\documentclass[12pt]{vbthesis}

\usepackage{amsfonts,url,fancyhdr,amsthm,amssymb,amsmath,amscd,epsfig,psfrag}
\usepackage[vcentermath]{youngtab}
\let\a=\alpha   \let\b=\beta    \let\g=\gamma    
           
   \let\l=\lambda  \let\m=\mu      
\let\n=\nu            \let\p=\pi           \let\s=\sigma 
                 
               \let\S=\Sigma 
             \let\D=\Delta

\def\bz{{\overline z}}

\def\cm{{\cal M}}  \def\co{{\cal O}}

\let\CW=\cw \let\CM=\cm \let\CO=\co \let\CF=\cf \let\CG=\cg \let\CL=\cl \let\CS=\cs

\def\IR{{\mathbb R}} \def\IC{{\mathbb C}} \def\IP{{\mathbb P}}
   
\def\IZ{{\mathbb Z}}

\let\top=\Diamond
\def\zm{z_{\rm min}}
\def\CY{Calabi-Yau}
\def\ipo{\hbox{\bf 0}}
\def\sst#1{{\scriptstyle #1}}
\def\xx{}  \def\yy{/}

\let\0=\over      \let\1=\vec      \def\2{{1\over2}}    \let\3=\ss
\let\4=\underline \let\5=\overline \let\6=\partial      \def\7#1{{#1}\llap{/}}
\def\8#1{{\textstyle{#1}}}         \def\9#1{{\ifmmode{\pmb{#1}}\else\bf#1\fi}}

          \def\({\left(}       \def\){\right)}

\def\ifundefined#1{\expandafter\ifx\csname#1\endcsname\relax}
\def\bye{\end{document}}   
\long\def\new#1\endnew{{\bf #1}}
\long\def\del#1\enddel{} 

\def\HS#1 {\hspace*{#1pt}} \def\VS#1 {\vspace*{#1pt}}
\def\BC{\begin{center}}    
\def\EC{\end{center}}

\def\mao#1{\mathop{\rm #1}\nolimits}  
       \def\Tr{\mao{Tr}}       
          
\def\Re{\mao{\hbox{\cal Re}}}   \def\Im{\mao{\hbox{\cal Im}}}

\let\and=\wedge

\let\bra=\langle        \let\ket=\rangle        \def\<#1\>{\bra #1 \ket}
\let\ni=\noindent

\def\rel#1 #2{\buildrel #1 \over {#2}}

          \def\({\left(}       \def\){\right)}

\def\rel#1 #2{\buildrel #1 \over {#2}}

\def\plb#1 #2 {Phys. Lett. {\bf B#1} #2 }
\def\phr#1 #2 {Phys. Rep. {\bf  #1} #2 }        
\def\npb#1 #2 {Nucl. Phys. {\bf B#1} #2 }
\def\aph#1 #2 {Ann. Phys. {\bf #1} #2 }         
\def\jmp#1 #2 {J. Math. Phys. {\bf #1} #2 }
\def\jgp#1 #2 {J. Geom. Phys. {\bf #1} #2 }
\def\prd#1 #2 {Phys. Rev. {\bf D#1} #2 }
\def\prl#1 #2 {Phys. Rev. Lett. {\bf #1} #2 }
\def\rmp#1 #2 {Rev. Mod. Phys.  {\bf #1} #2 }
\def\zpc#1 {Z. Phys. {\bf #1C} }
\def\cmp#1 #2 {Commun. Math. Phys. {\bf #1} #2 }
\def\cqg#1 #2 {Class.Quant.Grav. {\bf #1} #2 }
\def\mpl#1 {Mod. Phys. Lett. {\bf A#1} }
\def\cpc#1 {Computer Phys. Commun. {\bf #1} }   
\def\ijmp#1 {Int. J. Mod. Phys. {\bf A#1} }
\def\ijmpC#1 {Int. J. Mod. Phys. {\bf C#1} }
\def\atmp#1 {Adv. Theor. Math. Phys. {\bf #1} }

\newdimen\tableauside\tableauside=1.0ex
\newdimen\tableaurule\tableaurule=0.4pt
\newdimen\tableaustep
\def\phantomhrule#1{\hbox{\vbox to0pt{\hrule height\tableaurule width#1\vss}}}
\def\phantomvrule#1{\vbox{\hbox to0pt{\vrule width\tableaurule height#1\hss}}}
\def\sqr{\vbox{%
  \phantomhrule\tableaustep
  \hbox{\phantomvrule\tableaustep\kern\tableaustep\phantomvrule\tableaustep}%
  \hbox{\vbox{\phantomhrule\tableauside}\kern-\tableaurule}}}
\def\squares#1{\hbox{\count0=#1\noindent\loop\sqr
  \advance\count0 by-1 \ifnum\count0>0\repeat}}
\def\tableau#1{\vcenter{\offinterlineskip
  \tableaustep=\tableauside\advance\tableaustep by-\tableaurule
  \kern\normallineskip\hbox
    {\kern\normallineskip\vbox
      {\gettableau#1 0 }%
     \kern\normallineskip\kern\tableaurule}%
  \kern\normallineskip\kern\tableaurule}}
\def\gettableau#1 {\ifnum#1=0\let\next=\null\else
  \squares{#1}\let\next=\gettableau\fi\next}
\def\.#1 #2\>{\bibitem{#1}#2}

\def\BC{\begin{center}}
\def\EC{\end{center}}
\def\eeql#1 {\label{#1}\eeq}      \let\nn=\nonumber  
\def\beq{\begin{equation}}      \def\eeq{\end{equation}}        
\def\bea{\begin{eqnarray}}      \def\eea{\end{eqnarray}}
\def\eeal#1 {\label{#1}\eea} 

\def\refeq#1{(\ref{#1})}
\def\reffig#1{figure \ref{#1}}

\def\fig#1#2#3#4{\begin{figure}[htp]\begin{center}\includegraphics[width=#3]{#4}\caption{#2}\label{#1}\end{center}\end{figure}}

\def\lu{{\lambda_u}}
\def\lv{{\lambda_v}}

\Yinterspace{0.3ex plus 0.3ex}

\newtheorem{definition}{Definition}[chapter]
\newtheorem{proposition}[definition]{Proposition}
\newtheorem{theorem}[definition]{Theorem}
\newtheorem{corollary}[definition]{Corollary}
\newtheorem{example}[definition]{Example}
\newtheorem{lemma}[definition]{Lemma}
\newtheorem{remark}[definition]{Remark}
\newtheorem{conjecture}[definition]{Conjecture}

\title{Toric Geometry and\\[1ex] String Theory}
\author{Vincent Bouchard}
\college{Magdalen College}
\degree{Doctor of Philosophy}
\degreedate{July 2005}

\begin{document}

\maketitle

%sets the number of sectioning levels that get number and appear in the contents
\setcounter{secnumdepth}{4}
\setcounter{tocdepth}{4}

%sets the interline and interparagraphs spaces
\baselineskip=17pt plus 3pt\parskip=9pt

\bibliographystyle{../BibTex/utcaps}

\begin{dedication}
This thesis is dedicated to my parents.
\end{dedication}

\begin{acknowledgements}
First and foremost, I would like to thank Philip Candelas for his invaluable help throughout my four years in Oxford. I also owe special thanks to Bogdan Florea, Marcos Mari\~no and Harald Skarke for stimulating and enjoyable collaborations on the research topics discussed in this thesis. Thanks also to Lara Anderson, Xenia de la Ossa, Marco Gualtieri, Wen Jiang, Shabnam Kadir, Lionel Mason, David Skinner, Fonger Ypma and many others for interesting mathematical physics discussions. 

Many thanks to my parents, who have always supported me in my choices of life and encouraged me to be true and honest with myself. Alexandra, thank you so much for your love and patience; distance relationships are difficult, but you showed me that no matter what the distance is love always exists. Thanks also to my sister Maryse, for your continual support and friendship; and for your invitation to discover Burmese and Malian cultures!

I would like to thank all my friends in Oxford, especially James, Stefano, Silje, Lucy, Kezia, Owen, Rahul; you showed me that better worlds exist, and most importantly that we can start creating them ourselves, right here and right now. Thanks to the OSSTW, OSAN, Oxford and UK Indymedia, ZOMBIE, PGA, OCSET and OARC crews, for the inspiration and the fantastic work to make our communities a better place to live. Thanks also to Maarit, Will and Ralph, for the wonderful Sunday nights spent in your good company. Thanks to all my friends from Qu\'ebec, particularly David, Sylvie, Hendrick, F\'elix, Marc; it is amazing that our friendship is still as intense as ever! Finally, I must not forget to thank all the people I have played music with, for all these special moments where collective creation of music engendered unity through diversity...

I must also acknowledge the generous funding of the Rhodes Trust and of the National Science and Engineering Research Council of Canada, without which I could not have pursued my studies in Oxford.

\end{acknowledgements}

\begin{abstractseparate}
In this thesis we probe various interactions between toric geometry and string theory. First, the notion of a top was introduced by Candelas and Font as a useful tool to investigate string dualities. These objects torically encode the local geometry of a degeneration of an elliptic fibration. We classify all tops and give a prescription for assigning an affine, possibly twisted Kac-Moody algebra to any such top. Tops related to twisted Kac-Moody algebras can be used to construct string compactifications with reduced rank of the gauge group. Secondly, we compute all loop closed and open topological string amplitudes on orientifolds of toric \CY\ threefolds, by using geometric transitions involving $SO/Sp$ Chern-Simons theory, localization on the moduli space of holomorphic maps with involution, and the topological vertex. In particular, we count Klein bottles and projective planes with any number of handles in some \CY\ orientifolds. We determine the BPS structure of the amplitudes, and illustrate our general results in various examples with and without D-branes. We also present an application of our results to the BPS structure of the coloured Kauffman polynomial of knots.
\end{abstractseparate}

\begin{romanpages}
\baselineskip=17pt plus 1pt\parskip=1pt

\tableofcontents

\baselineskip=17pt plus 3pt\parskip=9pt
\listoffigures
\end{romanpages}

\chapter[Introduction]{Introduction}\label{intro}

Whether string theory has a deep r\^ole to play in our understanding of the universe or not, it has already achieved a great success in the recent decades: building bridges between various areas of mathematics and physics that were previously unrelated. Mathematicians and physicists have always worked together to search the meaning of Nature. But more than ever, both fields have become interlaced, to the extent that it is now difficult to trace a line between what was previously known as mathematics and physics. 

Mathematical physics, particularly string theory, has led to fascinating insights, both in mathematics and in physics. The aim of this thesis is to pursue this interaction between mathematics and physics, through the study of toric geometry and string theory.

Recently, toric geometry became an important part of the geometrical study of string theory. Toric varieties provide an elementary way to understand many abstract concepts of algebraic geometry. In a similar fashion, the use of toric varieties in string theory has led to deep insights on the nature of string theory itself, on its various internal dualities and on its numerous relations with other areas of mathematics and physics. Owing to its beauty and simplicity, toric geometry also gives the possibility to compute various non-trivial results in string theory that could not be calculated otherwise. 
 
In this thesis we shall focus on two particular interactions between toric geometry and string theory. On the one hand, we will continue the study of \CY\ hypersurfaces in toric varieties, using Batyrev's reflexive polytopes and their close cousins named `tops', and investigate their relations with dualities between compactifications of string theory. On the other hand, we will develop a novel approach to closed and open topological strings on orientifolds of toric \CY\ threefolds, using various mathematical devices; geometric transitions involving $SO/Sp$ Chern-Simons theory, the topological vertex, and unoriented localization techniques.

\bigskip\bigskip

In chapter \ref{toric}, we start by summarizing various results in toric geometry relevant for this thesis. The goal of this chapter is not to provide a complete introduction to toric geometry. Toric geometry is a well developed subject, and many thorough accounts already exist in the literature, such as William Fulton's very good {\it Introduction to Toric Varieties} \cite{Fulton:1993}.

Rather, this chapter aims at offering a gentle walk through the concepts of toric geometry that are essential for the rest of this thesis. We tried to formulate a pedagogical synthesis of the required background, starting from the definition of a toric variety and culminating with Batyrev's approach to hypersurfaces in toric varieties. The non-expert reader should be able to find his way through the sometimes abstract concepts of toric geometry, and in that sense it is self-contained. However, many proofs are omitted, and various interesting aspects of toric geometry are not discussed, for the sake of clarity and brevity.

Toric varieties may be approached from various points of view. The ``classical'' viewpoint makes heavy use of numerous concepts of algebraic geometry. In this approach, toric varieties are built by gluing together affine toric varieties in a certain way associated to a geometrical object called fan. Although this is the most common approach to toric varieties, we will not use it in this chapter. We prefer to follow other avenues that lead to the same results in a somewhat more intuitive way, at least from the point of view of string theory.

A second way to construct toric variety from a fan has been developed by Cox \cite{Cox:1993fz}. By associating homogeneous coordinates to the one-dimensional cones of the fan, this approach exploits an interesting similarity between toric varieties and (weighted) projective spaces. It is perhaps the simplest way to get a grip on toric varieties without requiring background knowledge in algebraic geometry. Therefore, we shall follow this path as far as possible.

Thus, we will describe toric varieties using the homogeneous coordinate technology. We will mainly focus on three dimensional toric varieties, in particular toric threefolds, which will be the focus of interest in the remaining chapters. However most of the content of chapter \ref{toric} can be generalized to higher dimensional varieties rather straightforwardly.

In string theory, we are often interested in \CY\ manifolds, which are a special kind of K\"ahler manifolds. \CY\ manifolds are defined in various ways in the literature; we will briefly discuss these definitions and their relations. In this thesis we will say that a manifold is \CY\ if it is K\"ahler and has vanishing first Chern class. We will not assume that it is compact; in fact, noncompact \CY\ manifolds will play a crucial r\^ole. One of our objectives is to implement the \CY\ condition in toric geometry. Hence, after having introduced some essential properties of toric threefolds, we will explore various formulations of the \CY\ condition for toric threefolds. In fact, we will show that toric \CY\ threefolds must be noncompact.

Using this property of toric \CY\ threefolds, we will introduce the concept of toric diagrams, which are two-dimensional graphs representing the degeneration loci of the fibers of a toric \CY\ threefold. We will spend some time describing the properties of these graphs, since they play an important r\^ole in chapters \ref{closed} and \ref{open}. To this end, we will have to develop another approach to toric varieties, using the symplectic quotient point of view. As most of these notions are not essential for this thesis, we will simply give the reader the necessary background to develop an intuitive understanding of the meaning of toric diagrams.

We then describe two important examples of toric \CY\ threefolds that will be studied in detail in chapters \ref{closed} and \ref{open}. For the case of the resolved conifold, we provide a complete analysis, using the various approaches referred to above.

Now, it may be disconcerting to learn that compact toric \CY\ threefolds do not exist. We promised that toric geometry leads to fascinating results in string theory, but compact toric \CY\ threefolds do not even exist, while in most realistic compactifications of string theory we need compact manifolds! There is fortunately a beautiful way to circumvent this limitation, due to Batyrev \cite{Batyrev:1994}. The fundamental idea is to construct \CY\ manifolds not as toric manifolds, but rather as compact hypersurfaces in compact toric varieties. In this way, the toric variety does not have to be \CY\, nor does the \CY\ manifold need to be toric. Batyrev's approach is to associate reflexive polytopes to \CY\ hypersurfaces in toric varieties. We describe, still using the homogeneous coordinate technique, this construction. 

We end this chapter by introducing a similar geometrical object, which we call top. The appearance of tops was first noticed in string duality applications of Batyrev's reflexive polytopes \cite{Candelas:1996su}. They were then generalized, and became interesting in their own rights. As they are at the centre of chapter \ref{tops}, we describe them in detail, both from a lattice and a toric point of view.

\bigskip\bigskip

In chapter \ref{tops} we study our first application of toric geometry to string theory. Toric geometry and Batyrev's construction \cite{Batyrev:1994} provide a very useful setup to study dualities between heterotic strings compactified on a \CY\ $n$-fold and F-theory (or type II) compactified on a \CY\ $(n+1)$-fold. In \cite{Candelas:1996su}, Candelas and Font used reflexive polyhedra to study the conjectured duality between the $E_8 \times E_8$ heterotic string compactified on the manifold $K3 \times T^2$ and the IIA string compactified on a \CY\ threefold. It was noticed, and later explained in \cite{Perevalov:1997vw}, that the affine Dynkin diagrams of nonabelian gauge groups occurring in type IIA and also in F-theory can be read off from the dual reflexive polyhedron corresponding to the \CY\ manifold used for compactification. The fibration stucture of the \CY\ manifold can be directly seen as a nesting structure of the reflexive polyhedron. The elliptic fibration structure of the $K3$ part of the \CY\ divides the three dimensional reflexive polyhedron corresponding to the $K3$ in two parts, a top and a bottom, separated by the two dimensional reflexive polygon of the fiber. The concept of top was then introduced as half of a reflexive polyhedron.

This was just the beginning of the story. The ideas of \cite{Candelas:1996su} were studied in detail in many other papers \cite{Berglund:1998ej,Candelas:1997pv,Candelas:1996ht,Candelas:1997eh,Candelas:1997jz,Candelas:1997pq,Hu:2000pr}. The \CY\ manifolds analyzed in these papers were $K3$ fibrations with an elliptically fibered $K3$ manifold where the elliptic fibration structure of the $K3$ carries over to the \CY\ manifold. These nested fibration structures can be seen explicitly in the toric diagrams as nestings of the corresponding reflexive polyhedra. The fan for the toric variety describing the base of the elliptic fibration is given by projecting the higher dimensional fan corresponding to the fibered \CY\ manifold along the two dimensions of the reflexive polygon that represents the fiber \cite{Kreuzer:1997zg}. Then the way the elliptic fibers degenerate along the curves in the base space can be found, in order to determine the enhanced gauge groups, by considering the preimage of the projection for each toric divisor in the base. The concept of top can now be generalized \cite{Candelas:1997pq} to the geometrical objects formed by the preimages of the corresponding toric divisors. These objects are three-dimensional lattice polyhedra with one facet containing the origin and the other facets at integral distance one from the origin. This definition implies that the facet containing the origin is a reflexive polygon. Note that this really generalizes the concept initially introduced by Candelas and Font, since the tops defined as half of a reflexive polyhedron have all the properties of the new tops, but the new tops cannot always be completed to reflexive polyhedra. Alternatively the more general definition can be seen as the description of a toric hypersurface that is an elliptic fibration over $\IC$ and its degeneration over $0$. This geometrical description was briefly explored in chapter \ref{toric}, where tops were first introduced from a purely geometrical perspective.

In this chapter we classify all the possible tops, using the general definition. In contrast to the case of reflexive polyhedra, there are infinitely many tops, even for each choice of one of the $16$ reflexive polygons as the facet containing the origin. We find that there is a precise prescription for assigning an affine Kac-Moody algebra to any top, in a way that involves the lengths of simple roots and the coefficients of the null root. Owing to this fact the classification of tops is related to that of affine Kac-Moody algebras. We also find one parameter families as well as sporadic cases. In addition, for each of the $16$ polygons there is also a family depending on $l-3$ integer parameters where $l$ is the number of lattice points of the polygon; these correspond to the $A_n^{(1)}$ series of affine Kac-Moody algebras in such a way that $n$ is a linear combination of the parameters. Each of the untwisted affine Kac-Moody algebras occurs quite a number of times, and in addition four of the six possible (families of) twisted algebras also occur. The tops featuring the latter are related in a very nice way to string compactifications with reduced rank, i.e. CHL strings \cite{Chaudhuri:1995fk} and their generalizations and duals.

This chapter is based on \cite{Bouchard:2003bu} which was written in collaboration with Harald Skarke.

\bigskip\bigskip

In chapter \ref{closed} we start inverstigating our second application of toric geometry to string theory. Geometric or large $N$ transitions relating open and closed topological strings have had a deep impact in the study of topological string theory. Since the original formulation of the duality for local conifold transitions in \cite{Gopakumar:1998ki}, they have been extended in various directions, leading to the first systematic solution of these models on noncompact, toric \CY\ threefolds through the topological vertex \cite{Aganagic:2003db} (see \cite{Marino:2004uf,Neitzke:2004ni} for a review).

The study of topological strings on \CY\ orientifolds was initiated in \cite{Sinha:2000ap}, where an orientifold of the geometric transition of \cite{Gopakumar:1998ki} relating the deformed and the resolved conifold was studied in detail, and continued in \cite{Acharya:2002ag} from the mirror $B$-model point of view. The geometric transition of \cite{Gopakumar:1998ki} can be extended to more general toric geometries \cite{Aganagic:2002qg,Aganagic:2001ug,Diaconescu:2002sf,Diaconescu:2002qf}. Accordingly, in this chapter we propose a generalization of the large $N$ correspondence of \cite{Sinha:2000ap} to a large class of orientifolds of toric \CY\ threefolds.

We find that the partition function of closed topological strings on the orientifold (including unoriented contributions and oriented contributions from the covering space) is equivalent in the large $N$ limit to the Chern-Simons partition function on the threefold after a geometric transition. The ${\IC \IP^1}$'s that were invariant under the involution, becoming ${\IR\IP^2}$'s in the orientifold, give $SO(N)$ --- or $Sp(N)$ --- Chern-Simons theory on the ${\bf S}^3$'s resulting from the geometric transition. One also has to add instanton contributions localized on the fixed locus of a torus action on the deformed geometry.

This is a highly non-trivial proposal, as more complicated orientifolds involve instanton contributions to the Chern-Simons partition function. Moreover, for more general orientifolds, the geometry of the covering space becomes quite different from the one of the resulting orientifold. It is not obvious at all that both the oriented and unoriented contributions to the closed topological strings partition function are encoded in the Chern-Simons setup. But it turns out to be true in the examples we consider.

We also find that the closed topological string amplitudes on the orientifolds of the type we describe below can be computed with the topological vertex introduced in \cite{Aganagic:2003db}, by using a prescription that takes into account the involution of the target. We explicitly prove that this prescription is equivalent to the large $N$ Chern-Simons dual. This prescription extends the general formalism of the topological vertex to include the case of orientifolds.

To test our result we compute the unoriented contributions on the closed topological strings side using the unoriented localization techniques developed in \cite{Diaconescu:2003dq}.  This computation does not rely on large $N$ duality at all, consequently providing an independent check of our proposal. In \cite{Sinha:2000ap} it was found that only unoriented maps with one crosscap contribute to the partition function. However, in the general case, we find that configurations with two crosscaps, that is Klein bottles, do contribute as well.

To make the proposal more concrete we focus on a particular geometry. We consider a noncompact \CY\ threefold $X$ whose compact locus consists of two compact divisors each isomorphic to a del Pezzo surface $dP_2$ and a rational $(-1,-1)$ curve that intersects both divisors transversely. The divisors do not intersect each other. This geometry was briefly described in chapter \ref{toric}. We will equip $X$ with a freely acting antiholomorphic involution $I$ and consider an orientifold of the theory obtained by gauging the discrete symmetry $\sigma I$, where $\sigma$ is an orientation reversal worldsheet diffeomorphism.

The partition function of the closed topological ${\bf A}$-model with this geometry as target space will sum both over maps from orientable worldsheets to $X$ (with the K\"ahler parameters identified by the involution set equal) as well as over non-orientable worldsheets to the orientifolded geometry.

The orientifolded geometry allows a local geometric transition that will be described in detail in section \ref{geometry}. This amounts to contracting two $\IC \IP^1$'s and a ${\IR\IP^2}$ and replacing them by three ${\bf S}^3$'s. We conjecture that the dual open string model will consist of a system of Chern-Simons theories supported on the three spheres, with $U(N_1)$ and  $U(N_2)$ groups on the spheres corresponding to the contracted $\IC \IP^1$'s and $SO(N_3)$ - or $Sp(N_3)$ - group on the sphere corresponding to the contracted $\IR\IP^2$. The new ingredient is that the Chern-Simons theories will be coupled by cylindrical instantons.

This chapter is based on \cite{Bouchard:2004iu} written in collaboration with Bogdan Florea and Marcos Mari\~no.

\bigskip\bigskip

In chapter \ref{open} we continue the study of topological string amplitudes on orientifolds of toric \CY\ threefolds initiated in chapter \ref{closed}. Our main goal is to extend the results of chapter \ref{closed} to open topological strings on orientifolds without fixed points. In other words, we consider orientifolds of noncompact \CY\ threefolds with D-branes. 

An important property of topological string amplitudes is that they have an integrality structure related to the counting of BPS states, as it was first realized by Gopakumar and Vafa \cite{Gopakumar:1998jq} in the case of closed string amplitudes. The integrality structure in the open case was studied in \cite{Ooguri:1999bv,Labastida:2000yw}. As a first step in our study of topological string amplitudes on orientifolds without fixed planes we analyze their BPS structure. What we find is that the total orientifold amplitude is the sum of an oriented amplitude (the untwisted sector) and an unoriented amplitude (the twisted sector) with different integrality properties. We explain how to compute the contribution of the twisted sector in the open case. We also spell out in detail the integrality properties of the twisted sector contributions. 

This integrality structure provides a strong requirement on open topological string amplitudes, and we check it explicitly on various examples involving orientifolds with D-branes. To compute these open string amplitudes we use the new vertex rule introduced in section \ref{topover}. We also compute the associated Gromov-Witten invariants using independent localization techniques developed in section \ref{localclosed} and in \cite{Diaconescu:2003dq}, and find perfect agreement with the results obtained with the vertex. 

One of the most interesting applications of the large $N$ duality between open and closed topological strings consists in the determination of structural properties of knot and link invariants related to the BPS structure of open topological strings. For example, from the results of \cite{Ooguri:1999bv,Labastida:2000yw} one can deduce structure theorems for the coloured HOMFLY polynomial of knots and links. The large $N$ duality on orientifolds now involves $SO(N)$ and $Sp(N)$ Chern-Simons theories. Therefore, the BPS structure of the amplitudes should lead to the determination of structural properties of a different type of knot and link invariant: the coloured Kauffman polynomial \cite{Kauffman:1990}. Although for arbitrary knots and links we cannot determine in detail the structure of the untwisted sector, we are able to derive general structural results for the coloured Kauffman polynomial. We test again these predictions on various examples involving torus knots. 

This chapter is based on \cite{Bouchard:2004ri}, written in collaboration with Bogdan Florea and Marcos Mari\~no.

\bigskip\bigskip

Finally, in chapter \ref{conclusion} we discuss the results of this thesis. We propose various avenues of research and future directions, and point out some possible extensions of our results to more complicated situations.

\chapter[Toric Geometry]{Toric Geometry}\label{toric}

In this chapter we explore various aspects of toric geometry relevant for this thesis. For good introductions to toric geometry the reader is referred to \cite{Fulton:1993,Hori:2003,Greene:1996cy}.

Toric varieties may be approached from various points of view. They can be described using fans and homogeneous coordinates, or viewed as symplectic manifolds, or correspondingly as the Higgs branch of the space of supersymmetric ground states of the gauged linear sigma model (GLSM), or even associated to convex polytopes in integral lattices. Perhaps the simplest approach is the homogeneous coordinate description \cite{Cox:1993fz}; therefore we shall proceed as far as possible using this approach. 

First, we will introduce toric varieties and rapidly describe how to extract a toric variety from a fan using homogeneous coordinates. We will then examine what the Calabi-Yau condition becomes for toric variety, and deduce that toric Calabi-Yau manifolds must be noncompact. Using this result, we shall introduce toric diagrams, and explain their meaning using the symplectic manifold point of view. We then give a few examples of toric Calabi-Yau threefolds that will be used in chapters \ref{closed} and \ref{open}.

In the remaining of this chapter we explain how compact Calabi-Yau manifolds may be obtained in toric geometry, namely as hypersurfaces in compact toric manifolds using Batyrev's well known reflexive polytopes \cite{Batyrev:1994}. We then introduce the concept of `top' --- which is the object of interest of chapter \ref{tops} --- and explain its meaning in toric geometry.

To understand toric geometry, we must dive into the abstract world of algebraic geometry. We will assume here a basic knowledge of algebraic geometry; a standard reference is \cite{Griffiths:1978}.

\section{Homogeneous Coordinates}\label{toricCY}

An interesting aspect of Cox's approach to toric geometry is that by using the homogeneous coordinate construction, toric varieties look very much like the usual complex (weighted) projective spaces. In fact, from that point of view we can understand toric varieties as an algebraic generalization of complex (weighted) projective spaces.

To start with, let us explain roughly what a toric variety is. Consider $\IC^m$ and an action by an algebraic torus $(\IC^*)^p$, $p < m$. We identify and then substract a subset $U$ that is fixed by a continuous subgroup of $(\IC^*)^p$, then safely quotient by this action to form
\beq\label{toricdef}
\cm = \( \IC^m \setminus U \) / (\IC^*)^p.
\eeq
$\cm$ is called a {\it toric variety}, as it still has an algebraic torus action by the group $(\IC^*)^{m-p}$ descending from the natural action of $(\IC^*)^m$ on $\IC^m$.

For instance, $\IC \IP^2$ is a toric variety. Indeed, a standard way of describing $\IC \IP^2$ is by embedding it into $\IC^3$:
\beq
\IC \IP^2 = (\IC^3 \setminus \{0\}) / (\IC^*),
\eeq
where the quotient is implemented by modding out by the equivalence relation
\beq
(x,y,z) \sim \l (x,y,z),
\eeq
where $\l \in \IC^*$. We see that this description of $\IC \IP^2$ satisfies the general definition of a toric variety given above.

We now describe how to extract toric varieties from a fan using the homogeneous coordinate approach developed by Cox \cite{Cox:1993fz}.

Let $M$ and $N$ be a dual pair of lattices, viewed as subsets of vector spaces $M_\IR=M\otimes_\IZ\IR$ and $N_\IR=N\otimes_\IZ\IR$. Let $(u,v)\to \langle u,v \rangle$ denote the pairings $M\times N\to \IZ$ and $M_\IR\times N_\IR \to \IR$.

\begin{definition}
A {\rm strongly convex rational polyhedral cone} $\s \in N_\IR$ is a set
\beq
s = \{a_1 v_1 + a_2 v_2 + \ldots + a_k v_k | a_i \geq 0 \}
\eeq
generated by a finite number of vectors $v_1,\ldots,v_k$ in $N$ such that $\s \cap (-\s) = \{ 0 \}$.
\end{definition}

Let us put words on this definition. Suppose the lattice $N$ is $n$-dimensional, that is $N \cong \IZ^n$. A convex rational polyhedral cone is a $n$ or lower dimensional cone in $N_{\IR}$, with the origin of the lattice as its apex, such that it is bounded by finitely many hyperplanes (`polyhedra'), its edges are spanned by lattice vectors (`rational') and it contains no complete line (`strongly convex').

A face of a cone $\s$ is either $\s$ itself or the intersection of $\s$ with one of the bounding hyperplanes.

\begin{remark}
In the remaining of this chapter we will refer to convex rational polyhedral cones simply as cones.
\end{remark}

\begin{definition}
A collection $\S$ of cones in $N_\IR$ is called a {\rm fan} if each face of a cone in $\S$ is also a cone in $\S$, and the intersection of two cones in $\S$ is a face of each.
\end{definition}

Now let $\Sigma$ be a fan in $N$. Let $\Sigma (1)$ be the set of one dimensional cones (or edges) of $\Sigma$. From now on we will focus on three dimensional toric varieties, or correspondingly on three dimensional lattices $M,N \simeq \IZ^3$.

Let $v_i$, $i=1,\dots,k$ be the vectors generating the one dimensional cones in $\Sigma(1)$, where $k=|\Sigma(1)|$. To each $v_i$ we associate an homogeneous coordinate $w_i \in \IC$. From the resulting $\IC^k$ we remove the set
\beq
Z_{\Sigma} = \bigcup_I \{(w_1,\dots,w_k):~w_i=0~\forall~i \in I\},
\eeq
where the union is taken over all sets $I \subseteq \{1,\dots,k\}$ for which $\{v_i:~i\in I\}$ does not belong to a cone in $\Sigma$. In other words, several $w_i$ are allowed to vanish simultaneously only if there is a cone such that the corresponding $v_i$ all belong to this cone.

Then the toric variety is given by 
\beq
\CM_{\Sigma}={\IC^k \setminus Z(\Sigma)\over G}
\eeql{toricvariety}
where $G$ is $(\IC^*)^{k-3}$ times a finite abelian group. For all the toric varieties we consider in this thesis the finite abelian group is trivial, so from now on we will omit it (see \cite{Skarke:1998yk} for an explanation of this group). The quotient by $(\IC^*)^{k-3}$ is implemented by taking equivalence classes with respect to the following equivalence relations among the coordinates $w_i$:
\beq
(w_1,\ldots, w_k) \sim (\l^{Q_a^1}w_1,\ldots,\l^{Q_a^k}w_k)
\eeql{equiv1}
with $\lambda \in \IC^*$ and $\sum_{i=1}^k Q_a^i v_i =0$. Among these relations, $k-3$ are independent. We choose the $Q_a^i$ such that they are integer and the greatest common divisor of the $Q_a^i$ with fixed $a$ is $1$.

Using this construction, it is easy to see that the complex dimension of a toric variety is always equal to the real dimension $n$ of the lattice $N \cong \IZ^n$.

\begin{figure}[htp]
\begin{center}
\psfrag{v_1}{$v_1$}
\psfrag{v_2}{$v_2$}
\psfrag{v_3}{$v_3$}
\includegraphics[width=5cm]{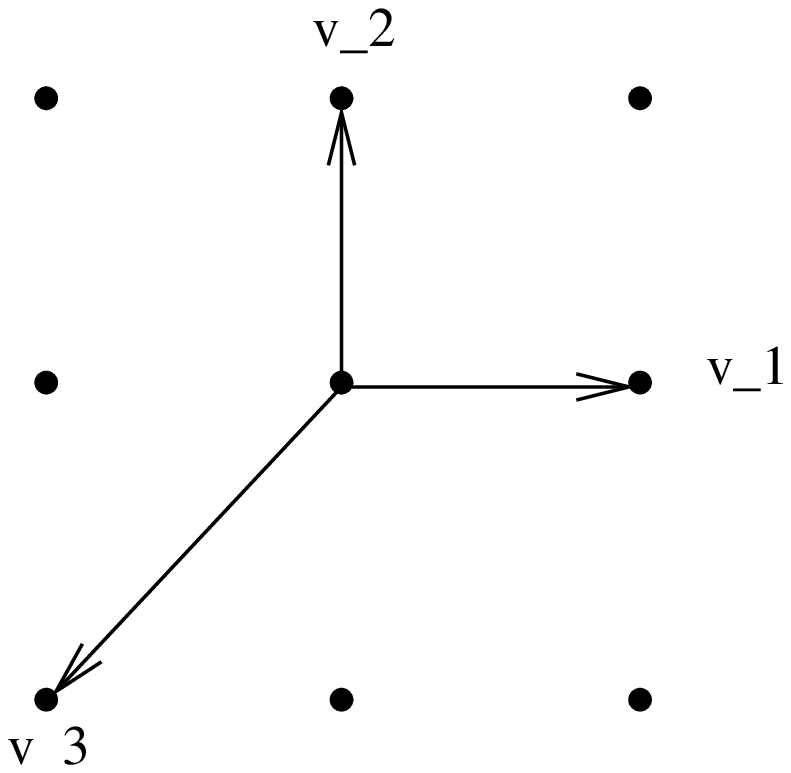}
\caption{The fan $\S$ of $\IC \IP^2$.}
\label{CP2}
\end{center}
\end{figure}

\begin{example}
Let us come back to the example of $\IC \IP^2$ (which is two dimensional rather than three dimensional, but easier to visualize as a first example). The fan is given in figure \ref{CP2}. There are three one-dimensional cones generated by the vectors $v_1=(1,0)$, $v_2=(0,1)$ and $v_3=(-1,-1)$, to which we associate the homogeneous coordinates $w_1$, $w_2$ and $w_3$ of $\IC^3$. The set $Z_{\S}$ is simply $\{ 0 \}$, and thus the toric variety is given by
\beq
\cm_{\S} = (\IC^3 \setminus \{0\}) / (\IC^*).
\eeq
Moreover, we have that $1(1,0)+1(0,1)+1(-1,-1)=(0,0)$, so the $\IC^*$ quotient is implemented by the equivalence relation $(w_1,w_2,w_3) \sim \l (w_1,w_2,w_3)$. This is the usual description of $\IC \IP^2$.
\end{example}

In toric geometry it is straightforward to know whether a toric variety is compact or not:

\begin{proposition}\label{compactness}
A toric variety $\cm_\S$ is compact if and only if its fan $\S$ fills $N_\IR$.
\end{proposition} 

The reader is referred to \cite{Fulton:1993} for a proof of this proposition.

\subsection{Toric Divisors}\label{toricdivisors}

In a toric variety there is a natural set of divisors called toric divisors.

\begin{definition}
Let $\cm_\S$ be a toric variety described by a fan $\S$. Associate a homogeneous coordinates $w_i$ to each vector $v_i$ generating the one dimensional cones of $\S$. The {\rm toric divisors} $D_i$ of $\cm_\S$ are the hypersurfaces defined by the equations $w_i=0$.
\end{definition}

Since we associated a homogeneous coordinates $w_i$ to each one dimensional cones $v_i$ in the fan $\S$ of $\cm_\S$, we can think of the vectors $v_i$ as corresponding to the toric divisors defined by $w_i=0$.

Similarly, higher dimensional cones of $\S$ correspond to lower dimensional algebraic subvarieties of $\cm_\S$.

In fact, it can be shown (see \cite{Fulton:1993}), by using methods very similar to those used for complex projective spaces, that the canonical bundle of $\cm_\S$ is given by
\beq
K_{\cm_\S} = \co(-\sum_i D_i).
\eeq
This result will be useful to determine whether a toric variety is \CY\ or not.

\section{Toric \CY\ Threefolds}\label{CYcondition}

First, let us briefly explain what a \CY\ manifold is.

\subsection{Definition of a \CY\ Manifold}

It was in 1954 that Calabi stated his conjecture \cite{Calabi:1956,Calabi:1957}, which was proved by Yau in 1976 \cite{Yau:1977,Yau:1978}. Given a compact K\"ahler manifold $M$ with $c_1=0$, the proof of the conjecture guarantees the existence of a Ricci-flat K\"ahler metric on $M$, that is a K\"ahler metric with zero Ricci form. Such a manifold is called a \CY\ manifold.

However, many different definitions of \CY\ manifolds exist in the literature; let us now list some of the most common definitions.

A \CY\ manifold of real dimension $2m$ is a compact K\"ahler manifold $(M,J,g)$: 
\begin{enumerate}
\item{with zero Ricci form,}
\item{with vanishing first Chern class,}
\item{with ${\rm Hol}(g) = SU(m)$ (or ${\rm Hol}(g) \subseteq SU(m)$),}
\item{with trivial canonical bundle,}
\item{that admits a globally defined and nowhere vanishing holomorphic $m$-form.}
\end{enumerate}

We will not study in detail these definitions and their interrelations in this thesis; for such an analysis the reader is referred to \cite{Bouchard:2005mo,Joyce:2000}. Let us simply say that strictly speaking, these definitions are all inequivalent.

An important point common to all these definitions is that the manifold is assumed to be {\it compact}. However, it is also possible to define {\it noncompact} (or local) \CY\ manifolds; by local or noncompact \CY\ manifolds we mean open neighbourhoods in compact \CY\ manifolds. These are very useful in many applications in physics, for instance in topological strings \cite{Marino:2004uf,Neitzke:2004ni}. They are also relevant in the study of geometric transitions \cite{Rossi:2004}. In fact, they will play a crucial r\^ole in the remaining of this thesis, especially in chapters \ref{closed} and \ref{open}.

In this thesis we will adopt the following definition of \CY\ manifolds.

\begin{definition}
A {\rm \CY\ manifold} is a K\"ahler manifold $(M,J,g)$ with trivial canonical bundle.
\end{definition}

This definition applies for both compact and noncompact \CY\ manifolds. The simplest noncompact \CY\ manifold is obviously $\IC^m$.

\subsection{\CY\ Manifolds in Toric Geometry}

We will now implement the \CY\ condition on toric threefolds.

It is a well known fact (see for instance \cite{Griffiths:1978,Bouchard:2005mo,Joyce:2000}) that to a divisor $D=\sum_i a_i N_i$ we can associate a line bundle with a meromorphic section such that the meromorphic section has a zero of order $a_i$ along $N_i$ if $a_i > 0$ and a pole of order $-a_i$ along $N_i$ if $a_i < 0$. The $N_i$ are irreducible hypersurfaces, that is hypersurfaces that cannot be written as the union of two hypersurfaces.

In the toric case, the toric divisors $D_i$ defined by $w_i=0$ are irreducible hypersurfaces. Therefore, using the above correspondence we see that the toric divisor $D_i$ is associated to a line bundle $\co(D_i)$ with a section $s$ that has a zero of order one along $D_i$; thus the section $s$ is simply $w_i$. Hence we see that each homogeneous coordinate $w_i$ is a section of the line bundle $\co(D_i)$ associated to the toric divisor $D_i$.

Now, if we consider a monomial  $w_1^{a_1} \cdots w_k^{a_k}$; for $a_i>0$, it has zeroes of order $a_i$ along $D_i$, while for $a_j<0$, it has poles of order $-a_j$ along $D_j$. Therefore it is a section of the line bundle $\co(\sum_i a_i D_i)$.

Let us now consider the case where $a_i = \< v_i,m\>$, $i=1,\ldots,k$ for some $m \in M$. Under the equivalence relations of the toric variety the monomial becomes
\beq
(\l^{Q_a^1} w_1)^{\< v_1,m \>} \cdots (\l^{Q_a^k} w_k)^{\< v_k,m \>} = \l^{\< \sum_{i=1}^k Q_a^i  v_i, m\>} w_1^{\< v_1,m \>} \cdots w_k^{\< v_k,m \>}.
\eeq
But since $\sum_{i=1}^k Q_a^i  v_i=0$, this monomial is invariant under the equivalence relations and therefore it is a true meromorphic function on our toric variety. This means that it must be a section of the trivial line bundle, i.e.
\beq\label{equivdiv}
\sum_{i=1}^k \< v_i,m \> D_i \sim 0~~{\rm for~any}~m \in M.
\eeq
Conversely, if $\sum_{i=1}^k a_i D_i \sim 0$, then there exists a $m\in M$ such that $a_i= \< v_i,m \>$ for all $i$.

Now, we know that a K\"ahler manifold is \CY\ if and only if its canonical class is trivial. We saw in section \ref{toricdivisors} that the canonical line bundle of a toric variety $\cm_{\S}$ is given by $K_{\CM_{\Sigma}} \cong \CO (-\sum_{i=1}^k D_i)$. Therefore the canonical bundle is trivial if and only if $\sum_{i=1}^k D_i \sim 0$. Using \refeq{equivdiv}, we see that this condition is equivalent to the existence of a $m \in M$ such that $\< v_i,m \>=1$ for all $i$, which leads to the following proposition.

\begin{proposition}\label{affine}
Let $\cm_\S$ be a toric manifold defined by a fan $\S$. $\cm_\S$ is \CY\ if and only if the vectors $v_i$ generating the one-dimensional cones of $\cm_\S$ all lie in the same affine hyperplane.
\end{proposition}

It is thus very easy to see whether a toric variety is \CY\ or not; in fact, it can be read off directly from the fan $\S$ of the toric variety.

A consequence of proposition \ref{affine} is the following:

\begin{corollary}
A toric \CY\ manifold is noncompact.
\end{corollary}

Since the $v_i$ lie in a hyperplane, $\Sigma$ does not fill $N_{\IR}$. Thus proposition \ref{compactness} tells us that $\cm_\S$ is noncompact.

This seems like a serious limitation of toric geometry, since in string theory we are often interested in compact \CY\ manifolds. However, we will see in section \ref{hyper} how to construct compact \CY\ manifolds in toric geometry.

The \CY\ condition can be rewritten in yet another equivalent form. In \refeq{equiv1} we defined the `charges' (the meaning of this name will become clear in section \ref{sympl}) $Q_a^i$ satisfying $\sum_{i=1}^k Q_a^i v_i =0$. Therefore $\sum_{i=1}^k Q_a^i \< v_i,m \> =0$ for any $m \in M$. In particular, there exists an $m \in M$ such that $\< v_i,m \>=1$ for all $i$ if and only if $\sum_{i=1}^k Q_a^i = 0$ for all $a$. But we showed that a toric manifold is \CY\ if and only if there exists and $m \in M$ such that $\< v_i,m \>=1$ for all $i$. Therefore, the condition can be restated as follows: 

\begin{proposition}
A toric manifold is \CY\ if and only if the charges $Q_a^i$ satisfy the condition $\sum_{i=1}^k Q_a^i = 0$ for all $a$.
\end{proposition}

This condition is also very simple to verify. We only have to check that the charges $Q^i_a$ given in the toric data describing the manifold add up to zero. Thus, if we are given a fan we simply check that the $v_i$ lie in an affine hyperplane, while if we are given the toric data we simply verify that the charges add up to zero.

To conclude this section we introduce a nice pictorial way of characterizing toric \CY\ threefolds. We showed that for toric \CY\ threefolds the $v_i$ lie in a two dimensional plane $P$. Therefore, we can draw the two dimensional graph ${\tilde \Gamma}$ given by the intersection of the plane $P$ and the fan $\Sigma$. ${\tilde \Gamma}$ determines completely the fan $\Sigma$ of a toric \CY\ threefold. Given ${\tilde \Gamma}$, we can draw a `dual' graph $\Gamma$ in the sense that the edges of ${\tilde \Gamma}$ are normals to the edges of $\Gamma$ and vice-versa. $\Gamma$ is called the {\it toric diagram} of a toric \CY\ threefold $\CM_{\Sigma}$. It represents the degeneration of the fibers of the torus fibration. We will describe in more details toric diagrams in section \ref{sympl}.

\begin{figure}[htp]
\begin{center}
\psfrag{(-1,-1)}{$(-1,-1)$}
\psfrag{(0,0)}{$(0,0)$}
\psfrag{(1,0)}{$(1,0)$}
\psfrag{(0,-1)}{$(0,-1)$}
\includegraphics[width=6cm]{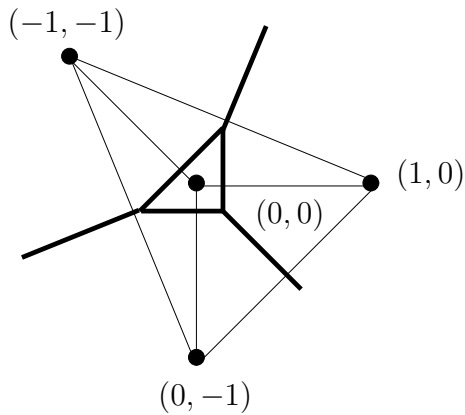}
\caption{The $\Gamma$ and $\tilde \Gamma$ graphs for $\CO(-3) \rightarrow \IC \IP^2$. The toric diagram $\Gamma$ is the normal diagram drawn in thick lines. The points $(\nu_i,1)$ give the fan $\Sigma$, where the $\nu_i$ are the vertices of $\tilde \Gamma$ and are shown in the figure.}
\label{p2graphs}
\end{center}
\end{figure}

Conversely, given a toric diagram $\Gamma$, it is straightforward to recover the fan $\Sigma$ of the toric \CY\ threefold. One first draws the dual graph $\tilde \Gamma$, and then defines the vectors $v_i = (\nu_i,1)$ where $\nu_i$ are the vertices of $\tilde \Gamma$. Because of the symmetries of a three dimensional lattice, the $v_i$ must be the generators of the edges of the fan $\Sigma$ of the toric \CY\ threefold $\CM_{\Sigma}$. Linear relations between the vectors $v_i$ give the charges $Q_a^i$. In other words, the fan $\Sigma$ is a three dimensional cone over the two dimensional graph $\tilde \Gamma$. An example of graphs $\Gamma$ and $\tilde \Gamma$ is given in figure \ref{p2graphs}.

\section{Toric Diagrams and Symplectic Quotients}\label{sympl}

In this section we describe the toric diagrams introduced above. To do so, we need to leave momentarily the homogeneous coordinates approach to toric varieties and see toric manifolds as symplectic quotients, or correspondingly as the Higgs branch of the space of supersymmetric vacua of the GLSM.

\subsection{Toric Manifolds as Symplectic Quotients}

Let $z_1,\dots,z_k$ be the coordinates of $\IC^k$. Let $\mu_a: \IC^k \to \IC$, $a=1,\dots,k-3$ be the $k-3$ moment maps defined by
\beq
\sum_{i=1}^k Q_a^i |z_i|^2 = \Re(t_a),
\eeql{moment}
where the $t_a$ are complex numbers. The $Q_a^i$ are the same charges that were introduced in \refeq{equiv1}. Therefore, the \CY\ condition imposes that $\sum_{i=1}^k Q_a^i=0$ for all $a$. We also consider the action of the group $G=U(1)^{k-3}$ on the coordinates defined by
\beq
z_j \rightarrow \exp (i Q_a^j \alpha_a ) z_j,~~~a=1,\dots,k-3.
\eeq
It turns out that
\beq
\CM = {\bigcap_{a=1}^{k-3} \mu^{-1} (\Re(t_a)) \over G}
\eeq
is a toric \CY\ threefold. The $k-3$ parameters $t_a$ are the complexified K\"ahler parameters of the \CY\ threefold.

Furthermore, since the charges $Q_a^i$ are the same as in \refeq{equiv1}, it is easy to recover the fan of $\CM$. One only has to find distinct vectors $v_i$ satisfying $\sum_{i=1}^k Q_a^i v_i =0 $; the $v_i$ generate the one dimensional cones of $\Sigma$. Moreover, since the \CY\ condition tells us that $\sum_{i=1}^k Q_a^i  =0$, we can choose (because of the symmetries of three dimensional lattices) vectors $v_i$ of the form $v_i = (\nu_i,1)$. The problem is then reduced to a two-dimensional problem which can easily be solved by inspection. We see that the charges $Q_a^i$ are the important data defining the toric \CY\ manifolds. This is usually called the {\it toric data} of the manifold.

This description of toric manifolds also arise in gauged linear sigma models. This is a two-dimensional $U(1)^{k-3}$ gauge theory with $k$ chiral superfields $\Phi_i$, whose scalar components are the $z_k$. The charges of the superfields $\Phi_i$ under the gauge group $U(1)^{k-3}$ are denoted by $Q_a^i$, $a=1,\dots,k-3$. This is why the $Q_a^i$ are generally called {\it charges}. It turns out that -- in the Higgs branch -- the supersymmetric ground states of the theory are parameterized by the so-called D-term equations modulo gauge equivalence, which are nothing but the moment maps $\mu_a$ defined in \refeq{moment}. In other words, the Higgs branch of the space of supersymmetric ground states of the GLSM is the toric variety $\CM$ defined above.

Now equipped with the description of toric \CY\ threefolds as symplectic quotients, let us come back to the toric diagrams introduced in section \ref{CYcondition}. There, we claimed that these diagrams encode the degeneration of the fibers of the manifold. This can be seen in two different ways: by looking at the threefold as a $T^3$ fibration or as a $T^2 \times \IR$ fibration. We will start with the first approach in section \ref{t3}, which is probably simpler. We will explore the second point of view in section \ref{t2}, using the topological vertex approach to toric \CY\ threefolds.

\subsection{$T^3$ Fibration}\label{t3}

We look at the threefolds as $T^3$ fibrations over three dimensional base manifolds with corners. Locally, we can introduce complex coordinates on the toric manifold: these are the $z_i$ introduced in \refeq{moment}. They are not all independent; for a threefold, there are $k-3$ relations between them given by the moment maps \refeq{moment}. Let us rewrite these coordinates as $z_j=|z_j| e^{i \theta_j}$, and introduce a new set of coordinates $\{ (p_1,\theta_1),\dots,(p_k,\theta_k)\}$, with $p_i\equiv |z_i|^2$, $i=1,\ldots,k$. The base of the threefold is then parameterized by the coordinates $p_i$, while the phases $\theta_i$ describe the fiber $T^3$. 

Since $|z_i|^2 \geq 0$, the coordinates $p_i$ satisfy $p_i \geq 0$. Therefore the boundaries of the base are where some of the coordinates $p_i$ vanish. But when $p_j=0$ the circle $|z_j|e^{i \theta_j}$ degenerates to a single point. Hence, the boundaries of the base correspond to degenerations of the corresponding fiber directions $\theta_j$. Geometrically, this means that the fiber degenerates in the direction given by the unit normal to the boundary.

To draw the toric diagram, we first use the moment maps \refeq{moment} to express the coordinates $p_j$, $j=4,\ldots,k$ in terms of the three coordinates $p_1$, $p_2$, $p_3$. Consequently, the boundary equations $p_j=0$, $j=4,\ldots,k$ become equations in the coordinates $p_1$, $p_2$ and $p_3$ involving the K\"ahler parameters $t_j$ of \refeq{moment}. In fact, each boundary equation gives a plane in the space generated by $p_1$, $p_2$ and $p_3$. The intersections of these planes are lines; they form the toric diagram of the toric variety, visualized as a three dimensional graph in the space generated by $p_1$, $p_2$ and $p_3$.

Hence, in this approach the toric diagram is simply the boundary of the three dimensional base parameterized by the $p_i$. There is a $T^3$ fiber over the generic point, which degenerates at the boundaries in a way determined by the unit normal. Thus, from this point of view toric diagrams should be visualized as three dimensional diagrams, encoding the degeneration of the $T^3$ fiber. It is perhaps simpler to understand this approach by working out a specific example.

\begin{figure}[htp]
\begin{center}
\psfrag{z1}{$p_1$}
\psfrag{z2}{$p_2$}
\psfrag{z3}{$p_3$}
\includegraphics[width=6cm]{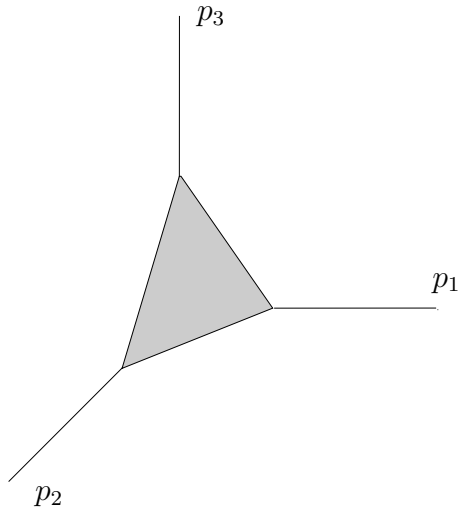}
\caption{Toric diagram $\Gamma$ of $\CO(-3) \rightarrow \IC \IP^2$ visualized as a three dimensional graph. It encodes the degeneration loci of the $T^3$ fiber.}
\label{p2graph3d}
\end{center}
\end{figure}

\begin{example}\label{exampcp2}
Let us find the toric diagram of $\CO(-3) \rightarrow \IC \IP^2$ from this point of view. This manifold is defined by the moment map $p_1+p_2+p_3-3p_4=t$, which we can use to express $p_4={1\over 3} (p_1+p_2+p_3-t)$. The boundary planes are then given by $p_1=0$, $p_2=0$, $p_3=0$ and $p_1+p_2+p_3=t$. The intersections of these planes give the toric diagram of $\CO(-3) \rightarrow \IC \IP^2$, which is drawn in figure \ref{p2graph3d}. We see that it is the same toric diagram as the one shown in figure \ref{p2graphs}, but visualized as a three dimensional graph. Note that from the fourth boundary equation one can see that the K\"ahler parameter $t$ controls the size of the $\IC \IP^2$, as it should be.
\end{example}

This is indeed an easy way to visualize the geometry of the manifold from the toric diagram; another example of this approach will be given in section \ref{examples}. However, it turns out that in many situations it is more enlightening to consider the manifold as a $T^2 \times \IR$ fibration, especially from the topological vertex perspective. Let us now describe this alternative viewpoint.

\subsection{$T^2 \times \IR$ Fibration}\label{t2}

In this language, a toric diagram $\Gamma$ is a two-dimensional graph which represents the degeneration locus of the $T^2 \times \IR$ fibration over the base $\IR^3$. Over a line in $\Gamma$ in the direction $(q,p)$, the cycle $(-q,p)$ of the $T^2$ fiber degenerates. 

To exhibit this structure, we will now follow the topological vertex approach to toric \CY\ threefolds developed by Aganagic, Klemm, Mari\~no and Vafa in \cite{Aganagic:2003db}. A good review is given in \cite{Marino:2004uf}.

The fundamental idea behind this approach is that toric \CY\ threefolds are built by gluing together $\IC^3$ patches. Therefore, the first step is to describe $\IC^3$ (which is the simplest noncompact toric \CY\ threefold) as a $T^2 \times \IR$ fibration and exhibit its degeneration locus in a two dimensional graph $\Gamma$, which turns out to be a trivalent vertex. Then, more general geometries are constructed by gluing together $\IC^3$ patches, which, in the toric diagram language, corresponds to gluing together trivalent vertices in a way specified by the toric data of the manifold. 

Conversely, given a toric \CY\ threefold, we can find a decomposition of the set of all coordinates into triplets that correspond to the decomposition of the threefold into $\IC^3$ patches. The moment maps \refeq{moment} relate the coordinates between the patches, therefore describing how the trivalent vertices corresponding to the $\IC^3$ patches are glued together to form the toric diagram of the manifold.

Let us start by describing $\IC^3$ from this point of view. Here we will only sketch the description; the details are given in \cite{Aganagic:2003db,Marino:2004uf}. Let $z_i$, $i=1,2,3$ be complex coordinates on $\IC^3$. Define the functions 
\bea
&&r_{\alpha}(z)=|z_1|^2-|z_3|^2, \nn \\ 
&&r_{\beta}(z)=|z_2|^2-|z_3|^2, \nn \\
&&r_{\g} (z)=\Im(z_1 z_2 z_3).
\eeal{hamil}
It turns out that these functions generate the fiber $T^2 \times \IR$. More specifically, $\IR$ is generated by $r_{\g}$ while the $T^2$ fiber is generated by the circle actions
\beq
\exp (i\a r_{\a} + i \b r_{\b}): (z_1,z_2,z_3) \rightarrow (e^{i \a} z_1, e^{i \b} z_2, e^{-i(\a+\b)} z_3).
\eeql{cycles}
The cycles generated by $r_{\a}$ and $r_{\b}$ are then respectively referred to as the $(0,1)$ and $(1,0)$ cycles.

We now describe the degeration loci of the fibers. We see from \refeq{hamil} and \refeq{cycles} that the $(0,1)$ cycle degenerates when $r_{\a} =0=r_{\g} $ and $r_{\b} \geq 0$, while the $(1,0)$ cycle degenerates when $r_{\a} \geq 0=r_{\g}$ and $r_{\b} = 0$. There is also a one-cycle parameterized by $\a+\b$ that degenerates when $r_{\a}-r_{\b} =0=r_{\g}$ and $r_{\a} \leq 0$.

\begin{figure}[htp]
\begin{center}
\psfrag{(0,1)}{$(0,1)$}
\psfrag{(-1,-1)}{$(-1,-1)$}
\psfrag{(1,0)}{$(1,0)$}
\includegraphics[width=5cm]{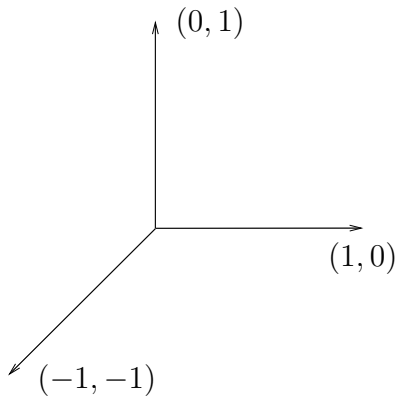}
\caption{Trivalent vertex associated to $\IC^3$, drawn in the $r_{\alpha}$-$r_{\b}$ plan. The vectors represent the generating cycles over the lines.}
\label{trivertex}
\end{center}
\end{figure}

The toric diagram is a planar graph that encodes the degeneration loci of the fibers. We can set $r_{\g}=0$ and draw the graph in the plane $r_{\a}-r_{\b}$. The graph consists in lines $p r_{\a}+q  r_{\b} = c$ where $c$ is a constant. Over this line the $(-q,p)$ cycle of the $T^2$ fiber degenerates (up to the equivalence $(q,p) \sim (-q,-p)$). For $\IC^3$, the degeneration loci can be represented as a toric diagram with lines defined by the equations $r_{\a}=0,~r_{\b}\geq 0$; $r_{\b}=0,~r_{\a} \geq 0$ and $r_{\a}-r_{\b} =0,~r_{\a} \leq 0$. Over these lines respectively the cycles $(0,1)$; $(-1,0) \sim (1,0)$ and $(1,1)$ degenerate. This gives the trivalent vertex associated to $\IC^3$, which is shown in figure \ref{trivertex}.

For more general geometries, we first find a decomposition of the set of coordinates $z_i$, $i=1,\dots,k$ into triplets of coordinates associated to the $\IC^3$ patches. We choose a patch and describe the functions $r_{\a}$ and $r_{\b}$ as above. It turns out that we can use these coordinates as global coordinates for the $T^2$ fiber in the $\IR^3$ base. As usual, we refer to the cycles $r_{\a}$ and $r_{\b}$ respectively as the $(0,1)$ and $(1,0)$ cycles. Using the moment maps \refeq{moment} defining the toric \CY\ threefold, we can find the action of the functions $r_{\a}$ and $r_{\b}$ on the other patches and therefore draw the toric diagram giving the degeneration loci of the $T^2$ fiber. An explicit example of this approach will be worked out in section \ref{examples}.

This decomposition of toric \CY\ threefolds into $\IC^3$ patches leads to a similar decomposition of topological string amplitudes on toric \CY\ threefolds into a basic building block associated to the trivalent vertex of the $\IC^3$ patches, which is called the {\it topological vertex}. By gluing together these topological vertices one can build topological string amplitudes on any toric \CY\ threefold. This beautiful property of topological amplitudes is the essence of the topological vertex approach developed in \cite{Aganagic:2003db}. We will not expand further on this subject, although we will use the topological vertex in various occasions in chapters \ref{closed} and \ref{open}. The interested reader is encouraged to go through the details of the construction in \cite{Aganagic:2003db}.

In the next section we illustrate these different approaches to toric \CY\ threefolds in specific examples.

\section{Examples}\label{examples}

We now describe two examples of toric \CY\ threefolds. The first example is the resolved conifold, namely $\CO(-1) \oplus \CO(-1) \rightarrow \IC \IP^1$. In this simple case, we illustrate in detail the different viewpoints explained in the previous sections. The second example is a more complicated geometry. It is a noncompact \CY\ threefold whose compact locus consists of two compact divisors each isomorphic to a del Pezzo surface $dP_2$\footnote{A del Pezzo surface $dP_n$, $n=0,\ldots,8$ is a complex two-dimensional Fano variety, that is $\IC \IP^2$ blown up in $n$ points.} and a rational $(-1,-1)$ curve that intersects both divisors transversely. We will give the toric data describing the manifold and draw the corresponding toric diagram.

\subsection{Resolved Conifold}

The resolved conifold $Y={\cal O}(-1) \oplus {\cal O}(-1) \rightarrow \IC {\IP}^1$  is a noncompact \CY\ threefold which admits a toric description given by the following toric data:
\beq
\begin{matrix}
 &  z_1 & z_2 & z_3 & z_4  \\
\IC^* & 1 & 1 & -1 & -1
\end{matrix}
\eeql{toricres}
The lines in this table give the charges $Q^i_a$ corresponding to the torus actions on the homogeneous coordinates $z_i$. We see that $\sum_i Q^i = 1+1-1-1=0$; therefore $Y$ is \CY. $Y$ is defined as the space obtained from
\beq
|z_1|^2 + |z_2|^2 -|z_3|^2 - |z_4|^2 =t
\eeql{rescon}
after quotienting by the $U(1)$ action specified by the charges in \refeq{toricres}.

We now find the fan $\Sigma$ describing $Y$. We have the relation $\sum_{i=1}^4 Q^i v_i = v_1+v_2-v_3-v_4=0$. We choose distinct vectors $v_i = (\nu_i,1)$ where the $\nu_i$ are two dimensional. A solution is $v_1 = (1,0,1)$, $v_2=(-1,0,1)$, $v_3=(0,1,1)$ and $v_4=(0,-1,1)$. These four vectors generate the four one dimensional cones of $\Sigma$.

\begin{figure}[htp]
\begin{center}
\psfrag{(0,1)}{$(0,1)$}
\psfrag{(0,-1)}{$(0,-1)$}
\psfrag{(-1,0)}{$(-1,0)$}
\psfrag{(1,0)}{$(1,0)$}
\includegraphics[width=7cm]{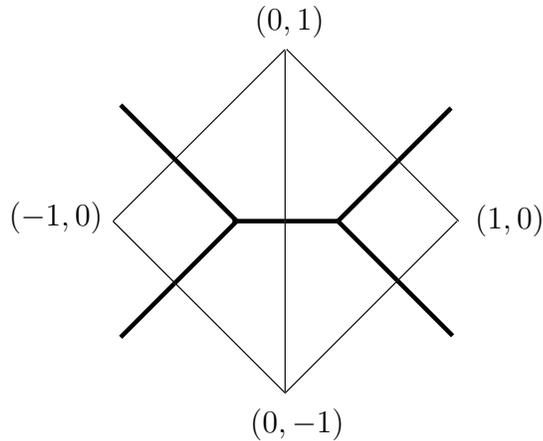}
\caption{The $\Gamma$ and $\tilde \Gamma$ graphs for $\CO(-1)\oplus \CO(-1) \rightarrow \IC \IP^1$. The toric diagram $\Gamma$ is the normal diagram drawn in thick lines. The points $(\nu_i,1)$ give the fan $\Sigma$, where the $\nu_i$ are the vertices of $\tilde \Gamma$ and are shown in the figure.}
\label{p1graphs}
\end{center}
\end{figure}

The two dimensional graph $\tilde \Gamma$ is given by the intersection of the plane $z=1$ and $\Sigma$. The vertices are $(1,0)$,$(-1,0)$,$(0,1)$ and $(0,-1)$. We can also draw the toric diagram, which is the dual graph $\Gamma$. They are shown in figure \ref{p1graphs}.

\begin{figure}[htp]
\begin{center}
\psfrag{(0,0,t)}{$(0,0,t)$}
\psfrag{p1}{$p_1$}
\psfrag{p2}{$p_2$}
\psfrag{p3}{$p_3$}
\includegraphics[width=7cm]{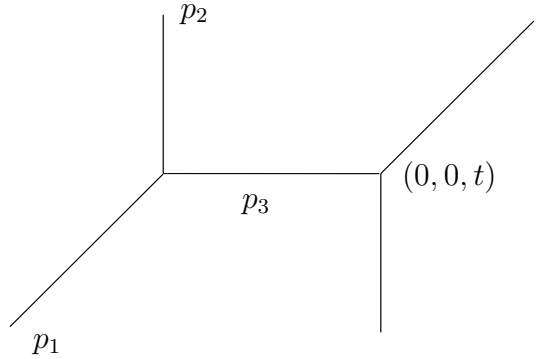}
\caption{Toric diagram $\Gamma$ of $\CO(-1)\oplus \CO(-1) \rightarrow \IC \IP^1$ visualized as a three dimensional graph. It encodes the degeneration loci of the $T^3$ fiber.}
\label{p1graph3d}
\end{center}
\end{figure}

If we look at the resolved conifold as a $T^3$ fibration, we have to understand the toric diagram $\Gamma$ as a three dimensional graph representing the base, where the $T^3$ fiber degenerates at the boundaries. The base is parameterized by the four coordinates $p_i \equiv |z_i|^2$ subject to the relation \refeq{rescon}. We can use \refeq{rescon} to eliminate $p_4$,
\beq
p_4=p_1+p_2-p_3-t.
\eeq 
Therefore, since $|z_i|^2 \geq 0$, the boundary equations of the toric base are given by
\bea
p_1 &=& 0, \nn\\
p_2 &=& 0, \nn\\
p_3 &=& 0, \nn\\
p_1+p_2-p_3 &=& t.
\eea
The intersections of these planes give the toric diagram of the resolved conifold shown in figure \ref{p1graphs}, but visualized as a three dimensional graph as in figure \ref{p1graph3d}. Note that as in example \ref{exampcp2}, by the fourth boundary equation above one can see that the K\"ahler parameter $t$ controls the size of the $\IC \IP^1$, as it should be.

We can also describe the resolved conifold as a $T^2 \times \IR$ fibration, using its decomposition into $\IC^3$ patches. We choose the first patch to be defined by $z_1 \neq 0$. Using \refeq{rescon} we can express $z_1$ in terms of the other coordinates, so the patch is parameterized by $(z_2,z_3,z_4)$. We define the functions
\bea
r_{\a}&=&|z_3|^2-|z_2|^2,\nn\\
r_{\b}&=&|z_4|^2-|z_2|^2.
\eeal{hamilp11}
This gives the usual trivalent graph of $\IC^3$.

\begin{figure}[htp]
\begin{center}
\psfrag{(0,1)}{$(0,1)$}
\psfrag{(1,0)}{$(1,0)$}
\psfrag{(-1,-1)}{$(-1,-1)$}
\psfrag{(1,1)}{$(1,1)$}
\psfrag{(0,-1)}{$(0,-1)$}
\psfrag{(-1,0)}{$(-1,0)$}
\psfrag{U1}{$U_1$}
\psfrag{U2}{$U_2$}
\includegraphics[width=8cm]{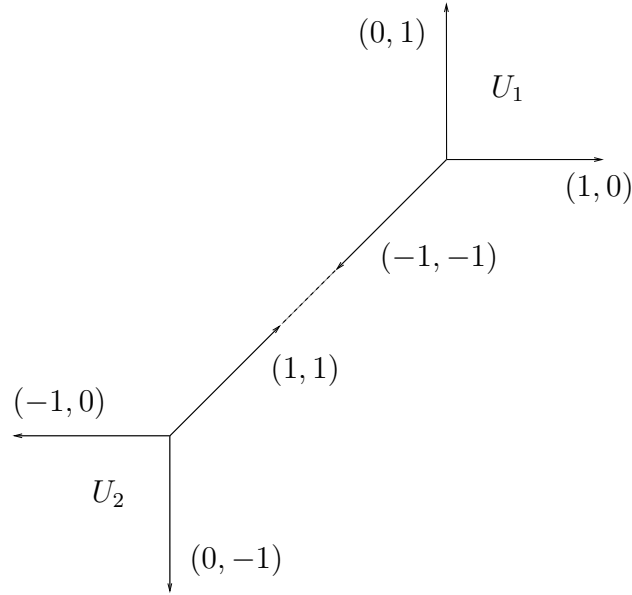}
\caption{Toric diagram of $\CO(-1)\oplus\CO(-1)\rightarrow\IC\IP^1$, drawn in the $r_{\alpha}$-$r_{\b}$ plan.  The vectors represent the generating cycles over the lines. The origin of the second patch $U_2$ is shifted to $(-t,-t)$.}
\label{p1graphtopo}
\end{center}
\end{figure}

The other patch is defined by $z_2 \neq 0$, therefore parameterized by $(z_1,z_3,z_4)$. Using \refeq{rescon} we can rewrite the functions \refeq{hamilp11} in terms of the coordinates on this patch:
\bea
r_{\a}&=&|z_1|^2-|z_4|^2-t,\nn\\
r_{\b}&=&|z_1|^2-|z_3|^2-t.
\eeal{hamilp12}
These functions generate the circle action
\beq
\exp (i\a r_{\a} + i \b r_{\b}): (z_1,z_3,z_4) \rightarrow (e^{i (\a+\b)} z_1, e^{-i \b} z_3, e^{-i\a} z_4).
\eeq
In this patch, the $(0,1)$ cycle degenerates when $r_{\a} \leq -t$ and $r_{\b} = -t$. The $(1,0)$ cycle degenerates when $r_{\a}=-t$ and $r_{\b} \leq -t$. The $(1,1)$ cycle degenerates when $r_{\a}-r_{\b}=0$ and $r_{\a} \geq -t$. Therefore, the graph associated to this patch is identical to the first one, although it is shifted such that its origin is at the point $(-t,-t)$. The two graphs are joined through the common edge given by $r_{\a}-r_{\b}=0$. $t$ gives the `length' of the internal edge, and correspondingly is the K\"ahler parameter associated to the $\IC \IP^1$. This gives the toric diagram of the resolved conifold shown in figure \ref{p1graphtopo}.

\subsection{Two $dP_2$'s Connected by a $\IC \IP^1$}

The next example is a noncompact \CY\ threefold $X$ whose compact locus consists of two compact divisors each isomorphic to a del Pezzo surface $dP_2$ and a rational $(-1,-1)$ curve that intersects both divisors transversely. The divisors do not intersect each other. This manifold is described by the following toric data:
\beq
\begin{matrix} 
 & z_1 & z_2 & z_3 & z_4 & z_5 & z_6 & z_7 & z_8 & z_9 & z_{10} \\
\IC^* & -1 & 1 & 1 & -1 & 0 & 0 & 0 & 0 & 0 & 0 \\
\IC^* & 1 & 0 & -1 & -1 & 0 & 1 & 0 & 0 & 0 & 0 \\
\IC^* & 1 & -1 & 0 & -1 & 1 & 0 & 0 & 0 & 0 & 0 \\
\IC^* & 0 & 0 & 0 & 1 & -1 & -1 & 1 & 0 & 0 & 0 \\
\IC^* & 0 & 0 & 0 & 0 & 1 & 0 & -1 & -1 & 0 & 1 \\
\IC^* & 0 & 0 & 0 & 0 & 0 & 1 & -1 & 0 & -1 & 1 \\
\IC^* & 0 & 0 & 0 & 0 & 0 & 0 & -1 & 1 & 1 & -1.
\end{matrix}
\eeql{toricXi}

\begin{figure}[htp]
\begin{center}
\includegraphics[width=7cm]{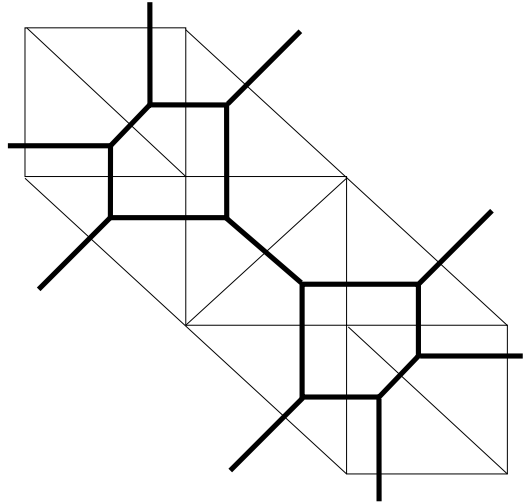}
\caption{The $\Gamma$ and $\tilde \Gamma$ graphs for the \CY\ threefold $X$ whose compact locus consists of two $dP_2$'s connected by a $\IC \IP^1$. The toric diagram $\Gamma$ is the normal diagram drawn in thick lines.}
\label{maingeograph}
\end{center}
\end{figure}

We see that the charges in each line add up to zero, hence $X$ is \CY. The toric diagram $\Gamma$ of $X$ and its dual $\tilde \Gamma$ are shown in figure \ref{maingeograph}. 

It is often useful \cite{Aganagic:2002qg} to consider a related Calabi-Yau threefold $\tilde X$ obtained from $X$ by flopping the two exceptional curves outside of the compact divisors. This is shown in figure \ref{toriccomm}.

This Calabi-Yau threefold --- and a simplified one where some of the K\"ahler parameters of $\tilde X$ are sent to infinity --- will be the focus of interest of chapter \ref{closed} and \ref{open}.

\section{Hypersurfaces in Toric Varieties}\label{hyper}

In the remaining of this chapter we explain how compact \CY\ manifolds may be obtained in toric geometry, namely as hypersurfaces in compact toric manifolds using Batyrev's well known reflexive polytopes \cite{Batyrev:1994}.

\subsection{Reflexive Polytopes}
 
In section \ref{toricCY} we described in details toric \CY\ threefolds. In particular, we showed that toric \CY\ threefolds are noncompact. However, from a string theory perspective, it is often desirable to consider compact \CY\ manifolds. Hence it seems that toric geometry is not a good setup for such geometries. 

Fortunately, there is a way to construct compact \CY\ manifolds in toric geometry, namely as compact hypersurfaces in compact toric varieties. Batyrev's reflexive polytopes \cite{Batyrev:1994} provide a very useful description of such compact \CY\ manifolds. The toric variety itself is not \CY, consequently it can be compact. Reflexivity of the polytopes then ensures that the compact hypersurface, which is not toric itself, is \CY.

An elementary introduction to these concepts and their applications to string theory and dualities can be found in \cite{Skarke:1998yk}. The following is partly based on the first sections of \cite{Bouchard:2003bu}.

As in section \ref{toricCY}, in the following we focus our attention on three-dimensional toric varieties, therefore leading to two-dimensional \CY\ hypersufaces, i.e. K3 surfaces. It is however straightforward to generalize the concepts to higher dimensional toric varieties.

\subsubsection{Lattice Description}

A polytope in $M_\IR$ is the convex hull of a finite number of points in $M_\IR$, and a polyhedron in $M_\IR$ is the intersection of finitely many half-spaces (given by inequalities $\langle u,v\rangle \geq c$ with some $v\in N_\IR$ and $c\in \IR$) in $M_\IR$. It is well known that any polytope is a polyhedron and any bounded polyhedron is a polytope. If a polyhedron $S\subset M_\IR$ contains the origin \ipo, its dual
\beq
S^*=\{v\in N_\IR : \langle u,v\rangle \geq -1 \hbox{ for all } u \in S\}.
\eeql{dualpoly}
is also a polyhedron containing \ipo, and $(S^*)^*=S$.

A lattice polytope in $M_\IR$ is a polytope with vertices in $M$. 

\begin{definition}
A polytope $\D\subset M_\IR$ containing \ipo\ is called {\rm reflexive} if both $\D$ and $\D^*$ are lattice polytopes. 
\end{definition}

This is equivalent to $\D$ being a lattice polytope whose bounding equations are of the form $\<u,v_i\>\ge -1$ with $v_i\in N$ (in coordinates, $\sum_j u_j v_{ij}\ge -1$ with integer coefficients $v_{ij}$). By convexity it is sufficient to consider only those equations corresponding to $v_i$ that are vertices of $\D^*$. In this way there is a duality between vertices of $\D^*$ and facets of $\D$; similarly, there are dualities between $p$ dimensional faces of $\D$ and $(n-p-1)$ dimensional faces of $\D^*$ (in three dimensions: between edges and dual edges).

An interior point $u$ of a reflexive polytope must satisfy $\langle u,v_i\rangle > -1$ for all $v_i$, so an interior lattice point must satisfy $\langle u,v_i\rangle \geq 0$. Thus if $u$ is an interior lattice point, then $nu$ is also an interior lattice point for any non-negative integer $n$. For $u\ne$ \ipo\ this would be in conflict with the boundedness of $\D$, implying that \ipo\ is the only interior lattice point.

\subsubsection{Toric Interpretation}\label{Kthree}

Given a three dimensional pair of reflexive polytopes $\D\in M_\IR$, $\D^*\in N_\IR$, a smooth K3 surface can be constructed in the following manner. Any complete triangulation of the surface of $\D^*$ defines a fan $\Sigma$ whose three dimensional cones are just the cones over the regular (i.e., lattice volume one) triangles. To any lattice point $p_i=(\bar x_i, \bar y_i, \bar z_i)$ on the boundary of $\D^*$ one can assign a homogeneous coordinate $w_i\in \IC$, with the rule that several $w_i$ are allowed to vanish simultaneously only if there is a cone such that the corresponding $p_i$ all belong to this cone. The equivalence relations among the homogeneous coordinates are given by
\beq
(w_1,\ldots, w_n) \sim (\l^{Q_a^1}w_1,\ldots,\l^{Q_a^k}w_k)\;\hbox{ for any }\l\in\IC^*
\eeql{equiv}
with any set of integers $Q_a^i$ such that $\sum Q_a^i p_i=0$; among these relations, $k-3$ are independent. This construction gives rise to a smooth compact three dimensional toric variety $\CM_{\Sigma}$ (smooth because the generators of every cone are also generators of $N$, compact because the fan fills $N_\IR$). The loci $w_i=0$ are the toric divisors $D_i$.

To any lattice point $q_j$ of $M$ we can assign a monomial $m_j=\prod_i w_i^{\<q_j,p_i\>+1}$; the exponents are non-negative as a consequence of reflexivity. The hypersurface defined by the zero-locus of a generic polynomial $P=\sum a_j m_j$ transforms homogeneously under \refeq{equiv} and can be shown to define a K3 hypersurface in $\CM_{\Sigma}$ (actually it defines a family of hypersurfaces depending on the coefficients $a_j$).

\begin{remark}
A good way to remember this construction is to notice that the polytope in $\D^* \in N_\IR$ gives the faN of the ambient toric variety, while the polytope in $\D \in M_\IR$ gives the MonoMials.
\end{remark}

\subsubsection{\CY\ Condition}

It was shown by Batyrev \cite{Batyrev:1994} that the hypersurface defined by the vanishing of a generic polynomial in the class determined by $\D$ is a smooth \CY\ manifold for $n \leq 4$, where $n$ is the dimension of the lattice $M$. For $n\leq3$ the underlying toric variety is smooth; in particular for $n=3$ the hypersurface describes a smooth K3 surface as explained above. For $n=4$ it may have point-like singularities, which are however missed by the generic hypersurface describing the \CY\ threefold.

Let us now explain why the hypersurface is a \CY\ manifold. Let a manifold $X$ be defined by the equation $P=0$ in a toric variety $\cm$. The polynomial $P$ defines a section of a line bundle (other sections are defined by different coefficients $a_j$). The divisor class of the line bundle can be read off from any monomial in $P$. Since the origin is always included in the polytopes, $P$ always includes the monomial $\prod_{i=1}^k w_i$, which corresponds to the divisor class $[\sum_{i=1}^k D_i]$. Thus, the polynomial $P$ determines a section of the anticanonical bundle of the toric variety $\cm$. 

Roughly speaking, on $X$ the section $P$ maps points of $X$ to $0$ in the fibers of the anticanonical bundle of $\cm$, since $X$ is defined as the zero-locus of $P$. Thus, $P$ serves as a coordinate near $X$, and in fact the normal bundle $N_X$ of $X$ is simply $K^*_\cm|_X$, since $P$ is a section of the anticanonical bundle of $\cm$. Thus, the exact sequence $0 \to T^{1,0}X \to T^{1,0} \cm|_X \to N_X \to 0$ becomes (this result is also known as the adjunction formula 1 --- see \cite{Griffiths:1978,Hubsch:1992})
\beq
0 \to T^{1,0}X \to T^{1,0} \cm|_X \to K^*_\cm|_X \to 0.
\eeq

Given any holomorphic vector bundle $B$ over $X$ of rank $k$ and any holomorphic sub-bundle $A$, one can always form the respective determinant bundles $\det B$ and $\det A$ which satisfy the identity $\det B = \det A \otimes \det (B / A)$. Using the above exact sequence, we can then write
\beq
\det T^{1,0} \cm|_X = \det T^{1,0} X \otimes \det K^*_\cm|_X.
\eeq
Now, using the definition of the anticanonical bundle as the determinant line bundle of the holomorphic tangent bundle and the fact that $\det K^*_\cm = K^*_\cm$ since $K^*_\cm$ is a line bundle, we find
\beq
K^*_\cm|_X = K^*_X \otimes K^*_\cm|_X,
\eeq
or equivalently
\beq
K_X = (K^*_\cm \otimes K_\cm)|_X,
\eeq
that is the canonical bundle $K_X$ of $X$ is trivial. If $X$ is smooth, which is guaranteed by reflexivity for $n \leq 4$, then $X$ is a \CY\ manifold.

\subsubsection{Fibration Structure}

Suppose the intersection of $\D^*$ with the plane $\bar z = 0$ gives a reflexive polygon. We may reinterpret $P$ as a polynomial in the $w_i$ for which $\bar z_i=0$, with coefficients depending on the remaining $w_i$, i.e. we are dealing with an elliptic curve parameterized by the $w_i$ for which $\bar z_i\ne 0$. The map $\CM_{\Sigma} \to \IC \IP^1$,
\beq
(w_1,\ldots, w_n) \;\to \; W=\prod_{i:\bar z_i\ne 0} w_i^{\bar z_i} 
\eeql{projmap}
is easily checked to be consistent with \refeq{equiv} and thus well defined. At any point of the $\IC \IP^1$ that is neither $0$ nor $\infty$ all the $w_i$ with $\bar z_i\ne 0$ are non-vanishing, and \refeq{equiv} can be used to set all except one of them to $1$. This gives the K3 surface the structure of an elliptic fibration.

\subsection[Tops]{Tops}\label{topsfirst}

We now describe `tops', the close cousins of Batyrev's reflexive polytopes.

\subsubsection[Lattice Description]{Lattice Description}

In \cite{Candelas:1996su} Candelas and Font considered reflexive polytopes whose intersections with a plane were themselves reflexive polygons; the fact that this intersection cuts the polytope into two parts (`top' and `bottom') gave rise to the notion of a `top' as half of a reflexive polytope in this sense. In \cite{Candelas:1997pq} this definition was generalized in the following way. 

\begin{definition}
A {\rm top} $\top\subset N_\IR$ is a lattice polytope such that one of its defining inequalities is of the form $\<u_0,v\>\ge 0$ and all others are of the form $\<u_i,v\>\ge -1$, with $u_i\in M$.
\end{definition}

We consider two tops to be isomorphic if they are related by a $GL(3,\IZ)$ transformation. This allows us to choose coordinates $(x,y,z)$ for $M$ and $M_\IR$ such that $u_0$ has coordinates $(0,0,1)$ (we will always make this choice whenever we work with specific coordinates). Then the inequality corresponding to the facet $F_0:=\{v\in \top:\<u_0,v\>=0\}$ is given by $\bar z\ge 0$ in terms of dual coordinates $(\bar x,\bar y,\bar z)$ for $N_\IR$. $F_0$ is bounded by the restrictions of the other inequalities to $\bar z=0$; as these are again of the type $\ldots \ge -1$ with integer coefficients, $F_0$ is a reflexive polygon. Thus the more general definition of a top indeed contains all the cases of \cite{Candelas:1996su}. A straightforward adaptation of the above argument about reflexive polytopes shows that a top has no interior lattice points.

The dual $\top^*\subset M_\IR$ of $\top$ is the polyhedron defined by the inequalities originating from the vertices of $\top$. The vertices $(\bar x_i,\bar y_i,0)$ of $F_0$ lead to inequalities of the form $x\bar x_i+y\bar y_i\ge -1$; we will refer to the corresponding facets as {\it vertical facets}. Thus $\top^*$ must be contained in a prism over $F_0^*$ (the dual of $F_0$ in the two dimensional sense). The remaining vertices of $\top$ have $\bar z>0$. The corresponding inequalities can be written as
\beq
z\bar z_i\ge -1-x\bar x_i-y\bar y_i
\eeql{nonvert}
implying that for every fixed $(x,y)\in F_0^*$ there is a minimal (but no maximal) value $z_{\rm min}(x,y)$ such that $(x,y,z)\in \top^*$ for all $z \ge z_{\rm min}(x,y)$. In this way we can view $\top^*$ as the result of `chopping off' the lower parts of an infinitely extended prism. Alternatively, we may see it as a `polytope' with one vertex $u_\infty=+\infty\, u_0$ at infinity, as it is the dual of $\top$ which may be defined by $\<u_i,v\>\ge -1$ for $i\ge 1$ and $\<\l u_0,v\>\ge -1$ for arbitrarily large positive $\l$. $\top^*$ has infinitely many interior lattice points $(0,0,z)$ with $z$ any non-negative integer.

The projection
\beq
\pi : \;\;\top^* \to F_0^*,\;\;(x,y,z)\to(x,y)
\eeql{proj}
takes vertices of $\top^*$ to lattice points of $F_0^*$. Conversely, every (finite) vertex of $\top^*$ is of the form $(x,y,\zm(x,y))$ where $(x,y)$ is a lattice point of $F_0^*$. This means that by specifying $F_0^*$ and $z_{\rm min}$ for each of its lattice points, we have specified $\top^*$ completely. The projections of non-vertical facets determine a partition of $F_0^*$. These facts afford a useful pictorial representation of $\top^*$ in terms of a picture of $F_0^*$ where every lattice point is labeled with the corresponding $z_{\rm min}$.

\begin{figure}[htp]
\begin{center}
\includegraphics[width=11cm]{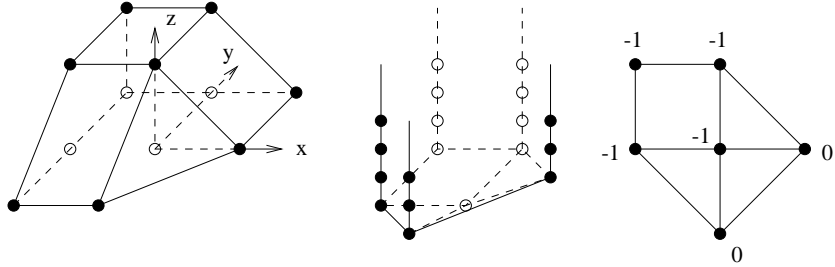}
\caption{A top, its dual and the minimal point notation.}
\label{fig:a3}
\end{center}
\end{figure}

\begin{example}
Suppose $\top$ is the convex hull of $(-1,-1,0)$, $(-1,1,0)$,
$(1,1,0)$, $(1,0,0)$, $(0,-1,0)$, $(0,0,1)$, $(-1,0,1)$, $(-1,1,1)$ and $(0,1,1)$, as shown in the first part of \reffig{fig:a3}. $\top$ has 7 facets corresponding to the inequalities

\bea
&&F_0:z\ge 0, \quad F_1:z\le 1, \quad F_2:x\ge -1, \quad F_3:\ y\le 1,\nn\\
&&F_4:x+z\le 1, \quad F_5:x-y+z\le 1, \quad F_6:z-y\le 1, 
\nn\eea
implying that $\top$ is indeed a top and that $\top^*$ has the vertices $(0,0,-1)$, $(1,0,0)$, $(0,-1,0)$, $(-1,0,-1)$, $(-1,1,-1)$, $(0,1,-1)$, in addition to the `vertex at infinity' $(0,0,\infty)$. Thus $F_0^*$ is the convex hull of $(1,0)$, $(0,-1)$, $(-1,0)$, $(-1,1)$, $(0,1)$ with $z_{\rm min}$ as shown in the third diagram of \reffig{fig:a3}.
\end{example}

\subsubsection[Toric Interpretation]{Toric Interpretation}

In section \ref{Kthree} we showed that three dimensional reflexive polytopes describe families of compact K3 surfaces in compact three dimensional toric varieties. Let us now find the toric interpretation of tops.

Let $\top$ be a top defined in $N_{\IR}$, and let $\top^* \subset M_{\IR}$ be its dual (notice that we lost the symmetry that was present in the case of reflexive polytopes, since the dual of a top is not a top). A complete triangulation of the surface of $\top$ defines a fan $\Sigma$, to which we associate a three dimensional toric variety $\CM_{\Sigma}$. As before, we can also assign a monomial $m_j=\prod_i w_i^{\<q_j,p_i\>+1}$ to any lattice point $q_j$ of $M$, and define a generic polynomial $P=\sum_j a_j m_j$. The equation $P=0$ defines a surface $\CS$ in $\CM_{\Sigma}$.

Although the description is very similar to the K3 case, there are many important differences. Since the fan associated to a top does not fill $N_{\IR}$, according to proposition \ref{compactness}  the toric variety is noncompact. The surface $\CS$ defined by $P=0$ is elliptically fibered as before, but it is now also noncompact. In fact, since there are no negative exponents in \refeq{projmap}, the base space of the fibration is now $\IC$ instead of $\IC \IP^1$.

In those cases where the top is half of a reflexive polytope, it encodes the geometry of the K3 away from the preimage of the point $\infty$. In addition we have an interpretation of a top in the case of an elliptically fibered higher dimensional \CY\ hypersurface in a toric variety. Here the polygon encoding the elliptic fiber is again an intersection of a reflexive polytope with a plane. The base space of the fibration is determined by projecting the fan along the two dimensions spanned by the polygon \cite{Kreuzer:1997zg}. Rays in this projected fan determine divisors in the base along which the fiber can degenerate; the inverse image of such a ray is again a top whose structure determines the generic type of degeneration over the intersection of a disc with the divisor. This may lead to tops with far more points than in a three dimensional reflexive polytope \cite{Candelas:1997pq}.

These tops and their associated toric descriptions possess very interesting mathematical properties. They also enter beautifully in the analysis of dualities between compactifications of string theory. In chapter \ref{tops} we investigate in detail the geometrical properties of tops and provide a complete classification of these objects, which leads to interesting results in string theory.

\chapter[The Classification of Tops and String Dualities]{The Classification of Tops and String Dualities}\label{tops}

In this chapter we study in detail the geometrical object introduced in section \ref{topsfirst}, namely tops. We start by continuing the investigation of the toric properties of tops that was started in section \ref{topsfirst}. This leads us to claim that not only tops but also elliptic fibration structures in general should be related to untwisted or twisted affine Kac-Moody algebras. Using Kodaira's classification of degenerations of elliptic fibrations \cite{Kodaira:1963}, we find a simple way to assign an affine Kac-Moody algebra to a top.

Then in section \ref{classification} we classify all the possible tops. We present here our classification scheme; the results are shown in appendix \ref{class}. Finally, we discuss our results with particular emphasis on the cases related to twisted algebras, first in terms of geometry (twisted algebras occur only for fibrations that allow orbifold actions) and then in terms of dualities between M-theory, F-theory or type II strings and heterotic strings or CHL type strings.

\section[Toric Properties and Affine Kac-Moody Algebras]{Toric Properties and Affine Kac-Moody Algebras}\label{kac}

In section \ref{topsfirst} we introduced tops and described some of their fundamental properties. In this section we will pursue further the analysis of their toric properties, which leads to a beautiful correspondence between tops and affine Kac-Moody algebras.

\subsection[Affine Kac-Moody Algebras]{Affine Kac-Moody Algebras}

Let us first recall some properties of tops introduced in section \ref{topsfirst}. Let $\top$ be a top defined in $N_{\IR}$, and let $\top^* \subset M_{\IR}$ be its dual. A complete triangulation of the surface of $\top$ defines a fan $\Sigma$, to which we associate a noncompact three dimensional toric variety $\CM_{\Sigma}$. We assign a monomial $m_j=\prod_i w_i^{\<q_j,p_i\>+1}$ to any lattice point $q_j$ of $M$, and define a generic polynomial $P=\sum_j a_j m_j$. The equation $P=0$ defines a noncompact surface $\CS$ in $\CM_{\Sigma}$.

The elliptic fibration structure of $\CS$ is exhibited as follows. The intersection of $\top$ with the plane $\bar z = 0$ gives a reflexive polygon (the 16 reflexive polygons are shown in figure \ref{fig:16pol}). We may reinterpret $P$ as a polynomial in the $w_i$ for which $\bar z_i=0$, with coefficients depending on the remaining $w_i$, i.e. we are dealing with an elliptic curve parameterized by the $w_i$ for which $\bar z_i\ne 0$. The map $\CM_{\Sigma} \to \IC$,
\beq
(w_1,\ldots, w_n) \;\to \; W=\prod_{i:\bar z_i\ne 0} w_i^{\bar z_i} 
\eeql{projmaptop}
is consistent with the equivalence relation \refeq{equiv}, hence well defined. Notice that in \refeq{projmaptop} there are no negative exponents, which is why it is a map to $\IC$ rather than $\IC \IP^1$ as in \refeq{projmap}. This gives the surface $\CS$ the structure of an elliptic fibration over $\IC$.

All of the interesting geometry happens at $W=0$. Let us start with a top that has only a single vertex $p_n$ at $\bar z =1$ and all other lattice points at $\bar z =0$. This leads to a hypersurface determined by a polynomial in $w_1,\ldots, w_{n-1}$ with coefficients that are power series in $W=w_n$ that start with a constant; each of these power series corresponds to a vertical edge in $\top^*$. In the generic case nothing special happens and we get a smooth elliptic curve at $W=0$. If we restrict some of the $W$ dependent coefficients to start at higher powers of $W$, this may lead to singularities. Now the non-vanishing coefficients correspond only to a subset ${\top'}^*$ of $\top^*$, and we can resolve the singularity by passing from $\top$ to $\top'$, which corresponds to a blow up.

\begin{figure}[h]
\begin{center}
\includegraphics[width=127mm]{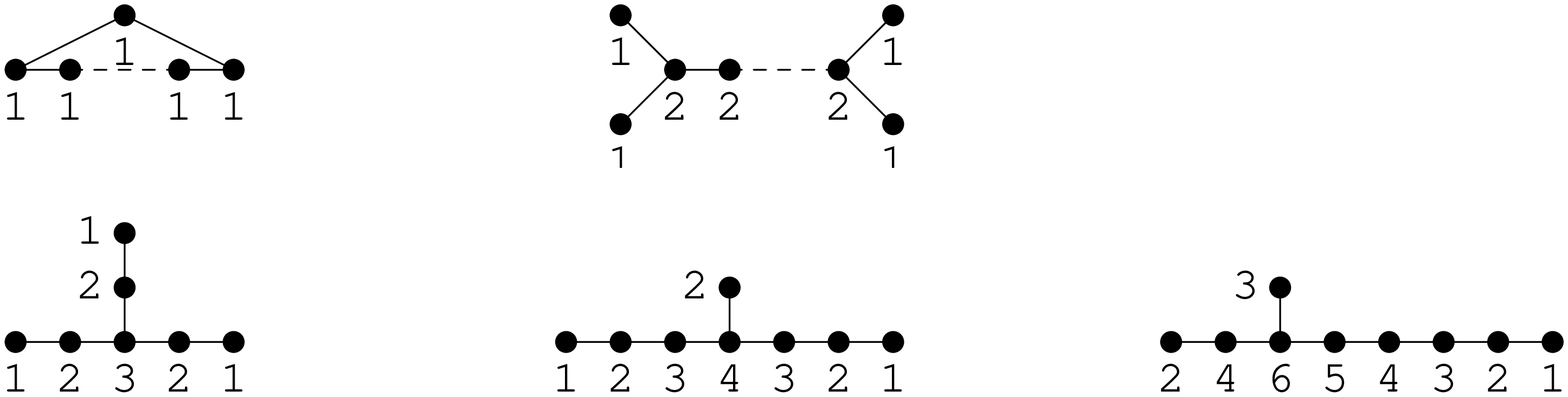}
\put(-350,100){$A_n^{(1)}$}
\put(-200,100){$D_n^{(1)}$}
\put(-360,40){$E_6^{(1)}$}
\put(-220,40){$E_7^{(1)}$}
\put(-55,40){$E_8^{(1)}$}
\caption{Dynkin diagrams of the self-dual untwisted ADE Kac-Moody algebras.}
\label{fig:ADE}
\end{center}
\end{figure}

An arbitrary top $\top$ always contains at least one lattice point at
$\bar z=1$. 
This can be seen by observing that a complete triangulation of the fan
leads to a triangulation of $\top$ in terms of tetrahedra of volume
one; such a tetrahedron with base at $\bar z=0$ must then have its
apex at $\bar z =1$.
Thus every top can be interpreted as the smooth resolution of the
singularity at $W=0$ of an elliptic fibration.
Such singularities and their resolutions were classified by Kodaira
\cite{Kodaira:1963}, resulting in the following picture (see also \cite{Barth:1984}).
Under certain assumptions fulfilled in the present case, the inverse
image of $W=0$ must consist either of a single (possibly singular)
curve or of a collection of smooth rational curves $C_i$ such that the
intersections of these curves obey $C_i\cdot C_j=-M_{ij}$ where $M$ is
the Cartan matrix of an untwisted Kac-Moody algebra of ADE type 
(these are precisely the self-dual affine Kac-Moody algebras); the
multiplicities of the $C_i$ are the coefficients of the null vector of
the algebra.
In other words, the intersection patterns and multiplicities are
respresented by the Dynkin diagrams with labels, as shown in
figure~\ref{fig:ADE}. 

Every toric divisor $D_i=\{w_i=0\}\subset \CM_{\S}$ may give rise via 
$D_i\cdot {\CS}=\sum_j C_{ij}$ to one or more curves $C_{ij}$ in $\CS$.
This type of intersection theory was worked out in detail for three 
dimensional reflexive polytopes in \cite{Perevalov:1997vw}, with the following results 
which also hold in the context of tops for the compact divisors with 
$\bar z > 0$.

\begin{figure}[h]
\begin{center}
\includegraphics[width=10cm]{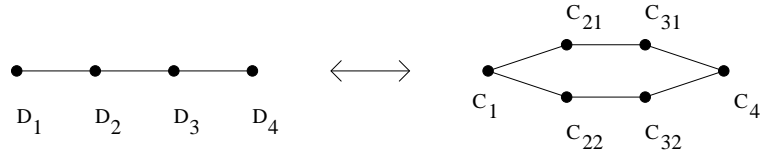}
\end{center}
\caption{An edge of $\top^*$ and the intersection pattern to which it corresponds.}
\label{fig:inters}
\end{figure}

A lattice point interior to a facet of $\top$ determines a divisor
$D_i$ that does not intersect $\CS$. A vertex gives rise to a single curve (rational for any vertex at $\bar z>0$ in a non-trivial top). A point interior to an edge corresponds to $l$ rational curves, where $l$ is the length of the dual edge. Each of the rational curves involved has self intersection $-2$. Mutual intersections occur only among neighbours along edges; in that case $D_i\cdot D_j \cdot {\CS}=l$. For example, an edge of length $3$ dual to an edge of length $2$ gives rise to an intersection pattern as indicated in figure~\ref{fig:inters}.

Here $D_1$ and $D_4$ come from vertices $v_1,v_4$ of $\top$ and $D_2$ and
$D_3$ from points interior to the edge $\5{v_1v_4}$.
$D_2$ and $D_3$ each give rise to two curves ($C_{21}, C_{22}$ and
$C_{31}, C_{32}$, respectively) in such a way that any two curves have
mutual intersection one if they are joined by an edge in the second
part of figure~\ref{fig:inters}.

We can use this information to predict the structure of the edge
diagram of the part of $\top$ with $\bar z \ge 1$.
If all the dual edges have length 1, it must have the structure of the
Dynkin diagram of an affine ADE algebra.
From (\ref{projmaptop}) it is clear that the multiplicity of a curve
$C_i=D_i\cdot {\CS}$ in $W=0$ is just $\bar z_i$, i.e. the Dynkin
labels encode the heights $\bar z$ of the lattice points.
If some of the dual edges have lengths $>1$, the edge diagram must be
the result of partially collapsing an ADE diagram,
as in reading figure~\ref{fig:inters} from right to left.
Each of the curves $C_{ij}$ originating from the same $D_i$ must have
multiplicity $\bar z_i$ in $W=0$.
There are two possibilities. 
If both vertices are at $\bar z > 0$, the uncollapsed diagram contains
a closed loop and the only possibility is the $A$-series.

\begin{figure}[h]
\begin{center}
\includegraphics[width=127mm]{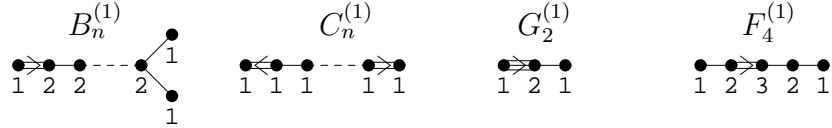}
\put(-320,40){$B_n^{(1)}$}
\put(-225,40){$C_n^{(1)}$}
\put(-150,40){$G_2^{(1)}$}
\put(-65,40){$F_4^{(1)}$}
\caption{Dynkin diagrams of the duals of untwisted non-simply laced Kac-Moody 
algebras.}
\label{fig:utnsl}
\end{center}
\end{figure}

The collapsed diagram must then look like the second one in figure
\ref{fig:utnsl}, where we use multiple lines and arrows to indicate
that we pass from a point giving rise to more than one curve to a vertex
associated with a single curve.
The other possibility for an edge whose dual has length $>1$ is that one 
of its vertices is at $\bar z =0$.
Then we
have to identify two or more ends of one of the $D$ or $E$ Dynkin diagrams.
Direct inspection shows that this is possible only for those ends
whose last point has a label of 1.
These are precisely the `extension points' if the ADE diagram is read
as the extended Dynkin diagram of an ADE Lie algebra.
If the folding procedure leaves at least one of these points
invariant, we may view this as folding an ordinary Dynkin diagram,
taking us from a self-dual simply laced algebra to the dual of a
non-simply laced one.
In this way we get $B_n$ from $D_{n+1}$, $C_n$ from $A_{2n-1}$, $G_2$
from $D_4$ (by a triple folding), and $F_4$ from $E_6$.

\begin{figure}[h]
\begin{center}
\includegraphics[width=127mm]{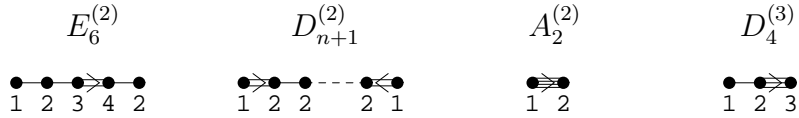}
\put(-320,40){$E_6^{(2)}$}
\put(-235,40){$D_{n+1}^{(2)}$}
\put(-145,40){$A_2^{(2)}$}
\put(-65,40){$D_4^{(3)}$}
\caption{Dynkin diagrams of the duals of twisted Kac-Moody algebras
that can be read off from tops.} 
\label{fig:twist}
\end{center}
\end{figure}

If all of the points with Dynkin label $1$ are involved in the folding
procedure, we pass from an untwisted Kac-Moody algebra to a twisted
one.
We find the possibilities $E_7^{(1)}\to E_6^{(2)}$, 
$D_{n+3}^{(1)}\to D_{n+1}^{(2)}$ (two foldings), 
$D_4^{(1)}\to A_2^{(2)}$ (quadruple folding) and 
$E_6^{(1)}\to D_4^{(3)}$ (triple folding); the resulting diagrams are
shown in figure~\ref{fig:twist}.
Our notation is the one used, for example, in \cite{Goddard:1986}.
Two further twisted algebras $A_{2r-1}^{(2)}$ and
$A_{2r}^{(2)}$ that come from foldings of $D$ diagrams that leave only the
central point invariant cannot occur in the context of tops.

\subsection[Higher Dimensional Geometries]{Higher Dimensional Geometries}

In applications to string theory we are often interested in higher dimensional 
geometries such that locally there is a product structure involving a 
neighbourhood of a degeneration of an elliptic fibration.
For example, the total space may be a higher dimensional elliptically fibered 
\CY\ space or a $K3$ bundle over ${\bf S}^1$.
Then it may happen that there is a closed loop such that over every 
neighbourhood in the loop we have one of Kodaira's degenerations, but upon 
going around the loop the exceptional curves get permuted.
Using the fact that the permuted curves intersect if and only if the original 
curves intersect and otherwise the same arguments as before (in particular,
matching of multiplicities), we see that a folding of an affine ADE diagram 
can be assigned in any such case, independently of whether we have a 
description in terms of a top.

\subsection[Elliptic Curve]{Elliptic Curve}

Before proceeding to the classification of tops, let us also discuss the toric
interpretation of the vertices of a top $\top$ in the plane $\bar z = 0$.
These are just the vertices of the polygon $F_0$.
In terms of the geometry of the elliptic curve determined by $F_0$ every 
such vertex $v$ gives rise to $l$ divisors (i.e., points) in the elliptic 
curve where $l$ is the length of the edge of $F_0^*$ dual to $v$.
In the context of $\top$ there are $l$ sections of the fibration for generic 
values of the coefficients.
If $\top$ is part of a three or higher dimensional reflexive polytope, 
$v$ determines a divisor $D$ in the corresponding \CY\ hypersurface 
that may be reducible or irreducible.
In the latter case this divisor projects to an $l$-fold cover of the base 
space of the fibration.
In the case of a three dimensional reflexive polytope, $D$ consists of $l$
curves in the corresponding K3 if $v$ is interior to an edge and is 
irreducible if $v$ is a vertex of the three dimensional polytope.

\section[Classification]{Classification}\label{classification}

\begin{figure}[!tbp]
\begin{center}
\includegraphics[width=12cm]{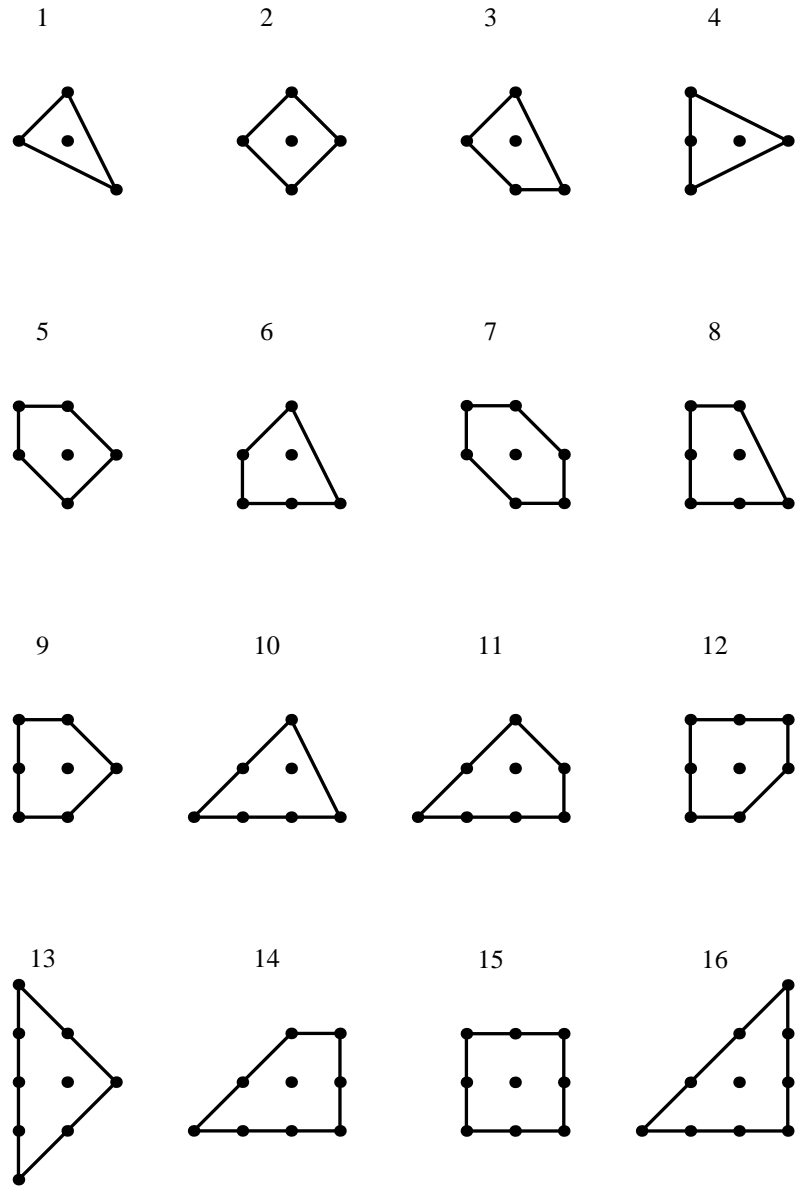}
\end{center}
\caption{The 16 two-dimensional reflexive polygons. 
The polygons $1,2,\ldots,6$ are respectively dual to the polygons
$16,15,\ldots,11$, and the polygons $7,\ldots,10$ are self-dual.}
\label{fig:16pol}
\end{figure}

The classification of reflexive polygons is well known \cite{Batyrev:1985,Koelman:1990}.
As we make extensive use of it, the resulting 16 polygons are shown in
figure \ref{fig:16pol}.
More recently, a general algorithm for classifying reflexive polytopes
was developed \cite{Kreuzer:1995cd,Kreuzer:2000qv,Skarke:1996hq} and successfully applied to the 
three \cite{Kreuzer:1998vb} and four dimensional \cite{Kreuzer:2000xy} cases.

The main idea of the algorithm of \cite{Kreuzer:1995cd,Kreuzer:2000qv,Skarke:1996hq} is to look for a set of
maximal polytopes that contain all others; these are dual
to minimal polytopes that do not contain any other reflexive
polytopes.
In the present case of classifying tops, we lose the symmetry between
the objects we are trying to classify and their duals.
Given that the duals are infinite, we will obviously look for maximal
objects among the duals and minimal objects among the tops themselves.
As every dual $\top^*$ of a top must be contained in a prism over one
of the 16 polygons of figure~\ref{fig:16pol}, it is natural to treat
these prisms as the maximal polyhedra.

We have already drastically reduced the $GL(3,\IZ)$ group of
isomorphisms between tops by demanding that $F_0$ lie in the plane
$\bar z =0$.
A further reduction comes from making a specific choice of coordinates
for $F_0$.
The remaining freedom is in the subgroup $G$ of $GL(3,\IZ)$ that fixes 
$F_0$ (not necessarily pointwise).
The elements of $G$ that fix every point of $F_0$ form a normal
subgroup $G_0\simeq \IZ^2$ of $G$; elements of $G_0$ act via
\beq 
(\bar x,\, \bar y,\, \bar z)\to 
(\bar x + a \bar z,\, \bar y +b \bar z,\, \bar z),
\quad (x,\,y,\,z)\to (x,\,y,\,z-ax-by) \quad 
\hbox{with } a,b\in\IZ.
\eeql{actg0}
The quotient $G/G_0$ can be identified with the subgroup of
$GL(2,\IZ)$ that fixes $F_0$; as it must take vertices to vertices and
keep the order (up to reversion), $G/G_0$ must be a subgroup of the
dihedral group of order $2n$ of the $n$-gon $F_0$.
The freedom in $G$ can then be eliminated by using (\ref{actg0}) to fix
$z_{\rm min}$ for two lattice points at the boundary of $F_0^*$ and
dealing with the remaining $G/G_0$ freedom by direct inspection.

The boundary point $b_0:=(0,0,z_0)\in \top^*$ below \ipo\ (with
$z_0:=\zm(0,0)$) is invariant under the transformation (\ref{actg0}).
It must belong to one or more facets of the type (\ref{nonvert}), so 
$1/z_0$ must be a negative integer.
Conversely, for every point $p=(\bar x, \bar y, \bar z)\in \top$,
$\bar z$ is invariant under (\ref{actg0}), and $\<p,b_0\>\ge -1$
implies $\bar z \le -1/z_0$. 
The vertices of $\top$ at $\bar z =-1/z_0$ are dual to the facets of
$\top^*$ that contain $b_0$.

\begin{lemma}\label{lemmab0}
 If $b_0=(0,0,-1)$, then every non-vertical facet of $\top^*$ contains $b_0$.
\end{lemma}

$b_0=(0,0,-1)$ implies that $\top$ is bounded by 
$\bar z \le 1$, so any vertex of $\top$ must have either $\bar z =0$,
corresponding to a vertical facet of $\top^*$, or $\bar z=1$,
corresponding to a facet that contains $b_0=(0,0,-1)$.

We are now in a position to enumerate the cases relevant to the
classification. 

\ni
{\bf Case 0:} $\top^*$ has a single non-vertical facet dual to a
vertex of $\top$ at $\bar z = 1$.\\[2mm]
We can use (\ref{actg0}) to have this vertex at $(0,0,1)$ and thus
$\zm=-1$ everywhere.
This trivial case exists for every choice of $F_0$.

\ni
{\bf Case 1:} $b_0$ is a vertex of $\top^*$.\\[2mm]
This implies $b_0=(0,0,-1)$, dual to a facet $F_1$ of $\top$
corresponding to $\bar z \le 1$.
According to lemma \ref{lemmab0}, the structure of $\top^*$ is determined by a
partition of $F_0^*$ in the style of cutting a cake.
Equations of non-vertical facets take the form (\ref{nonvert}) with
$\bar z=1$, implying that $\zm(x,y)$ is integer whenever $x$ and $y$
are integer.
We have seen an example in figure~\ref{fig:a3}; this example should
also serve as a useful background for the following discussion.

Consider three consecutive lattice points $p_{i-1}, p_i, p_{i+1}$
along the boundary of $F_0^*$.
It is easily checked that they fulfill $p_{i-1}+p_{i+1}=(2-l_i)\,p_i$ where
$l_i$ is the length (in lattice units) of the edge of $F_0$ dual to $p_i$
(with $l_i=0$ if $p_i$ is not a vertex).
The facets of $\top^*$ whose projections contain the triangles
$b_0\,p_{i-1}\,p_i$ and $b_0\,p_i\,p_{i+1}$, respectively, are dual to
vertices $v_{i-1}, v_i\in F_1$, with $v_{i-1}=v_i$ if 
$p_{i-1}, p_i, p_{i+1}$ belong to the projection of a single
non-vertical facet. 
One can calculate that $v_{i-1},v_i$ have lattice distance 
\beq 
\zm(p_{i-1})+(l_i-2)\zm(p_i)+\zm(p_{i+1})+l_i; 
\eeql{latdist}
non-negativity of this expression is just the local convexity condition.
The circumference in lattice units of the polygon $F_1$ is the sum 
$\sum_i l_i (\zm(p_i)+1)$ of these expressions.

All possible cases can be enumerated by choosing an integer $\zm$ for
every lattice point at the boundary of $F_0^*$, subject to consistency
with convexity, i.e. non-negativity of (\ref{latdist}) at each lattice
point of $F_0^*$;
to ensure that $b_0$ is a vertex, we also need that at least three of
these expressions are positive.
The freedom in (\ref{actg0}) can be eliminated, for example, by
putting two adjacent boundary points at $z=-1$.

\begin{figure}[h]
\begin{center}
\includegraphics[width=12cm]{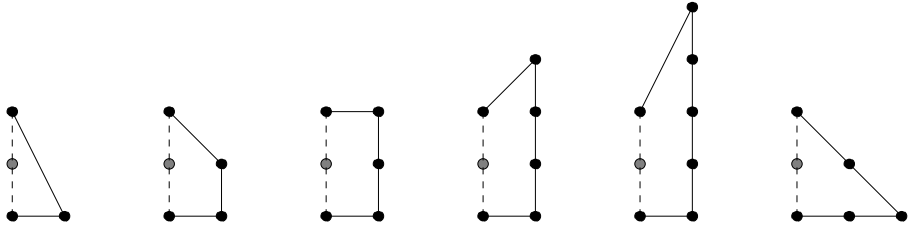}
\caption{Possible halves of reflexive polygons.}
\label{fig:dcen}
\end{center}
\end{figure}

\ni
{\bf Case 2:} $b_0$ lies on a line connecting two lattice points $p_1,p_2$
of $\top^*$.\\[2mm]
Then $b_0=(p_1+p_2)/2$, so $2z_0$ must be integer, i.e. 
$z_0\in\{-1,-1/2\}$.
The projection of $\5{p_1p_2}$ divides $F_0^*$ into two halves.
By inspection of figure~\ref{fig:16pol} we see that either half must
look, up to automorphisms of the two dimensional lattice, like one of
the possibilities shown in figure~\ref{fig:dcen} 
(without loss of generality, we assume that $\5{p_1p_2}$ is at $x=0$).

\ni
{\bf a)} $z_0=-1$: Because of lemma \ref{lemmab0} there are no more than two
non-vertical facets corresponding to the two halves with $x\le 0$ and
$x\ge0$. 
We can use (\ref{actg0}) to put either of them, but not both, at $z=-1$.
If we choose, say, $z=-1$ for the facet at $x\ge0$, then the facet at 
$x\le 0$ must be the transform of a facet at $z=-1$ by (\ref{actg0})
with $b=0$; convexity implies $a\ge 0$.
This gives a one parameter family that starts with the case of a
single facet at $z=-1$ (case 0).
For $a\ge 1$, $\top$ has precisely two vertices at $\bar z=1$ whose
distance is $a$. 

\ni
{\bf b)} $z_0=-1/2$: There are one or two vertices of $\top$ at
$\bar z=2$, dual to the facet(s) containing $b_0$.
All other vertices of $\top$ must be at $\bar z=0$ or $1$.
We can again treat the halves separately and find that after using up
the freedom given by (\ref{actg0}) there are only finitely many cases
for each diagram of figure~\ref{fig:dcen}, corresponding to partitions
such that the facet containing $b_0$ is dual to a vertex at $\bar z=2$
and all other facets correspond to $\bar z=1$.
Here the freedom of applying (\ref{actg0}) to one of the halves leads
to families such that the edge at $\bar z=2$ has length $2a$ or $2a+1$.

\begin{figure}[h]
\begin{center}
\includegraphics[width=14cm]{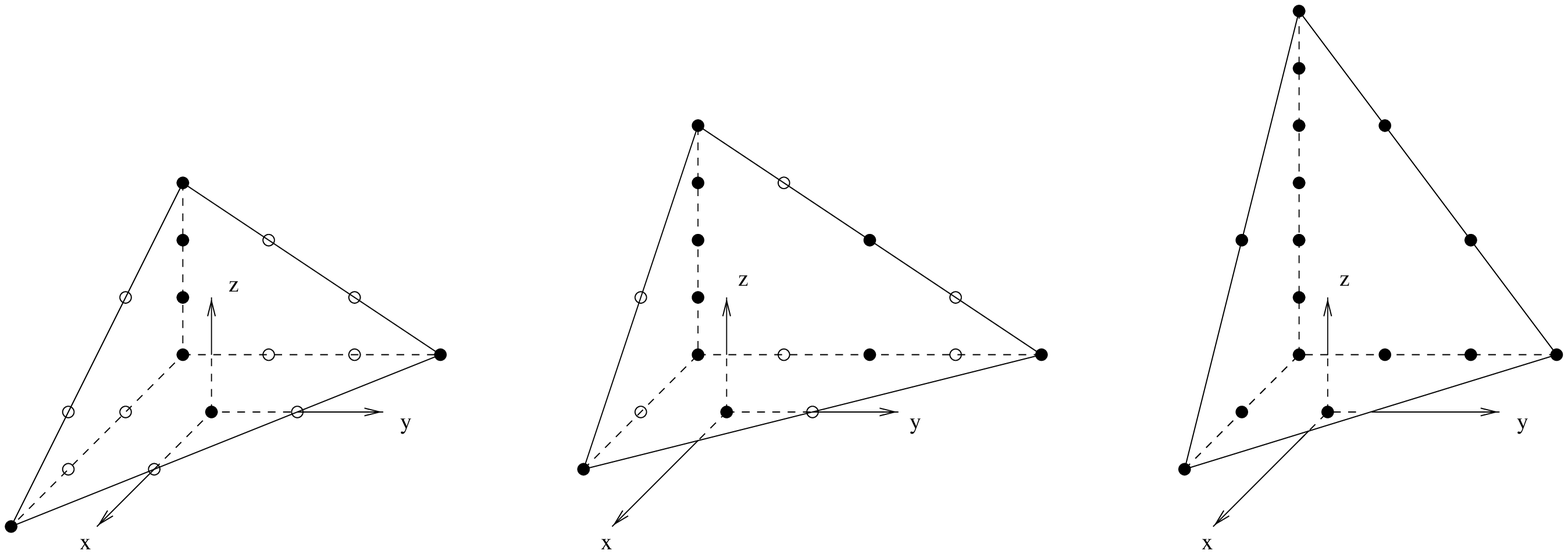}
\caption{Three maximal tops (the significance of $\bullet$ vs. $\circ$
will be explained in section \ref{bullet}).}\label{fig:e6e7e8}
\end{center}
\end{figure}

\ni
{\bf Case 3:} Neither of the above.\\[2mm]
Then $b_0$ must be interior to a facet of $\top^*$ in such a way that
cases 0 and 2 do not apply.
In particular, after triangulating this facet $b_0$ must be interior to
one of the triangles $v_1\,v_2\,v_3$.
Applying $\p$, we see that $(0,0)$ is interior to $\D\subseteq F_0^*$
where $\D$ is the triangle with vertices $\p(v_i)$.
Every two dimensional lattice polytope whose only interior lattice
point is the origin is reflexive, so $\D$ must be one of the
triangles occurring in figure~\ref{fig:16pol} (numbers 1, 4, 10, 13, 16).
The linear relations among the vertices of these triangles imply
\bea v_1+v_2+v_3=3b_0&& \hbox{ for the triangles 1, 16},\\
   v_1+v_2+2v_3=4b_0&& \hbox{ for the triangles 4, 13},\\
   v_1+2v_2+3v_3=6b_0&&\hbox{ for triangle 10},
\eea
so $-1/z_0$ must divide one of the numbers $3,4,6$.
We can dismiss the following possibilities.\\
$z_0=-1$ implies case 0 by the lemma,\\
$z_0=-1/2$ is possible for triangles 4, 10, 13, but easily
seen to correspond to case 2,\\
$z_0=-1/3$ for triangles 10 or 16 can be reduced to triangle 1,\\
$z_0=-1/4$ for triangle 13 can be reduced to triangle 4.\\[2mm]
This leaves us with\\
{\bf a)} $z_0=-1/3$ for triangle 1,\\
{\bf b)} $z_0=-1/4$ for triangle 4,\\
{\bf c)} $z_0=-1/6$ for triangle 10.\\[2mm]
In each of these cases $\p(v_2)$ and $\p(v_3)$
generate the two dimensional lattice, so we can use (\ref{actg0}) to
put $v_2$ and $v_3$ at $z=0$ which forces $v_1$ to be at $z=-1$.
Let us denote by $P$ the prism over $\D$ cut off at the
$v_1v_2v_3$--plane. 
Then $P\subseteq \top^*$ implies $\top\subseteq P^*$, but $P^*$ is a
top with a finite number of lattice points. 
In other words, every top containing points at $\bar z >2$ is
contained in one of the three tops shown in figure~\ref{fig:e6e7e8}.
This implies that once a choice of $F_0^*$ as one of the 16 polygons
and (if possible) of $\D\subseteq F_0^*$ has been made, there is
only a finite number of consistent possibilities of assigning $\zm$ to
the remaining lattice points in $F_0^*$. 

The classification of all possible tops is then straightforward.
All that has to be done is to examine each of the 16 polygons of
figure~\ref{fig:16pol} with respect to all possibilities of applying
one of the cases 0, 1, 2, 3, taking care to avoid overcounting
wherever there are non-trivial automorphisms of $F_0^*$.
A complete list is given in appendix \ref{class}.

\section{Geometrical Interpretation}\label{bullet}

The results of the classification of course confirm the predictions made 
in section \ref{kac} on the structure of the edge diagram of a top at $\bar z \ge 1$.
By combining the arguments of the previous sections it is clear that
case 1 of our classification corresponds to $A_n^{(1)}$ algebras,
with 
\beq n+1=\sum_i l_i (\zm(p_i)+1) \eeq 
in the notation used around (\ref{latdist}).
Case 2a leads to $C_n^{(1)}$ diagrams where $n$ is just the parameter $a$ 
used there, 
and case 2b to $D_n^{(1)}$ and its folded versions where $n-4$ is 
the length of the edge at $\bar z = 2$, i.e. $2a$ or $2a+1$.
Cases 3a,b,c correspond to $E_6^{(1)}$, $E_7^{(1)}$, $E_8^{(1)}$
and their folded versions, respectively.

\subsection{Twisted Algebras: Geometry}

There is, however, a great difference between the occurrences of
untwisted and twisted Kac-Moody algebras as edge diagrams.
While each of the untwisted algebras occurs quite a number of times
and five of the reflexive polygons ($F_0^*$ one of 10, 11, 13, 14, 16
of figure~\ref{fig:16pol}) feature every possible untwisted algebra,
twisted algebras are quite rare.
Each of the diagrams of $E_6^{(2)}$, $A_2^{(2)}$ and $D_4^{(3)}$
occurs only once, and the members of the $D^{(2)}$ series occur twice.

There are only three pairs $(F_0,F_0^*)$ leading to diagrams of
twisted algebras, namely $(1,16)$, $(2,15)$ and $(4,13)$.
These reflexive pairs of polygons are quite special in several ways.
They are the only dual pairs where $F_0^*$ has no edge of length one;
this implies that no vertex of $F_0$ corresponds to a single section.
Moreover each of theses polygons is reflexive on two distinct
lattices, in such a way that the dual is the same polygon on the other
lattice. 
More precisely, polygon 1 is the same as polygon 16 on a sublattice of
index 3, and polygons 2 and 4 are the same as their respective duals on
sublattices of index 2.
We find that whenever twisted algebras occur, the corresponding tops
can be understood as coming from a restriction to a sublattice,
suitably extended in the third dimension.

Consider again the first top in figure~\ref{fig:e6e7e8}.
On the full lattice (with points both of the type $\circ$ and
$\bullet$) the edge diagram is an $E_6^{(1)}$ Dynkin diagram, but if
we consider the sublattice of index $3$ determined only by points shown
as $\bullet$, we find the diagram of $D_4^{(3)}$.
This is the only twisted algebra coming from the pair $(1,16)$.

\begin{figure}[h]
\begin{center}
\includegraphics[width=14cm]{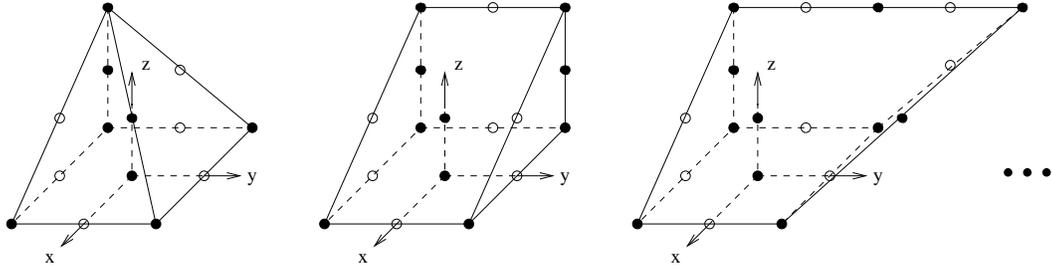}
\caption{A family of tops over squares.}
\label{fig:square}
\end{center}
\end{figure}

The pair $(2,15)$ of reflexive squares gives rise to a family of
$D^{(2)}$ algebras as shown in figure~\ref{fig:square}.
Note, however, that the twisted diagrams $D_{i+3}^{(2)}$ with
$i=0,1,2,\ldots$ come from sublattice versions of tops corresponding
to $D_{2i+4}^{(1)}$, i.e. not the diagrams whose foldings produce the
twisted ones.

The situation is most intricate for the pair $(4,13)$ of reflexive
triangles.
In the second picture of figure~\ref{fig:e6e7e8}, passing
to the index $2$ sublattice indicated by $\bullet$ means that we get an
$E_6^{(2)}$ diagram from an $E_7^{(1)}$ diagram.
\begin{figure}[h]
\begin{center}
\includegraphics[width=14cm]{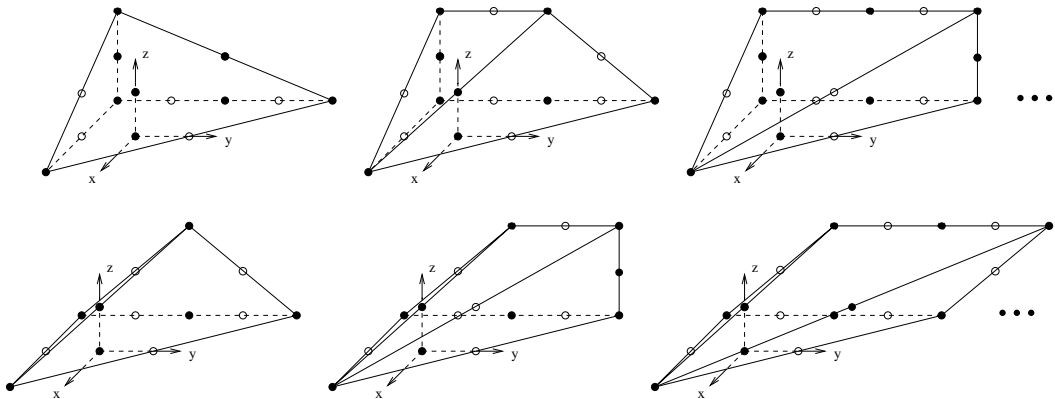}
\caption{Two families of tops over the dual pair $(4,13)$.}
\label{fig:p112}
\end{center}
\end{figure}
Now consider the tops shown in figure~\ref{fig:p112}.
In the finer lattice including the points $\circ$, each member of the
first family shown in the upper row is isomorphic to the member of the
second family directly below; 
the transformation is (\ref{actg0}) with $a=0$, $b=1$, resulting in
tilting the top along the $y$ direction.
The corresponding algebras are $B_3^{(1)}$ and $D_{2i+4}^{(1)}$ with
$i\in\{1,2,\ldots\}$. 
In the coarser $\bullet$ lattice, however, this isomorphism is lost
as it would correspond to half integer parameters in (\ref{actg0}).
The first row gives a family of twisted diagrams $D_{i+3}^{(2)}$ with
$i\in\{0,1,2,\ldots\}$, as in the case of the square.
The second row gives another $D_{i+3}^{(2)}$ family for $i\ge 1$ with
a special case for $i=0$ where we have an $A_2^{(2)}$ diagram.
As an additional subtlety, the tops for odd $i$ are nevertheless
isomorphic, namely by an isomorphism that acts on $F_0$ by
swapping the two vertices of the long edge.
For this reason we have listed these tops in table~\ref{tab:results13}
in the appendix as two families of type $D_{2i+3}^{(2)}$ and one
family of type $D_{2i+4}^{(2)}$.

The fact that the tops giving rise to twisted algebras all correspond 
to pairs of lattices has the following interpretation.
Passing from a coarser lattice $N$ to a finer lattice $N'$ in a way that 
is compatible with the structure of the fan means passing from a toric 
variety to a quotient of this variety by a finite abelian group isomorphic 
to $N'/N$.
In the present cases this group is just $\IZ_3$ or $\IZ_2$.
So the varieties corresponding to diagrams related to twisted algebras
allow group actions in such a way that taking the quotient would 
take us to the variety defined by the finer lattice.

\section{String Theory Interpretation}

Finally let us discuss the relation between tops and string theory.
The simplest case is that of taking an elliptic K3 surface whose toric 
polytope consists of two tops.
Blowing down all divisors except two at $\bar z = \pm 1$ will result in
the occurrence of two ADE singularities (at $W=0/\infty$) corresponding 
to the unfolded diagram.
Compactifying M-theory on such a space leads to a theory where the 
generic abelian gauge group is enhanced in such a way that the corresponding 
ADE groups arise.
The same groups also arise in compactifications of type IIA and F-theory 
without background fields.

If a top is part of a diagram describing an elliptically fibered Calabi-Yau
threefold or fourfold, the generic local geometry is a product of a 
neighbourhood $U$ in $\IC$ or $\IC^2$ with the two dimensional geometry 
featuring the ADE pattern;
every exceptional rational curve gives rise to a divisor $U\times \IC \IP^1$.
Globally several of these divisors may correspond to a single irreducible 
divisor (this is the non-split case of \cite{Bershadsky:1996nh}).
Clearly this can happen only if the different $\IC \IP^1$'s all come from the 
same lattice point in the toric diagram.
In this case there are special loci in the base space over which some or all 
of the $\IC \IP^1$'s coincide. 
Monodromy around these loci will interchange the $\IC \IP^1$'s.
Upon compactifications of the same theories as above this results in non 
simply laced gauge groups \cite{Aspinwall:1996nk}, again of the type determined by the toric diagram.

These constructions are conjectured to be dual to compactifications 
of heterotic strings;
in particular the K3 compactifications are dual to toroidal heterotic 
compactifications with maximal rank of the gauge group.
However, the heterotic moduli space also contains components of reduced 
rank of the gauge theory \cite{Chaudhuri:1995fk,Chaudhuri:1995bf}.
These require non-standard IIA and F-theory compactifications as duals.
In the IIA case these are compactifications on orbifolds in the presence
of a non-trivial RR background \cite{Schwarz:1995bj}; 
this has an M-theory lift where the orbifolding also acts by a shift on the 
$S^1$ in the eleventh dimension.
In terms of F-theory a non-generic monodromy group together with non-trivial
$B_{\m\n}$ flux in the underlying IIB theory is required \cite{Bershadsky:1998vn};
F-theory duals were also considered in \cite{Park:1996it,Berglund:1998va}.
We will now see how these requirements are met by the tops that give rise 
to twisted Kac-Moody algebras; that these algebras should play a role
on the heterotic side was predicted in \cite{Lerche:1997rr}.

\subsection{Twisted Algebras: String Theory}

Consider once more the second diagram of figure~\ref{fig:e6e7e8}.
To be specific let us assume that it is part of the reflexive polytope that 
is obtained by adding the `bottom' that is the reflection of the top through 
the $xy$-plane (this reflexive polytope was also considered in \cite{Berglund:1998va},
but our discussion will be different).
Passing from the space determined by the $\bullet$ diagram to its $\IZ_2$
orbifold means that we get singularities that may be resolved by blowing
up along the divisors indicated by $\circ$.
In a standard IIA compactification passing to the orbifold means that one 
loses as many non-algebraic cycles as there are $\circ$ cycles and the rank
of the gauge group remains 24.
In the compactification with RR background the $\circ$ cycles do not 
contribute and we end up with reduced rank.
By counting with the right multiplicities (one for every $\circ$ point 
in an edge except $(0,1,0)$ which has multiplicity two) we get a rank 
reduction of eight as it should be \cite{Chaudhuri:1995fk,Chaudhuri:1995dj}.
The same reflexive polytope allows for a second fibration structure
with the fiber determined by the polygon at $x+y=0$.
Now the orbifolding acts by changing the trivial top at $x+y>0$ to the $B_3^{(1)}$ top and the $C_6^{(1)}$ top at $x+y<0$ to the $D_{10}^{(1)}$ top.
This is the toric description of the involution discussed in section 2 of 
\cite{Bershadsky:1998vn}.

In a similar manner the diagrams of figures \ref{fig:square} and 
\ref{fig:p112} can be used to construct theories with rank reduction of 
eight.
Taking the first top in figure~\ref{fig:e6e7e8} together with its mirror 
image, we get a K3 with orbifold group $\IZ_3$. 
According to \cite{Chaudhuri:1995dj} this should lead to a rank reduction of 12;
with the same counting as before this is indeed confirmed.

The gauge groups that we get are again non simply laced with a mechanism 
very similar to the one we encountered before.
In the M-theory picture we compactify on $(K3\times S^1)/G$ which is a smooth 
non-trivial bundle over $S^1$ with fiber the original $K3$.
So locally over a neighborhood in $S^1$ we get all the $\IC \IP^1$'s of the 
untwisted diagram but upon going around the $S^1$ they are permuted.
For obtaining gauge groups we blow down the toric divisors corresponding
to all points of a top except for one at $\bar z =1$.
The collapsed cycles belong to ordinary ADE Dynkin diagrams (the heights
play no role here) that are folded by the permutations.
In this way $E_6^{(2)}$, $D_{n+1}^{(2)}$, $A_2^{(2)}$ and $D_4^{(3)}$
give rise to the groups $F_4$, $B_n$, $A_1$ and $G_2$, respectively.

Tops from the families of figures \ref{fig:square} and \ref{fig:p112} that do 
not fit into
three dimensional reflexive polytopes should play a role in theories dual to 
non-toroidal compactifications of CHL strings \cite{Kachru:1997bz}.

\chapter[Closed Topological Strings on Orientifolds]{Closed Topological Strings on Orientifolds}\label{closed}

In this chapter we study closed topological strings on orientifolds of toric \CY\ threefolds. We compute all loop topological string amplitudes, using geometric transitions involving $SO$/$Sp$ Chern-Simons theory, localization on the moduli space of holomorphic maps with involution, and the topological vertex.

We start by summarizing general results for {\bf A}-model topological strings on an orientifold. Then we describe in some details the geometric transition on which we will focus in this chapter. The geometry was presented in section \ref{examples}. In section \ref{ampl} we compute explicitely the Chern-Simons amplitude obtained after the geometric transition. Then we present the unoriented localization computation, and show that it gives exactly the same contributions for the one and two crosscaps instanton configurations. We then propose our prescription based on the topological vertex in section \ref{topover}, proving its equivalence to the Chern-Simons computation.

\section{{\bf A}-model Topological Strings on an Orientifold}\label{amodel}

\subsection{Type IIA Superstrings and Topological Strings on an Orientifold}

It is a well known fact that, when type IIA theory is compactified on a
\CY\ manifold $X$, the resulting
four dimensional theory is ${\cal N}=2$ supergravity with $h^{1,1}(X)$ vector multiplets
$t_i$.
The ${\cal N}=2$ prepotential that governs the effective action of the vector multiplets,
$F_0(t_i)$, can be
computed by the genus zero free energy of the
{\bf A}-model topological strings with the \CY\ as target space (see
\cite{Hori:2003} for a review of topological strings and related issues). Higher genus
free energies $F_g(t_i)$ of the topological string theory also play a r\^ole in the
four dimensional supergravity theory, and compute higher curvature
couplings involving the graviphoton \cite{Bershadsky:1993cx, Antoniadis:1993ze}.

One way to break ${\cal N}=2$ supersymmetry down to ${\cal N}=1$ is
to consider an orientifold of the theory.
The orientifold is defined by combining an involution symmetry $I$ on
the \CY\ $X$ with a diffeomorphism $\sigma$ on the worldsheet $\Sigma$.
In the context of type IIA superstrings, the orientifold is only well defined if
the involution is anti-holomorphic.
Furthermore, the worldsheet diffeomorphism has to be orientation reversal
\cite{Acharya:2002ag,Diaconescu:2003dq,Brunner:2003zm}. The resulting theory in four dimensions
has ${\cal N}=1$ supersymmetry, and $h_-^{1,1}(X)$ out of the $h^{1,1}(X)$ ${\cal N}=2$
vector multiplets
become ${\cal N}=1$ chiral multiplets in four dimensions,
where $h_-^{1,1}(X)$ is the number of
harmonic $(1,1)$ forms on $X$ which have $-1$ eigenvalue under $I$ (see \cite{Brunner:2003zm} for
a description of the spectrum of massless modes in four dimensions).

These considerations hold in the context of {\bf A}-model topological strings as well:
{\bf A}-model topological strings possess a worldsheet orientation reversal
symmetry when accompanied with an
anti-holomorphic involution of the target space \cite{Acharya:2002ag}.
It is thus possible to consider {\bf A}-model topological strings on an orientifold
defined as above. The twisted sector of the topological string amplitude on the
orientifold includes amplitudes for unoriented Riemann surfaces\footnote{The attentive reader may have noticed that this nomenclature is slightly strange, as by definition a Riemann surface must be orientable. But as is conventional in the string theory literature, an {\it unoriented (or non-orientable) Riemann surface} means a non-orientable surface which results from the action of an orientation reversal diffeomorphism on a Riemann surface.}. Recall that
a closed, non-orientable
Riemann surface is characterized by its genus $g$ and by the
number of crosscaps $c$, which can be one or two (crosscaps can be
traded for handles when the number of crosscaps is higher than two).
For example, the surface with $g=0$ and $c=1$ is the real projective plane
$\IR \IP^2$, while the surface with $g=0$ and $c=2$ is the Klein bottle.
It was shown in \cite{Acharya:2002ag} that the superpotential of the $h_-^{1,1}(X)$
chiral multiplets is given by the $\IR\IP^2$ amplitude of the topological
string theory. As far as we know, the topological amplitudes
involving more handles or crosscaps do not have an interpretation in the ${\cal N}=1$
supergravity theory.

Generally speaking, one could consider type IIA superstrings on a noncompact
orientifold, with D-branes and
orientifold planes \cite{Acharya:2002ag}. In this chapter we only consider type IIA superstrings
without D-branes or orientifold planes. This means
that the anti-holomorphic involution must have no fixed points. Moreover,
as the parent theory has no D-branes,
to compute the superpotential we only need to consider {\bf A}-model
closed topological strings. Open topological string amplitudes corresponding to orientifolds with D-branes shall be considered in chapter \ref{open}.

\subsection{Structure of the Topological String Amplitudes}\label{structopclosed}

Roughly speaking, the free energy of {\bf A}-model closed topological strings
counts the number of
holomorphic maps from the worldsheet to the target space, weighted by a
factor of $e^{-A}$ where $A$ is the area of the embedded curve.
In the context of orientifolds, the partition function of topological strings
sums over holomorphic maps in two different sectors:
the ``untwisted'' and the ``twisted'' sectors. The former consists of
usual holomorphic maps from orientable worldsheets to the covering space,
i.e. the noncompact \CY\ threefold without the involution. The latter
consists of equivariant maps
$f: \Sigma \rightarrow X$ satisfying the equivariance
condition
\beq
f\circ \sigma = I \circ f,
\eeql{equivar}
where $I$ is the anti-holomorphic involution acting on $X$, and
$\sigma: \Sigma \rightarrow \Sigma$ is the orientation
reversal diffeomorphism of the
Riemann surface which is needed in order to construct the orientifold
action. Notice that, if $\Sigma$ has genus zero, the action of
$\sigma$ is given by $z \rightarrow -1/{\bar z}$. The relevant maps in the
twisted sector are then the maps which
are compatible with the orientation reversal diffeomorphism on the worldsheet
and the anti-holomorphic involution on the target space, and
descend to holomorphic maps from non-orientable worldsheets to the orientifold.

The structure of the total free energy of the {\bf A}-model is then
\beq
\CF(X/I, g_s)= \CF(X/I, g_s)_{\rm untwisted} + \CF(X/I, g_s)_{\rm unor},
\eeql{strucfclosed}
where $g_s$ is the string coupling constant. In this equation,
$\CF(X/I, g_s)_{\rm untwisted}$ is the contribution of the untwisted sector, and
$\CF( X/I, g_s)_{\rm unor}$ is the contribution of the twisted sector.

One of the most important results of topological string theory is the fact that 
topological string amplitudes have an integrality, or BPS structure, which expresses 
them in terms of numbers of BPS states.

In the case of closed topological strings on Calabi-Yau threefolds, the BPS structure 
was obtained by Gopakumar and Vafa in \cite{Gopakumar:1998jq}. Let us denote by $F_g(t)$ the topological string 
free energy at genus $g$, where $t$ denotes the set of K\"ahler parameters of the 
Calabi-Yau threefold $X$, 
and let 
\beq
\CF(t,g_s)=\sum_{g=0}^{\infty} g_s^{2g-2}F_g(t) 
\eeql{totalfree}
be the total free energy. Then, one has the following structure result:
\beq
\CF(t,g_s)=\sum_{d=1}^{\infty}\sum_{g=0}^{\infty}\sum_{\beta} 
{1\over d} { n^g_Q \over (q^{d\over 2} - q^{-{d\over 2}})^{2-2g}}
e^{-d Q \cdot t}.
\eeql{gv}
where $q=e^{i g_s}$, the sum over $Q$ is over two-homology classes in $X$, and $n^g_Q$ (the so-called Gopakumar-Vafa 
invariants) are {\it integers}. The factor $(q^{d\over 2} - q^{-{d\over 2}})^{2g}$ comes from computing 
a signed trace over the space of differential forms on a Riemann surface of genus $g$, while the factor 
$(q^{d\over 2} - q^{-{d\over 2}})^{-2}$ comes from a Schwinger computation \cite{Gopakumar:1998jq}. 

We thus find that the untwisted sector has the following structure:
\beq
  \CF(X/I, g_s)_{\rm untwisted}={1\over 2}\CF(X, g_s)={1\over 2}\sum_{d=1}^{\infty}\sum_{g=0}^{\infty}\sum_{Q} 
{1\over d} { n^g_Q \over (q^{d\over 2} - q^{-{d\over 2}})^{2-2g}}
e^{-d Q\cdot t}.
\eeql{orcont}
Here, $\CF(X, g_s)$ is the free energy of the covering $X$ of $X/I$,
after suitably identifying
the K\"ahler classes in the way prescribed by the involution $I$, and
we have written it in terms of Gopakumar-Vafa invariants $n^g_Q $ \cite{Gopakumar:1998jq}.

The unoriented contribution in \refeq{strucfclosed} comes from holomorphic maps from
closed non-orientable Riemann surfaces to the orientifold $X/I$.
The Euler characteristic of a closed Riemann surface of genus $g$ and $c$ crosscaps
is $\chi = -2g+2-c$ where $c$ is the number of crosscaps. We then have
\beq
\CF(X/I, g_s)_{\rm unor}= \CF(X/I, g_s)_{\rm unor}^{c=1}+ \CF(X/I, g_s)_{\rm unor}^{c=2},
\eeql{strucunclosed}
which corresponds to the contributions of one and two crosscaps. Following
the arguments in \cite{Gopakumar:1998jq} we expect the structure
\bea
\CF(X/I, g_s)_{\rm unor}^{c=1}&=&
\pm \sum_{d \, \, {\rm odd}}\sum_{g=0}^{\infty}\sum_{Q }  n^{g,c=1}_Q
{1\over d}(q^{d\over 2} - q^{-{d\over 2}})^{2g-1}e^{-d Q\cdot t},\nn\\
\CF(X/I, g_s)_{\rm unor}^{c=2}&=&
\sum_{d \, \, {\rm odd}} \sum_{g=0}^{\infty}\sum_{Q }
n^{g,c=2}_Q {1\over d}(q^{d\over 2} - q^{-{d\over 2}})^{2g}e^{-d Q\cdot t},
\eeal{strucun}
where $n^{g,c}_Q$ are integers. The $\pm$ sign in the $c=1$ free energy is due to
the following:
the target space anti-holomorphic involution does not fully specify the unoriented part
of the free energy
on the orientifold, since we have to make a choice for the sign of the crosscaps. Depending
on this choice, we will have
the two different signs for $c=1$. This corresponds to the choice of $SO$ or $Sp$ group in the gauge
theory dual. This remaining choice is also easily understood on the mirror symmetric
side \cite{Acharya:2002ag}. For
the conifold, the
{\bf B}-model mirror symmetric description involves two orientifold 5-planes. The two
choices of signs for crosscap states correspond
on the mirror symmetric side to the two following choices for the charges of the
$O5$-planes: $+-$ and $-+$ \cite{Acharya:2002ag}. A similar story holds for more complicated orientifolds. Notice as well that the sum
over multicoverings $d$ in \refeq{strucun} is only over {\it odd} integers. In the case of
$c=1$ this follows from an elementary geometric argument, since there are no even
multicoverings (see \cite{Sinha:2000ap,Acharya:2002ag}). For $c=2$ there is no such a simple argument, but
our explicit computations both in Chern-Simons theory and in localization of unoriented
instantons indicate that only odd multicoverings contribute.

\section{Geometric Transitions} \label{geometry}

\subsection{Orientifold of the Resolved Conifold and its Geometric Transition}

In \cite{Sinha:2000ap} it was proposed that in the large $N$ limit, closed topological strings on the
orientifold of the conifold are
dual to $SO(N)$/$Sp(N)$ Chern-Simons theory on ${\bf S}^3$, where the choice of gauge group is
related to the choice of sign for the crosscaps.
Since this is the starting point for our discussion, let us review in some
detail the results of \cite{Sinha:2000ap}.

We start with a theory of topological
open strings on the deformed conifold defined by $w_1 w_4 - w_2 w_3=\mu$. The conifold
contains an ${\bf S}^3$, and if we wrap $2N$ branes on the
three-sphere, the spacetime description of the
open topological string theory is Chern-Simons theory on ${\bf S}^3$
with gauge group $U(2N)$ and at level $k$ (the level is related to the open string
coupling constant). We
now consider the following involution of the geometry
\beq
I:(w_1, w_2, w_3, w_4) \rightarrow (\bar w_4, -\bar w_3, -\bar w_2, \bar w_1)
\eeql{stinvclosed}
that leaves the
${\bf S}^3$ invariant. The string field
theory for the resulting open strings is now
Chern-Simons theory with gauge group $SO(N)$ or $Sp(N)$, depending on the choice of
orientifold action on the gauge group. The total free energy of the
Chern-Simons theory with gauge group $SO/Sp$ can be written as
\beq
\CF=-\log S_{00}^{SO(N)/Sp(N)} = {1\over 2} \sum_{d=1}^{\infty} {1 \over d} {e^{-dt}\over (q^{d\over2}
-q^{-{d\over 2}})^2} \mp  \sum_{d \, {\rm odd}}
 {1 \over d} {e^{-dt/2}\over q^{d\over2} -q^{-{d\over 2}}},
\eeql{sopcs}
where the $\mp$ sign corresponds to $SO/Sp$, respectively. In \refeq{sopcs}, $q=e^{i g_s}$, with
\beq
g_s= { 2\pi \over k + y},
\eeql{stringcouclosed}
and $y$ is the dual Coxeter of the gauge group, which is $N-2$ for $SO(N)$ and
$N+1$ for $Sp(N)$.
The parameter $t$ in \refeq{sopcs} is the
't Hooft parameter, given by
\beq
t=(N\mp 1) g_s,
\eeql{thooft}
for $SO/Sp$, respectively.

\begin{figure}[htp]
\begin{center}
\psfrag{SO(N)/Sp(N)}{$SO(N)/Sp(N)$}
\psfrag{RP2}{$\IR \IP^2$}
\includegraphics[width=5.0in]{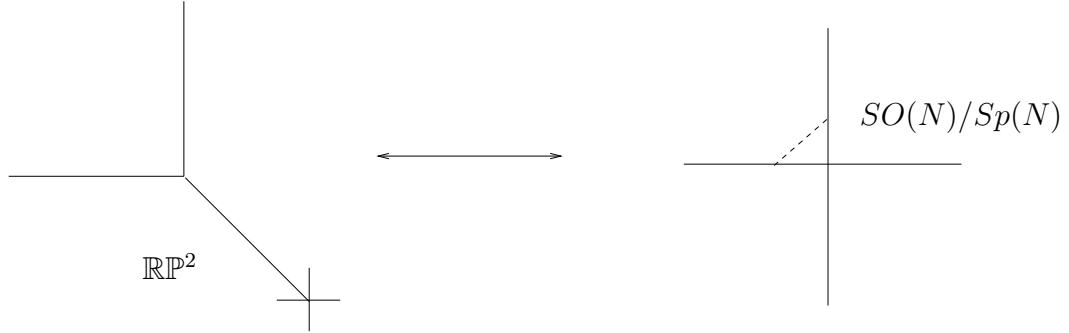}
\caption{Geometric transition for the orientifold of the conifold. The cross in the figure to the 
left represents an $\IR\IP^2$ obtained by quotienting a $\IC \IP^1$ by the involution $I$, and the dashed line in the figure on the right represents an ${\bf S}^3$ with $SO/Sp$ gauge group.}
\label{orienticon}
\end{center}
\end{figure}

In the usual geometric transition of \cite{Gopakumar:1998ki}, the dual to the deformed conifold is the
resolved conifold $Y={\cal O}(-1) \oplus {\cal O}(-1) \rightarrow {\IC \IP}^1$. This \CY\ threefold was described in detail in section \ref{examples}.

The involution \refeq{stinvclosed} of the deformed conifold maps to the anti-holomorphic
involution of $Y$ defined by:
\beq
I: (z_1, z_2, z_3, z_4) \rightarrow ({\overline z}_2, -{\overline z}_1,
{\overline z}_4, -{\overline z}_3).
\eeql{orres}
It is easy to see that $Y/I$ contains a single ${\IR\IP}^2$ obtained from the
quotient of the ${\IC \IP}^1$ of $Y$ by $I$. We will represent the quotient of
the resolved conifold by that involution in terms of the toric diagram depicted 
in figure \ref{orienticon}.

The free energy of the $SO/Sp$ Chern-Simons theory gives the total free
energy of closed strings propagating on $Y/I$.
The first term in \refeq{sopcs} gives the oriented contribution, while the
second term gives the unoriented contribution, and
they have the structure explained in \refeq{orcont} and \refeq{strucunclosed}. Notice that
in the case of the unoriented contribution we have
\beq
n_{Q=1/2}^{g=0,c=1}=\mp 1
\eeql{gvunor}
depending on the choice of sign for the crosscaps, and all the remaining
Gopakumar-Vafa invariants vanish. In particular, the contribution
of Riemann surfaces with two crosscaps is zero. As we will see, in more general 
cases there are two crosscaps contributions.
The above prediction of the large $N$ transition for the free energy was
checked in \cite{Acharya:2002ag} against mirror symmetry, and in \cite{Diaconescu:2003dq} against
localization computations for unoriented Gromov-Witten theory.

\subsection{Our Main Example}

In this paper we want to generalize the open/closed string duality studied in
\cite{Sinha:2000ap} to more general orientifolds. We will mainly focus on the noncompact
\CY\ manifold $X$
described in section \ref{examples}. 

 The compact locus consists of two divisors that are each isomorphic to a del Pezzo surface $dP_2$ and
a rational $(-1,-1)$ curve that intersects both divisors transversely. Note that the two compact divisors do
not intersect. We consider a real torus action on $X$ given by:
\beq
e^{i\phi}\cdot (z_1,z_2,\ldots,z_{10})\rightarrow
(e^{i\lambda_1\phi}z_1,e^{i\lambda_2\phi}z_2,\ldots,e^{i\lambda_{10}\phi}z_{10}).
\eeql{tact}
We now define the anti-holomorphic involution as follows:
\bea
&&I:(z_1,z_2,z_3,z_4,z_5,z_6,z_7,z_8,z_9,z_{10})\rightarrow\nn\\
&&~~~~~~~~~~~~~~~~~~~~~~(\bz_{10},\bz_8,\bz_9,\bz_7,-\bz_6,\bz_5,-\bz_4,\bz_2,\bz_3,\bz_1).
\eeal{antinv}
The subtorus of \refeq{tact} that is compatible with the involution is defined by the following constraints on the weights
\beq
\lambda_1+\lambda_{10}=0,~~\lambda_2+\lambda_8=0,~~\lambda_3+\lambda_9=0,~~\lambda_4+\lambda_7=0,~~
\lambda_5+\lambda_6=0.
\eeql{constr}
Imposing these constraints does not enlarge the set of invariant curves.

\begin{figure}[htp]
\begin{center}
\psfrag{dP}{$dP_2$}
\psfrag{P1}{$\IC \IP^1$}
\psfrag{P2}{$\IC \IP^2$}
\psfrag{RP}{$\IR \IP^2$}
\psfrag{Flops}{Flops}
\psfrag{Anti-holomorphic}{Anti-holomorphic involution}
\includegraphics[width=15cm]{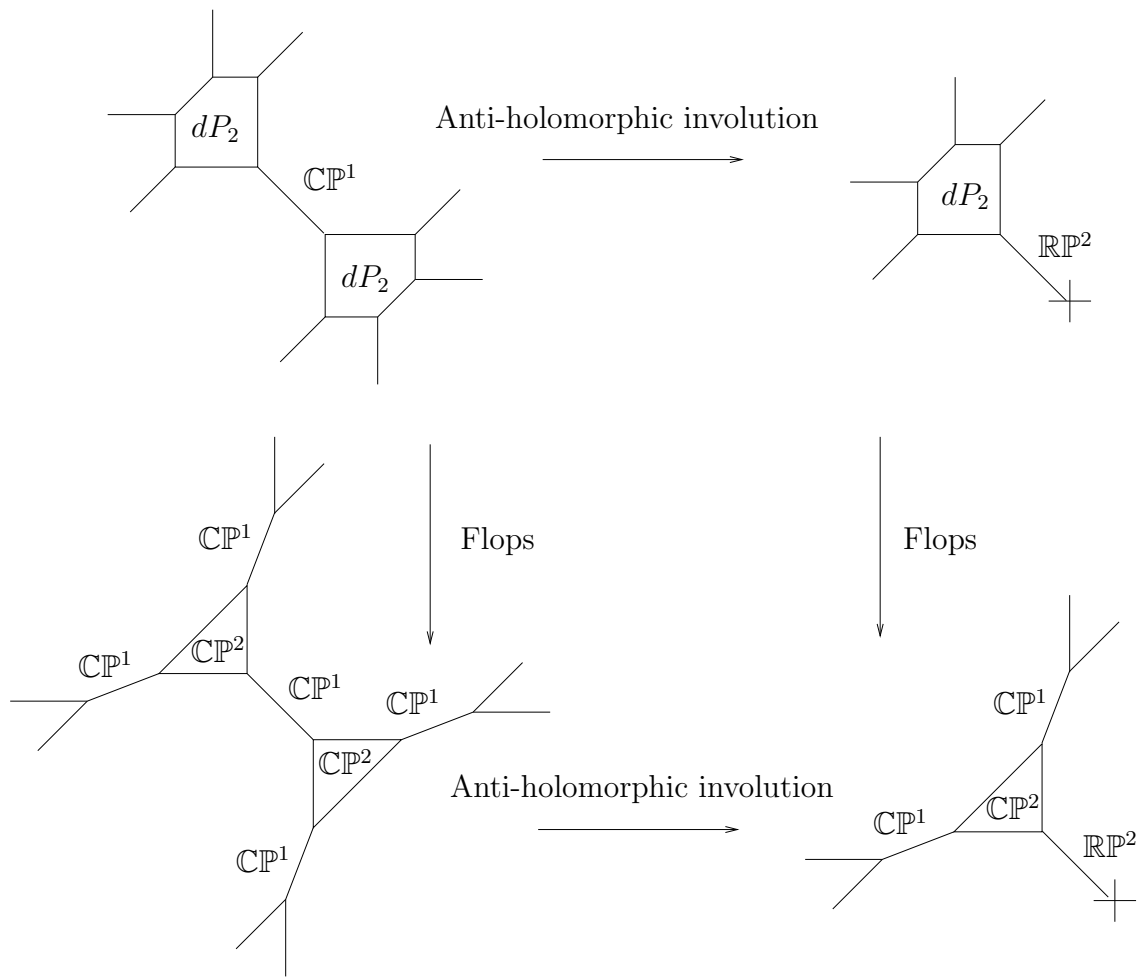}
\caption{The geometry on the closed topological strings side. The orientifolding action acts from left to right, while flopping the ${\IC \IP^1}$'s acts top-down. The ${\IR\IP^2}$ is represented by a cross at the end of the toric leg.}
\label{toriccomm}
\end{center}
\end{figure}

It is often useful \cite{Aganagic:2002qg} to consider a related \CY\ threefold $\widetilde X$ obtained from $X$ by flopping
the two exceptional curves outside of the compact divisors. The ``commuting square'' of geometries (where the arrows
correspond either to flopping or to quotienting by the anti-holomorphic involution) is presented in figure \ref{toriccomm}.

\begin{figure}[htp]
\begin{center}
\psfrag{U(N1)}{$U(N_1)$}
\psfrag{U(N2)}{$U(N_2)$}
\psfrag{SO(N3)/Sp(N3)}{$SO(N_3)/Sp(N_3)$}
\psfrag{P1}{$\IC \IP^1$}
\psfrag{P2}{$\IC \IP^2$}
\psfrag{RP}{$\IR \IP^2$}
\includegraphics[width=15cm]{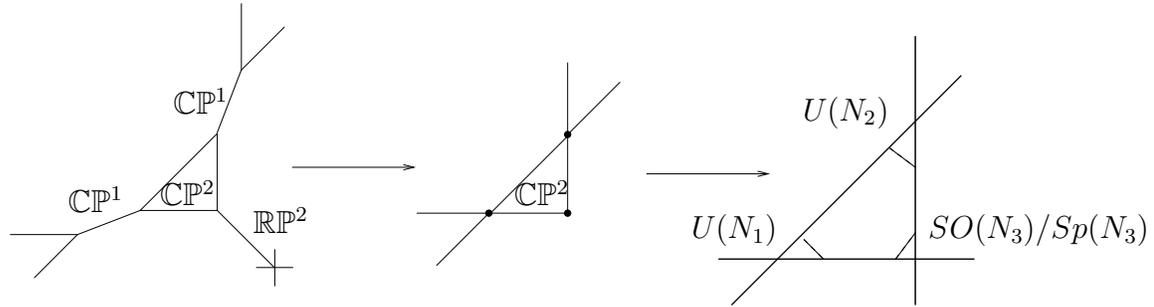}
\caption{The geometric transition. The two ${\IC \IP^1}$'s and the ${\IR\IP^2}$ of the left figure are shrunk to singular points in the middle diagram, and then deformed into three ${\bf S}^3$.}
\label{transition}
\end{center}
\end{figure}

We can now follow the logic in \cite{Aganagic:2002qg,Aganagic:2001ug,Diaconescu:2002sf,Diaconescu:2002qf} and consider a
geometric transition in which each of the resolved conifolds (or their orientifolds)
that exist locally in the geometry are replaced by
deformed conifolds (or their orientifolds). In the above example, this means that
we contract two $\IC \IP^1$'s and a $\IR\IP^2$ and we replace them
with three spheres carrying $U(N)$ and $SO(N)/Sp(N)$ Chern-Simons theories, respectively.
The transition is represented
in figure \ref{transition}. In the next section we will see how to obtain the closed string amplitudes
in the orientifold from Chern-Simons theory.

\section{Closed String Amplitudes from Chern-Simons Theory}\label{ampl}

\subsection{Results from Chern-Simons Theory with Classical Gauge Groups}\label{CSnotation}

As we will see in a moment,
in order to compute the free energies of topological strings on orientifolds via geometric
transitions we have to compute the Chern-Simons invariants of the
unknot and the Hopf link of linking number $+1$ in arbitrary representations
of $U(N)$, $SO(N)$ and $Sp(N)$. Some general results from Chern-Simons theory and the topological vertex have been regrouped in appendix \ref{schurapp}.
In this section we will denote $q=e^{i g_s}$ and
\beq
\lambda=q^{N+a},
\eeql{lam}
where
\beq
a=\begin{cases}
0\,\,\,\,\,\,\,\,\,\,\,\,\,\,\,\,\,\,\,\,\,\,\,\,\,\,\, {\rm for
\,\,\,\,} U(N), \\
-1\,\,\,\,\,\,\,\,\,\,\,\,\,\,\,\,\,\,\,\,\,\, {\rm for
\,\,\,\,} SO(N),\\
1\,\,\,\,\,\,\,\,\,\,\,\,\,\,\,\,\,\,\,\,\,\,\,\,\,\,\, {\rm for
\,\,\,\,} Sp(N).
\end{cases}
\eeql{avaluesclosed}
Notice that the 't Hooft parameter of the classical gauge groups
can be written as
\beq
t=(N+a)g_s
\eeql{thooftgen}
therefore $\lambda=e^t$.
For an arbitrary gauge group $G$ it is a well known result that the Chern-Simons invariant of the unknot in an arbitrary 
representation $R$ is given by the so-called {\it quantum dimension} of $R$ \cite{Witten:1988hf}:
\beq
 \CW_R = {S_{0R} \over S_{00}} = \dim_q R,
\eeql{invunknot}
where $S_{0R}$, $S_{00}$ are entries of the $S$ matrix of the Wess-Zumino-Witten model with the corresponding gauge group and at level 
$k$ (recall that $k$ is related to the string coupling constant by \refeq{stringcouclosed}). The general definition of the quantum dimension is given in \refeq{qdgeneral}.

The expression \refeq{qdgeneral} can be written more explicitly for the different classical gauge groups.
Let $R$ be a representation corresponding to a Young tableau with row lengths $\{ \mu_i \}_{i=1,...,d(\mu)}$, with 
$\mu_1 \geq \mu_2 \geq ...$
and where $d(\mu)$ denotes the number of rows. Then the quantum dimension of a representation $R$ of $U(N)$ reads 
(see for example \cite{Marino:2001re})
\beq
\dim_q^{U(N)} R = \prod_{1 \leq i < j \leq d(\mu)} {[\mu_i-\mu_j+j-i] \over [j-i]} \prod_{i=1}^{d(\mu)} {\prod_{v=-i+1}^{\mu_i-1}
[v]_{\lambda} \over \prod_{w=1}^{\mu_i} [w-i+d(\mu)]},
\eeql{qdUN}
where the quantum number $[x]$ is defined in \refeq{qnumber}, and we defined
\beq
[x]_{\lambda} = \lambda^{1/2} q^{x/2} -\lambda^{-1/2} q^{-x/2},
\eeql{qnumberii}
and $\lambda=q^{N}$ for $U(N)$ representations.

We can also find explicit expressions for the quantum dimensions of $SO(N)$ and $Sp(N)$ representations
\bea
\dim_q^{SO(N)} R &=& \prod_{1 \leq i < j \leq d(\mu)} {[\mu_i-\mu_j+j-i] [\mu_i + \mu_j+1-i-j]_{\lambda} 
\over [j-i][1-i-j]_{\lambda}}\nn\\
&&\times \prod_{i=1}^{d(\mu)}{[\mu_i-i]_{\lambda}^{SO(N)} \over [-i]_{\lambda}^{SO(N)}} \prod_{v=1}^{\mu_i}
{ [\mu_i+1-i-v-d(\mu)]_{\lambda}
\over  [v-i+d(\mu)]},\nn\\
\dim_q^{Sp(N)} R &=& \prod_{1 \leq i < j \leq d(\mu)} {[\mu_i-\mu_j+j-i] [\mu_i + \mu_j+1-i-j]_{\lambda} \over 
[j-i][1-i-j]_{\lambda}} \nn\\
&&\times \prod_{i=1}^{d(\mu)}{[1-i]_{\lambda}^{Sp(N)} [2\mu_i-2i+1]_{\lambda} \over [1-i+\mu_i]_{\lambda}^{Sp(N)} 
[1-2i]_{\lambda}} \nn\\
&&\times \prod_{v=1}^{\mu_i}{ [\mu_i+1-i-v-d(\mu)]_{\lambda} \over  [v-i+d(\mu)]},
\eeal{qdSON}
where we defined
\beq
[x]_{\lambda}^{\rm SO(N)} = {\lambda}^{1/4} q^{{1 \over 4} (2x+1)} -{\lambda}^{-1/4} q^{-{1\over 4}(2x+1)},
\eeql{qnumberiii}
and
\beq
[x]_{\lambda}^{\rm Sp(N)} = {\lambda}^{1/4} q^{{1 \over 4} (2x-1)} -{\lambda}^{-1/4} q^{-{1\over 4}(2x-1)},
\eeql{qnumberiv}
with $\lambda=q^{N+a}$ which leads to $\lambda=q^{N-1}$ for $SO(N)$ and $\lambda=q^{N+1}$ for $Sp(N)$. Using \refeq{qdSON} 
one can show that
\beq
\dim_q^{Sp(N)} R = (-1)^{\ell(R)} \dim_q^{SO(-N)} R^T,
\eeql{SOSpreln}
where $R^T$ is the transposed or conjugate 
representation, related to $R$ by exchanging rows with columns, $SO(-N)$ is meant in the sense of analytic
continuation,
and $\ell(R)$ is the number of boxes of the Young tableau.
This relation is part of the ``$SO(N)=Sp(-N)$'' equivalence \cite{Cvitanovic}. A relation similar to \refeq{SOSpreln} holds
for usual dimensions \cite{Ramgoolam:1993hh}.

Using \refeq{qdUN} and \refeq{qdSON} one can also infer the following formula for quantum dimensions of representations of $SO(N)$ 
and $Sp(N)$ in terms of quantum dimensions of representations of $U(N)$:
\beq
\dim_q^{SO(N)/Sp(N)} R = \sum_{Q=Q^t}(-1)^{1/2(\ell(Q)\mp r(Q))} \dim_q^{U(N)} (R/Q),
\eeql{qdSONUN}
where the skew quantum dimension is defined, as in \refeq{skewschur}, by
\beq
\dim_q^{U(N)} (R/Q)=\sum_{R'} N_{R' Q}^R  \dim_q^{U(N)} R'
\eeql{skewq}
and $N_{R_1 R_2}^R$ are the usual Littlewood-Richardson coefficients that appear in the tensor product of $U(N)$ representations:
$R_1 \otimes R_2 =\sum_R N_{R_1 R_2}^R R$. In \refeq{qdSONUN} the sum is over self-conjugate representations, i.e. representations
that are equal to their transpose, and starts with the trivial representation: $\{ \cdot, \yng(1), \yng(2,1),
\yng(2,2), \yng(3,1,1), ... \}$. 
$r(Q)$ denotes the rank of $Q$, which is defined as the number of boxes in the leading diagonal of the Young tableau \cite{Littlewood:1940}. The
$-$ sign is for $SO(N)$ representations while the $+$ sign is for $Sp(N)$ representations.

As we will see in the following sections, the relations between quantum dimensions of representations of
$SO(N)$ and $Sp(N)$ \refeq{SOSpreln} and \refeq{qdSONUN} are responsible for the fact that partition functions of $SO(N)$
and $Sp(N)$ differ only by an overall sign in front of the unoriented contributions with an odd number of crosscaps,
which leads to the interpretation that they correspond to different choices of sign for the crosscap states. 
Basically, the first term in the sum of the right hand side of \refeq{qdSONUN}
is responsible for oriented contributions to the partition functions, so they are the same for $SO(N)$, $Sp(N)$ and
$U(N)$ gauge groups. The other terms in the sum are responsible for unoriented contributions
to the partition function, and the difference of sign in the exponent of the $(-1)$ factor leads to a relative minus
sign between unoriented contributions with an odd number of crosscaps of the $SO(N)$ and $Sp(N)$ partition functions.

Another important ingredient we will need is the framing of knots and links \cite{Witten:1988hf}. Given a knot invariant
in representation $R$, we can change its framing
by $p$ units (where $p$ is an integer) if we multiply it by
\beq
(-1)^{\ell(R) p}q^{p C_R/2}
\eeql{framing}
where $C_R$ is the quadratic Casimir of the representation $R$. The quadratic Casimirs have the following expressions 
for the different classical gauge groups:
\beq
C_R=\kappa_R + (N+a) \ell(R),
\eeql{crvalues}
where $a$ is given by \refeq{avaluesclosed}, and
\beq
\kappa_R = \sum_i \mu_i (\mu_i - 2 i+1).
\eeql{kappaquantity}
The framing factor can then be written as
\beq
(-1)^{\ell(R) p} \lambda^{p \ell(R)/2}q^{p \kappa_R/2}.
\eeql{framingl}
The sign in \refeq{framing} is not standard in the context of Chern-Simons theory, but as shown in
\cite{Aganagic:2001nx,Marino:2001re}, it is crucial in the context of topological string theory in order to guarantee integrality
properties in the resulting amplitudes. To incorporate a change of framing in a link, we just change the framings
of each of its components according to the rule \refeq{framing} as well.

In our computations we will also need the invariants of Hopf links with linking number $+1$. For arbitrary gauge group $\CG$,
the invariant of the Hopf link with linking number $+1$ is given by the normalized 
inverse $S$ matrix \cite{Witten:1988hf}, and it can be written in terms of quantum dimensions as (see for example \cite{Guadagnini:1990uw})
\beq
\CW_{R_1 R_2} = {S_{R_1 R_2}^{-1} \over S_{00}} = \sum_{R \in R_1 \otimes R_2}  q^{{1\over2}(C_R-C_{R_1}-C_{R_2})} \dim_q R,
\eeql{hopflinkgeneral}
where the sum is over all representations $R$ occurring in the decomposition of the tensor product of $R_1$ and $R_2$.
In the $U(N)$ case, we can replace the Casimir operators $C_{R_i}$ appearing in \refeq{hopflinkgeneral} by $\kappa_{R_i}$, since
$\ell(R)=\ell (R_1)+\ell (R_2)$ in the decomposition of a tensor product of irreducible representations of $U(N)$.
However this relation between the number of boxes of Young tableaux does not hold in the $SO(N)$ and $Sp(N)$ cases. We thus find
\bea
\CW_{R_1R_2}^{U(N)} &=&  \sum_{R} N_{R_1 R_2}^R  q^{{1\over2}(\kappa_R-\kappa_{R_1}-\kappa_{R_2})} \dim_q^{U(N)} R,\nn\\
\CW_{R_1R_2}^{SO(N)/Sp(N)} &=& \sum_{R } M_{R_1 R_2}^R
\lambda^{{1\over2}(\ell(R)-\ell(R_1)-\ell(R_2))} q^{{1\over2}(\kappa_R-\kappa_{R_1}-\kappa_{R_2})}\nn\\
&&~~~~~~~~~~~~~~~~~~\times \dim_q^{SO(N)/Sp(N)} R,
\eeal{hopflink}
where we have denoted by $M_{R_1 R_2}^R$ the tensor product coefficients for irreducible representations of
$SO(N)$ and $Sp(N)$, which turn out to be the same for $SO(N)$ and $Sp(N)$.

To compute \refeq{hopflink} we need the values of $M_{R_1 R_2}^R$, in other words, we have to
decompose any tensor product of $SO(N)$ or $Sp(N)$ representations
into a sum of irreducible representations. This can be done with a technique first developed by Littlewood in \cite{Littlewood:1940}.
Let us first consider $SO(N)$ representations. Let $[R]$ be
the character of the representations $R$, as a function of the eigenvalues of an $SO(N)$ matrix,
and let $\{R\}$ be the Schur function of these eigenvalues labeled by the same representation.
One can prove the following formulae \cite{Littlewood:1940}:
\bea
[R]&=&\{R\} + \sum_{R_1 \in \{\delta\}} (-1)^{\ell(R_1)/2} N_{R_1R_2}^R \{ R_2 \},\nn\\
\{R\}&=&[R]+\sum_{R_1 \in \{\gamma\}}  N_{R_1R_2}^R [R_2],
\eeal{characSON}
where $\{\delta\}$ and $\{\gamma\}$ are subsets of Young tableaux that we describe in appendix \ref{young}.
By using these relations one can express each character $[R][R']$ in the product as a sum of Schur functions, then multiply these
with the usual Littlewood-Richardson coefficients, and finally rexpress the Schur functions in terms of a sum of characters by the
second equation of
\refeq{characSON}. For example,
\bea
[\yng(2)][\yng(1)]&=&(-1+\{\yng(2)\})(\{\yng(1)\})\nn\\
&=&\{\yng(3)\}+\{\yng(2,1)\}-\{\yng(1)\}\nn\\
&=&[\yng(3)]+[\yng(1)]+[\yng(2,1)]+[\yng(1)]-[\yng(1)]=[\yng(1)]+[\yng(3)]+[\yng(2,1)],
\eeal{examptensor}
where the Young tableaux are associated to irreducible representations of $SO(N)$.
To compute the decompositions for $Sp(N)$ representations, one only has to replace the
subsets $\{\delta\}$ and $\{\gamma\}$ respectively by the subsets $\{ \beta \}$ and $\{ \alpha \}$, which are also explained 
in appendix \ref{young} \cite{Ramgoolam:1993hh}. Using this technique one can decompose any tensor products of $SO(N)$ and $Sp(N)$ representations into a 
sum of irreducible representations, which is needed in the computation of expectation values of Hopf links using \refeq{hopflink}. 
One finds that the decomposition of tensor products is always the same for $SO(N)$ and $Sp(N)$ representations, justifying 
our claim above.

The procedure we have described turns out to be rather involved, and fortunately there is a more direct way of computing 
$M_{R_1 R_2}^R$ through the following formula \cite{King:1971,Fulton:1997}:
\beq
M_{R_1 R_2}^R = \sum_{Q,T,U} N_{QT}^{R_1} N_{QU}^{R_2} N_{TU}^R,
\eeql{kingfor}
which expresses these coefficients in terms of usual Littlewood-Richardson coefficients. This formula allows to easily
compute the invariants of Hopf links for $SO/Sp$ gauge groups for any pair of representations.

As shown in \cite{Aganagic:2002qg,Aganagic:2003db}, the Hopf link invariant $\CW_{R_1R_2}^{U(N)}$ plays a crucial r\^ole in the computation of
oriented string amplitudes. It is a Laurent polynomial in $\lambda^{1\over 2}$ whose highest power is $\lambda^{(\ell(R_1) 
+ \ell(R_2))/2}$:
\beq
\CW_{R_1 R_2}^{U(N)}=\lambda^{(\ell(R_1) + \ell(R_2))/2} W_{R_1 R_2}(q)+ \cdots,\eeql{leadinghopf}
where the dots refer to terms with lower powers of $\lambda$.
The leading part of $\CW_{R_1 R_2}^{U(N)}$, which we have denoted by $W_{R_1 R_2}$, can be
computed in terms of Schur polynomials in an infinite number of
variables (see for example \cite{Aganagic:2003db,Okounkov:2003sp,Eguchi:2003sj} for more details):
\beq
W_{R_1 R_2}(q)= s_{R_2}(x_i=q^{-i+{1\over 2}})s_{R_1}(x_i=q^{\mu_i^{R_2}-i+{1\over 2}}),
\eeql{hopfschurclosed}
where $\{ \mu_i^{R_2} \}_{i=1, \cdots, d(\mu^{R_2})}$ is the partition corresponding to $R_2$. 
We will also denote $W_R=W_{R \cdot}=s_R (x_i=q^{-i+{1\over 2}})$.
By looking at the formula in \refeq{hopflink} for $\CW_{R_1R_2}^{SO(N)/Sp(N)}$, one can see that it is a
Laurent polynomial in $\lambda^{1\over 2}$, whose highest power is also $\lambda^{(\ell(R_1) + \ell(R_2))/2}$, and
which has the same leading coefficient $W_{R_1 R_2}(q)$.

More results about Schur functions and their relations to Chern-Simons invariants and the topological vertex are presented in appendix \ref{schurapp}. In the following we will often use this point of view, especially in chapter \ref{open}.

\subsection{Computation of Open String Amplitudes}\label{compopen}

\begin{figure}[htp]
\begin{center}
\psfrag{M1:U(N1)}{$M_1$: $U(N_1)$}
\psfrag{M2:U(N2)}{$M_2$: $U(N_2)$}
\psfrag{M3:SO(N3)/Sp(N3)}{$M_3$: $SO(N_3)/Sp(N_3)$}
\psfrag{1}{$1$}
\psfrag{0}{$0$}
\psfrag{rc1}{$r_{c1}$}
\psfrag{rc2}{$r_{c2}$}
\psfrag{rc3}{$r_{c3}$}
\includegraphics[width=4.5in]{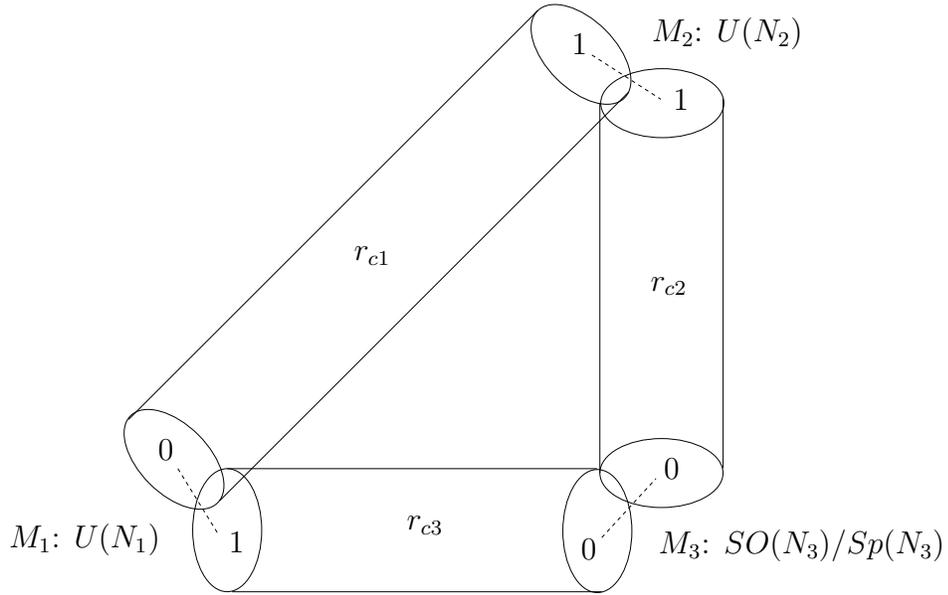}
\caption{The deformed geometry. $M_i$ $i=1,2,3$ are the three spheres and
$r_{ci}$ are the K\"ahler parameters of the cylindrical instantons.
The gauge groups of the Chern-Simons theories on the spheres and the framings of the unknots are also indicated.}
\label{CSsetup}
\end{center}
\end{figure}

We will now follow the results in \cite{Aganagic:2002qg,Diaconescu:2002sf,Diaconescu:2002qf} to compute the open topological string amplitudes
in the geometry described in section \ref{geometry},
which is shown in figure \ref{CSsetup}. There are $N_i$ D-branes wrapped around the three ${\bf S}^3$'s $M_i$, $i=1,2,3$.
This geometry is similar to the one considered for example in \cite{Aganagic:2002qg}; the main difference being that one
of the spheres in our geometry, more precisely $M_3$, is left invariant by the anti-holomorphic involution,
thus leading to a $SO(N)$ or $Sp(N)$ Chern-Simons theory.

For open strings with both ends on the same ${\bf S}^3$, the dynamics is described by a Chern-Simons theory as usual.
For $M_1$ and $M_2$, the Chern-Simons theories respectively have gauge groups $U(N_1)$ and $U(N_2)$, while for $M_3$
it has gauge group $SO(N_3)$ or $Sp(N_3)$. However, there are also cylindrical open string instantons coupling the
Chern-Simons theories on different spheres \cite{Witten:1992fb}. Schematically, the path integral becomes
\beq
Z=\int \prod_{i=1}^3 {\cal D} A_i \, e^{\sum_{i=1}^3 S_i^{CS}(A_i) + S_{inst}},
\eeql{PI}
where $S_i^{CS} (A_i)$, $i=1,2,3$ are the Chern-Simons actions for the three ${\bf S}^3$'s. The instanton sector,
$S_{inst}$, can be computed by using localization (as in \cite{Diaconescu:2002sf,Diaconescu:2002qf}) or by using the techniques of \cite{Aganagic:2002qg}.
We will follow here the procedure of \cite{Aganagic:2002qg}. As explained there,
the bifundamental strings stretching between two three sphere ${\bf S}^3$'s give a massive complex scalar field, with
mass proportional to the complexified K\"ahler parameter $r_c$ corresponding to the ``distance'' between the two spheres.
After integrating out this scalar field one finds an operator which corresponds to a primitive annulus of size $r_c$ 
together with its multicovers. The boundaries of the annulus are on the two three-spheres between which the 
bifundamental strings are stretched. These cylindrical instantons and the geometry are shown in figure \ref{CSsetup}. 
Inserting one operator for each cylindrical instanton we find
\beq
e^{S_{inst}}= \CO(U_3,U_1) \CO(V_1, V_2) \CO(U_2,V_3),
\eeql{Finst}
where we have defined the holonomy variables
\beq
U_i=P \exp \oint_{\Xi_i} A_i,~~~V_i = P \exp \oint_{\Gamma_i} A_i,~~~i=1,2,3,
\eeql{holonomy}
and the $\Xi_i,\Gamma_i$, $i=1,2,3$ are the boundary components of the cylindrical instantons, which are unknots in the
corresponding three-spheres. The operators in \refeq{Finst} are given by
\beq
\CO(A,B, r_c)= \sum_{R}\Tr_R A e^{-\ell(R) r_c} \Tr_R B,
\eeql{operatorR}
where the sum is over all representations, including the trivial one.

The careful reader may note that the operator \refeq{operatorR} is only equivalent to the
usual operator \cite{Ooguri:1999bv,Aganagic:2002qg}
\beq
\exp \sum_{n=1}^{\infty} {e^{-n r_c} \over n} \Tr A^n  \Tr B^{n}
\eeql{operatorOV}
in the $U(N)$ case. In the more general case where the gauge group is $SO(N)$ or $Sp(N)$,
the two operators are not equivalent. It turns out that \refeq{operatorR} is the good operator
to use; it would be interesting to investigate further why this is so.

We can now write the total free energy $\CF=-\log Z$ (with $Z$ given in \refeq{PI}) as
\beq
\CF=\sum_{i=1}^3 \CF(M_i) + {\cal F}_{inst},
\eeql{freeP}
where $\CF(M_i)$ are the free energies of the Chern-Simons theories in the spheres $M_i$,
$i=1,2,3$, and ${\cal F}_{inst}$ is:
\beq
\CF_{inst}= - \ln \biggl\{
\sum_{R_1,R_2,R_3}  e^{-\sum_{i=1}^3 \ell(R_i) r_{ci}} K_{R_3 R_1} (\CL_1) K_{R_1 R_2} (\CL_2) K_{R_2 R_3} (\CL_3)
\biggr\},
\eeql{Finstii}
where $\CL_i$ is the link formed by the knots $(\Xi_i,\Gamma_i)$ and
\bea
K_{R_3R_1} (\CL_1)&=&{\langle R_{3} | V_{M_{1}} | R_1 \rangle \over Z_{M_{1}}} ,\nn\\
K_{R_1R_2} (\CL_2)&=&{\langle R_{1} | V_{M_{2}} | R_2 \rangle \over Z_{M_{2}}},\nn\\
K_{R_2R_3} (\CL_3)&=&{\langle R_{2} | V_{M_{3}} | R_3 \rangle \over Z_{M_{3}}} .
\eeal{Wdef}
It was shown in \cite{Aganagic:2002qg} (using our notation as in figure \ref{CSsetup}) that
\beq
V_{M_1} = TS^{-1},~~~V_{M_2} = S T^{-1} S,~~~ V_{M_3}=S^{-1},
\eeql{Vmatrix}
which means that the three links $\CL_i$, $i=1,2,3$ are Hopf links with linking number $+1$ and that the framings 
are as follows:
$(\Gamma_1, \Xi_3, \Gamma_3)$ are canonically framed, i.e. with framings $(0,0,0)$, while $(\Xi_1,\Xi_2,\Gamma_2)$
have framings $(1,1,1)$, as shown in figure \ref{CSsetup}. We can thus write
\bea
K_{R_3 R_1} (\CL_1)&=&  (-1)^{\ell(R_3)} q^{\kappa_{R_3} \over 2} {S_{R_3 R_1}^{-1} \over S_{00}}=(-1)^{\ell(R_3)} 
q^{\kappa_{R_3} \over 2} \CW_{R_3R_1},\nn\\
K_{R_1 R_2} (\CL_2)&=&  (-1)^{\ell(R_1)+ \ell(R_2)} q^{{1\over 2}(\kappa_{R_1}+\kappa_{R_2})} {S_{R_1 R_2}^{-1} 
\over S_{00}} =
(-1)^{\ell(R_1)+ \ell(R_2)} q^{{1\over 2}(\kappa_{R_1}+\kappa_{R_2})} \CW_{R_1 R_2}\nn\\
K_{R_2 R_3} (\CL_3)&=& {S_{R_2 R_3}^{-1} \over S_{00}} =\CW_{R_1 R_3},
\eeal{Ws}
where the $\lambda$ dependent pieces of \refeq{framingl} have been absorbed in a redefinition of $r_{ci}$. Therefore 
\refeq{Finstii} becomes

\bea
\CF_{inst}&=& - \ln \biggl\{
 1+ \sum_{R_1,R_2,R_3} (-1)^{\sum_{i=1}^3 l_i} e^{-\sum_{i=1}^3 \ell(R_i) r_{ci}} q^{{1 \over 2}(\kappa_{R_1}+\kappa_{R_2}
+\kappa_{R_3})}\nn\\
&&\times \CW_{R_3R_1}(\CL_1) \CW_{R_1R_2} (\CL_2) \CW_{R_2R_3} (\CL_3) \biggr\},
\eeal{Finstiii}
where we singled out the term coming from $R_1,R_2,R_3=\cdot$, i.e. the three representations being the trivial representation.

\subsection{Duality Map and Closed String Amplitudes}

Let us first recall the variables we have defined so far. We first defined the Chern-Simons variables $q=e^{i g_s}$
and $\lambda_i=q^{N_i+a_i}$, with $g_s = {2 \pi \over k_i+y}$ being the same for the three theories. We denote the 
three K\"ahler parameters of the cylindrical instantons by $r_{ci}$, $i=1,2,3$ and the three 't Hooft 
parameters of the different gauge groups by $t_i$. To compare the amplitudes on both sides of the duality, we have to relate the open string parameters $t_i$ and $r_{ci}$ 
to the following closed string parameters: $t$, which is the K\"ahler parameter of ${\IC \IP}^2$, and $s_i$, $i=1,2,3$, 
which are the K\"ahler parameters of the two ${\IC \IP}^1$'s and the ${\IR\IP}^2$. The duality map reads
\bea
&&t = r_{c1} - {t_1+t_2 \over 2} =  r_{c2} - {t_2 + t_3 \over 2}= r_{c3} - {t_1 + t_3 \over 2},\nn\\
&&t_1 = s_1,~~~t_2 = s_2,~~~t_3 = s_3 .
\eeal{duality}
Let now $q_i$ be $q_i=e^{-s_i}=e^{-t_i}$, $i=1,2$, $Q=e^{-s_3}=e^{-t_3}$ and let $\ell$ be $\ell(R_1)+\ell(R_2)+\ell(R_3)$. We can rewrite the
open string partition function \refeq{Finstiii} using \refeq{duality}:
\bea
\CF_{inst}&=& -\ln \biggl\{  1+ \sum_{\ell} (-1)^{\ell} e^{-\ell t} q^{{1 \over 2}(\kappa_{R_1}+\kappa_{R_2}+\kappa_{R_3})} 
q_1^{\ell(R_1)+
\ell(R_3) \over 2} q_2^{\ell (R_1) +\ell(R_2) \over 2} Q^{\ell(R_2)+\ell(R_3) \over 2} \nn\\
&&\times \CW_{R_3 R_1}^{U(N)} \CW_{R_1 R_2}^{U(N)} \CW_{R_2 R_3}^{SO(N)/Sp(N)}  \biggr\},
\eeal{Finstiv}
where the Hopf link invariants in the last line are evaluated at $\lambda=q_i^{-1}$, $i=1,2$, and $\lambda=Q^{-1}$ respectively.
Notice that the leading power of $\lambda$ in $\CW_{R_1 R_2}^{U(N)}$ and in $\CW_{R_1 R_2}^{SO(N)/Sp(N)}$ is in both
cases $\lambda^{(\ell(R_1) + \ell(R_2))/2}$, therefore the above expression for $\CF_{inst}$ gives a power series in $q_i$ and $Q$ with
positive integer coefficients, as it should.
We can now expand the logarithm to find
\beq
\CF_{inst}=\sum_{\ell=1}^{\infty} Z_{\ell}^{(c)} e^{-\ell t},
\eeql{Finstv}
where the connected coefficient $Z_{\ell}^{(c)}$ are given by
\beq
Z_{\ell}^{(c)}=\sum_{1 \leq d \leq \ell} {(-1)^{d+1} \over d} \sum_{m_1+m_2+...+m_d=\ell} Z_{m_1} Z_{m_2} \cdots Z_{m_d}.
\eeql{connected}
These coefficients give the instanton partition function order by order in the K\"ahler parameter $e^{-t}$. Using the
formulae given above for Hopf link invariants with classical gauge groups, we can explicitly compute
the coefficients $Z_{\ell}^{(c)}$.
The contributions independent of the K\"ahler parameter $t$ are given by the sum of Chern-Simons free energies on ${\bf S}^3$
$\sum_{i=1}^3 \CF_{CS}(M_i)$,
which have already been computed in \cite{Sinha:2000ap, Gopakumar:1998ki}.

\subsection{The Oriented Contribution}

\begin{figure}[btp]
\begin{center}
\psfrag{R2(0)}{$R_2^{(0)}$}
\psfrag{R4(0)}{$R_4^{(0)}$}
\psfrag{R3(-2)}{$R_3^{(-2)}$}
\psfrag{R1(-2)}{$R_1^{(-2)}$}
\psfrag{R5(-2)}{$R_5^{(-2)}$}
\psfrag{R'2(0)}{$R_2^{'(0)}$}
\psfrag{R'4(0)}{$R_4^{'(0)}$}
\psfrag{R'3(-2)}{$R_3^{'(-2)}$}
\psfrag{R'1(-2)}{$R_1^{'(-2)}$}
\psfrag{R'5(-2)}{$R_5^{'(-2)}$}
\psfrag{R(0)}{$R^{(0)}$}
\includegraphics[width=4.5in]{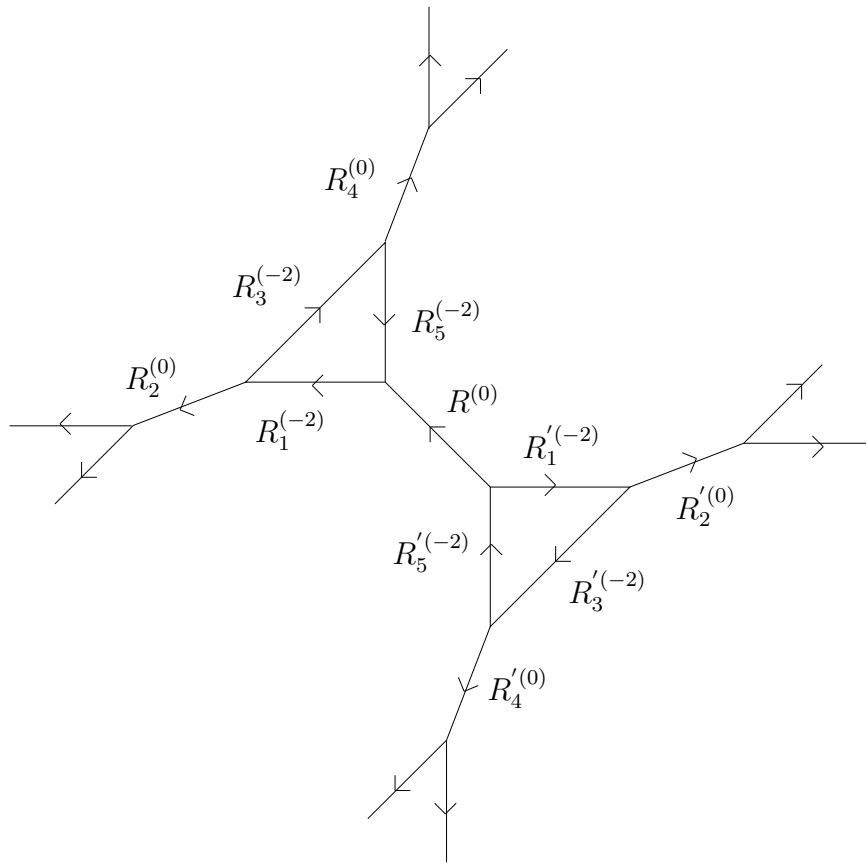}
\caption{The geometry in the topological vertex formalism. In brackets next to the representations are the framings in the corresponding propagator.}
\label{topologicalfig}
\end{center}
\end{figure}

As we explained in section \ref{amodel}, $\CF_{inst}$ contains contributions due to oriented and to unoriented instantons.
In order to compute the closed, unoriented string amplitudes we have to subtract the oriented part, which we must compute
independently. The covering space $X$ is the \CY\ manifold depicted in figure \ref{topologicalfig}. The oriented amplitude 
can be computed using the topological vertex formalism \cite{Aganagic:2003db}. Using the formulae and gluing rules explained in \cite{Aganagic:2003db} we find
\beq
Z(X)=\sum_{R} \CO_{R} (t, q_1, q_2) \CO_{R^T} (t_, q_1, q_2) (-1)^{l(R)} Q^{l(R)},
\eeql{Ztopological}
where $q_i=e^{-s_{i}}$, $i=1,2$ and $Q=e^{-s_3}$. $t$ is the K\"ahler parameter of the ${\IC \IP}^2$, $s_{1}$ and $s_{2}$ are the K\"ahler parameters of the
two ${\IC \IP}^1$'s attached to the ${\IC \IP}^2$, and $s_{3}$ is the K\"ahler parameter of the ${\IC \IP}^1$ between the two ${\IC \IP}^2$'s.
Notice that we have identified the K\"ahler parameters in the way prescribed by the involution.
In \refeq{Ztopological} we introduced the operator
\bea
\CO_{R} (t,q_1,q_2) &=& \sum_{R_i} C_{R R_5 R_1^T} C_{R_1 R_3^T R_2^T} C_{R_3 R_5^T R_4^T} C_{R_2 \cdot \cdot} C_{R_4 \cdot \cdot} (-1)^{\sum_{i} \ell(R_i)} \nn\\
&& \times q^{\kappa(R_1)+\kappa(R_3)+\kappa(R_5)}
e^{-(\ell(R_1)+\ell(R_3)+\ell(R_5))t} q_1^{\ell(R_2)} q_2^{\ell(R_4)},
\eeal{Optopological}
where $C_{R_i R_j R_k}$ is the topological vertex amplitude, which can be expressed in terms of the quantities \refeq{hopfschurclosed}:
\beq
C_{R_1 R_2 R_3}=\sum_{Q_1,Q_3, Q} N_{Q Q_1}^{~~R_1} N_{Q
Q_3}^{~~R_3^T}
q^{\kappa_{R_2}/2+\kappa_{R_3}/2}
{W_{R_2^T Q_1}W_{R_2 Q_3}\over W_{R_2}}.
\eeql{topvertex}
More results about the topological vertex and its relation to Chern-Simons invariants and Schur functions are given in appendix \ref{schurapp}.

Using \refeq{Ztopological} we can express again the free energy as a sum over connected coefficients
\beq
\CF (X)=-\log Z(X) = 
\sum_{\ell,\ell_1, \ell_2, \ell_3} Z_{\ell,\ell_1,\ell_2,\ell_3}^{(c)} q_1^{\ell_1} q_2^{\ell_2} Q^{\ell_3} e^{-\ell t}.
\eeql{Ftopological}
The free energy computed in \refeq{Finstv} should equal, according to \refeq{strucfclosed},
\beq
\CF_{inst}= {1\over 2}\CF(X) + \CF(X/I,g_s)_{unor},
\eeql{frees}
where $\CF(X)$ is given in \refeq{Ftopological}. This determines the unoriented part, which should have
the structure given in \refeq{strucunclosed}. We will
encode the resulting oriented and unoriented Gopakumar-Vafa invariants in the
following generating functionals
\bea
\CF_d^{g,0}&=&{1 \over 2} \sum_{d_1,d_2,d_Q} n^{g,0}_{d,d_1,d_2,d_Q} q_1^{d_1} q_2^{d_2} Q^{d_Q},\nn\\
\CF_d^{g,1}&=&\sum_{d_1,d_2,d_Q} n^{g,1}_{d,d_1,d_2,d_Q} q_1^{d_1} q_2^{d_2} Q^{d_Q/2},\nn\\
\CF_d^{g,2}&=&\sum_{d_1,d_2,d_Q} n^{g,2}_{d,d_1,d_2,d_Q} q_1^{d_1} q_2^{d_2} Q^{d_Q},
\eeal{functionalsi}
where $d$ is the degree in $e^{-t}$, and the superscripts $g,c$ with $c=0,1,2$ denote the genus and the number of
crosscaps, respectively. Of course, $c=0$ is the oriented contribution obtained from 
\refeq{Ftopological} (multiplied by the factor of $1/2$), and
in the second equation of \refeq{functionalsi} $d_Q$ must be odd. In order to compute these functionals,
we have to remove multicoverings according to \refeq{orcont} and \refeq{strucunclosed}.
It is important to note that the requirement that the partition function satisfies the good integrality properties 
leading to \refeq{functionalsi} is highly non-trivial.

\begin{figure}[htp]
\begin{center}
\psfrag{R2}{$R_2$}
\psfrag{R1}{$R_1$}
\psfrag{R3}{$R_3$}
\psfrag{R}{$R$}
\psfrag{RP2}{$\IR \IP^2$}
\includegraphics[width=1.9in]{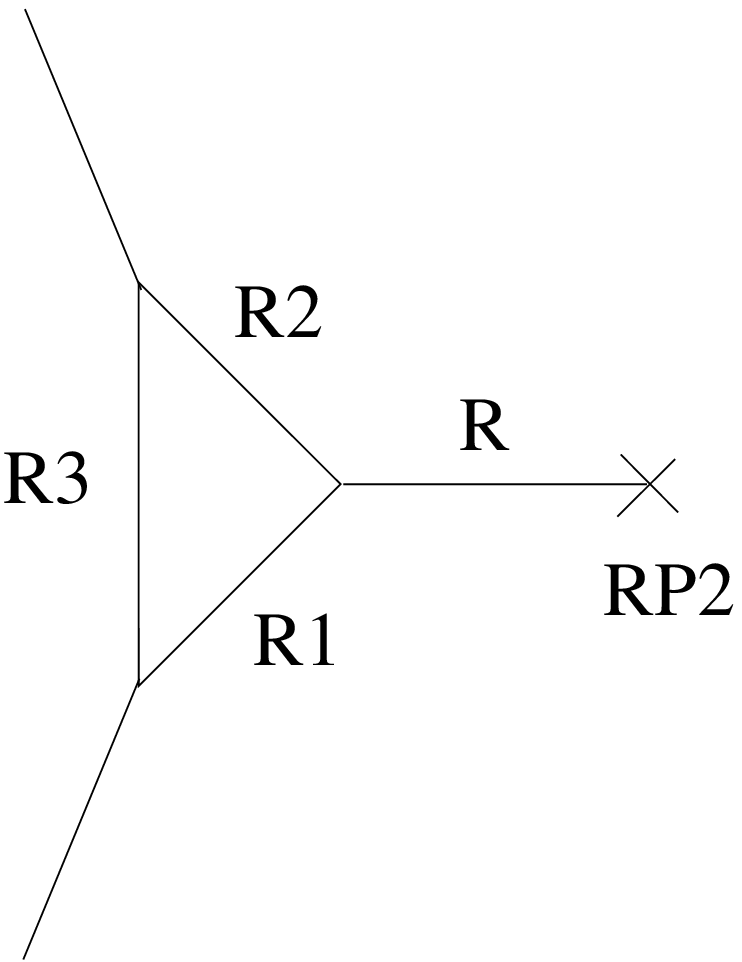}
\caption{Toric diagram for local ${\IC \IP}^2$ attached to an $\IR\IP^2$.}
\label{ptworptwo}
\end{center}
\end{figure}

We present the results for the functionals given by \refeq{functionalsi} in appendix \ref{resultfunct}. For the sake of brevity,
here we present the results only in the limiting case where we take the volumes of the two ${\IC \IP}^1$'s attached
to the ${\IC \IP}^2$ to infinity, as in \cite{Aganagic:2002qg}. We thus obtain the answer for the simpler geometry whose toric diagram
is depicted in figure \ref{ptworptwo}. This geometry already captures all the interesting features of the unoriented and
oriented generating functionals.

To take this limit, one can directly consider the generating functionals \refeq{functionalsi} and set $q_{1,2}=0$,
which corresponds to sending the two K\"ahler parameters to infinity. One can also obtain this limit by taking
the leading piece of the $U(N)$ Hopf link invariants in \refeq{Finstiv}, in the sense explained in \refeq{leadinghopf}. The
free energy of this geometry is just:
\beq
\CF= -\ln \biggl\{  1+ \sum_{\ell} (-1)^{\ell} e^{-\ell t} q^{{1 \over 2}(\kappa_{R_1}+\kappa_{R_2}+\kappa_{R_3})}
Q^{\ell(R_2)+\ell(R_3) \over 2} W_{R_3 R_1} W_{R_1 R_2} \CW_{R_2 R_3}^{SO(N)/Sp(N)}  \biggr\}.
\eeql{prtwo}
The result can now be encoded in the simpler generating functionals
\bea
\CF_d^{g,0}&=&{1 \over 2} \sum_{d_Q} n^{g,0}_{d,d_Q} Q^{d_Q},\nn\\
\CF_d^{g,1}&=&\sum_{d_Q} n^{g,1}_{d,d_Q} Q^{d_Q/2},\nn\\
\CF_d^{g,2}&=&\sum_{d_Q} n^{g,2}_{d,d_Q} Q^{d_Q},
\eeal{functionalslimit}
with the same restrictions as for \refeq{functionalsi}. We present here the all genus results we obtain up to degree
$6$ in $e^{-t}$. At this order $n_{d,d_Q}^{g,c}=0~\forall~g \geq 11$ (all the invariants $n_{d,d_Q}^{g,c}$ with $d \leq 6$ 
that are not shown in the tables are understood to be zero). The results in tables \ref{tabfunct1}--\ref{tabfunct16} correspond to $Sp(N)$ gauge group; 
to obtain the $SO(N)$ result it suffices to change the sign of the invariants with $c=1$.

\begin{table}[p]
\begin{center}
\begin{footnotesize}
\begin{tabular}{|c|rrrrrr|}\hline
$c=0$&$d_Q=0$&$1$&$2$&$3$&$4$&$5$
\\
\hline
$d=0$&$0$&$1$&$0$&$0$&$0$&$0$
\\
$1$&$6$&$-4$&$0$&$0$&$0$&$0$
\\
$2$&$-12$&$14$&$-2$&$0$&$0$&$0$
\\
$3$&$54$&$-84$&$30$&$0$&$0$&$0$
\\
$4$&$-384$&$725$&$-392$&$51$&$0$&$0$
\\
$5$&$3390$&$-7540$&$5434$&$-1368$&$84$&$0$
\\
$6$&$-34128$&$87776$&$-79198$&$29466$&$-4040$&$124$
\\
\hline
\end{tabular}
\caption{Invariants $n_{d,d_Q}^{0,0}$ at genus 0, up to $d=6$.}
\label{tabfunct1}
\end{footnotesize}
\end{center}
\end{table}

\begin{table}[p]
\begin{center}
\begin{footnotesize}
\begin{tabular}{|c|rrrrrr|}\hline
$c=1$&$d_Q=1$&$3$&$5$&$7$&$9$&$11$
\\
\hline
$d=0$&$1$&$0$&$0$&$0$&$0$&$0$
\\
$1$&$-2$&$0$&$0$&$0$&$0$&$0$
\\
$2$&$5$&$-3$&$0$&$0$&$0$&$0$
\\
$3$&$-32$&$30$&$-4$&$0$&$0$&$0$
\\
$4$&$286$&$-369$&$112$&$-5$&$0$&$0$
\\
$5$&$-3038$&$5016$&$-2410$&$328$&$-6$&$0$
\\
$6$&$35870$&$-72150$&$47554$&$-11528$&$819$&$-7$
\\
\hline
\end{tabular}
\caption{Invariants $n_{d,d_Q}^{0,1}$ at genus 0, up to $d=6$.}
\label{tabfunct1x}
\end{footnotesize}
\end{center}
\end{table}

\begin{table}[p]
\begin{center}
\begin{footnotesize}
\begin{tabular}{|c|rrrr|}\hline
$c=2$&$d_Q=2$&$3$&$4$&$5$
\\
\hline
$d=3$&$1$&$0$&$0$&$0$
\\
$4$&$-11$&$2$&$0$&$0$
\\
$5$&$131$&$-66$&$7$&$0$
\\
$6$&$-1690$&$1460$&$-333$&$12$
\\
\hline
\end{tabular}
\caption{Invariants $n_{d,d_Q}^{0,2}$ at genus 0, up to $d=6$.}
\label{tabfunct2}
\end{footnotesize}
\end{center}
\end{table}

\begin{table}[p]
\begin{center}
\begin{footnotesize}
\begin{tabular}{|c|rrrrr|}\hline
$c=0$&$d_Q=0$&$1$&$2$&$3$&$4$
\\
\hline
$d=3$&$20$&$-18$&$0$&$0$&$0$
\\
$4$&$-462$&$612$&$-168$&$0$&$0$
\\
$5$&$8904$&$-15210$&$7380$&$-930$&$0$
\\
$6$&$-161896$&$336636$&$-228532$&$56536$&$-3851$
\\
\hline
\end{tabular}
\caption{Invariants $n_{d,d_Q}^{1,0}$ at genus 1, up to $d=6$.}
\label{tabfunct3}
\end{footnotesize}
\end{center}
\end{table}

\begin{table}[p]
\begin{center}
\begin{footnotesize}
\begin{tabular}{|c|rrrrr|}\hline
$c=1$&$d_Q=1$&$3$&$5$&$7$&$9$
\\
\hline
$d=3$&$-9$&$7$&$0$&$0$&$0$
\\
$4$&$288$&$-315$&$59$&$0$&$0$
\\
$5$&$-6984$&$9954$&$-3630$&$282$&$0$
\\
$6$&$152622$&$-269501$&$145467$&$-25672$&$1014$
\\
\hline
\end{tabular}
\caption{Invariants $n_{d,d_Q}^{1,1}$ at genus 1, up to $d=6$.}
\label{tabfunct3x}
\end{footnotesize}
\end{center}
\end{table}

\begin{table}[p]
\begin{center}
\begin{footnotesize}
\begin{tabular}{|c|rrrr|}\hline
$c=2$&$d_Q=2$&$3$&$4$&$5$
\\
\hline
$d=4$&$-6$&$0$&$0$&$0$
\\
$5$&$201$&$-55$&$1$&$0$
\\
$6$&$-5180$&$3180$&$-438$&$2$
\\
\hline
\end{tabular}
\caption{Invariants $n_{d,d_Q}^{1,2}$ at genus 1, up to $d=6$.}
\label{tabfunct4}
\end{footnotesize}
\end{center}
\end{table}

\begin{table}[p]
\begin{center}
\begin{footnotesize}
\begin{tabular}{|c|rrrrr|}\hline
$c=0$&$d_Q=0$&$1$&$2$&$3$&$4$
\\
\hline
$d=4$&$-204$&$216$&$-24$&$0$&$0$
\\
$5$&$10860$&$-15444$&$5154$&$-276$&$0$
\\
$6$&$-388044$&$690273$&$-365536$&$60235$&$-1800$
\\
\hline
\end{tabular}
\caption{Invariants $n_{d,d_Q}^{2,0}$ at genus 2, up to $d=6$.}
\label{tabfunct5}
\end{footnotesize}
\end{center}
\end{table}

\begin{table}[p]
\begin{center}
\begin{footnotesize}
\begin{tabular}{|c|rrrrr|}\hline
$c=1$&$d_Q=1$&$3$&$5$&$7$&$9$
\\
\hline
$d=4$&$108$&$-103$&$9$&$0$&$0$
\\
$5$&$-7506$&$9474$&$-2567$&$95$&$0$
\\
$6$&$329544$&$-521400$&$231550$&$-29010$&$554$
\\
\hline
\end{tabular}
\caption{Invariants $n_{d,d_Q}^{2,1}$ at genus 2, up to $d=6$.}
\label{tabfunct5x}
\end{footnotesize}
\end{center}
\end{table}

\begin{table}[p]
\begin{center}
\begin{footnotesize}
\begin{tabular}{|c|rrr|}\hline
$c=2$&$d_Q=2$&$3$&$4$
\\
\hline
$d=4$&$-1$&$0$&$0$
\\
$5$&$146$&$-18$&$0$
\\
$6$&$-8296$&$3520$&$-274$
\\
\hline
\end{tabular}
\caption{Invariants $n_{d,d_Q}^{2,2}$ at genus 2, up to $d=6$.}
\label{tabfunct6}
\end{footnotesize}
\end{center}
\end{table}

\begin{table}[p]
\begin{center}
\begin{footnotesize}
\begin{tabular}{|c|rrrrr|}\hline
$c=0$&$d_Q=0$&$1$&$2$&$3$&$4$
\\
\hline
$d=4$&$-30$&$28$&$0$&$0$&$0$
\\
$5$&$7344$&$-9094$&$2036$&$-30$&$0$
\\
$6$&$-581706$&$913220$&$-381934$&$40728$&$-408$
\\
\hline
\end{tabular}
\caption{Invariants $n_{d,d_Q}^{3,0}$ at genus 3, up to $d=6$.}
\label{tabfunct7}
\end{footnotesize}
\end{center}
\end{table}

\begin{table}[p]
\begin{center}
\begin{footnotesize}
\begin{tabular}{|c|rrrrr|}\hline
$c=1$&$d_Q=1$&$3$&$5$&$7$&$9$
\\
\hline
$d=4$&$14$&$-12$&$0$&$0$&$0$
\\
$5$&$-4519$&$5133$&$-977$&$11$&$0$
\\
$6$&$447502$&$-642780$&$233460$&$-19781$&$139$
\\
\hline
\end{tabular}
\caption{Invariants $n_{d,d_Q}^{3,1}$ at genus 3, up to $d=6$.}
\label{tabfunct7x}
\end{footnotesize}
\end{center}
\end{table}

\begin{table}[p]
\begin{center}
\begin{footnotesize}
\begin{tabular}{|c|rrr|}\hline
$c=2$&$d_Q=2$&$3$&$4$
\\
\hline
$d=5$&$58$&$-2$&$0$
\\
$6$&$-8489$&$2352$&$-90$
\\
\hline
\end{tabular}
\caption{Invariants $n_{d,d_Q}^{3,2}$ at genus 3, up to $d=6$.}
\label{tabfunct8}
\end{footnotesize}
\end{center}
\end{table}

\begin{table}[p]
\begin{center}
\begin{footnotesize}
\begin{tabular}{|c|rrrrr|}\hline
$c=0$&$d_Q=0$&$1$&$2$&$3$&$4$
\\
\hline
$d=5$&$2772$&$-3084$&$424$&$0$&$0$
\\
$6$&$-580800$&$821490$&$-270708$&$17600$&$-36$
\\
\hline
\end{tabular}
\caption{Invariants $n_{d,d_Q}^{4,0}$ at genus 4, up to $d=6$.}
\label{tabfunct9}
\end{footnotesize}
\end{center}
\end{table}

\begin{table}[p]
\begin{center}
\begin{footnotesize}
\begin{tabular}{|c|rrrrr|}\hline
$c=1$&$d_Q=1$&$3$&$5$&$7$&$9$
\\
\hline
$d=5$&$-1542$&$1599$&$-191$&$0$&$0$
\\
$6$&$407661$&$-536973$&$157255$&$-8372$&$13$
\\
\hline
\end{tabular}
\caption{Invariants $n_{d,d_Q}^{4,1}$ at genus 4, up to $d=6$.}
\label{tabfunct9x}
\end{footnotesize}
\end{center}
\end{table}

\begin{table}[p]
\begin{center}
\begin{footnotesize}
\begin{tabular}{|c|rrr|}\hline
$c=2$&$d_Q=2$&$3$&$4$
\\
\hline
$d=5$&$12$&$0$&$0$
\\
$6$&$-5862$&$976$&$-15$
\\
\hline
\end{tabular}
\caption{Invariants $n_{d,d_Q}^{4,2}$ at genus 4,  up to $d=6$.}
\label{tabfunct10}
\end{footnotesize}
\end{center}
\end{table}

\begin{table}[p]
\begin{center}
\begin{footnotesize}
\begin{tabular}{|c|rrrr|c|rrrr|}\hline
$c=0$&$d_Q=0$&$1$&$2$&$3$&$c=1$&$d_Q=1$&$3$&$5$&$7$
\\
\hline
$d=5$&$540$&$-552$&$36$&$0$&$d=5$&$-276$&$265$&$-15$&$0$
\\
$6$&$-393714$&$509724$&$-130496$&$4684$&$6$&$254310$&$-309962$&$71523$&$-2141$
\\
\hline
\end{tabular}
\caption{Invariants $n_{d,d_Q}^{5,c}$ at genus 5, with $c=0,1$, up to $d=6$.}
\label{tabfunct11}
\end{footnotesize}
\end{center}
\end{table}

\begin{table}[p]
\begin{center}
\begin{footnotesize}
\begin{tabular}{|c|rrr|}\hline
$c=2$&$d_Q=2$&$3$&$4$
\\
\hline
$d=5$&$1$&$0$&$0$
\\
$6$&$-2758$&$245$&$-1$
\\
\hline
\end{tabular}
\caption{Invariants $n_{d,d_Q}^{5,2}$ at genus 5, up to $d=6$.}
\label{tabfunct11x}
\end{footnotesize}
\end{center}
\end{table}

\begin{table}[p]
\begin{center}
\begin{footnotesize}
\begin{tabular}{|c|rrrr|c|rrrr|}\hline
$c=0$&$d_Q=0$&$1$&$2$&$3$&$c=1$&$d_Q=1$&$3$&$5$&$7$
\\
\hline
$d=5$&$42$&$-40$&$0$&$0$&$d=5$&$-20$&$18$&$0$&$0$
\\
$6$&$-180780$&$216960$&$-41904$&$696$&$6$&$108440$&$-123342$&$21630$&$-302$
\\
\hline
\end{tabular}
\caption{Invariants $n_{d,d_Q}^{6,c}$ at genus 6, with $c=0,1$, up to $d=6$.}
\label{tabfunct12}
\end{footnotesize}
\end{center}
\end{table}

\clearpage

\begin{table}[p]
\begin{center}
\begin{footnotesize}
\begin{tabular}{|c|rr|}\hline
$c=2$&$d_Q=2$&$3$
\\
\hline
$d=5$&$0$&$0$
\\
$6$&$-868$&$34$
\\
\hline
\end{tabular}
\caption{Invariants $n_{d,d_Q}^{6,2}$ at genus 6, up to $d=6$.}
\label{tabfunct12x}
\end{footnotesize}
\end{center}
\end{table}

\begin{table}[p]
\begin{center}
\begin{footnotesize}
\begin{tabular}{|c|rrrr|c|rrrr|}\hline
$c=0$&$d_Q=0$&$1$&$2$&$3$&$c=1$&$d_Q=1$&$3$&$5$&$7$
\\
\hline
$d=6$&$-55076$&$61896$&$-8532$&$44$&$d=6$&$30948$&$-33110$&$4156$&$-18$
\\
\hline
\end{tabular}
\caption{Invariants $n_{d,d_Q}^{7,c}$ at genus 7, with $c=0,1$, up to $d=6$.}
\label{tabfunct13}
\end{footnotesize}
\end{center}
\end{table}

\begin{table}[p]
\begin{center}
\begin{footnotesize}
\begin{tabular}{|c|rr|}\hline
$c=2$&$d_Q=2$&$3$
\\
\hline
$d=6$&$-174$&$2$
\\
\hline
\end{tabular}
\caption{Invariants $n_{d,d_Q}^{7,2}$ at genus 7, up to $d=6$.}
\label{tabfunct13x}
\end{footnotesize}
\end{center}
\end{table}

\begin{table}[p]
\begin{center}
\begin{footnotesize}
\begin{tabular}{|c|rrr|c|rrr|c|r|}\hline
$c=0$&$d_Q=0$&$1$&$2$&$c=1$&$d_Q=1$&$3$&$5$&$c=2$&$d_Q=2$
\\
\hline
$d=6$&$-10620$&$11268$&$-992$&$d=6$&$5634$&$-5710$&$458$&$d=6$&$-20$
\\
\hline
\end{tabular}
\caption{Invariants $n_{d,d_Q}^{8,c}$ at genus 8, with $c=0,1,2$, up to $d=6$.}
\label{tabfunct14}
\end{footnotesize}
\end{center}
\end{table}

\begin{table}[p]
\begin{center}
\begin{footnotesize}
\begin{tabular}{|c|rrr|c|rrr|c|r|}\hline
$c=0$&$d_Q=0$&$1$&$2$&$c=1$&$d_Q=1$&$3$&$5$&$c=2$&$d_Q=2$
\\
\hline
$d=6$&$-1170$&$1180$&$-50$&$d=6$&$590$&$-570$&$22$&$d=6$&$-1$
\\
\hline
\end{tabular}
\caption{Invariants $n_{d,d_Q}^{9,c}$ at genus 9, with $c=0,1,2$, up to $d=6$.}
\label{tabfunct15}
\end{footnotesize}
\end{center}
\end{table}

\begin{table}[p]
\begin{center}
\begin{footnotesize}
\begin{tabular}{|c|rr|c|rr|}\hline
$c=0$&$d_Q=0$&$1$&$c=1$&$d_Q=1$&$3$
\\
\hline
$d=6$&$-56$&$54$&$d=6$&$27$&$-25$
\\
\hline
\end{tabular}
\caption{Invariants $n_{d,d_Q}^{10,c}$ at genus 10, with $c=0,1$, up to $d=6$.}
\label{tabfunct16}
\end{footnotesize}
\end{center}
\end{table}

\clearpage

\section{Unoriented Localization}\label{localclosed}

As explained in section \ref{amodel}, to compute the full partition function of closed topological strings on the geometry 
before the geometric transition, we have to sum both over holomorphic maps from orientable Riemann surfaces to the 
\CY\ space $X$ as well as maps from non-orientable worldsheets to the orientifolded space $X/I$.

In \cite{Diaconescu:2003dq} it was developed a method for summing unoriented world-sheet instantons for
closed topological strings based on localization with respect to a torus action on a moduli space of equivariant holomorphic 
maps. Although in \cite{Diaconescu:2003dq} this moduli space has not been constructed, a computational definition for its virtual fundamental cycle 
was given. Concretely, this reduces to enumerating all fixed loci under an induced torus action on the moduli space and 
assigning a local contribution to each component of the fixed locus using an equivariant version of the localization 
theorem of \cite{Graber:1999}. Moreover, in \cite{Diaconescu:2003dq} it was shown that the fixed loci can be represented in terms of Kontsevich graphs \cite{Kontsevich:1994na} 
with involution. 

This method does not rely on large $N$ duality, and therefore may provide an independent check of our large $N$ duality proposal 
for orientifolds. Namely, we can employ the localization techniques of \cite{Diaconescu:2003dq} to compute one crosscap and two crosscaps contributions to the
full closed topological string partition function on the orientifolded geometry before the geometric transition.

We can use the computation in \cite{Diaconescu:2003dq} to confirm the one crosscap invariants for low degree and genus obtained 
from the Chern-Simons computation. There, it was computed the unoriented free
energy for a ${\IC \IP^2}$ with a ${\IR\IP^2}$
attached. This is exactly the limiting geometry for which we presented our results in tables \ref{tabfunct1}--\ref{tabfunct16}, related to the full 
geometry of section \ref{amodel} by sending the two K\"ahler parameters of the ${\IC \IP^1}$'s of the full geometry to infinity. 
In our variables, the result of \cite{Diaconescu:2003dq} reads
\bea
\CF&=&{1 \over g_s} (Q^{1/2}-2 e^{-t} Q^{1/2} + 5 e^{-2t} Q^{1/2}+...+{1 \over 9} Q^{3/2} -
3 e^{-2t} Q^{3/2}+{268 \over 9} e^{-3t} Q^{3/2}+...)\nn\\
&&+g_s ({1 \over 24} Q^{1/2}-{1\over 12}e^{-t} Q^{1/2} + ...).
\eeal{localization}
By expanding $q=e^{i g_s}$ in powers of $g_s$, it is straightforward to show that the contributions with $c=1$ 
in tables \ref{tabfunct1}-\ref{tabfunct16} are in agreement with \refeq{localization}.

In the following we will compute some Klein bottle amplitudes using unoriented localization.
We will find agreement with the Chern-Simons and with the topological vertex computations. We will perform the computations in the \CY\ geometry $\widetilde X$. In the patch $\{z_1\neq 0,
z_7\neq 0,z_{10}\neq 0\}$
we introduce local coordinates
\beq
z={z_1^3 z_4\over z_7z_{10}^3},\quad u={z_6z_7z_{10}^2\over z_1},\quad v={z_5z_7z_{10}^2\over z_1}.
\eeql{locor}
Using \refeq{constr} we obtain the weights of the local coordinates
\beq
\l_z=6\l_1+2\l_4,\quad\l_u=-3\l_1-\l_4+\l_6,\quad\l_v=-3\l_1-\l_4+\l_5.
\eeql{locw}
Note that the compatibility of the involution with the torus action implies $\l_z+\l_u+\l_v=0$.

We will denote the contributions of the fixed loci by $C_{\chi,d,h}$, where $\chi$ is the Euler characteristic of the 
unoriented source Riemann surface and $d$ and $h$ are the degrees of the map with respect to the $\IR\IP^2$ and hyperplane 
class of $\IC \IP^2$ respectively.

\subsection{Unoriented Localization at $2$ Crosscaps and Degree $2$ $\IR\IP^2$}

The computation at degree $0$ hyperplane class has been performed in \cite{Diaconescu:2003dq}. Let us recall the graphs
and their contributions.

\fig{unverz}{Two crosscaps and no hyperplane at degree $2$ $\IR\IP^2$.}
{2.5in}{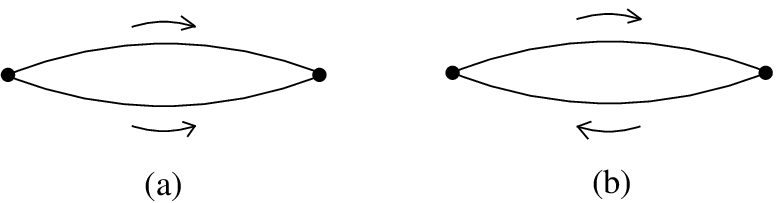}

Note that in case $(b)$ the anti-holomorphic involution exchanges the two components of the source curve. In
\cite{Diaconescu:2003dq} it has been postulated that such an operation will introduce an additional minus sign. Therefore the
contributions of the two graphs are
\beq
C_{0,2,0}^{(a)}={\l_u\l_v\over 4\l_z^2},\qquad C_{0,2,0}^{(b)}=-{\l_u\l_v\over 4\l_z^2}.
\eeql{tcnh}

Let us consider now the degree $1$ hyperplane class configurations. The graphs allowed are presented in figure \ref{unvervb}.

\fig{unvervb}{Two crosscaps and one hyperplane at degree $2$ $\IR\IP^2$. Mirror pairs are $\{(a),(c)\}$ and $\{(b),(d)\}$
respectively.}
{3.0in}{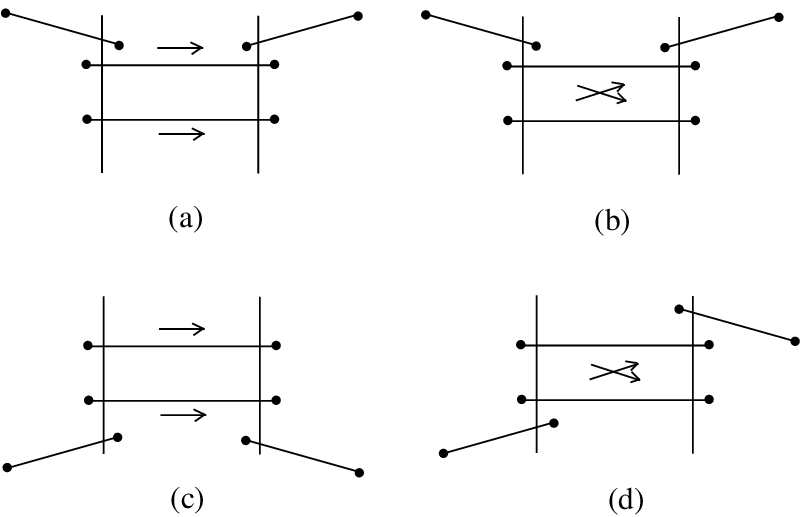}

The allowed configurations are obtained by performing bubblings at the nodes of the graphs in figure \ref{unverz} and
inserting degree $1$ hyperplane graphs; we will call such configurations type ${\rm I}$ graphs. These come in pairs, each one 
admits a mirror graph. From now on, we will draw a single graph for each mirror pair. The contributions of the above configurations 
are given by
\beq
C_{0,2,1}^{(a)}={\l_v^2\over 2\l_z^2},\quad  C_{0,2,1}^{(b)}=-{\l_v^2\over 2\l_z^2},
\quad C_{0,2,1}^{(c)}={\l_u^2\over 2\l_z^2},\quad  C_{0,2,1}^{(d)}=-{\l_u^2\over 2\l_z^2}
\eeql{consdego}
where we have used again the sign rule postulated in \cite{Diaconescu:2003dq}. The graph contributions add up to zero.

The discussion is similar at degree $2$ hyperplane class. The type ${\rm I}$ graphs appearing cancel in pairs due to the same sign rule 
as above. There also appear new configurations, which we will call type ${\rm II}$ graphs, and which we present in figure \ref{unverbi}.  

\fig{unverbi}{Two crosscaps and two hyperplanes at degree $2$ $\IR\IP^2$: type ${\rm II}$ graphs.}
{3.2in}{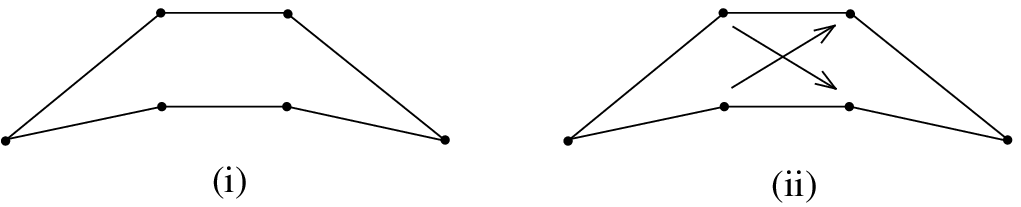}

Their contributions are given by 
\beq
C_{0,2,2}^{(i)}={(\l_u-2\l_v)(\l_v-\l_u)\over 2\l_v^2},\quad C_{0,2,2}^{(ii)}=-{(\l_u-2\l_v)(\l_v-\l_u)\over 2\l_v^2}
\eeql{consdegoi}
and therefore they cancel due to the same sign rule that we used previously.

At degree $3$ hyperplane class, we obtain again pairs of graphs of type ${\rm I}$ and type ${\rm II}$ that
cancel each other. In figure \ref{unver} we draw some new type ${\rm II}$ graphs whose analogues at higher $\IR\IP^2$ degree will play an important 
r\^ole.

\fig{unver}{Two crosscaps and three hyperplanes at degree $2$ $\IR\IP^2$: type ${\rm II}$ graphs.}
{3.2in}{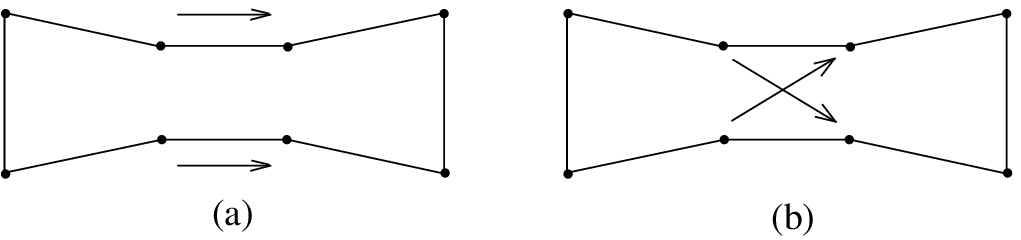}

The contributions of the two graphs of figure \ref{unver} are: $C^{(a)}_{0,2,3}=-C^{(b)}_{0,2,3}=1$. To conclude,
we obtain that up to degree $3$ hyperplane class, the $2$ crosscaps degree $2$ $\IR\IP^2$ Gromov-Witten invariants vanish. 
In fact, this will be true at any hyperplane class degree.

\subsection{Unoriented Localization at $2$ Crosscaps and Degree $4$ $\IR\IP^2$}

\fig{unverii}{Two crosscaps and no hyperplane at degree $4$ $\IR\IP^2$.}
{5.2in}{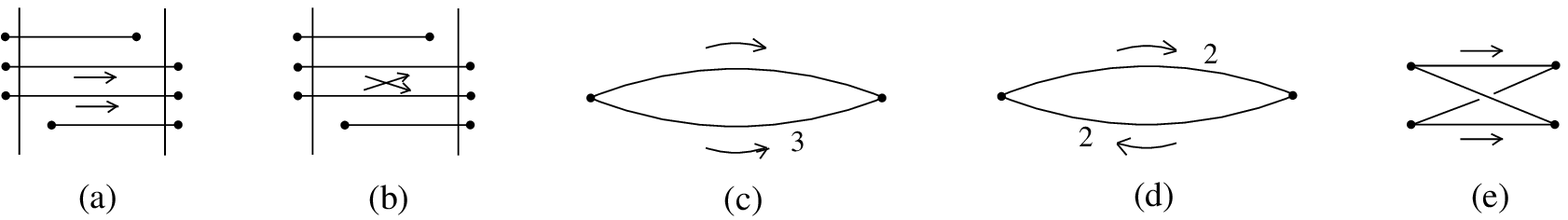}

At degree $0$ hyperplane class this computation has been performed in \cite{Diaconescu:2003dq} . We list the graphs in figure \ref{unverii}. The contributions of the graphs are:
\bea
&& C_{0,4,0}^{(a)}={1\over 2}{\l_u^2\l_v^2\over \l_z^4}, \qquad C_{0,4,0}^{(b)}=-{1\over 2}{\l_u^2\l_v^2\over \l_z^4},
\qquad C_{0,4,0}^{(c)}=
{1\over 8}{\l_u\l_v(2\l_z^2-9\l_u\l_v)\over \l_z^4}, \nn\\
&& C_{0,4,0}^{(d)}=-{1\over 4}{\l_u\l_v(\l_z^2-4\l_u\l_v)\over \l_z^4},
\qquad C_{0,4,0}^{(e)}={1\over 8}{\l_u^2\l_v^2\over \l_z^4}.
\eeal{nohyp}
Note that $C_{0,4,0}^{(a)}+C_{0,4,0}^{(b)}=0$ and $C_{0,4,0}^{(c)}+C_{0,4,0}^{(d)}+C_{0,4,0}^{(e)}=0$.

At degree $1$ hyperplane class there appear new configurations, which we will call type ${\rm III}$ graphs; they are obtained by adding 
to the first two graphs in figure \ref{unverii} degree $1$ hyperplane lines as shown in figure \ref{unveriii}.

\fig{unveriii}{Two crosscaps and one hyperplane at degree $4$ $\IR\IP^2$: type ${\rm III}$ graphs.}
{3.7in}{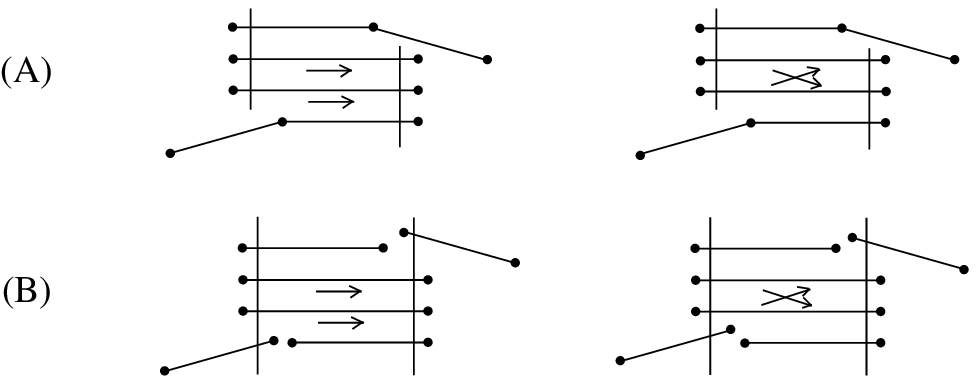}

\fig{unverivclosed}{Two crosscaps and one hyperplane at degree $4$ $\IR\IP^2$: type ${\rm I}$ graphs.}
{5.0in}{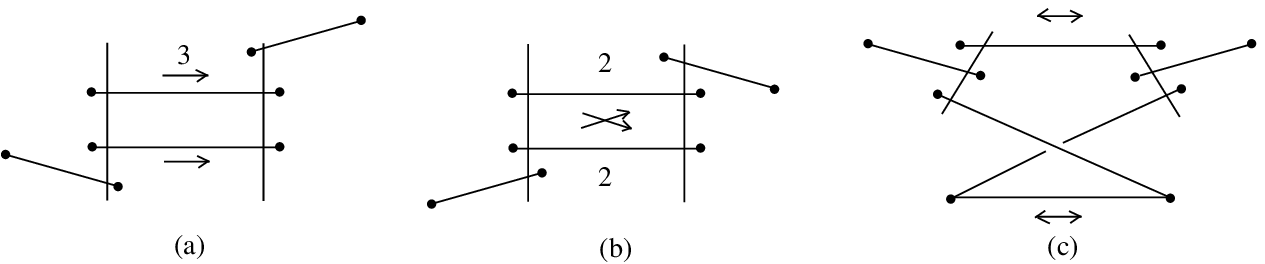}

Using again the sign rule in \cite{Diaconescu:2003dq}, the two graphs in each line of figure \ref{unveriii} add up to zero. We now turn 
to type ${\rm I}$ graphs; they
are presented in figure \ref{unverivclosed} and their contributions are:
\bea
&&C_{0,4,1}^{(a)}={1\over 2}(2-9{\l_u\l_v\over\l_z^2})({\l_u^2\over\l_z^2}+{\l_v^2\over\l_z^2}),~~~~~
C_{0,4,1}^{(b)}=(-1+4{\l_u\l_v\over\l_z^2})({\l_u^2\over\l_z^2}+{\l_v^2\over\l_z^2}),\nn\\
&&C_{0,4,1}^{(c)}={\l_u\l_v\over 2\l_z^2}({\l_u^2\over\l_z^2}+{\l_v^2\over\l_z^2}).
\eea
It is easy to check that $C_{0,4,1}^{(a)}+C_{0,4,1}^{(b)}+C_{0,4,1}^{(c)}=0$. This is in fact the same
cancellation that took place at degree $0$ hyperplane class between the contributions of the corresponding
three graphs. Again, we see that at degree $1$ hyperplane class there is nothing essentially new compared to
degree $0$ hyperplane class.

Let us now consider the case of degree $2$ hyperplane class. We can split the allowed configurations in graphs
of type ${\rm I}$ and ${\rm III}$ above. Configurations of type ${\rm III}$ are built by starting with the graphs $(a)$ and $(b)$
in figure \ref{unverii} and further adding in all possible ways degree $2$ graphs in $\IC \IP^2$. They
will always cancel in pairs. Configurations of type ${\rm I}$ are constructed by starting with
the graphs $(c)$, $(d)$ and $(e)$ in figure \ref{unverii}, performing a bubbling at a pair of identified nodes and inserting degree $2$ graphs
in $\IC \IP^2$. The contributions of the graphs with degree $2$ multicoverings of one of the hyperplane sections cancel as before; there also 
appear configurations as in figure \ref{unvervclosed}. However, their contributions also add up to zero, and this will be true for any quartet of 
type ${\rm I}$ graphs as in figure \ref{unvervclosed}. 

\fig{unvervclosed}{Two crosscaps and two hyperplanes at degree $4$ $\IR\IP^2$: type ${\rm I}$ graphs.}
{6.0in}{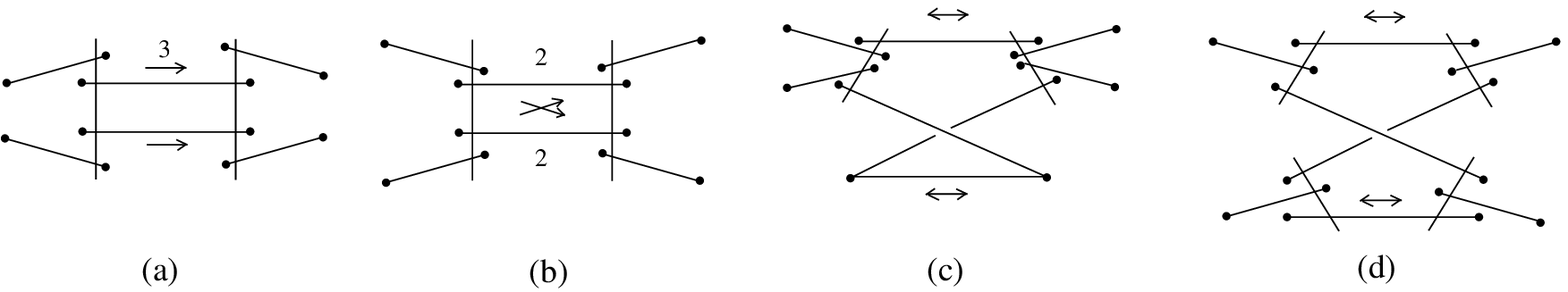}

\fig{unverviclosed}{Two crosscaps and two hyperplanes at degree $4$ $\IR\IP^2$: type ${\rm II}$ graphs.}
{5.0in}{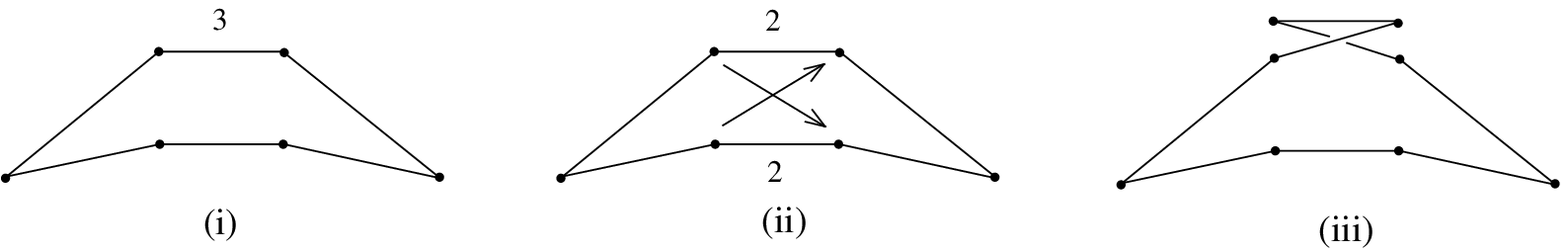}

There are also type ${\rm II}$ graphs, which are constructed by starting with the graphs $(c)$, $(d)$ and $(e)$ in figure \ref{unverii}. A triplet 
of such graphs is presented in figure \ref{unverviclosed}. The total contribution of these three graphs is 
\beq
C_{0,4,2}^{(i)}+C_{0,4,2}^{(ii)}+C_{0,4,2}^{(iii)}=-{(\l_u-\l_v)^2(\l_u+\l_v+\l_z)Q_9(\l_u,\l_v,\l_z)\over 4
\l_u^2\l_v^2\l_z^2(2\l_u+\l_z)^2(2\l_v+\l_z)^2(3\l_u+\l_z)(3\l_v+\l_z)},
\eeql{newtypeii}
where $Q_9(\l_u,\l_v,\l_z)$ is a degree $9$ homogeneous polynomial in $\l_u,\l_v,\l_z$. We recall that consistency 
of the anti-holomorphic involution 
with the torus action implies $\l_u+\l_v+\l_z=0$, and therefore the sum of the graphs in figure \ref{unverviclosed} is zero. This will also be true 
for the other possible triplet of type ${\rm II}$ graphs. 
We conclude that up to degree $2$ hyperplane class, the two crosscaps degree $4$ $\IR\IP^2$ Gromov--Witten invariants vanish.

At degree $3$ hyperplane class there appear all three types of graphs. We claim that the 
type ${\rm I}$ and ${\rm III}$ graphs sum up to zero, as above. Besides sets of type ${\rm II}$ graphs that have analogues at lower 
degree hyperplane class, and whose contributions add up to zero in a similar fashion, at degree $3$ hyperplane class there also are new 
collections of graphs. Such a set is presented in figure \ref{unveriVV}.

\fig{unveriVV}{Two crosscaps and three hyperplanes at degree $4$ $\IR\IP^2$: type ${\rm II}$ graphs.}
{5.0in}{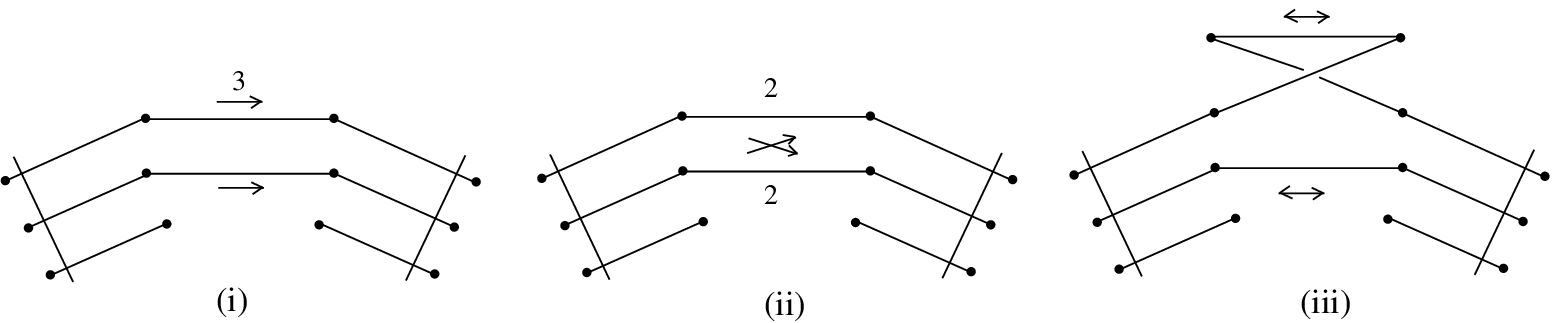}

The total contribution of the above three graphs is 
\beq
C_{0,4,3}^{(i)}+C_{0,4,3}^{(ii)}+C_{0,4,3}^{(iii)}={\l_u(\l_u+\l_v+\l_z)(\l_u-\l_v)^2(\l_u-2\l_v)^2Q_3(\l_u,\l_v,\l_z)\over 2
\l_z^2\l_v^4(2\l_v+\l_z)^2(3\l_v+\l_z)},
\eeq
where $Q_3(\l_u,\l_v,\l_z)$ is a degree $3$ homogeneous polynomial in $\l_u,\l_v,\l_z$. But $\l_u+\l_v+\l_z=0$, and these graphs 
sum up to zero.

\fig{unveriV}{Two crosscaps and three hyperplanes at degree $4$ $\IR\IP^2$: type ${\rm II}$ graphs.}
{5.0in}{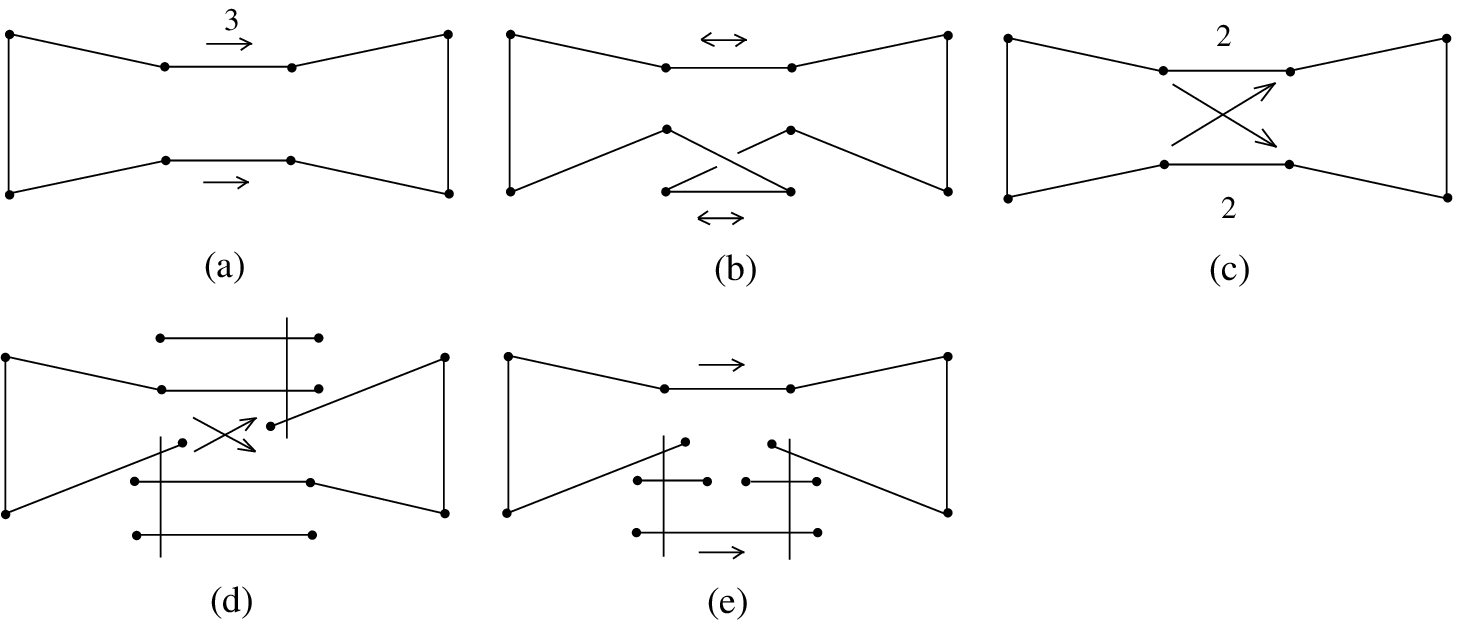}

However, at degree $3$ hyperplane class there is a unique set of type ${\rm II}$ graphs, presented in figure \ref{unveriV}, whose total 
contribution does not vanish. The contributions of these graphs are given by
\bea
&&C_{0,4,3}^{(a)}=1-3{\l_u\l_v\over\l_z^2},\quad C_{0,4,3}^{(b)}={\l_u\l_v\over\l_z^2},\quad C_{0,4,3}^{(c)}=2{\l_u\l_v\over\l_z^2},\nn\\
&&C_{0,4,3}^{(d)}=-1+2{\l_u\l_v\over\l_z^2},\quad C_{0,4,3}^{(e)}=1-2{\l_u\l_v\over\l_z^2}.
\eeal{disapp}
We see that the sum of the above expressions is equal to $1$, which is the Gromov-Witten invariant $n_{3,2}^{0,2}$ of table \ref{tabfunct2}. It is 
straightforward to perform now a similar computation but taking also into account the two $(-1,-1)$ curves that are transversal to the $\IC \IP^2$. 
The result is that at degree $3$ hyperplane class we obtain the following contribution to the free energy from $2$ crosscap configurations
\beq
{\cal F}^{0,2}_3=Q^2-q_1Q^2-q_2Q^2+q_1q_2Q^2.
\eeql{degftc}
This is in agreement with the Chern-Simons theory result presented in appendix \ref{resultfunct}.

\section{Topological Vertex Computation}\label{topover}

Using large $N$ duality, it was recently proposed \cite{Aganagic:2003db} that the free energy of
closed topological strings on a toric
manifold can be computed using a cubic field theory, namely a topological vertex
and gluing rules. In this section we present
a prescription to compute all genus topological string amplitudes
on orientifolds with an external
``${\IR\IP}^2$ leg" by using the topological vertex formalism. We will also explicitly
show that this prescription is equivalent to the large $N$ dual Chern-Simons computation.

\subsection{General Prescription}

\begin{figure}[htp]
\begin{center}
\psfrag{R}{$R$}
\psfrag{RP2}{$\IR \IP^2$}
\includegraphics[width=2in]{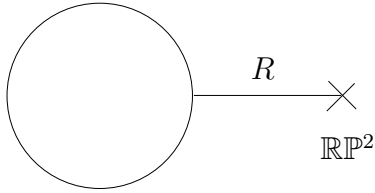}
\caption{Toric diagram for the quotient $X/I$ of a local, toric \CY\ manifold
with a single ${\IR\IP}^2$.}
\label{vertexfigclosed}
\end{center}
\end{figure}

Consider a quotient $X/I$ of a local, toric \CY\ manifold $X$ by an involution $I$
which can be represented as in figure \ref{vertexfigclosed}. We have a
bulk geometry, represented by the blob, attached to an ${\IR\IP}^2$ through an
edge associated to the representation $R$. Let us denote by
${\cal O}_{R}$ the amplitude
for the blob with the external leg. We propose
the following formula for the total partition function:
\beq
Z=\sum_{R=R^T} \CO_R  Q^{\ell(R)/2}
(-1)^{{1\over 2}(\ell(R)\mp r(R))}
\eeql{vertextotalclosed}
where the sum is over all self-conjugate
representations $R$. Here $r(R)$ denotes the rank of $R$, as
in \refeq{qdSONUN}, and $Q=e^{-s}$ is the exponentiated
K\"ahler parameter corresponding to the ${\IR\IP}^2$. The $\mp$ sign is correlated with the
choice of $\pm$ sign for the crosscaps, and corresponds to the $SO/Sp$ gauge duals,
respectively.

The prescription \refeq{vertextotalclosed} comes from the action of the involution $I$ on the partition function
on the covering space, which is given by
\beq
Z=\sum_{R} \CO_R (t_i) \CO_{R^T} (t_i) Q^{\ell(R)}
(-1)^{\ell(R)}.
\eeql{covering}
where the K\"ahler parameters have been identified in the way prescribed by the involution. 
The involution $I$ maps one half of the toric diagram onto the other half, reversing the orientation of the middle leg. 
The resulting partition function is the one given by \refeq{vertextotal}.

We are presently investigating in more details the origin of \refeq{vertextotalclosed}. Having a clear understanding of this formula will 
probably allow us to define a similar prescription for involutions with a fixed locus, like the $I^+$ of \cite{Acharya:2002ag}.

The restriction to self-conjugate representations may appear surprising at first sight. But in the topological vertex formalism, inverting the orientation of one edge sends $R$ to its transpose $R^T$ (and also introduces a factor of $(-1)^{\ell (R)}$). Therefore, since the ${\IR\IP}^2$ leg is unoriented, its partition function must sum only over self-conjugate representations, which are the only representations consistent with the involution $I$.

It is interesting to note that the formula \refeq{vertextotalclosed} is very similar to the formula for quantum dimensions of $SO/Sp$ gauge
group in terms of $U(N)$ quantum dimensions \refeq{qdSONUN}. Both formulas share the constraint $R=R^T$ and the factor of
$(-1)^{{1\over 2}(\ell(R)\mp r(R))}$. This gives a geometrical argument, from the topological vertex formalism, for the appearance of $SO/Sp$ gauge groups on the Chern-Simons side.

\subsection{Examples}

We now consider two examples of the above prescription. This will lead us to prove an identity between the topological vertex and $SO/Sp$ Chern-Simons expectation values of Hopf links.

\subsubsection{Orientifold of the Resolved Conifold}

The simplest example is the orientifold of the resolved conifold first considered in \cite{Sinha:2000ap}, which we reviewed in section \ref{geometry}.
In that case, the toric diagram is very simple and has been drawn in the left hand side of 
figure \ref{orienticon}. The rule \refeq{vertextotalclosed} gives:
\beq
F= -\log \biggl\{ \sum_{R=R^T}   C_{\cdot \cdot R} Q^{\ell(R)/2}
(-1)^{{1\over 2}(\ell(R)\mp r(R))} \biggr\} .
\eeql{firstex}
This should equal the free energy of Chern-Simons on the sphere for the gauge groups
$SO/Sp$ \refeq{sopcs}.

Using \refeq{topvertexschur}, \refeq{identityone} and \refeq{littlewood}, we can rewrite \refeq{firstex} as
\bea
F&=&-\log \biggl\{ \sum_{R=R^T} s(Q^{1/2}q^{-\rho}) (-1)^{{1\over 2}(\ell(R)\pm r(R))}\biggr\}\nn\\
&=&-\log \biggl\{ \prod_{i=1}^{\infty} (1\mp Q^{1/2} q^{i-1/2}) \prod_{1\leq i < j < \infty} (1-Q q^{i+j-1}) \biggr\}\nn\\
&=&-\log \biggl\{ {\prod_{i=1}^{\infty} (1\mp Q^{1/2} q^{i-1/2})^{1/2}  \prod_{i,j,=1}^{\infty} (1-Q q^{i+j-1})^{1/2}\over \prod_{i=1}^{\infty} (1\pm Q^{1/2} q^{i-1/2})^{1/2}} \biggr\}
\eeal{trivialii}
Using \refeq{expident} we find that it becomes
\bea
F &=& {1 \over 2} \sum_{n=1}^{\infty} {Q^{n} \over n (q^{n/2}-q^{-n/2})^2} \pm {1 \over 2} \sum_{n=1}^{\infty} {((-1)^n -1) Q^{n/2} \over n (q^{n/2}-q^{-n/2})}\nn\\
&=&  {1 \over 2} \sum_{n=1}^{\infty} {Q^{n} \over n (q^{n/2}-q^{-n/2})^2} \mp\sum_{n=1,~n~{\rm odd}}^{\infty} {Q^{n/2} \over n (q^{n/2}-q^{-n/2})} 
\eeal{trivialiii}
which is exactly $-\log S_{00}^{\rm{SO(N)/Sp(N)}}$ as given by \refeq{sopcs}.

\subsubsection{Local ${\IC \IP}^2$ Attached to ${\IR\IP}^2$}\label{localcp2}

The second example to consider is the local ${\IC \IP}^2$ attached to a single
${\IR\IP}^2$, whose toric diagram is drawn in figure \ref{ptworptwo}, and which was
discussed before from the point of view of
geometric transitions. The amplitude for this geometry is given by
\refeq{vertextotalclosed} with
\beq
{\cal O}_{R}= \sum_{R_i} q^{\sum_i \kappa_{R_i}} (-1)^{\sum_i \ell(R_i)}
 C_{\cdot R_3 R_1^T} C_{\cdot R_2 R_3^T} C_{R_1 R_2^T R}e^{-t \sum_i \ell(R_i)},
\eeql{orrp}
and $t$ is the K\"ahler parameter of the local ${\IC \IP}^2$. If we now compare
this expression to the one obtained by geometric transition in \refeq{prtwo}, we find that
both amplitudes are equal if
\beq
{1\over S_{00}^{SO(N)/Sp(N)}}\sum_{R=R^T}   C_{R_1 R_2^T R} Q^{\ell(R)/2}
(-1)^{{1\over 2}(\ell(R)\mp r(R))} = q^{-{\kappa_{R_2}\over 2}}
Q^{{1\over 2}(\ell(R_1) + \ell (R_2))}
{\cal W}^{SO(N)/Sp(N)}_{R_1 R_2},
\eeql{magicclosedstate}
where we have taken into account that the partition function of the geometry in figure \ref{ptworptwo} also
includes a $t$-independent piece which equals $S_{00}^{SO(N)/Sp(N)}$.
The r.h.s. of \refeq{magicclosedstate} involves the Hopf link invariant for the gauge groups $SO/Sp$,
where we put $\lambda=Q^{-1}$. Notice that $Q^{{1\over 2}(\ell(R_1) + \ell (R_2))}
{\cal W}^{SO(N)/Sp(N)}_{R_1 R_2}$ is a polynomial in $Q^{1\over 2}$, while
the l.h.s. of \refeq{magicclosedstate} is {\it a priori} an infinite series in $Q^{1\over 2}$.

\subsection{A Theorem Relating the Topological Vertex and $SO/Sp$ Chern-Simons Invariants}

We can now prove the following theorem that relates the topological vertex to $SO/Sp$ Chern-Simons invariants of the Hopf link.

\begin{theorem}\label{thetheo}
Let $X/I$ be the quotient of a local, toric \CY\ manifold $X$ by an involution $I$ without fixed points, which can be represented as in figure \ref{vertexfigclosed}. The topological vertex formula for the total partition function given by \refeq{vertextotalclosed} is equivalent to the total partition function obtained by large $N$ $SO/Sp$ geometric transition.
\end{theorem}

We saw in section \ref{localcp2} that in the case of the local $\IC \IP^2$ attached to $\IR \IP^2$, the theorem is proved if the identity \refeq{magicclosedstate} holds. In fact, it is straightforward to show that if \refeq{magicclosedstate} holds then the topological vertex rule given in the theorem to compute amplitudes on orientifolds agrees with the results of large $N$ $SO/Sp$ transitions for all the geometries of the form depicted in figure \ref{vertexfigclosed}. Therefore, to prove the theorem we only have to prove the following lemma.

\begin{lemma}\label{theidentity}
Let $C_{R_2^T R R_1}$ be the topological vertex of \cite{Aganagic:2003db} defined in \refeq{topvertexapp} and \refeq{topvertexschur}, 
${\cal W}^{SO(N)/Sp(N)}_{R_1 R_2}$ be the $SO$/$Sp$ Chern-Simons expectation value 
of the Hopf link with linking number $+1$ (after setting the Chern-Simons variable $\lambda$ 
defined in \refeq{lam} to be $\lambda=Q^{-1}$), 
and $S_{00}^{SO(N)/Sp(N)}$ be the partition function of $SO$/$Sp$ Chern-Simons theory on ${\bf S}^3$. Then the following identity holds:
\beq
{1\over S_{00}^{SO(N)/Sp(N)}}\sum_{R=R^T}   C_{R_1 R_2^T R} Q^{\ell(R)/2}
(-1)^{{1\over 2}(\ell(R)\mp r(R))} = q^{-{\kappa_{R_2}\over 2}}
Q^{{1\over 2}(\ell(R_1) + \ell (R_2))}
{\cal W}^{SO(N)/Sp(N)}_{R_1 R_2}.
\eeql{magicclosed}
\end{lemma}

We will only consider here the $Sp$ case for the sake of clarity, but the proof for the $SO$ case is similar.

Let us start by considering \refeq{magicclosed} for $R_1=\cdot$ being the trivial representation and $R_2=\mu$ being any representation. We must show that
\beq
{1 \over S_{00}^{{\rm Sp(N)}}} \sum_{R=R^T} C_{\cdot \mu^T R} Q^{{1 \over 2} (\ell(R)-\ell(\mu))} q^{\kappa_{\mu} \over 2} (-1)^{{1\over 2}(\ell(R)+r(R))}= \CW_{\mu}^{\rm Sp(N)}.
\eeql{trivonei}
Using \refeq{topvertexschur} and \refeq{identityone} the l.h.s can be rewritten as
\beq
{1 \over S_{00}^{{\rm Sp(N)}}} s_{\mu} (Q^{-1/2} q^{\rho}) \sum_{R=R^T} (-1)^{{1\over 2}(\ell(R)-r(R))} s_R (Q^{1/2} q^{-\ell(\mu)-\rho}).
\eeql{trivoneii}
From \refeq{littlewood}, the first line of \refeq{trivialii} and the definition of $W_R (q) = s_R (q^\rho)$ in terms of q-numbers (see for instance eq. (7.5) of \cite{Aganagic:2003db}), we find, after some algebra:
\bea
&&Q^{-\ell(\mu)/2} \prod_{1\leq i< j \leq d(\mu)} {[l_i+l_j-i-j+1]_{Q^{-1}} [l_i-l_j+j-i] \over [-i-j+1]_{Q^{-1}} [j-i]} \nn\\
&&\times \prod_{i=1}^{d(\mu)} {[1-i]_{Q^{-1}}^{\rm Sp(N)} [2l_i-2i+1]_{Q^{-1}} \over [l_i + 1-i]_{Q^{-1}}^{\rm Sp(N)} [-2i+1]_{Q^{-1}}} \prod_{v=1}^{l_i} Q^{1/2} {[l_i-i-v-d(\mu)+1]_{Q^{-1}} \over [v-i+d(\mu)]}
\eeal{trivoneiii}
where $d(\mu)$ is the number of rows of $\mu$, and we used the $q$-numbers defined in \refeq{qnumber}, \refeq{qnumberii} and \refeq{qnumberiv}. One can see that the two factors of $Q$ cancel out of \refeq{trivoneiii}, and the remaining expression is exactly the definition of the $Sp(N)$ quantum dimension of $\mu$ for $\lambda = Q^{-1}$, as given by \refeq{qdSON}. But $\CW_{\mu}^{\rm Sp(N)} = \dim_q^{\rm Sp(N)} \mu (\lambda=Q^{-1})$. Therefore \refeq{trivonei} is proved.

We are now in position to prove \refeq{magicclosed} in the general case, namely we have to show that
\beq
{1\over S_{00}^{Sp(N)}}\sum_{R=R^T}   C_{R_1 R_2^T R} Q^{\ell(R)/2}
(-1)^{{1\over 2}(\ell(R) + r(R))} = q^{-{\kappa_{R_2}\over 2}}
Q^{{1\over 2}(\ell(R_1) + \ell (R_2))}
{\cal W}^{Sp(N)}_{R_1 R_2}.
\eeql{magicii}
Let us first rewrite the Hopf link expectation value in terms of quantum dimensions, using \refeq{hopflink}. The r.h.s. becomes:
\beq
\sum_{\mu,\lambda_1,\lambda_2,\lambda_3} N^{R_1}_{\lambda_1 \lambda_2} N^{R_2}_{\lambda_1 \lambda_3} N^{\mu}_{\lambda_2 \lambda_3} q^{{1\over2}(\kappa_{R_1}-\kappa_{\mu})}
Q^{{1\over 2}(\ell(\mu))} {\cal W}^{Sp(N)}_{\mu},
\eeql{hopfquant}
where we expressed the $Sp$ tensor product coefficients in terms of Littlewood-Richardson coefficients using \refeq{kingfor}. We can now rewrite the r.h.s. using \refeq{trivonei} and \refeq{topvertexschur} as 
\bea
{1\over S_{00}^{Sp(N)}}&& \sum_{R=R^T} Q^{{1\over2}\ell(R)} (-1)^{{1\over 2}(\ell(R)+ r(R))}s_{R^T} (q^{\rho})  q^{{1\over2}(\kappa_{R_1}+\kappa_{R})}\nn\\
&&\times \sum_{\mu,\lambda_1,\lambda_2,\lambda_3} N^{R_1}_{\lambda_1 \lambda_2} N^{R_2}_{\lambda_1 \lambda_3} N^{\mu}_{\lambda_2 \lambda_3} s_{\mu^T}(q^{\ell(R)+\rho})
\eeal{rhsschur}
The sum in the second line can be explicitely evaluated by using \refeq{identityone}, \refeq{identitytwo}, the definition of skew Schur functions \refeq{skewschur} and the fact that $s_{R_1} (x) s_{R_2} (x) = \sum_{R} N^{R}_{R_1 R_2} s_R (x)$:
\beq
\sum_{\mu,\lambda_1,\lambda_2,\lambda_3} N^{R_1}_{\lambda_1 \lambda_2} N^{R_2}_{\lambda_1 \lambda_3} N^{\mu}_{\lambda_2 \lambda_3} s_{\mu^T}(q^{\ell(R)+\rho})
 =\sum_{\lambda_1} s_{R_1^T/\lambda_1^T} (q^{\ell(R)+\rho}) s_{R_2^T/\lambda_1^T} (q^{\ell(R) + \rho})
\eeql{sumeval}
Inserting \refeq{sumeval} in \refeq{rhsschur} gives (using the fact that $R=R^T$):
\bea
{1\over S_{00}^{Sp(N)}} && \sum_{R=R^T} Q^{{1\over2}\ell(R)} (-1)^{{1\over 2}(\ell(R) + r(R))} \nn\\
 &&\times {\left[ q^{{1\over2}(\kappa_{R_1}+\kappa_{R})}s_{R^T} (q^{\rho}) \sum_{\lambda_1}s_{R_2^T/\lambda_1} (q^{\ell(R^T) + \rho}) s_{R_1^T/\lambda_1} (q^{\ell(R)+\rho})  \right]}
\eeal{rhsfinal}
The term in brackets is exactly the definition of $C_{R_2^T R R_1}$ in terms of Schur functions (see \refeq{topvertexschur}). Therefore \refeq{rhsfinal} is equal to the l.h.s. of \refeq{magicii} and lemma \ref{theidentity} is proved.

As we explained lemma \ref{theidentity} implies theorem \ref{thetheo}; thus we have just shown that the topological vertex prescription \refeq{vertextotalclosed} is equivalent to computations using large $N$ $SO/Sp$ transitions. We will use extensively the prescription \refeq{vertextotalclosed} and the identity \refeq{magicclosed} in chapter \ref{open}.

\chapter[Open Topological Strings on Orientifolds]{Open Topological Strings on Orientifolds}\label{open}

In this chapter we study open topological strings on orientifolds of toric Calabi-Yau threefolds. In other words, we consider orientifolds of toric Calabi-Yau threefolds with D-branes.

We first explore the BPS content of open topological string amplitudes on orientifolds and formulate their structural properties. We then compute explicitely the amplitudes for various examples in section \ref{examplesopen}: the $SO/Sp$ framed unknot, the $SO/Sp$ framed Hopf link, and an outer brane in 
$\IC \IP^2$ attached to $\IR \IP^2$. The independent localization computations we provide for all these examples corroborate our methods 
and proposals. In section \ref{kauffman} we apply our results to formulate structural properties of the coloured Kauffman polynomial.

\section{Open Topological String Amplitudes in Orientifolds}\label{structure}

\subsection{BPS Structure of Topological String Amplitudes}

We saw in section \ref{structopclosed} that closed topological string amplitudes have an integrality structure which expresses them in terms of numbers of BPS states. The aim of this section is to formulate the BPS structure of open topological string amplitudes. First, let us briefly review the known results for open strings. 

The structure of the open string amplitudes was found in \cite{Ooguri:1999bv,Labastida:2000yw} and is much more delicate than in the closed string case -- which was explored in section \ref{structopclosed}.  
To define an open string amplitude we have to specify boundary conditions through a set of submanifolds of $X$, $S_1, 
\cdots, S_L$. To each of these submanifolds we associate a source $V_{\ell}$, $\ell=1, \cdots, L$, which is 
a $U(M)$ matrix. The total partition function is given by
\beq
Z(V_1, \cdots, V_L)=\sum_{R_1, \cdots, R_L}Z_{(R_1, \cdots, R_L)}
\prod_{\alpha=1}^L {\rm Tr}_{R_{\alpha}} V_{\alpha},
\eeql{partf}
where $R_{\alpha}$ denote representations of $U(M)$ and we are considering
the limit $M\rightarrow \infty$. 
The amplitudes $Z_{(R_1, \cdots, R_L)}$ can be computed in the noncompact, toric case by using 
the topological vertex \cite{Aganagic:2003db}. According to the correspondence proposed in \cite{Ooguri:1999bv}, they are given 
in some cases by invariants of links whose components are coloured by representations $R_1, \cdots, R_L$. 
The free energy is defined as usual by 
\beq
\CF(V_1, \cdots, V_L)= - \log \, Z(V_1, \cdots, V_L)
\eeql{freeopen}
and is understood as a series in traces of $V$ in different representations.   
We define the generating function
$f_{(R_1, \cdots, R_L)} (q, \lambda)$ through the
following equation:
\beq
{\cal F}(V)= - \sum_{n=1}^\infty \sum_{R_1, \cdots, R_L}
{1\over n} f_{(R_1, \cdots, R_L)} (q^n, {\rm e}^{-n t})
\prod_{\alpha=1}^L {\rm Tr}_{R_\alpha}V_{\alpha}^n
\eeql{genfunctopen}
The main result of \cite{Labastida:2000yw} is that 
$f_{(R_1, \cdots, R_L)} (q, e^{-t})$ is given by:
\bea
f_{(R_1, \cdots, R_L)}(q, e^{-t})&=&(q^{1\over 2}-q^{-{1 \over 2}})^{L-2}\sum_{g\ge 0} \sum_{\beta}
\sum_{R'_1, R_1'' \cdots, R'_L, R_L''}  \prod_{\alpha=1}^L
c_{R_{\alpha}\,R_{\alpha}'\,R_{\alpha}''}S_{R_{\alpha}'}(q)
\nn\\ 
&&~~~~~~~~~~\times N_{(R_1'', \cdots, R_L''),g,\beta}(q^{1\over 2}-q^{-{1\over 2}})^{2g} {\rm e}^{-\beta\cdot t}.
\eeal{fr}
In this formula $R_{\alpha},R_{\alpha}',R_{\alpha}''$ label
representations of the symmetric group
$S_\ell$, which can be labeled
by a Young tableau with a total of $\ell$ boxes.
$c_{R\,R'\,R''}$ are the Clebsch-Gordon coefficients
of the symmetric group, and the monomials $S_R (q)$ are defined as
follows. If $R$ is a hook representation
\beq
\yng(6,1,1,1,1)
\eeql{hookii}
with $\ell$ boxes in total, and with $\ell-d$ boxes in the first row, then
\beq
S_R (q)=(-1)^d q^{ -{\ell -1 \over 2}+d} ,
\eeql{expsr}
and it is zero otherwise.
Finally, $N_{(R_1, \cdots, R_L),g,\beta}$ are {\it integers}
associated to open string amplitudes. They compute the net number
of BPS domain walls of charge $\beta$ and spin $g$ transforming
in the representations $R_{\alpha}$ of $U(M)$, where we are using
the fact that representations of $U(M)$ can also be labeled by Young
tableaux. It is also useful to introduce a generating functional for these
degeneracies as in \cite{Labastida:2000yw}:
\beq
{\widehat f}_{(R_1, \cdots, R_L)}(q, e^{-t})=
\sum_{g \ge 0}\sum_\beta N_{(R_1, \cdots, R_L),g,\beta}
(q^{{1\over 2}}-q^{-{1 \over 2}})^{2g+L-2}{\rm e}^{-\beta\cdot t}.
\eeql{tildefrlinks}
We then have the relation:
\beq
 f_{(R_1, \cdots, R_L)}(q,e^{-t})=
\sum_{R_1', \cdots, R'_L}\prod_{\alpha=1}^L 
M_{R_{\alpha} R'_{\alpha}}(q)
{\widehat f}_{(R_1, \cdots, R_L)}(q,e^{-t} ),
\eeql{relafslinks}
where the matrix $M_{R R'}(q)$ is given by
\beq
M_{R R'}(q)= \sum_{R''} c_{R\,R'\,R''}S_{R''}(q)
\eeq
and it is symmetric and invertible \cite{Labastida:2000yw}. 
The $f_{(R_1, \cdots, R_L)}$ introduced in \refeq{conj} can be extracted
from $Z_{(R_1, \cdots, R_L)}$ through a procedure spelled out in detail in
\cite{Labastida:2000zp,Labastida:2001ts,Labastida:2000yw}. One has, for example,
\beq
f_{\yng(1) \yng(1)}=
Z_{\yng(1) \yng(1)}-Z_{\yng(1) \cdot} Z_{\cdot \yng(1)},
\eeql{exm}
where $\cdot$ denotes the trivial representation. 
As it was emphasized in \cite{Labastida:2000zp,Labastida:2001ts,Labastida:2000yw}, this structure result has interesting 
consequences for knot theory, since it implies a series of integrality results for 
knot and link invariants. We will come back to this issue in section \ref{kauffman}.

\subsection{BPS Structure of Topological Strings on Orientifolds}\label{bpsstructorient}

We want to understand now the corresponding BPS structure of open topological string 
amplitudes on orientifolds without fixed points, like the ones considered in \cite{Sinha:2000ap,Acharya:2002ag}. 
In chapter \ref{closed} the closed case was studied in detail, in the noncompact case, by using large $N$ 
transitions and the topological vertex. In particular, the BPS structure was formulated in section \ref{structopclosed} Let us now recall some of the results that we found.

Let us denote by $X/I$ the orientifold obtained by 
an involution on $X$. The total free energy has in this case the 
structure
\beq
\CF(X/I, g_s)= {1\over 2}\CF(X, g_s) + \CF(X/I, g_s)_{\rm unor},
\eeql{strucf}
where $g_s$ is the string coupling constant. In the r.h.s. of this equation, 
the first summand is the contribution of the untwisted sector,
and it involves the free energy $\CF(X, g_s)$ of the covering $X$ of $X/I$, 
after suitably identifying
the K\"ahler classes in the way prescribed by the involution $I$. This piece of 
the free energy has the expansion given by \refeq{orcont}. The second 
summand, that we call the unoriented part $\CF( X/I, g_s)_{\rm unor}$, is the contribution of the twisted sector, 
and involves the counting of holomorphic maps from
closed non-orientable Riemann surfaces to the orientifold $X/I$.
The Euler characteristic of a closed Riemann surface of genus $g$ and $c$ crosscaps
is $\chi = -2g+2-c$ where $c$ is the number of crosscaps. We then have
\beq
\CF(X/I, g_s)_{\rm unor}= \CF(X/I, g_s)_{\rm unor}^{c=1}+ \CF(X/I, g_s)_{\rm unor}^{c=2},
\eeql{strucunchap5}
which corresponds to the contributions of one and two crosscaps. Following
the arguments in \cite{Gopakumar:1998jq} we predict the following structure
\bea
\CF(X/I, g_s)_{\rm unor}^{c=1}&=&
\pm \sum_{d \, \, {\rm odd}}\sum_{g=0}^{\infty}\sum_{\beta }  n^{g,c=1}_\beta
{1\over d}(q^{d\over 2} - q^{-{d\over 2}})^{2g-1}e^{-d \beta\cdot t},\nn\\
\CF(X/I, g_s)_{\rm unor}^{c=2}&=&
\sum_{d \, \, {\rm odd}} \sum_{g=0}^{\infty}\sum_{\beta }
n^{g,c=2}_\beta {1\over d}(q^{d\over 2} - q^{-{d\over 2}})^{2g}e^{-d \beta\cdot t},\nn\\
\eeal{strucuntwo}
where $n^{g,c}_\beta$ are integers. The $\pm$ sign in the $c=1$ free energy is due to
the two different choices for the sign of the crosscaps, and the restriction to 
$d$ odd comes, in the case of $c=1$, from the geometric absence of even 
multicoverings. In the $c=2$ case this was concluded from examination of 
different examples. The structure results in \refeq{strucf}, \refeq{strucunchap5} and \refeq{strucuntwo} were tested in chapter \ref{closed} through detailed 
computations in noncompact geometries.  

We now address the generalization to open string amplitudes in orientifolds. We first consider 
for simplicity the case of a single boundary condition in the orientifold $X/I$ associated to a topological 
D-brane wrapping a submanifold 
$S$. As in the closed string case, the total open string amplitude will have a contribution from 
untwisted sectors, and a contribution from twisted sectors. We will then write
\beq
\CF(V)={1\over 2} \CF_{\rm or} (V) + \CF_{\rm unor}(V),
\eeql{freen}
The contribution from the untwisted sector, $F_{\rm or} (V)$, 
involves the covering geometry, which will be given by $X$, the submanifold $S$, and its image under the involution 
$I(S)$. In other words, the covering amplitude will involve now {\it two} different sets of D-branes, in general. The 
covering geometry with two sets of branes has the total partition function
\beq
Z_{\rm cov}(V_1, V_2)=\sum_{R_1, R_2} {\cal C}_{R_1 R_2} {\rm Tr}_{R_1}V_1 \, {\rm Tr}_{R_2} V_2,
\eeql{covam}
where $V_1$, $V_2$ are the sources corresponding to $S$ and $I(S)$ and represent open string 
moduli. Since the two D-branes in $S$ and $I(S)$ are related by an involution, the two-brane 
amplitude in \refeq{covam} is symmetric under their exchange, {\it i.e.} we have 
\beq
{\cal C}_{R_1 R_2}={\cal C}_{R_2 R_1}.
\eeql{sime}
In order to obtain ${\cal F}_{\rm or}(V)$ we have to make the identification of {\it both} closed and open 
string moduli under the involution $I$. This means identifying the K\"ahler parameters that appear in 
${\cal C}_{R_1 R_2}$ (the closed background) but also setting $V_1=V_2=V$ (the open background). We then find
\beq
Z_{\rm or}(V)=\sum_R Z_R^{\rm or} {\rm Tr}_R V
\eeql{zcov}
where
\beq
Z_R^{\rm or}=\sum_{R_1, R_2} N^R_{R_1 R_2} {\cal C}_{R_1 R_2} =\sum_{R'} {\cal C}_{R/R' R'}.
\eeql{zror}
Here we have used that
\beq
{\rm Tr}_{R_1}V \, {\rm Tr}_{R_2} V =\sum_{R}N^R_{R_1 R_2} \, {\rm Tr}_{R} V
\eeql{tensorp}
and $N^R_{R_1 R_2}$ are tensor product coefficients. In \refeq{zror} we also used these 
coefficients to define skew coefficients with labels $R/R'$, as in \refeq{skewschur}. 
If we denote ${\cal C}_R \equiv {\cal C}_{R\cdot}$, we 
have for example
\beq
Z_{\yng(1)}^{\rm or} =2\, {\cal C}_{\yng(1)},\quad
Z_{\yng(2)}^{\rm or} = 2 \, {\cal C}_{\yng(2)} +
{\cal C}_{\yng(1) \yng(1)},\quad
Z_{\yng(1,1)}^{\rm or} =2{\cal C}_{\yng(1,1)}+
{\cal C}_{\yng(1) \yng(1)}.
\eeql{listz}
It turns out that the quantities $Z^{\rm or}_R$ defined in this way 
have the integrality properties of a one-brane 
amplitude, as it should. One finds, for example,
\beq
{\widehat f}_{\yng(1)}^{\rm or} =2 \widehat f^{\rm cov}_{\yng(1) \cdot},\quad
{\widehat f}_{\yng(2)}^{\rm or} = 2 \widehat f^{\rm cov}_{\yng(2) \cdot}-{1\over
q^{1\over 2} -q^{-{1\over 2}}} \widehat f^{\rm cov}_{\yng(1) \yng(1)}, \quad
{\widehat f}_{\yng(1,1)}^{\rm or} =2 \widehat f^{\rm cov}_{\yng(1,1) \cdot}-{1\over
q^{1\over 2} -q^{-{1\over 2}}} \widehat f^{\rm cov}_{\yng(1) \yng(1)}. \quad
\eeql{listf}
In these equations, the superscript ``${\rm cov}$'' refers to quantities computed from the two-brane 
amplitude ${\cal C}_{R_1 R_2}$ according to the general rules for open string amplitudes in the usual, oriented 
case explained above. One can easily verify from the integrality properties of $\widehat f_{R_1 R_2}$ as a
2-brane amplitude that indeed
${\widehat f}_R^{\rm or}$ has the integrality properties of a one-brane
amplitude. In fact, using the identity
\beq
\sum_{R', R_1', R_2'} M_{R R'}^{-1} N^{R'}_{R_1' R_2'} M_{R'_1 R_1} M_{R'_2 R_2}= 
{1\over q^{-{1\over 2}}- q^{1\over 2}} N^R_{R_1 R_2}
\eeql{idef}
we can write
\beq
{\widehat f}_R^{\rm or}=\sum_{R_1 R_2} N^R_{R_1 R_2} {\widehat f}^{\rm cov}_{R_1 R_2},
\eeql{covint}
where we put ${\widehat f}_{R \cdot}\equiv (q^{-{1\over 2}}- q^{1\over 2}) {\widehat f}_R$. 

We would like to determine now the structural properties of $\CF_{\rm unor}(V)$. This is indeed very easy. 
The analysis of \cite{Labastida:2000yw} to determine the structural properties of $F(V)$ in the usual oriented case was based 
on an analysis of the Hilbert space associated to an oriented Riemann surface $\Sigma_{g,\ell}$ with $\ell$ 
holes ending on $S$ and in the relative homology class $\beta \in H_2(X,S)$. 
The relevant Hilbert space turns out to be
\beq
{\rm Sym}\bigl(F^{\otimes \ell} \otimes H^*(J_{g,\ell}) \otimes
H^*({\cal M}_{g,\ell, \beta})\bigr)
\eeql{hilbert}
where $J_{g,\ell}={\bf T}^{2g+\ell-1}$ is the Jacobian of $\Sigma_{g,\ell}$, $F$ is a 
copy of the fundamental representation of the gauge group, ${\cal M}_{g,\ell, \beta}$ is the 
moduli space of geometric deformations of the Riemann surface inside the Calabi--Yau manifold, and ${\rm Sym}$ 
means that we take the completely symmetric piece with respect to permutations of the 
$\ell$ holes. Since the bulk of the Riemann surface is not relevant for the action of 
the permutation group, we can
factor out the cohomology of the Jacobian ${\bf T}^{2g}$. The projection onto the
symmetric piece can easily be done using the Clebsch-Gordon
coefficients $c_{R\,R'\,R''}$ of the permutation group $S_{\ell}$ \cite{Fulton:1997}, 
and one finds 
\beq
\sum_{R\,R'\,R''} c_{R\,R' \, R''} {\bf
S}_R(F^{\otimes \ell})\otimes{\bf S}_{R'}(H^*(({\bf
S}^1)^{\ell-1})) \otimes {\bf S}_{R''}(H^*({\cal M}_{g,\ell,\beta}))
\eeql{decomhilbert}
where ${\bf S}_R$ is the Schur functor that projects onto
the corresponding subspace. The space ${\bf S}_R(F^{\otimes
\ell})$ is nothing but the vector space underlying the irreducible
representation $R$ of $U(M)$. ${\bf S}_{R'}(H^*(({\bf
S}^1)^{\ell-1}))$ gives the hook Young tableau, and the Euler
characteristic of ${\bf S}_{R''}(H^*({\cal M}_{g,\ell, \beta}))$ is
the integer invariant $N_{R'',g,\beta}$. Therefore, the above
decomposition corresponds very precisely to \refeq{fr} (here we are considering 
for simplicity the one-brane case). 

In the case of an {\it unoriented} Riemann surface, the above argument goes through, 
with the only difference that now the Jacobian is $J_{g,c,\ell}={\bf T}^{2g-1+\ell +c}$, 
where $c=1,2$ denotes the number of crosscaps. Therefore, the analysis of the cohomology associated 
to the boundary is the same. We then conclude that
\beq
\CF_{\rm unor}(V)= - \sum_R \sum_{d \, {\rm odd}}{1\over d}f_R^{\rm unor} (q^d, e^{-d t})
{\rm Tr}_R V^d,
\eeql{freeunor}
and using again \refeq{relafslinks} one can obtain new functions
\beq
\widehat f^{\rm unor}_R=\sum_{R'} M^{-1}_{R R'} f_{R'}^{\rm unor}
\eeql{unorinv}
with 
contributions from one and two crosscaps: 
\beq
{\widehat f}^{\rm unor}_R={\widehat f}^{c=1}_R+ (q^{1\over 2} -q ^{-{1\over 2}}){\widehat f}^{c=2}_R,
\eeql{ccont}
and we finally have
\beq
\widehat f_R^c (q, e^{-t})=\sum_{g,\beta} N^c_{R,g,\beta} (q^{1\over 2} -q ^{-{1\over 2}})^{2g}
e^{-\beta\cdot t}.
\eeql{cs}
Each crosscap contributes then an extra factor of $q^{1\over 2} -q^{-{1\over 2}}$, as in the closed case.

In real life, what one computes is the total amplitude in the l.h.s. of \refeq{freen}, in terms of 
\beq
{\cal F}(V)=- \log Z(V)=- \log \Bigr( \sum_R Z_R {\rm Tr}_R \, V \Bigl),
\eeql{totalf}
and one wants to find the unoriented contribution to the amplitude after subtracting the oriented 
contribution. The above formulae give a precise prescription to compute $f_R^{\rm 
unor}$. The results one finds, up to three boxes, 
are the following:
\bea
f^{\rm unor}_{\yng(1)}&=&Z_{\yng(1)}-{\cal C}_{\yng(1)}, \nn\\
f^{\rm unor}_{\yng(2)}&=&Z_{\yng(2)}-{1\over 2}Z_{\yng(1)}^2 -{\cal C}_{\yng(2)} +{1\over 2}
{\cal C}_{\yng(1)}^2 
-{1\over 2}f^{\rm cov}_{\yng(1) \yng(1)}, \nn\\
f^{\rm unor}_{\yng(1,1)}&=&Z_{\yng(1,1)}- {1\over 2}Z_{\yng(1)}^2 -{\cal C}_{\yng(1,1)} +
{1\over 2}{\cal C}_{\yng(1)}^2-{1\over 2}f^{\rm cov}_{\yng(1) \yng(1)},
\eeal{threeres}
and
\bea
f^{\rm unor}_{\yng(3)}&=&Z_{\yng(3)}-Z_{\yng(2)} Z_{\yng(1)} +{1 \over 3} Z^3_{\yng(1)}-{\cal C}_{\yng(3)}
+{\cal C}_{\yng(2)} {\cal C}_{\yng(1)} -{1 \over 3} {\cal C}^3_{\yng(1)} \nn\\
&&-{1\over 3} f^{\rm unor}_{\yng(1)}(q^3, Q^3)-f^{\rm cov}_{\yng(2) \yng(1)},\nn\\
f^{\rm unor}_{\yng(2,1)}&=&Z_{\yng(2,1)}-Z_{\yng(2)} Z_{\yng(1)} -Z_{\yng(1,1)} Z_{\yng(1)} 
+{2 \over 3} Z^3_{\yng(1)}-{\cal C}_{\yng(2,1)}+{\cal C}_{\yng(2)} {\cal C}_{\yng(1)} +{\cal C}_{\yng(1,1)} 
{\cal C}_{\yng(1)} +{2 \over 3} {\cal C}^3_{\yng(1)}\nn\\ 
&&+{1\over 3} f_{\yng(1)}^{\rm unor} (q^3,Q^3) - {1\over 2}
(f^{\rm cov}_{\yng(2) \yng(1) } +f^{\rm cov}_{\yng(1,1) \yng(1)}),\nn\\
f^{\rm unor}_{\yng(1,1,1)}&=&Z_{\yng(1,1,1)}-Z_{\yng(1,1)} Z_{\yng(1)} +{1 \over 3} Z^3_{\yng(1)}
-{\cal C}_{\yng(1,1,1)}-{\cal C}_{\yng(1,1)} {\cal C}_{\yng(1)} +{1 \over 3} {\cal C}^3_{\yng(1)}\nn\\
&&-{1\over 3} f^{\rm unor}_{\yng(1)}(q^3, Q^3)-f^{\rm cov}_{\yng(1,1) \yng(1) }.
\eeal{threeresi}

The above considerations are easily extended to the case in which we have $L$ sets of 
D-branes in the orientifold geometry. The covering 
amplitude involves now $2L$ D-branes, and reads
\beq
Z_{\rm cov}=\sum_{R_1, S_1, \cdots, R_L, S_L} {\cal C}_{R_1 S_1 \cdots R_L S_L} {\rm Tr}_{R_1}V_1{\rm Tr}_{S_1}W_1  
\cdots {\rm Tr}_{R_L}V_1{\rm Tr}_{S_L} 
W_L.
\eeql{covaL}
The oriented amplitude is obtained by identifying the moduli in pairs under $I$, and is given by
\beq
Z_{Q_1 \cdots Q_L}^{\rm or}=\sum_{R_i, S_i} N^{Q_1}_{R_1 S_1} \cdots N^{Q_L}_{R_L S_L} {\cal C}_{R_1 S_1 \cdots R_L S_L} .
\eeql{zrorL}
The equations \refeq{freen}, \refeq{freeunor} and \refeq{cs} generalize in an obvious way, but now we have
\beq
\widehat f_{(R_1 \cdots R_L)}^c 
(q, e^{-t})=\sum_{g,\beta} N^c_{(R_1, \cdots, R_L),g,\beta} (q^{1\over 2} -q ^{-{1\over 2}})^{2g+L-1}
e^{-\beta\cdot t},
\eeql{linkc}
where the extra $L-1$ factors of $q^{1\over 2} -q ^{-{1\over 2}}$ have the same origin as in \refeq{tildefrlinks}.

\section{Examples of Open String Amplitudes} \label{examplesopen}

In this section we study in detail some examples and verify the above formulae for the 
unoriented part of the free energy. In order to do that, we have to compute the total amplitudes 
$Z_R$ in orientifold geometries. These amplitudes can be obtained in three ways: by using the unoriented 
localization methods of developed in section \ref{localclosed} and in \cite{Diaconescu:2003dq}, by using mirror symmetry \cite{Acharya:2002ag}, and by using Chern-Simons theory 
and the topological vertex. For the examples of open string amplitudes studied in this section 
we will use the topological vertex of \cite{Aganagic:2003db}, which can be adapted to the orientifold case as was proposed in section \ref{topover}, and 
also localization. We first summarize very briefly the results of section \ref{topover} on the topological vertex on orientifolds, 
and then we study in detail three examples. Finally we check some of the topological vertex results with 
unoriented localization.   

\subsection{The topological vertex on orientifolds}

Let us consider a quotient $X/I$ of a local, toric Calabi-Yau manifold $X$ by an involution $I$
without fixed points, as was represented in figure \ref{vertexfigclosed}. Theorem \ref{thetheo} tells us that the following topological vertex formula for the total partition function is equivalent to the large $N$ Chern-Simons result:
\beq
Z=\sum_{S=S^T} \CO_S  Q^{\ell(S)/2}
(-1)^{{1\over 2}(\ell(S)\mp r(S))}
\eeql{vertextotal}
where the sum is over all self-conjugate
representations $S$. In section \ref{topover} the above prescription was used to compute closed 
string amplitudes. Moreover, to prove Theorem \ref{thetheo}, we proved the following identity, which was stated in Lemma \ref{theidentity}:
\beq
{1\over S_{00}^{SO(N)/Sp(N)}}\sum_{R=R^T}   C_{R_1 R_2^T R} Q^{\ell(R)/2}
(-1)^{{1\over 2}(\ell(R)\mp r(R))} = q^{-{\kappa_{R_2}\over 2}}
Q^{{1\over 2}(\ell(R_1) + \ell (R_2))}
{\cal W}^{SO(N)/Sp(N)}_{R_1 R_2},
\eeql{magic}
 where $C_{R_2^T R R_1}$ is the topological vertex of \cite{Aganagic:2003db} defined in \refeq{topvertexapp} and \refeq{topvertexschur}, 
${\cal W}^{SO(N)/Sp(N)}_{R_1 R_2}$ is the $SO$/$Sp$ Chern-Simons expectation value 
of the Hopf link with linking number $+1$ (after setting the Chern-Simons variable $\lambda$ 
defined in \refeq{lam} to be $\lambda=Q^{-1}$), 
and $S_{00}^{SO(N)/Sp(N)}$ is the partition function of $SO$/$Sp$ Chern-Simons theory on ${\bf S}^3$. In the 
examples that follow we will use \refeq{vertextotal} to compute open string amplitudes on orientifolds, 
making use as well of the identity \refeq{magic}.

\subsection{The $SO/Sp$ framed unknot}

\begin{figure}[htp]
\begin{center}
\psfrag{R}{$R$}
\psfrag{RP2}{$\IR \IP^2$}
\includegraphics[width=2truein]{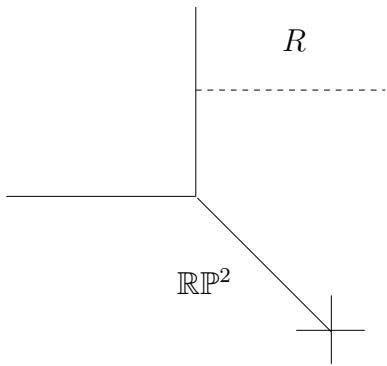}
\caption{A D-brane in an outer leg of the orientifold of 
the resolved conifold.}
\label{dorienti}
\end{center}
\end{figure}

We start again with the simplest non-trivial Calabi-Yau orientifold, namely the orientifold of the 
resolved conifold. This geometry was studied in section \ref{geometry}.

Let us now put a D-brane in an {\it outer} leg of the orientifold geometry. In the oriented case, the open string 
amplitude labelled by $R$ is computed by the Chern-Simons invariant of 
the framed unknot with gauge group $U(N)$ (see for example \cite{Marino:2004uf,Marino:2001re}). 
We want to study now the unoriented case. In order to extract the unoriented 
string amplitudes, we have to compute both the total amplitudes $Z_R$ and the 
covering amplitudes ${\cal C}_{R_1 R_2}$. Let us start analyzing the covering amplitude. 
 
\begin{figure}[htp]
\begin{center}
\psfrag{R1}{$R_1$}
\psfrag{R2}{$R_2$}
\includegraphics[width=3truein]{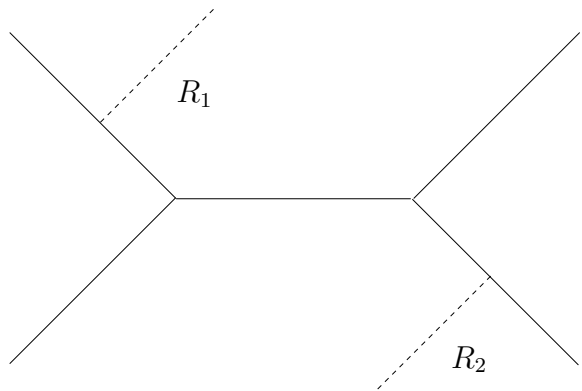}
\caption{The covering configuration contains two D-branes.}
\label{dcover}
\end{center}
\end{figure}

The covering geometry involves both the original D-brane and its image under the 
involution $I$, and a simple analysis shows that we have to consider {\it two} D-branes in opposite legs as 
depicted in figure \ref{dcover}. The amplitude for this two-brane configuration can be 
computed by using the topological vertex of \cite{Aganagic:2003db} (see appendix \ref{schurapp} for a list of 
useful formulae and properties of the vertex). A simple application of the rules in \cite{Aganagic:2003db} gives
\beq
{\cal C}_{R_1 R_2}={1 \over Z_{{\IP}^1}}\sum_R C_{R_1 R \cdot}C_{R^T \cdot R_2} (-Q)^{\ell(R)}={1 \over Z_{{\IP}^1}}
\sum_{R} W_{R_1 R^T}
W_{R R_2} (-Q)^{\ell(R)}
\eeql{cfirst}
where $Q=e^{-t}$, $Z_{\IP^1}$ is the partition function of the resolved conifold
\beq
Z_{\IP^1} = \prod_{k=1}^{\infty} (1-Q q^k)^k,
\eeql{zconif}
and the quantities $W_{R_1 R_2}$ are defined in \refeq{hopfschur}. The above quotient of series 
can be computed in a closed way by using the techniques of \cite{Eguchi:2003sj,Iqbal:2003ix}, and in fact one obtains two 
equivalent expressions.
The first expression is
\beq
{\cal C}_{R_1 R_2} =W_{R_1} W_{R_2} \prod_k (1- q^k Q)^{C_k (R_1, R_2)}
\eeql{csecond}
where the coefficients $C_k (R_1, R_2)$ are given by \refeq{expcoeff} or \refeq{coeffs}. Notice that \refeq{csecond} is a 
Laurent polynomial in $q^{\pm {1 \over 2}}$ and a
{\it polynomial} in $Q$. There is, however, a second expression for ${\cal C}_{R_1 R_2}$ which involves skew quantum dimensions as defined in \refeq{skewq}.
The derivation uses the representation of the vertex in terms of skew Schur functions given in \refeq{topvertexschur}.
Define first, as in \refeq{invunknot}:
\beq
{\cal W}_R=  ({\rm dim}_q^{U(N)} R)(\lambda=Q^{-1}),
\eeql{renquan}
where $\lambda$ is again the Chern-Simons variable \refeq{lam}, and the quantum dimension is defined in \refeq{qdgeneral} (an explicit formula was given in \refeq{qdUN}). 
Then we have, after using \refeq{identitythree},
\beq
{\cal C}_{R_1 R_2}= q^{{ \kappa_{R_1} + \kappa_{R_2} \over 2}}Q^{\ell (R_1)+ \ell(R_2)\over 2}
\sum_R (-1)^{\ell (R)} {\cal W}_{R_1^T/R} {\cal W}_{R_2^T/R^T}.
\eeql{cthree}

We now compute the total amplitude for the configuration 
depicted in figure \ref{dorienti}. To do this we can use \refeq{vertextotal}, 
where ${\cal O}_S$ is now an open string amplitude given by $C_{\cdot RS}$. One finds
\bea
Z_R&=&{1 \over Z_{X/I}} \sum_{R'=R'^T}C_{\cdot R R'} Q^{\ell (R')/2}(-1)^{{1\over 2}(\ell(R')\mp r(R'))}\nn\\
&=&q^{\kappa_R\over 2} {\cal W}_{R^T}^{SO/Sp},
\eeal{fone}
where we have used the formula \refeq{magic} to express the amplitude in terms of $SO/Sp$ quantum dimensions. 
We then see that the {\it total} brane amplitude in figure \ref{dorienti} is given by the Chern-Simons invariant of 
an unknot for gauge groups $SO/Sp$. To obtain the unoriented piece of this amplitude, we have 
to subtract the covering contribution, which involves a nontrivial combination of quantum dimensions for $U(N)$. 
For a {\it framed} D-brane one should simply change
\bea
Z_R &&\rightarrow (-1)^{\ell (R)p } q^{p\kappa_R\over 2}Z_R, \nn\\
{\cal C}_{R_1 R_2} &&\rightarrow (-1)^{(\ell (R_1)+ \ell(R_2))p } q^{p{\kappa_{R_1}
+\kappa_{R_2}\over 2}}{\cal C}_{R_1 R_2},
\eeal{framingch}
since in the covering configuration one has to put
the same framing in both legs, by symmetry.

We can now compute $f^{\rm unor}_R$ by using the results of 
the previous section. We will present explicit results only up to 
three boxes. The first thing one finds is that $f^{c=2}_R$ vanishes at this 
order in $R$. For $f^{c=1}_R$ one finds (we present here the results for $SO(N)$; for $Sp(N)$ one only has to change the overall sign of the $c=1$ contributions):
\bea
\widehat f^{c=1}_{\yng(1)}&=& (-1)^p Q^{1/2},\nn\\
\widehat f^{c=1}_{\yng(2)}&=&{q^{1 - p}\,\left( 1 - q^{p} - q^{1 + p} + 
    q^{1 + 2\,p} \right) \,Q^{1/2}\,\left( -1 + Q \right)\over (q -1)^2 (q +1)},\nn\\
\widehat f^{c=1}_{\yng(1,1)}&=&q^{-p} {\left( 1 - q^{1 + p} - q^{2 + p} + 
    q^{3 + 2\,p} \right) \,Q^{1/2}\,\left( -1 + Q \right)\over (q -1)^2 (q +1)},\nn
\eea
and
\bea
\widehat f^{c=1}_{\yng(3)}&=&{ (-1)^p q^{2-3p} (-1+q^p) (-1+q^{1+p}) Q^{1/2} (-1+Q) \over(-1+q)^4 (1+q)^2 (1+q^2+q^4)} \nn\\
&&\times [-q+q^{2p}+q^{1+p}(-1-q+2q^{p}-q^{2p})+q^{2(1+p)}(2+q-q^{p}-q^{2p})\cr
&&~~~+ Q( 1+q^p-q^{2p}+q^{1+p}(1-2q^{p})+q^{2(1+p)}(-2-q+q^{p})\nn\\
&&~~~~~~~~~~~+q^{3(1+p)}(1+q^{p}))], \nn\\
\widehat f^{c=1}_{\yng(2,1)}&=& {(-1)^p q^{-3p} (-1+q^{1+p}) Q^{1/2} \over(-1+q)^4 (1+q)^2 (1+q^2+q^4)} \nn\\
&&\times [q (-1+q^p) (1+q^p+q^{1+p}(1-2q^{p})+q^{2(1+p)}(-2-2q)\nn\\
&&~~~~~~~~~~~+q^{3(1+p)}(1+q)+q^{4(1+p)})\nn\\
&&~~~-Q (1+q) (-1 +q^{1+p}(-1 +3q^{p})+ q^{2(1+p)}(2+3 q - 3 q^{p})\nn\\
&&~~~~~~~~~~~+q^{3(1+p)}(-2 -3q^{2})+q^{4(1+p)}+q^{5(1+p)})\nn\\
&&~~~+ Q^2 (-1 +q^{1+p}(-1+2q^{p}) +q^{2(1+p)}(2+3q+q^{2}-q^{p} ) \nn\\
&&~~~~~~~~~~~+ q^{3(1+p)}(-3-2q-2q^{2}) +q^{5+4p}+q^{6+5p})],\nn\\
\widehat f^{c=1}_{\yng(1,1,1)}&=&{(-1)^p q^{-1-3p} (1-q^{1+p}-q^{2+p}+q^{3+2p}) Q^{1/2}(-1+Q) \over(-1+q)^4 (1+q)^2 (1+q^2+q^4)} \nn\\
&&\times [-q (1+q^{1+p}(1+q-q^{p})+q^{2(1+p)}(-2-2q-q^{2})\nn\\
&&~~~~~~~~~~~+q^{3(1+p)}(1+q)+q^{5+4p}))\nn\\
&&~~~+Q(1+q^{1+p}(1+q)+q^{2(1+p)}(-1-2q-2q^{2}-q^{3})\nn\\
&&~~~~~~~~~~~+q^{3(1+p)}(q^{2}+q^{3})+q^{7+4p})].\nn
\eea
One can indeed check that, for any integer $p$, the above polynomials are 
of the form predicted in \refeq{cs} (they are polynomials in $(q^{1/2}-q^{-1/2})^2$ 
with integer coefficients).

\subsection{$\IC \IP^2$ Attached to ${\IR \IP^2}$}

\begin{figure}[htp]
\begin{center}
\psfrag{R2(-2)}{$R_2^{(-2)}$}
\psfrag{R3(-2)}{$R_3^{(-2)}$}
\psfrag{R1(-2)}{$R_1^{(-2)}$}
\psfrag{R'2(-2)}{$R_2^{'(-2)}$}
\psfrag{R'3(-2)}{$R_3^{'(-2)}$}
\psfrag{R'1(-2)}{$R_1^{'(-2)}$}
\psfrag{R(0)}{$R^{(0)}$}
\psfrag{R}{$R$}
\psfrag{S}{$S$}
\psfrag{S1}{$S_1$}
\psfrag{S2}{$S_2$}
\psfrag{Orientifold}{Orientifold}
\psfrag{Covering geometry}{Covering geometry}
\includegraphics[width=5.5truein]{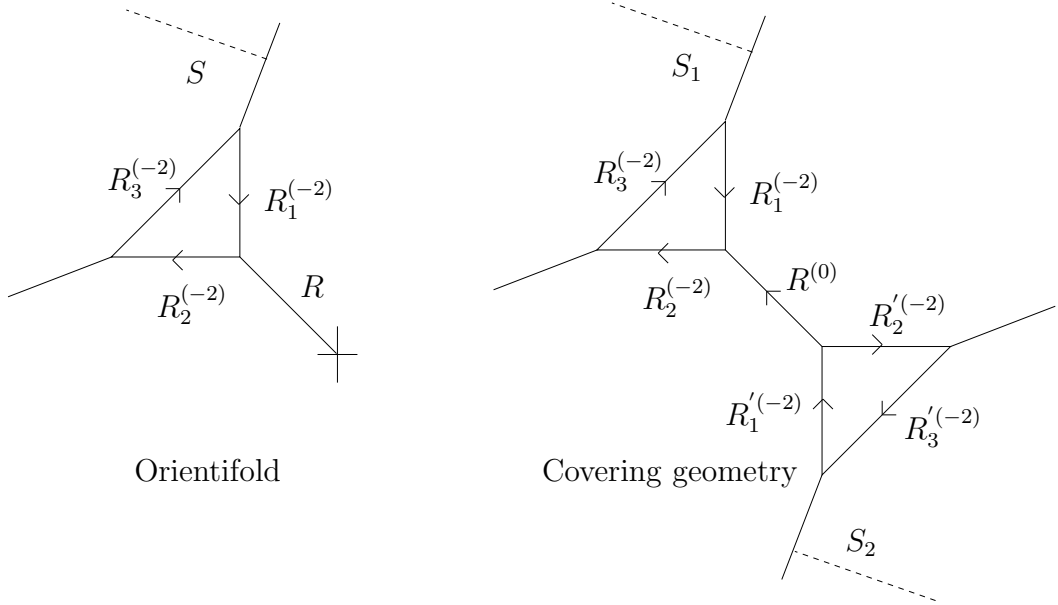}
\caption{A D-brane in an outer leg of the orientifold of the two ${\IC \IP^2}$'s connected by a ${\IC \IP^1}$.}
\label{geounorptwo}
\end{center}
\end{figure}

The next example we consider is the orientifold studied in chapter \ref{closed} and in \cite{Diaconescu:2003dq}, with a D-brane located in an outer leg. In this case, the covering space consists of two ${\IC \IP^2}$'s connected by a ${\IC \IP^1}$, with two 
D-branes in opposite legs  (the geometry is shown in figure \ref{geounorptwo}). Let us now define the following operator corresponding to the ${\IC \IP^2}$ with an outer D-brane:
\beq
\CO_{R S} = \sum_{R_i} q^{\sum_i \kappa_{R_i}} (-1)^{\sum_i \ell (R_i)} C_{S R_3 R_1^T} C_{\cdot R_2 R_3^T} C_{R_1 R_2^T R} e^{-t \sum_i \ell(R_i)}
\eeq
where $S$ is the representation attached to the D-brane. Using the topological vertex rules we can write,
for arbitrary framing $p$ ($Z_{\rm closed}$ is the amplitude without D-branes): 
\beq
{\cal C}_{S_1 S_2}= {1 \over Z_{\rm closed}^{\rm cov}} \sum_{R}(-1)^{p(\ell (S_1) + \ell (S_2))} q^{{p \over 2} (\kappa_{S_1}+\kappa_{S_2})} \CO_{R S_1} \CO_{R^T S_2} (-Q)^{\ell(R)}
\eeq
and
\beq
Z_{\rm closed}^{\rm cov} = 1 + \sum_{R} \CO_{R \cdot} \CO_{R^T \cdot} (-Q)^{\ell(R)},
\eeql{Ptwoclosed}
where in the last equation we have singled out the term where all the representations are trivial. As we do not have a closed expression for 
${\cal C}_{S_1 S_2}$ we have to evaluate $Z_{\rm or}$ order by order in $Q$ and $e^{-t}$.

Let us compute now $Z_S$ by using the topological vertex rules for orientifolds developed in section \ref{topover}. We find that
\beq
Z_S = {1 \over Z_{\rm closed}} (-1)^{p \ell(S)} q^{p \kappa_S \over 2} \sum_{R=R^T} \CO_{R S} Q^{\ell(R)/2} (-1)^{{1\over 2} (\ell(R) \mp r(R))}
\eeq
where
\beq
Z_{\rm closed} = 1+ \sum_{R=R^T} \CO_{R \cdot} Q^{\ell(R)/2} (-1)^{{1\over 2} (\ell(R) \mp r(R))}.
\eeql{Ptwoclosedii}
Using the results in the previous section, we can compute the functions ${\widehat f_S^c} (q, Q, e^{-t})$. 
We find the following results at low order, for arbitrary framing $p$ (again we present the results for $SO(N)$):
\bea
\widehat f^{c=1}_{\yng(1)}&=& (-1)^p \Bigl[ -Q^{1/2} e^{-t} + 4 Q^{1/2} e^{-2 t} -(35 + 8  \, z)Q^{1/2} e^{-3t} \nn\\
&& \,\,\, \,\,\,\,\,\, + (400 + 344\, z 
+ 112 \, z^2 + 13 \, z^3)Q^{1/2} e^{-4t} -2  Q^{3/2} e^{-2 t} \nn\\
&& \,\,\,\,\,\,\,\,\,+ (30 + 6  \, z)Q^{3/2} e^{-3t} - (488 + 359\, z + 104 \, z^2 + 11 \, z^3)Q^{3/2} e^{-4t}  \nn\\
&& \,\,\,\,\,\,\,\,\,- 3\, Q^{5/2} e^{-3t}+ (132 + 59\, z + 8 \, z^2 )Q^{5/2} e^{-4t} + \cdots \Bigr],\nn\\
\widehat f^{c=2}_{\yng(1)}&=& -(-1)^p \Bigl[Q^2 e^{-3t} -(15 + 7 \, z + z^2) Q^2 e^{-4t} + 2\, Q^4 e^{-4t}+ \cdots\Bigr] ,\nn\\
\widehat f^{c=1}_{\yng(2)}&=& {q^{-p+1} (-1+q^{p}) (-q+q^{p}) \over (q-1)^2 (q+1)} Q^{1/2} e^{-t} 
- {3  q^{-p+1} (-1+q^{p})^2 \over (q-1)^2} Q^{1/2} e^{-2t} + \cdots, \nn\\
\widehat f^{c=2}_{\yng(2)}&=&{q^{-p+1} (q^p-1)^2 \over (q-1)^2} Q^2 e^{-3t} + \cdots,\nn\\
\widehat f^{c=1}_{\yng(1,1)}&=& {q^{-p+1} (q^p-1) (q^{1+p}-1) \over (q-1)^2 (q+1)} Q^{1/2} e^{-t}-{3 q^{-p} (q^{1+p}-1)^2 \over (q-1)^2} Q^{1/2} e^{-2t} 
+ \cdots, \nn\\
\widehat f^{c=2}_{\yng(1,1)}&=&{q^{-p} (q^{1+p}-1)^2 \over (q-1)^2} Q^2 e^{-3t} + \cdots ,\nn
\eea
where $z\equiv (q^{1\over 2}- q^{-{1\over 2}})^2$. 

By comparing with \refeq{cs}, it is easy to see that the results above have the expected polynomial form with integer coefficients for any $p$. In contrast to the 
example above of a D-brane in the orientifold of the conifold, in this example there are nonzero amplitudes with an even number of 
crosscaps. 

\subsection{$SO$/$Sp$ Hopf Link Invariant}

\begin{figure}[htp]
\begin{center}
\psfrag{R(0)}{$R^{(0)}$}
\psfrag{R}{$R$}
\psfrag{S}{$S$}
\psfrag{S1}{$S_1$}
\psfrag{S2}{$S_2$}
\psfrag{P1}{$P_1$}
\psfrag{P2}{$P_2$}
\psfrag{P3}{$P_3$}
\psfrag{P4}{$P_4$}
\psfrag{Orientifold}{Orientifold}
\psfrag{Covering geometry}{Covering geometry}
\includegraphics[width=5.5truein]{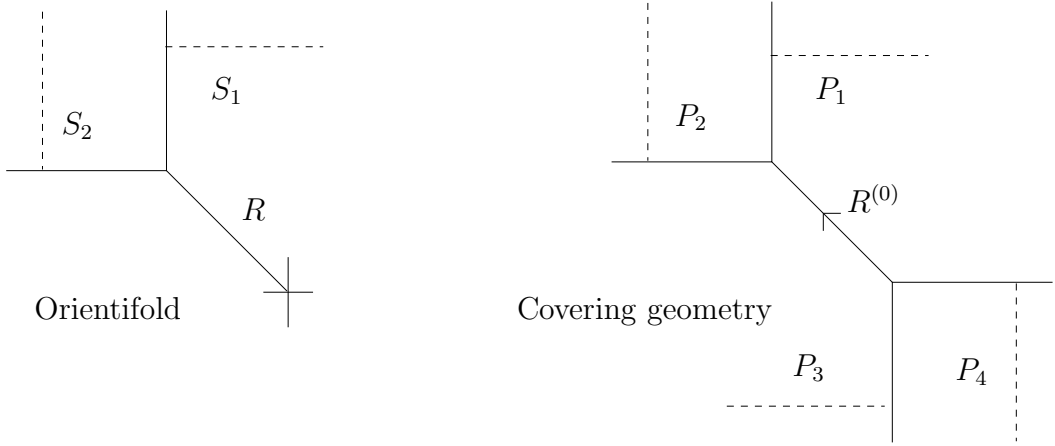}
\caption{Two adjacent D-branes in the outer legs of the orientifold of 
the resolved conifold.}
\label{geohopf}
\end{center}
\end{figure}

 Our third and final example is the orientifold of the resolved conifold with two adjacent D-branes in the outer legs. The covering geometry now involves 
four sets of D-branes in the outer legs of the resolved conifold, oppositely identified by the involution. The geometry is shown in figure \ref{geohopf}.

Using the topological vertex, we find for the covering amplitude (for arbitrary framings $p_1$ and $p_2$):
\bea
{\cal C}_{P_1 P_2 P_3 P_4} &=&{1 \over Z_{\IP^1}} \sum_R (-1)^{p_1 (\ell(P_1)+\ell(P_3)) + p_2 (\ell(P_2)+\ell(P_4))} \nn\\
&&\times q^{{p_1 \over 2} (\kappa_{P_1}+\kappa_{P_2})+{p_2 \over 2} (\kappa_{P_2}+\kappa_{P_4})} (-Q)^{\ell(R)} C_{R^T P_1 P_2} C_{R P_3 P_4}
\eeal{covampiii}
To obtain the oriented amplitude from \refeq{covampiii} we have to identify the moduli $P_1$ ($P_2$) with $P_3$ ($P_4$) as explained in \refeq{zrorL}. 
We can rewrite \refeq{covampiii} by using the expression of the topological vertex in terms of Schur functions \refeq{topvertexschur}:
\bea
{\cal C}_{P_1 P_2 P_3 P_4} &=&{1 \over Z_{\IP^1}} q^{{1\over2}(\sum_{i=1}^4 \kappa_{R_i})} s_{R_1^T} (q^{\rho}) s_{R_3^T} (q^{\rho}) \sum_{\eta_1, \eta_2} (-Q)^{\ell(\eta_1)} s_{R_2^T/\eta_1} (q^{\ell(R_1)+\rho}) s_{R_4^T/\eta_2} (q^{\ell(R_3)+\rho})\nn\\
&& \times \sum_{R} s_{R^T/\eta_1} (-Q q^{\ell(R_1^T)+\rho}) s_{R/\eta_2} (q^{\ell(R_3^T)+\rho}).
\eeal{covampschur}
By using the identities \refeq{identitytwo}, \refeq{identitythree} and \refeq{identityfour} we finally obtain that
\bea
&&{\cal C}_{P_1 P_2 P_3 P_4}= q^{{1\over2}(\sum_{i=1}^4 \kappa_{R_i})} 
s_{R_1^T} (q^{\rho}) s_{R_3^T} (q^{\rho}) \prod_k (1-Q q^k)^{C_k (R_1^T, R_3^T)}~~~~~~~~~~~~~~~~~~~~~~~~~~\nn\\
&&~~~~~\times \sum_{\eta} (-Q)^{\ell(\eta)} s_{R_2^T/\eta^T} (q^{\ell(R_1)+\rho}, Q q^{-\ell(R_3^T)-\rho}) s_{R_4^T/\eta} (q^{\ell(R_3)+\rho}, Q q^{-\ell(R_1^T)-\rho}),
\eeal{covampfinal}
where we defined the functions
\beq
s_{R/Q} (x,y) = \sum_{\eta} s_{R/\eta} (x) s_{\eta/Q} (y),
\eeql{schurdouble}
and the coefficients $C_k (R_1^T, R_3^T)$ are defined in \refeq{expcoeff} or \refeq{coeffs}. Notice that \refeq{covampfinal} is a polynomial in $Q$.

Now that we have our final expression for the covering amplitude, let us look at the full amplitude. The vertex rules for orientifolds tell us that, for the amplitude where there are two D-branes in the outer legs, one has (for arbitrary framings $p_1$ and $p_2$)
\bea
Z_{S_1 S_2} &=&{1 \over Z_{X/I}} \sum_{R=R^T}(-1)^{p_1 \ell(S_1)+p_2 \ell(S_2)} q^{{1 \over 2}(p_1 \kappa_{S_1}+p_2 \kappa_{S_2})} C_{S_1 S_2 R} Q^{\ell (R)/2}(-1)^{{1\over 2}(\ell(R)\mp r(R))} \nn\\
&=&(-1)^{p_1 \ell(S_1)+p_2 \ell(S_2)} q^{{1 \over 2}(p_1 \kappa_{S_1}+p_2 \kappa_{S_2})}q^{{1 \over 2} \kappa_{S_2}} Q^{{1 \over 2} (\ell(S_1)+\ell(S_2))} \CW_{S_1 S_2^t}^{SO(N)/Sp(N)},
\eeal{amptwofinal}
where we used again \refeq{magic}. This time, we see that the {\it total} amplitude of the two D-brane configuration in the orientifold of the conifold is given by the $SO/Sp$ Chern-Simons invariants of the Hopf link.

By substracting the oriented piece from the unoriented amplitude, and using the results of the previous section, we can compute the $N^{c}_{(S_1, S_2), g, \beta}$ integer invariants through the $\widehat f_{S_1 S_2}^c$ functions. As noted in \refeq{tildefrlinks} we now expect a slightly different structure for the $\widehat f_{S_1 S_2}^c$ functions than the one given by \refeq{cs}, since $L=2$. Namely, we expect
\beq
\widehat f_{S_1 S_2}^c = \sum_{g, \beta} N^{c}_{(S_1, S_2), g, \beta} (q^{1\over 2} - q^{-{1\over 2}})^{2g+1} Q^{\beta}.
\eeql{fhattwo}
with the $\widehat f_{S_1 S_2}^c$ functions defined as in \refeq{ccont}.

We obtain the following results for $SO(N)$:
\bea
{\widehat f}_{\yng(1) \yng(1)}^{c=1} &=& (-1)^{p_1+p_2} Q^{1/2} (1-Q) (q^{1/2}-q^{-1/2}),\nn\\
{\widehat f}_{\yng(2) \yng(1)}^{c=1} &=&{(-1)^{2p_1+p_2} Q^{1/2} (1-Q) q^{-p_1-1/2} \over q-1} (-q+2q^{1+p_1}-q^{1+2p_1} \nn\\
&&~~~~~~~~~~~~~~~~~~~~~~~~~~~~~~~~~~~~~~~~~~~~~~~~+Q(q^2 +q^{2p_1} -2q^{1+p_1})),~~~~~~~~~~~~~\nn\\
{\widehat f}_{\yng(1,1) \yng(1)}^{c=1}&=&{(-1)^{2p_1+p_2} Q^{1/2} (1-Q) q^{-p_1-1/2} \over q-1}(-(-1+q^{1+p_1})^2 + Q q (-1 + q^{p_1})^2).
\eeal{resulthopf}
It is straightforward to show that for any fixed framings $p_1$ and $p_2$ the $\widehat f$ functions \refeq{resulthopf} have the structure predicted by \refeq{fhattwo} with integer invariants $N^{c}_{(S_1, S_2), g, \beta}$. Up to the order $\ell(S_1)+\ell(S_2) =3$ the contributions with two crosscaps vanish.

\subsection{Localization Computations}

In the previous subsections we found many open BPS invariants using the topological vertex prescription of section \ref{topover} and the structure predictions of section \ref{structure}. As far as we are aware these invariants have never been computed before. Therefore it would be nice to have an independent check of our results which does not rely on large $N$ duality.

In section \ref{localclosed} and in \cite{Diaconescu:2003dq} localization techniques were defined to compute closed unoriented Gromov-Witten invariants of Calabi-Yau orientifolds. In this section we will extend these techniques to the case of {\it open} unoriented Gromov-Witten invariants, therefore providing an alternative and independent way to compute the invariants of the previous subsections.

In order to compare our results with localization computations we have to extract open Gromov-Witten invariants from the $f$ polynomials. First let us recall the definition of the $f$ functions \refeq{conj}:
\beq
{\cal F}^c (V_1, \dots, V_L) = - \sum_{d=1}^\infty \sum_{R_1, \cdots, R_L}
{1\over d} f_{(R_1, \cdots, R_L)}^c (q^d, {\rm e}^{-d t})
\prod_{\alpha=1}^L {\rm Tr}_{R_\alpha}V_{\alpha}^d
\eeql{deffunct}
where we added the superscript $c$ for the number of crosscaps. As usual, we can also work in the $\vec{k}$ basis. In this basis the free energy reads (see \cite{Labastida:2001ts}):
\beq
{\cal F}^c (V_1, \dots, V_L) = - \sum_{\{ \vec{k}^{\alpha} \}} W_{(\vec{k}^{(1)}, \dots, \vec{k}^{(L)})}^{({\rm conn}),c} \prod_{\alpha}^L {1 \over z_{\vec{k}^{(\alpha)}}} \Upsilon_{\vec{k}^{(\alpha)}} (V_{\alpha}),
\eeql{freekbasis}
where we defined the connected vevs $W_{(\vec{k}^{(1)}, \dots, \vec{k}^{(L)})}^{({\rm conn}),c}$, 
and $z_{\vec{k}} = \prod_m k_m ! m^{k_m}$. Since $q=e^{i g_s}$, we can expand the r.h.s. of \refeq{freekbasis} in $g_s$. We find a series with the structure \cite{Labastida:2001ts}:
\bea
{\cal F}^c (V_1, \dots, V_L) &=& \sum_{g=0}^{\infty} i^{\sum_{\alpha=1}^L |{\vec{k}^{(\alpha)}}|+c} F_{g,(\vec{k}^{(1)}, \dots, \vec{k}^{(L)})}^c g_s^{2g-2+c+\sum_{\alpha=1}^L |{\vec{k}^{(\alpha)}}|} \Upsilon_{\vec{k}^{(\alpha)}} (V_{\alpha}) \nn\\
&&=- {\left(  \prod_{\alpha}^L {1 \over z_{\vec{k}^{(\alpha)}}} \right)} W_{(\vec{k}^{(1)}, \dots, \vec{k}^{(L)})}^{({\rm conn}),c} \Upsilon_{\vec{k}^{(\alpha)}} (V_{\alpha}),
\eeal{fgwfunct}
where $F^c_{g,(\vec{k}^{(1)}, \dots, \vec{k}^{(L)})}$ is the generating functional for open Gromov--Witten invariants at genus $g$, with $c$ crosscaps and fixed boundary conditions given by $(\vec{k}^{(1)}, \dots, \vec{k}^{(L)})$. The factor of $i^{\sum_{\alpha=1}^L |{\vec{k}^{(\alpha)}}|+c}$ is necessary to compare Chern-Simons (or topological vertex) results with localization computations \cite{Marino:2001re}. Thus, we see that to extract open Gromov--Witten invariants we have to compute the connected vevs 
$W_{(\vec{k}^{(1)}, \dots, \vec{k}^{(L)})}^{({\rm conn} ),c}$ from the $f$ functions. Such a relation has been found in \cite{Labastida:2001ts}:
\beq
W_{(\vec{k}^{(1)}, \dots, \vec{k}^{(L)})}^{({\rm conn}),c} = \sum_{d| \vec{k}^{(\alpha)},~d~{\rm odd}} d^{\sum_{\alpha} | \vec{k}^{(\alpha)}|-1} \sum_{\{ R_{\alpha}\}} \prod_{\alpha=1}^{L} \chi_{R_{\alpha}} ( C(\vec{k}^{(\alpha)}_{1/d})) f_{(R_1, \dots, R_L)}^c (q^d, {\rm e}^{-dt}),
\eeql{relvevf}
where $C(\vec{k})$ is the conjugacy class associated to a vector $\vec{k}$, which has $k_j$ cycles of length $j$, and $\chi_{R}$ is the character of the symmetric group $S_{\ell}$. In \refeq{relvevf} the vector $\vec{k}_{1/d}$ is defined as follows. Fix a vector $\vec{k}$, and consider all the positive integers $d$ that satisfy the following condition: $d|j$ for every $j$ with $k_j \neq 0$. When this happens, we will say that ``$d$ divides $\vec{k}$", and we will denote this as $d|\vec{k}$. We can then define the vector $\vec{k}_{1/d}$ whose components are $(\vec{k}_{1/d})_i = k_{di}$. In \refeq{relvevf} the integer $d$ has to divide all the vectors $\vec{k}^{(\alpha)}$, $\alpha=1,\dots,L$. Note that in \refeq{relvevf} the sum is only over $d$ {\it odd}: this is because in the unoriented case only odd multicovers contribute.

Using \refeq{fgwfunct} and \refeq{relvevf} one can find expressions for the generating functionals of open Gromov-Witten invariants in terms of $f$ functions. Let us define the all genera generating functionals for open Gromov-Witten invariants with $c$ crosscaps and fixed boundary conditions given by $(\vec{k}^{(1)}, \dots, \vec{k}^{(L)})$:
\beq
F_{(\vec{k}^{(1)}, \dots, \vec{k}^{(L)})}^c = \sum_{g=0}^{\infty} F_{g,(\vec{k}^{(1)}, \dots, \vec{k}^{(L)})}^c g_s^{2g-2+c+\sum_{\alpha=1}^L |{\vec{k}^{(\alpha)}}|}.
\eeql{allgene}
For configurations with one representation ($L=1$), one finds
\bea
&&F^c_{(1,0,\dots)} =  i^{1-c} f^c_{\yng(1)}, ~~~F^c_{(2,0,\dots)} = {i^{-c} \over 2} (f^c_{\yng(2)}+f^c_{\yng(1,1)}),~~~F^c_{(0,1,0,\dots)}= {i^{1-c} \over 2} (f^c_{\yng(2)}-f^c_{\yng(1,1)})\nn\\
&&~~~F^c_{(3,0,\dots)}=-{i^{1-c} \over 6} (f^c_{\yng(3)}+2 f^c_{\yng(2,1)}+f^c_{\yng(1,1,1)}),~~~~F^c_{(1,1,0,\dots)}={i^{-c} \over 2} (f^c_{\yng(3)}-f^c_{\yng(1,1,1)})\nn\\
&&~~~~~F^c_{(0,0,1,0,\dots)}={i^{1-c} \over3}(f^c_{\yng(3)}- f^c_{\yng(2,1)}+f^c_{\yng(1,1,1)}+ f^c_{\yng(1)} (q^3, {\rm e}^{-3t})),
\eeal{genfunct}
For configurations with two representations ($L=2$), one finds
\bea
&&F^c_{((1,0,\dots),(1,0,\dots))} = i^{-c}f^c_{\yng(1) \yng(1)},~~~~~~~~F^c_{((2,0,\dots),(1,0,\dots))} = -{i^{1-c} \over2}(f^c_{\yng(2) \yng(1)}+f^c_{\yng(1,1) \yng(1)}),\nn\\
&&~~~~~~~~~~~~~~~F^c_{((0,1,0,\dots),(1,0,\dots))} = {i^{-c}\over2}(f^c_{\yng(2) \yng(1)}-f^c_{\yng(1,1) \yng(1)}).
\eeal{genfuncttwo}

Using the above formulae, we can compute the $F^c_{(\vec{k}^{(1)}, \dots, \vec{k}^{(L)})}$ generating functionals and put them in the form 
of \refeq{allgene} by expanding in $g_s$. This will extract the open Gromov-Witten invariants from our previous results. 

\subsubsection{The $SO/Sp$ Framed Unknot}

The topological vertex gives the following results:
\bea
F_{(1,0,0,...)}^{c=1} &=& (-1)^p Q^{1/2},\nn\\
F_{(2,0,0,...)}^{c=1}&=& {{1\over 2}\left[(1+p)^2 Q^{1/2} (1-Q)\right]} g_s  - {{1 \over 48}\left[  (1+p)^2 (1+4p+2p^2) Q^{1/2} (1-Q)\right]} g_s^3 +\dots,\nn\\
F_{(0,1,0,...)}^{c=1}&=& {\left[(1+p) Q^{1/2} (1-Q)\right]}-{{1 \over 24} \left[(3+11p+12p^2+4p^3) Q^{1/2} (1-Q)\right]} g_s^2 +\dots,\nn\\
F_{(3,0,0,\dots)}^{c=1}&=& { {1\over6}\left[(-1)^p Q^{1/2} (1+p)^3  (1+3p-6Q(1+p)+Q^2 (5+3p))\right]} g_s^2 + \dots,\nn\\
F_{(1,1,0,\dots)}^{c=1} &=& {\left[ (-1)^p Q^{1/2} (1+p)^2 (1+2p - 4Q(1+p)+Q^2(3+2p)) \right]} g_s + \dots,\nn\\
F_{(0,0,1,\dots)}^{c=1} &=&{ {1\over6}\left[ (-1)^p Q^{1/2}(3(1+p)(2+3p+Q^2(4+3p))-2Q(8+18p+9p^2))\right]}+\dots~.\nn
\eea

In order to compare with the localization computation, we introduce first some notation. We will consider the following real torus 
action on the resolved conifold $X$:
\beq
e^{i\phi}\cdot(z_1,z_2,z_3,z_4)=(e^{i\lambda_1\phi}z_1,e^{i\lambda_2\phi}z_2,e^{i\lambda_3\phi}z_3,e^{i\lambda_4\phi}z_4).
\eeq
The weights of the torus action on the local coordinates $z=z_1/z_2,u=z_2z_3,v=z_2z_4$ are given by $\lambda_z=\lambda_1-\lambda_2,
\lambda_u=\lambda_2+\lambda_3,\lambda_v=\lambda_2+\lambda_4$ respectively. Note that from the compatibility of the torus action with the 
anti-holomorphic involution it follows that $\lambda_u+\lambda_v+\lambda_z=0$. Now we can present the localization results:
\bea
F_{(1,0,0,...)}^{c=1} &=& Q^{1\over 2},\nn\\
F_{(2,0,0,...)}^{c=1}&=& {1\over 2}\left[\left({a\over a-1}\right)^2 Q^{1\over 2}(1-Q)\right] g_s-{1 \over 48}\left[{a^2(a^2+2a-1)\over (a-1)^4} 
Q^{1\over 2}(1-Q)\right] g_s^3 +\dots,\nn\\
F_{(0,1,0,...)}^{c=1}&=& \left[{a\over a-1} Q^{1\over 2}(1-Q)\right] -{1 \over 24}\left[{a(a+1)(3a-1)\over (a-1)^3} 
Q^{1\over 2}(1-Q)\right] g_s^2 +\ldots,\nn\\
F_{(3,0,0,\dots)}^{c=1}&=& -{ {1\over6}\left[Q^{1/2} \left(a\over a-1\right)^3  \left( {a+2\over a-1}-{6a \over a-1}Q+\ldots\right)\right]}
g_s^2 + \dots,\nn\\
F_{(1,1,0,\dots)}^{c=1}&=& -{\left[Q^{1/2} \left(a\over a-1\right)^2 \left( {a+1\over a-1}-{4 a \over a-1}Q+\ldots\right)\right]}
g_s + \dots,\nn\\
F_{(0,0,1,\dots)}^{c=1}&=& -{ {1\over6}\left[Q^{1/2}\left( {3 a(2a+1)\over (a-1)^2}-{2(8a^2+2a-1) \over (a-1)^2}Q +\ldots\right)
\right]} + \dots~.\nn
\eea
where $a=-\lambda_v/\lambda_z$. After making the substitution $a=1+{1\over p}$, we find that the above results coincide with the 
expressions obtained from the vertex computation up to factors of $\pm (-1)^p$. The sign difference is due to 
different choice of conventions between the vertex and the localization computations. 
As an example, we present in figure \ref{unveriv} the graphs contributing to the unoriented open 
Gromov--Witten invariant for genus $1$ maps with degree $3$ $\IR\IP^2$ and winding vector $(0,1,0,...)$, as well as their contributions.

\fig{unveriv}{One crosscap, genus $1$ and three crosscaps, genus $0$ at degree $3$ $\IR\IP^2$, and winding vector $(0,1,0,...)$.}
{5.2in}{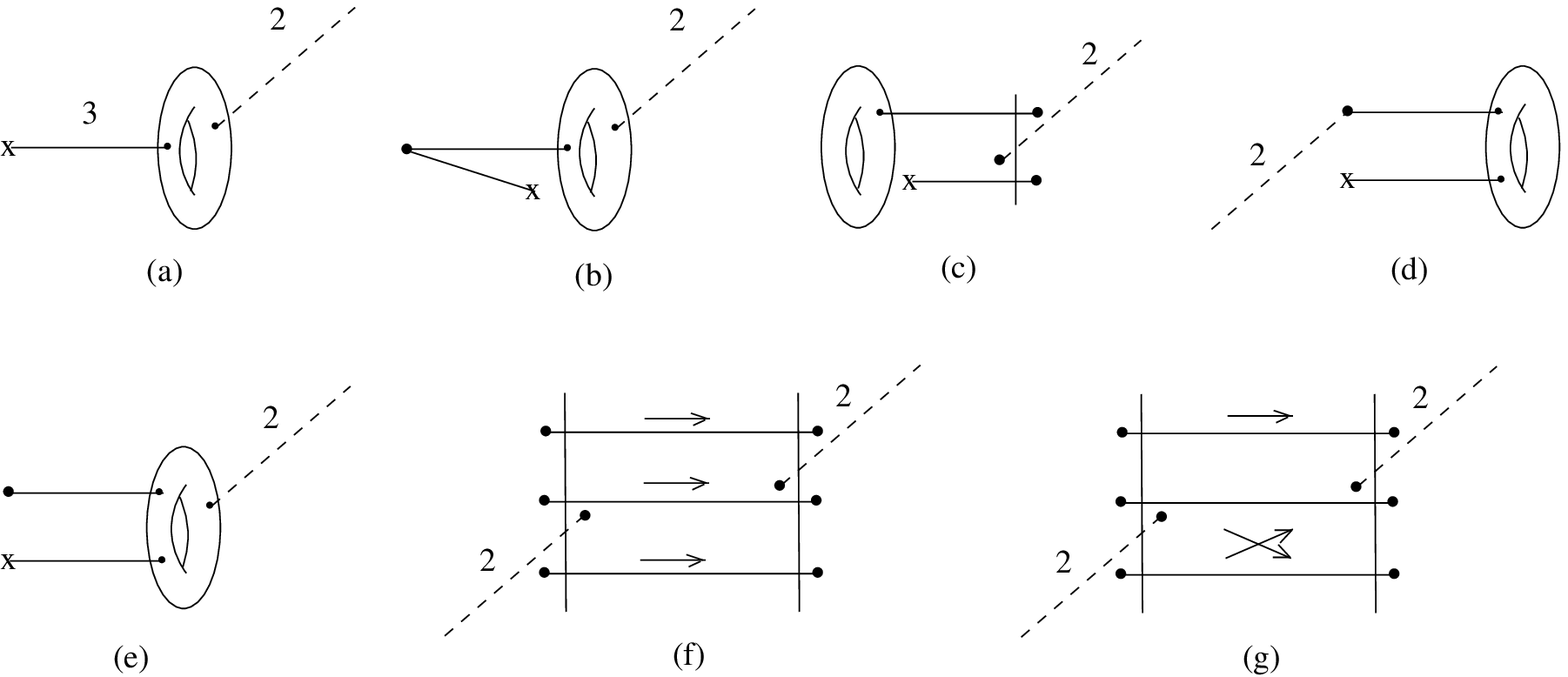}

The contributions of these graphs are computed according to the rules explained in \cite{Diaconescu:2003dq}. We obtain:
\bea
&& C_{(0,1,0,...)}^{(1,3),(a)}={(\lu-2\lv)(2\lu-\lv)\lv(\lu+2\lv)(\lambda_u^3+6\lambda_u^2\lv+\lu\lambda_v^2+2\lambda_v^3)
\over 48\lambda_u^3\lambda_z^4},\nn\\
&& C_{(0,1,0,...)}^{(1,3),(b)}=-{(\lu-2\lv)\lambda_v^2(\lu+2\lv)\over 48\lambda_u^2\lambda_z^2},\quad C_{(0,1,0,...)}^{(1,3),(c)}
={\lambda_v^2(\lu+2\lv)
\over 24\lu\lambda_z^2},\nn\\
&& C_{(0,1,0,...)}^{(1,3),(d)}=-{\lambda_v^2(2\lu+\lv)(\lu+2\lv)\over 24\lambda_z^4},\quad C_{(0,1,0,...)}^{(1,3),(e)}={\lambda_v^3
(\lu+2\lv)(\lambda_u^3-2\lu\lambda_v^2-2\lambda_v^3)\over 12\lambda_u^3 \lambda_z^4},\nn\\
&& C_{(0,1,0,...)}^{(1,3),(f)}=-{(\lu+2\lv)\lambda_v^3(\lu-2\lv)\over 6\lu\lambda_z^4},\quad  
C_{(0,1,0,...)}^{(1,3),(g)}={(\lu+2\lv)\lambda_v^3(\lu-2\lv)\over 2\lu\lambda_z^4}.\nn
\eea
which sum up to ${1 \over 24} {a(a+1)(3a-1) \over (a-1)^4}$.

\subsubsection{$\IC \IP^2$ Attached to $\IR\IP^2$}

The topological vertex gives the following results:
\bea
F_{(1,0,...)}^{c=1}&=& (-1)^p Q^{1/2} e^{-t} [-1 -2(-2+Q)e^{-t} +(-35+30 Q-3Q^2) e^{-2t}\nn\\
&&+ 4(100-122Q+33Q^2)e^{-3t}+\dots]+\dots, \nn\\
F_{(1,0,...)}^{c=2}&=& -(-1)^p Q^2 e^{-3t} {\left[ 1+(-15+2Q^2) e^{-t}+\dots \right]} g_s +\dots, \nn\\
F_{(2,0,...)}^{c=1}&=& {1 \over 2} Q^{1/2} e^{-t} {\left[-p^2 + (3 + 6p + 6p^2) e^{-t} +\dots \right]} g_s+\dots, \nn\\
F_{(2,0,...)}^{c=2}&=& -{1 \over 2} Q^2 e^{-3t} {\left[ 1+2p+2p^2+\dots \right]} g_s^2 + \dots,\nn\\
F_{(0,1,0,...)}^{c=1}&=&Q^{1/2} e^{-t} {\left[ -p + (3+6p)e^{-t}+\dots \right]} + \dots, \nn\\
F_{(0,1,0,...)}^{c=2}&=& -Q^2 e^{-3t} {\left[ 1+2p +\dots \right]} g_s  + \dots~.\nn
\eea
For the localization computations, we will use the same notation as in section \ref{localclosed}. We present below some of the localization 
computations we performed. First, we obtain
\beq
F_{(2,0,...)}^{c=1}={1\over 2} Q^{1/2} e^{-t} {\left[-\left({\lv-\lu\over\lv-2\lu}\right)^2 + \left({3 (2 \lambda_u^2-2\lu\lv + \lambda_v^2)\over 
(2\lu-\lv)^2}\right) e^{-t} +\dots \right]} g_s 
+\dots~.\nn
\eeq

\fig{unverv}{One crosscap graphs at degree $1$ $\IR\IP^2$, degree $2$ hyperplane and winding vector $(2,0,...)$.}
{5.5in}{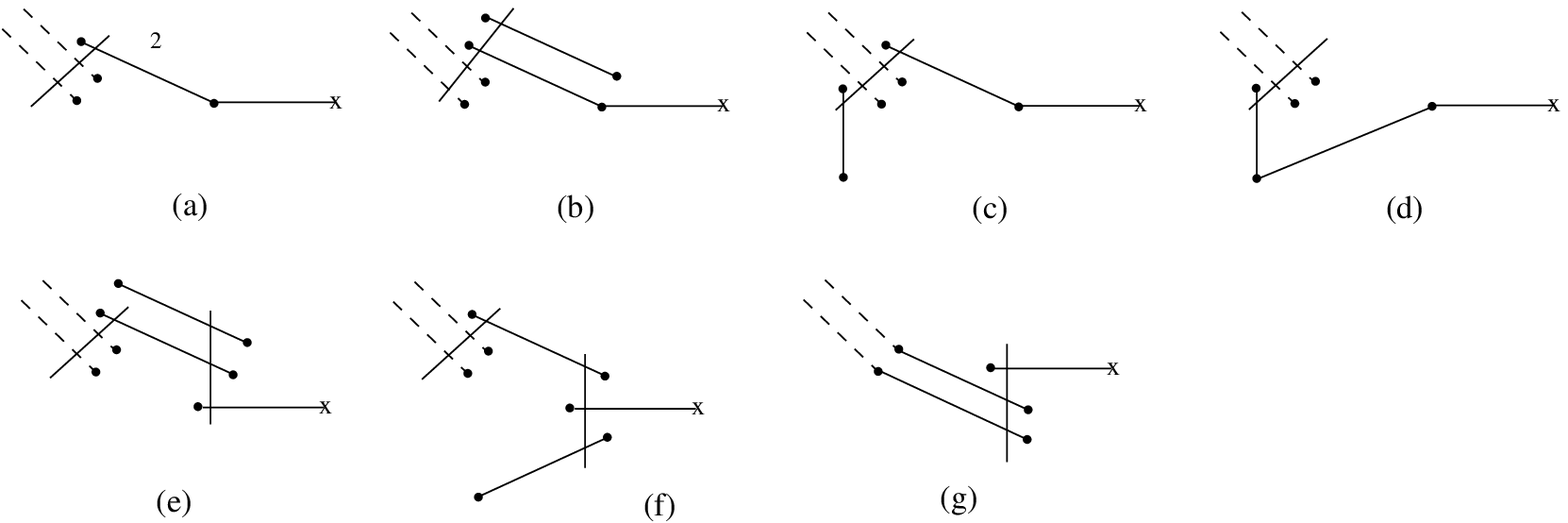}

We present in figure \ref{unverv} the graphs contributing at degree $2$ hyperplane class in the expression above. Their contributions are:
\bea
&& C_{(-1,1,2)}^{(2,0,0,...),(a)}={\lv(\lu-\lv)^2(3\lu-2\lv)\over \lambda_u^2(\lv-2\lu)^2},\quad 
C_{(-1,1,2)}^{(2,0,0,...),(b)}={(\lu-\lv)^4\over \lambda_u^2(\lv-2\lu)^2},\nn\\
&&  C_{(-1,1,2)}^{(2,0,0,...),(c)}={2\lambda_u^2-2\lu\lv+\lambda_v^2\over 2(\lv-2\lu)^2},\quad
 C_{(-1,1,2)}^{(2,0,0,...),(d)}={\lambda_u^2\over 2(\lv-2\lu)^2},\nn\\
&& C_{(-1,1,2)}^{(2,0,0,...),(e)}={\lambda_v^2(\lu-\lv)^2\over 2\lambda_u^2(\lv-2\lu)^2},\quad
C_{(-1,1,2)}^{(2,0,0,...),(f)}={(\lu-\lv)^2\over 2(\lv-2\lu)^2}, \quad C_{(-1,1,2)}^{(2,0,0,...),(g)}={\lambda_v^2\over 2\lambda_u^2}.\nn
\eea
which sum up to ${1 \over 2} \({3 (2 \lambda_u^2-2\lu\lv + \lambda_v^2)\over 
(2\lu-\lv)^2} \)$.

In the expressions above, the subscript of the contributions is $(\chi,d_1,d_2)$ where $\chi$ is the unoriented genus of the 
closed component of the map and $d_1$ and $d_2$ are the $\IR\IP^2$ and hyperplane degrees respectively. Then, for the same winding vector, 
at $2$ crosscaps we obtain 
\beq
F_{(2,0,...)}^{c=2}=-{1\over 2}Q^2e^{-3t}\left[{2\lambda_u^2-2\lu\lv+\lambda_v^2\over (\lv-2\lu)^2}+\dots\right]+\dots~.\nn
\eeq

\fig{unvervi}{Two crosscaps graphs at degree $4$ $\IR\IP^2$, degree $3$ hyperplane and winding vector $(2,0,...)$.}
{5.2in}{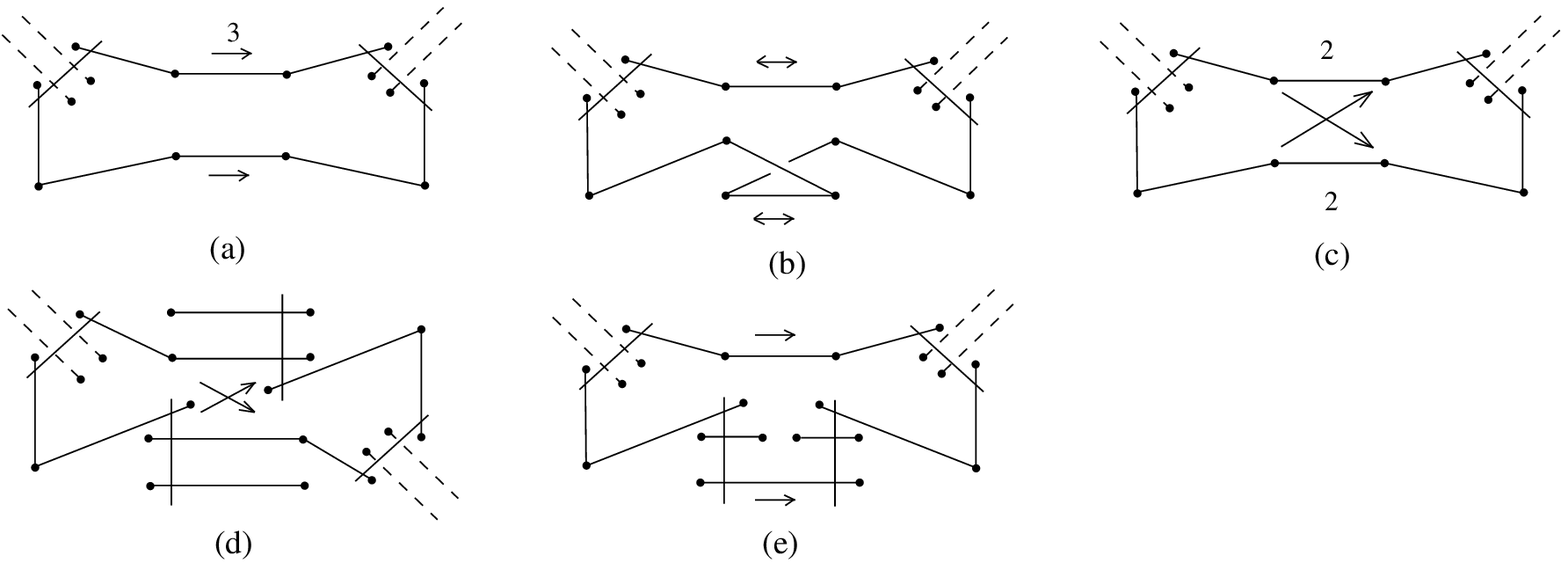}

The two crosscaps configurations were discussed at length in section \ref{localclosed}. The graphs come in sets and there is a single set such that 
the sum of the contributions of the corresponding graphs does not vanish. That set is presented in figure \ref{unvervi}. The contributions of the graphs are
\bea
&&C_{(0,4,3)}^{(2,0,0,...),(a)}=-{(\lambda_u^2-\lu\lv+\lambda_v^2)(2\lambda_u^2-2\lu\lv+\lambda_v^2)\over 2\lambda_z^2(\lv-2\lu)^2},\nn\\
&&C_{(0,4,3)}^{(2,0,0,...),(b)}={1\over 2}C_{(0,4,3)}^{(2,0,0,...),(c)}=
-{\lu\lv(2\lambda_u^2-2\lu\lv+\lambda_v^2)\over 2\lambda_z^2(\lv-2\lu)^2},\nn\\
&&C_{(0,4,3)}^{(2,0,0,...),(d)}=-C_{(0,4,3)}^{(2,0,0,...),(e)}=
{(\lambda_u^2+\lambda_v^2)(2\lambda_u^2-2\lu\lv+\lambda_v^2)\over 2\lambda_z^2(\lv-2\lu)^2}.\nn
\eea
which sum up to $-{1\over 2} \({2\lambda_u^2-2\lu\lv+\lambda_v^2\over (\lv-2\lu)^2}\)$.

We also obtain
\bea
&&F_{(0,1,0,...)}^{c=1}=Q^{1/2}e^{-t}\left[{\lv-\lu\over \lv-2\lu}-3\left({\lv\over \lv-2\lu}\right)e^{-t}+\dots\right]+\dots,\nn\\
&&F_{(0,1,0,...)}^{c=2}=Q^2e^{-3t}\left[\left({\lv\over \lv-2\lu}\right)+\dots\right]+\dots~.\nn
\eea
We note that for this geometry we obtain agreement with the vertex computation if we set $p=-{\lv-\lu\over\lv-2\lu}$.

\subsubsection{$SO/Sp$ Hopf Link Invariant}

The results obtained from the topological vertex are:
\bea
F_{((1,0,...),(1,0,\dots))}^{c=1}&=& (-1)^{p_1+p_2} Q^{1/2} (1-Q) g_s - {1 \over 24} (-1)^{p_1+p_2}Q^{1/2} (1-Q) g_s^3 + \dots, \nn\\
F_{((2,0,...),(1,0,\dots))}^{c=1}&=& {1\over 2}(-1)^{p_2} Q^{1/2} {\left[1+2p_1+2p_1^2-2Q(1+2p_1^2)+Q^2(1-2p_1+2p_1^2) \right]} g_s^2 + \dots, \nn\\
F_{((0,1,0,...),(1,0,\dots))}^{c=1}&=& (-1)^{p_2} Q^{1/2} {\left[1+2p_1 -4Q p_1-Q^2(1-2p_1) \right]} g_s + \dots~.\nn
\eea
The localization results are:
\bea
F_{((1,0,...),(1,0,\dots))}^{c=1}&=& Q^{1/2} (1-Q) g_s - {1 \over 24}Q^{1/2} (1-Q) g_s^3 + \dots, \nn\\
F_{((2,0,...),(1,0,\dots))}^{c=1}&=& -{1\over 2}Q^{1/2} \bigg[{\lambda_u^2+2\lu\lv+2\lambda_v^2\over \lambda_u^2}-2Q \left(\lu^2+2\lv^2 \over \lu^2 \right) \nn\\
&&~~~~~~~~~~+ Q^2 \left(\lambda_u^2-2\lu\lv+2\lambda_v^2\over \lambda_u^2 \right) \bigg] g_s^2 + \dots, \nn\\
F_{((0,1,0,...),(1,0,\dots))}^{c=1}&=& Q^{1/2} {\left[{\lu+2\lv\over\lu} -4Q \left(\lv \over \lu \right) -Q^2 \left(\lu-2 \lv \over \lu \right) \right]} g_s+\dots~.\nn
\eea
To obtain agreement with the vertex result for this D-brane configuration, we need to set $p_1={\lv\over\lu}$. These computations 
offer strong evidence of the equivalence between the vertex computation and the localization on the moduli space of stable open unoriented 
maps.

\section{Application: the BPS Structure of the Coloured Kauffman 
Polynomial}\label{kauffman}

One of the most interesting applications of the above results is the 
determination of the BPS structure of the coloured Kauffman polynomial. 
In contrast to the results obtained for orientifolds of toric geometries above, 
we will not be able to give a full determination of all quantities involved for 
arbitrary knots, but we can still formulate some interesting structural 
properties of the knot polynomials similar to those explained in \cite{Ooguri:1999bv,Labastida:2000yw,Labastida:2000zp,Labastida:2001ts}. We will first recall the results for the coloured HOMFLY polynomial, and then 
we will state and illustrate the results for the coloured Kauffman polynomial. 

\subsection{Chern-Simons Invariants and Knot Polynomials}

Let us consider Chern-Simons theory on ${\bf S}^3$ with gauge group $G$. 
The natural operators in this theory were introduced in section \ref{compopen}. They consist in the holonomies of the gauge 
connection around a knot ${\cal K}$, 
\beq
W_R^{\cal K}(A)={\rm P} \, \exp \oint_{\cal K} A.
\eeql{wloop}
If we now consider a link ${\cal L}$ with components ${\cal K}_{\alpha}$, $\alpha=1,
\cdots, L$, the correlation function
\beq
W_{R_1 \cdots R_L}^G({\cal L})=\langle W^{{\cal K}_1}_{R_1}\cdots
W^{{\cal K}_L}_{R_L}\rangle
\eeql{vevwilson}
defines a topological
invariant of the link ${\cal L}$. In this equation 
the bracket denotes a normalized vacuum expectation value, and we have 
indicated the gauge group $G$ as a superscript. It is well known \cite{Witten:1988hf} that 
Chern-Simons produces in fact invariants of {\it framed} links, but in the following 
we will consider knots 
in the so-called standard framing (see \cite{Marino:2004uf,Guadagnini:1990uw} for a review of these topics). 
The correlation functions \refeq{vevwilson} turn out to be rational functions 
of the variables $q^{\pm 1/2}$, $\lambda^{\pm 1/2}$. The variables $q$ and $\lambda$ are defined as in section \ref{CSnotation}.

The vacuum expectation values of Wilson loops are related to link invariants 
obtained from quantum groups \cite{Witten:1988hf}:

1) If $G=U(N)$ and $R_1=\cdots =R_L=\yng(1)$, then
\beq
W^{U(N)}_{\yng(1) \cdots \yng(1)}({\cal L})=
\lambda^{{\rm lk}({\cal L})}
\biggl( {\lambda^{1\over 2} -\lambda^{-{1\over 2}}
\over q^{1\over 2} -q ^{-{1\over 2}}} \biggr) P_{\cal L}(q, \lambda)
\eeql{homflyrel}
where $P_{\cal L}(q, \lambda)$ is the HOMFLY polynomial of ${\cal L}$ \cite{Freyd:1985dx}, 
and ${\rm lk}({\cal L})$ is its linking number.

2) If $G=SO(N)$ and $R_1=\cdots =R_L=\yng(1)$, then
\beq
W^{SO(N)}_{\yng(1) \cdots \yng(1)}({\cal L})=
\lambda^{{\rm lk}({\cal L})}
\biggl(1+ {\lambda^{1\over 2} -\lambda^{-{1\over 2}}
\over q^{1\over 2} -q ^{-{1\over 2}}} \biggr) F_{\cal L}(q, \lambda)
\eeql{kaufmannrel}
where $F_{\cal L}(q, \lambda)$ is the Kauffman polynomial of ${\cal L}$ \cite{Kauffman:1990}.

We will call $W_{R_1 \cdots R_L}^{U(N)}({\cal L})$ and $W_{R_1 \cdots R_L}^{SO(N)}({\cal L})$ 
the coloured HOMFLY and Kauffman polynomials of ${\cal L}$, respectively. Note that there is 
a slight abuse of 
language here, since these Chern-Simons correlation functions are not polynomials, but rather rational 
functions.

\subsection{BPS Structure: Statement and Examples}
  
In \cite{Ooguri:1999bv}, Ooguri and Vafa extended the duality of \cite{Gopakumar:1998ki} between Chern-Simons theory on ${\bf S}^3$ and 
topological strings on the resolved conifold by incorporating the correlation functions \refeq{vevwilson}. 
We will consider the case of knots, although everything we will say has a straightforward 
generalization to links. The results of \cite{Ooguri:1999bv} are the following: first, to any knot ${\cal K} \in {\bf S}^3$ one can associate 
a Lagrangian submanifold $S_{\cal K}$ in the resolved conifold. Moreover, 
the generating functional of knot invariants 
\beq
Z_{U(N)}(V)=\sum_R  W_R^{U(N)}({\cal K}) {\rm Tr}_R \, V
\eeql{genun}
where $V$ is a $U(M)$ matrix, is the partition function for open topological strings propagating on the resolved 
conifold and with Dirichlet boundary conditions associated to $S_{\cal K}$ 
(after some appropriate analytic continuation). 
Equivalently, we consider $M$ branes wrapping $S_{\cal K}$, where $M$ is the 
rank of $V$, and compute the partition function of topological string theory in this 
D-brane background. Since open string amplitudes have the BPS structure explained in \refeq{genfunctopen} and \refeq{fr}, 
this leads to structure results for the knot invariants $W_R^{U(N)}({\cal K})$ (which play the 
r\^ole of $Z_R$). This is explained in detail in \cite{Labastida:2000yw,Labastida:2000zp,Labastida:2001ts}. 

The large $N$ duality of \cite{Gopakumar:1998ki} can be generalized by considering an orientifold of the 
two geometries involved in the geometric transition, namely the resolved and the deformed conifold \cite{Sinha:2000ap}, which was described in detail in section \ref{geometry}. The string field
theory for the resulting open strings is Chern-Simons theory with gauge group $SO(N)$ or $Sp(N)$, depending on the choice of
orientifold action on the gauge group. The orientifold action 
on the resolved conifold is given by \refeq{orres}. It then follows from the results of 
\cite{Ooguri:1999bv} and the orientifold action considered in \cite{Sinha:2000ap} that the Chern-Simons generating functional 
\beq
Z_{SO/Sp}(V)=\sum_R W_R^{SO/Sp}({\cal K}) {\rm Tr}_R \, V, 
\eeql{cssogen}
where $V$ is again a $U(M)$ matrix, is the total partition function for open strings propagating 
on the orientifold of the resolved conifold with $M$ branes wrapping $S_{\cal K}$. In 
particular, the logarithm of \refeq{cssogen} will have the structure explained in \refeq{freen}, where the 
oriented contribution is obtained by considering a covering geometry 
with both $S_{\cal K}$ and its image under the involution \refeq{orres}, 
$I(S_{\cal K})$. We can then translate the structure results presented 
in section \ref{structure} into structure results for the coloured Kauffman polynomial. 

The main problem in making this translation precise is that, given an arbitrary knot ${\cal K}$, 
we lack a precise prescription to compute the contribution of the covering amplitude. 
The covering amplitude ${\cal C}_{R_1 R_2}$ is defined as the oriented amplitude in the covering 
geometry in the presence of two sets of branes wrapping $S_{\cal K}$ and $I(S_{\cal K})$, 
with representations $R_1$, $R_2$, respectively. If one of the representations is trivial, 
we recover the oriented amplitude in the presence of $S_{\cal K}$, therefore ${\cal C}_{R}=W_R^{U(N)}({\cal K})$. 
But in the general case it is not obvious how to determine ${\cal C}_{R_1 R_2}$. Although there are 
proposals for the geometry of the Lagrangian submanifolds $S_{\cal K}$ \cite{Labastida:2000yw,Taubes:2001wk}, a  
direct Gromov--Witten computation of the corresponding open string amplitudes seems to be very difficult. One 
possible way of determining ${\cal C}_{R_1 R_2}$ would be to translate it into a pure knot-theoretic computation 
in the context of Chern-Simons theory, but we have not found a completely satisfactory solution 
to this problem yet. 

Although we do not know how to compute the covering amplitude for an arbitrary knot, we can 
still extract the $\widehat f_{R}^{c=1}$ amplitudes from the knowledge of $W_{R}^{SO(N)}({\cal K})$. 
This goes as follows. Let us define the rational functions $g_R(q,\lambda)$ through the 
following equation
\beq
\log Z_{SO}(V)=\sum_R \sum_{d \, {\rm odd}}{1\over d}g_R (q^d, \lambda^d)
{\rm Tr}_R V^d,
\eeql{gR}
and define as well
\beq
\widehat g_R (q,\lambda)=\sum_{R R'} M^{-1}_{R R'} (q)g_R (q, \lambda).
\eeql{hgR}
Clearly, since we are not substracting the covering piece in the l.h.s. of \refeq{gR}, we cannot expect 
much structure for $\widehat g_R (q,\lambda)$. However, one has that
\beq
\widehat f_R^{c=1}(q, \lambda)={1\over 2} \bigl(\widehat g_R (q,\lambda^{1\over 2})-
(-1)^{\ell(R)}\widehat g_R (q,-\lambda^{1\over 2})\bigr).
\eeql{frcone}
This follows from parity considerations. The invariants $W_R^{U(N)}({\cal K})$ have powers of $\lambda^{1\over 2}$ of the form 
$\ell(R) + 2k$, while $W_R^{SO(N)}({\cal K})$ have powers of $\lambda^{1\over 2}$ both of the form $\ell(R) + 2k$ and $\ell(R) + 2k+1$. 
The first ones correspond to both oriented and $c=2$ contributions, while the last ones correspond to $c=1$ 
contributions. Also, the covering contribution ${\cal C}_{R_1 R_2}$ (being an oriented amplitude) contains only powers 
in $\lambda^{1\over 2}$ of the form $\ell(R_1) + \ell(R_2) + 2k$. It is now easy to see from the 
results in section \ref{bpsstructorient} that $\widehat f_R^{c=1}$ does not involve at all the covering contributions, and can be 
computed solely from the $SO(N)$ invariants, precisely in the way specified by \refeq{frcone}. We can then formulate 
the following conjecture concerning the structure of the coloured Kauffman polynomial:

\begin{conjecture}\label{ourconj}
Let $\widehat g_R (q, \lambda)$ be defined in terms of the coloured Kauffman polynomial 
by \refeq{gR} and \refeq{hgR}. Then, we have that
\beq
{1\over 2} \bigl(\widehat g_R (q,\lambda^{1\over 2})-
(-1)^{\ell(R)}\widehat g_R (q,-\lambda^{1\over 2})\bigr)=\sum_{g,\beta} N^{c=1}_{R,g,\beta} 
(q^{1\over 2} -q^{-{1\over 2}})^{2g} \lambda^\beta,
\eeql{conj}
where $N^{c=1}_{R,g,\beta}$ are {\rm integer} numbers. They are 
BPS invariants corresponding to unoriented open string amplitudes with one crosscap.    
\end{conjecture}
 
In the case of $W_{\yng(1)}^{SO(N)}(q,\lambda)$, which is the unnormalized Kauffman polynomial, we can be slightly 
more precise, since we know that ${\cal C}_{\yng(1)}(q,\lambda)=W_{\yng(1)}^{U(N)}(q,\lambda)$, which is 
the unnormalized HOMFLY polynomial. We then deduce that
\beq
W_{\yng(1)}^{SO(N)}(q,\lambda)-W_{\yng(1)}^{U(N)}(q,\lambda) = \sum_{g,\beta} N^{c=1}_{\yng(1),g,\beta} 
(q^{1\over 2} -q^{-{1\over 2}})^{2g} \lambda^\beta + 
\sum_{g,\beta} N^{c=2}_{\yng(1),g,\beta} (q^{1\over 2} -q^{-{1\over 2}})^{2g+1} \lambda^\beta.
\eeql{HOMK}
On the other hand, it follows from integrality of the oriented amplitudes that
\beq
W_{\yng(1)}^{U(N)}(q,\lambda)=\sum_{g\ge 0} p^H_g (\lambda) (q^{1\over 2} -q^{-{1\over 2}})^{2g-1},
\eeql{expH}
where $p^H_g (\lambda)$ is an odd polynomial in $\lambda^{\pm {1\over 2}}$. Thus the following corollary follows from conjecture \ref{ourconj}:

\begin{corollary} 
In the case where $R=\yng(1)$, the structure of the unnormalized Kauffman polynomial is given by
\beq
W_{\yng(1)}^{SO(N)}(q,\lambda)=\sum_{b\ge 0} p^K_b (\lambda) (q^{1\over 2} -q^{-{1\over 2}})^{b-1},
\eeql{expHii}
where $p^K_b (\lambda)$ is an odd (even) polynomial in $\lambda^{\pm {1\over 2}} $ for $b$ even (odd). Moreover, 
\beq
p_0^K(\lambda)=p_0^H(\lambda).
\eeql{equalHK}
\end{corollary}

This structural prediction turns out to be a well-known 
result in the theory of the Kauffman polynomial, see for example \cite{Lickorish:1998}, page 183. 
One can easily compute $N^{c=1,2}_{\yng(1),g,\beta}$ for various knots by computing the 
corresponding Kauffman polynomial. For example, the results of \cite{Labastida:1995kf} imply that 
\beq
N^{c=2}_{\yng(1),g,\beta}=0
\eeql{twovan}
for all torus knots.  
 
Let us now turn to checks of conjecture \ref{ourconj} for different knots and higher representations. 
The simplest case is of course the unknot, but 
this case has been already checked in section \ref{examplesopen} (indeed, in the case of the unknot we know even how 
to compute the covering amplitude for arbitrary representations). In order to test the conjecture, we 
have to compute the invariants $W_{R}^{SO(N)}({\cal K})$ for arbitrary $R$. 
A class of nontrivial knots where this is feasible are torus knots. In the case of $U(N)$ invariants, this 
was done in \cite{Labastida:2000zp} by using the formalism of knot operators \cite{Labastida:1990bt} and the results of \cite{Labastida:1993xg}. 
For $SO(N)$, the formalism of knot operators was used in \cite{Labastida:1995kf} to compute invariants in the fundamental 
representation, but this has not been generalized to higher representations. For torus knots of the form $(2,m)$, 
however, one can use the results of \cite{RamaDevi:1992dh} to write down a formula for the invariants in any 
representation of any gauge group. The formula reads as follows:
\beq
{\cal W}_{R}^G({\cal K}_{(2,m)}) =\sum_{S \in R \otimes R} ({\rm dim}_q S) (c_S(R, R))^m
\eeql{twominv}
where
\beq
c_S(R_1, R_2)= \epsilon_{R_1 R_2}^S q^{{C_{R_1} + C_{R_2} \over 2} -{C_S\over 4}}
\eeql{coeff}
In this equation, $C_R$ is the quadratic 
Casimir 
\beq
C_R=\kappa_R + \ell(R) (N+a),
\eeql{quadcas}
where $a$ is given in \refeq{avaluesclosed}, and $\epsilon_{R_1 R_2}^S$ is a sign which counts whether $S$ appears 
symmetrically or antisimmetrically in the tensor product $R_1 \otimes R_2$. 
In cases where $S$ appears with no multiplicity, there is an explicit expression for this sign 
given by \cite{Naculich:1992uf}
\beq
\epsilon_{R_1 R_2}^S= (-1)^{\rho\cdot(\Lambda_1 + \Lambda_2 -\Lambda_S)},
\eeql{sign}
where $\Lambda_1$, $\Lambda_2$, $\Lambda_S$ are the 
highest weights of to the representations 
$R_1$, $R_2$, $S$, respectively. 
Using \refeq{twominv} one can easily compute the invariants of the 
$(2,m)$ torus knots in the $SO(N)$ case, and extract $g_R$ (hence $N_{R,g,\beta}^{c=1}$) for various 
representations. In all cases we have found agreement with the above 
conjecture. We now present some results for the BPS invariants for the simplest torus knot, the $(2,3)$ knot or trefoil knot, 
for representations up to three boxes. We present the results up to two boxes in tables \ref{trefoil1}--\ref{trefoil3} (in the tables the representations $R$ are denoted by their row lengths $(\m_1,\ldots,\m_{d(\m)})$). The results with three boxes are listed in appendix \ref{trefoilapp}, namely in tables \ref{trefoil4}--\ref{trefoil6}. All the invariants that are not shown in the tables are understood to be zero.

\begin{table}[htp]
\begin{center}
\begin{footnotesize}
\begin{tabular}{|c|rrr|}\hline
&$\beta=1$&$2$&$3$
\\
\hline
$g=0$&$3$&$-3$&$1$
\\
$1$&$1$&$-1$&$0$
\\
\hline
\end{tabular}
\caption{BPS invariants $N_{(1),g,\beta}^{c=1}$ for the trefoil knot.}
\label{trefoil1}
\end{footnotesize}
\end{center}
\end{table}

\begin{table}[htp]
\begin{center}
\begin{footnotesize}
\begin{tabular}{|c|rrrrr|}\hline
&$\beta=3/2$&$5/2$&$7/2$&$9/2$&$11/2$
\\
\hline
$g=0$&$8$&$-39$&$69$&$-53$&$15$
\\
$1$&$6$&$-61$&$146$&$-126$&$35$
\\
$2$&$1$&$-37$&$128$&$-120$&$28$
\\
$3$&$0$&$-10$&$56$&$-55$&$9$
\\
$4$&$0$&$-1$&$12$&$-12$&$1$
\\
$5$&$0$&$0$&$1$&$-1$&$0$
\\
\hline
\end{tabular}
\caption{BPS invariants $N_{(2),g,\beta}^{c=1}$ for the trefoil knot.}
\label{trefoil2}
\end{footnotesize}
\end{center}
\end{table}

\begin{table}[htp]
\begin{center}
\begin{footnotesize}
\begin{tabular}{|c|rrrrr|}\hline
&$\beta=3/2$&$5/2$&$7/2$&$9/2$&$11/2$
\\
\hline
$g=0$&$16$&$-69$&$111$&$-79$&$21$
\\
$1$&$20$&$-146$&$307$&$-251$&$70$
\\
$2$&$8$&$-128$&$366$&$-330$&$84$
\\
$3$&$1$&$-56$&$230$&$-220$&$45$
\\
$4$&$0$&$-12$&$79$&$-78$&$11$
\\
$5$&$0$&$-1$&$14$&$-14$&$1$
\\
$6$&$0$&$0$&$1$&$-1$&$0$
\\
\hline
\end{tabular}
\caption{BPS invariants $N_{(1,1),g,\beta}^{c=1}$ for the trefoil knot.}
\label{trefoil3}
\end{footnotesize}
\end{center}
\end{table}

Although we have focused in this section on the case of knots, it is straightforward 
to extend the conjecture above to the case of links, and extract the $c=1$ piece from the 
$SO$ Chern-Simons invariants. Framed knots can be also considered by using exactly the 
same rules that are used for $U(N)$ invariants \cite{Marino:2001re}. In fact, these simple generalizations of our results have recently been worked out explicitely in \cite{Borhade:2005pw}, where many more checks of conjecture \ref{ourconj} have also been performed.

\chapter{Conclusions and Future Directions}\label{conclusion}

In this thesis we explored various applications of toric geometry to string theory. We focused on two particular fields of research deeply rooted in concepts of toric geometry.

Our first theme relied on the construction of Calabi-Yau threefolds as hypersurfaces in toric variety, using Batyrev's reflexive polytopes. One of the main applications of such an approach has been to dualities between compactifications of heterotic strings and type II string theory (or F-theory). Among other things, it was conjectured that the enhanced gauge group of the heterotic compactifications can be read off directly from a geometrical object that was called {\it top}.

In chapter \ref{tops} we classified mathematically all these tops. Moreover, we derived a simple prescription to assign an affine Kac-Moody algebra to each top. Some tops lead to twisted affine Kac-Moody algebras. These ``twisted tops" may be used to build compactifications with reduced rank, which should be dual to CHL strings.

Although the duality between F-theory (or type II string theory) and heterotic strings has been conjectured quite a while ago, even the exact form of the conjecture is still unclear. Many concurrent proposals exist in the literature, applying to different situations and types of geometries. It is unclear if these proposals are all equivalent, and many subtle but important details have been glossed over. It would be interesting to use our results of chapter \ref{tops} to try to formulate rigorously this duality, and consequently shed light on its mathematical properties and physical implications. 

Our second application concerned topological strings on orientifolds of toric Calabi-Yau threefolds. We found how to compute closed --- in chapter \ref{closed} --- and open --- in chapter \ref{open} --- topological string amplitudes on Calabi-Yau orientifolds, by using geometric transitions involving $SO/Sp$ Chern-Simons theory, the topological vertex formalism and a suitable extension to include orientifolds, and localization techniques. We found the general structure of the twisted and untwisted contributions, determined the BPS structure of the corresponding amplitudes, and checked our results in various examples. This allowed us to extract BPS invariants counting higher genus curves with one and two crosscaps.

We want to remark that, although our main testing ground has been orientifolds of (noncompact) toric Calabi-Yau threefolds, with or without noncompact D-branes, the general results about the structure and integrality properties of the amplitudes should be valid in general.

This work can be extended in various ways. First of all, it would be very interesting to consider Calabi-Yau orientifolds in which the involution has fixed loci, like for example the ones considered in \cite{Acharya:2002ag}. This would lead to topological strings on orientifolds with fixed planes. In this case, the geometric transition of \cite{Sinha:2000ap} is no longer useful and one has to find other ways of implementing a Chern-Simons dual description. In the context of the topological vertex formalism, we should find the right prescription to deal with fixed point loci, by using perhaps the group-theoretic results of \cite{Littlewood:1940} for $SO/Sp$.

It would also be very important to clarify some issues that appeared in the orientifolds we studied here. For example, one would like to have a more detailed derivation of the multicovering formulae for amplitudes involving two crosscaps and of the choice of annulus operator we made, as well as a more rigorous justification of the localization techniques we used.

On a different note, one of the motivations behind the study of open topological string amplitudes developed in chapter \ref{open} was to extend the results of \cite{Labastida:2000yw,Labastida:2001ts} on the BPS structure of the coloured HOMFLY polynomial to the coloured Kauffman polynomial of knots and links. Although our general structural results on open string amplitudes on orientifolds give a first principles answer to this problem, as it has been made clear in the analysis of the framed unknot and the Hopf link, we have not been able to determine the covering contribution for arbitrary knots. This is an important open issue that one should resolve in order to obtain a complete picture of the correspondence between enumerative geometry and knot invariants implied by large $N$ dualities.

Furthermore, in this thesis we showed that the predictions obtained from the topological vertex and its extension in the unoriented case agree with unoriented localization computations. However, it would be very interesting to derive a more general and precise correspondence between these two approaches, following the lines of the mathematical treatment of the vertex given in \cite{Diaconescu:2003qa,Li:2004uf}, and maybe connect the unoriented Gromov-Witten theory sketched here and in \cite{Diaconescu:2003dq} with a moduli problem involving ideal sheaves, generalizing in this way the results of \cite{Maulik:2003,Maulik:2004}. 

Finally, various new points of view on topological string theory have recently been found. Among others, a derivation using integrable hierarchies has been proposed \cite{Aganagic:2003qj}, and fascinating relations to the quantum foam \cite{Okounkov:2003sp,Iqbal:2003ds,Saulina:2004da,Okuda:2004mb,Halmagyi:2005vk} and to BPS black holes \cite{Ooguri:2004zv,Aganagic:2004js,Vafa:2004qa} have been discovered. These results also lead to the conjectured existence of a {\it topological M-theory} \cite{Nekrasov:2004vv,Dijkgraaf:2004te,Pestun:2005rp,Smolin:2005gu,deBoer:2005pt}, in which topological strings on orientifolds may play a r\^ole \cite{deBoer:2005pt}. It would be very interesting to investigate how topological strings on orientifolds and our extension of the topological vertex formalism relate to these various approaches.

\appendix
\chapter{Useful Formulae}

\section{Subsets of Young Tableaux}\label{young}

To compute the tensor product decomposition of irreducible representations of $SO(N)$ and $Sp(N)$ using Littlewood's technique as 
explained in \refeq{characSON}, we had to use four different subsets of Young tableaux: $\{ \delta \}$ and $\{ \gamma \}$ for $SO(N)$, and 
$\{ \beta \}$ and $\{ \alpha \}$ for $Sp(N)$. These four sets are defined as follows \cite{Littlewood:1940}.

$\{\delta \}$ is the set of all partitions into even parts only: $\{ \yng(2), \yng(4), \yng(2,2), ... \}$.

$\{\beta \}$ is the set of all partitions such that there are an even number of parts of any given magnitude: $\{\yng(1,1), \yng(2,2), 
\yng(1,1,1,1), ... \}$.

To define the two remaining sets we have to use the Frobenius notation \cite{Ramgoolam:1993hh,Littlewood:1940}. In this notation, a Young tableau is described by an 
array of pair of numbers. The number of pairs is equal to the number of boxes in the leading diagonal of the tableau; the upper number of the 
pair is the number of boxes to the right and the lower number is the number of boxes below. For example, the Young tableau $\tableau{3 3 1}$ 
is described in the Frobenius notation by $\begin{pmatrix} 2 & 1 \\ 2 & 0 \\ \end{pmatrix}$.

Using this notation we can define the two remaining sets. Consider Young tableaux defined in the Frobenius notation by
\beq
\begin{pmatrix}
 a_1 & a_2 & a_3 & a_4 & ... \\ b_1 & b_2 & b_3 & b_4 & ... \\
\end{pmatrix}.
\eeql{frobgen}

$\{ \gamma \}$ is the set of Young tableaux such that $a_i = b_i+1~~ \forall~~ i$: $\{ \yng(2), \yng(3,1), \yng(4,1,1),... \}$.

$\{ \alpha \}$ is the set of Young tableaux such that $a_i+1 = b_i~~ \forall~~ i$: $\{ \yng(1,1), \yng(2,1,1), \yng(3,1,1,1),...\}$.

Note that $\{ \beta \}$ and $\{ \alpha \}$ are respectively related to $\{ \delta \}$ and $\{ \gamma \}$ by taking the
transpose of the representations, where by transpose we mean exchanging rows and columns.

\section{The Topological Vertex, Chern-Simons Invariants and Schur Functions}\label{schurapp}

In this appendix we will list some useful identities of Schur functions and their relations to the unknot and Hopf link invariants. For a more 
detailed discussion of Schur functions see for example \cite{Fulton:1997,Macdonald:1995}. Applications of these results to topological 
string computations can be found in \cite{Okounkov:2003sp,Eguchi:2003sj,Hollowood:2003cv,Iqbal:2004ne}.

Let $R$ be a partition associated to a Young tableau. Let $\ell(R)$ be the number of boxes of the Young tableau and $l_i (R)$ be the number of boxes in the $i$-th row. We define the quantity
\beq
W_R (q) = s_R (q^\rho)
\eeql{unknotschur}
where $s_R (q^\rho)$ is the Schur function with the substitution $s_R( x_i = q^{-i + 1/2})$, where $i$ runs from $1$ to $\infty$. 
$W_R (q)$ is the leading order of the $U(N)$ quantum dimension $\dim_q^{U(N)} R$ (in the sense defined in \cite{Aganagic:2002qg}). Also, using Weyl's formula, we can write  a general formula 
for quantum dimensions of a group $G$:
 \beq
\dim_q^G R = \prod_{\alpha \in \Delta_+} {[(\Lambda_R + \rho, \alpha)] \over [(\rho, \alpha)]},
\eeql{qdgeneral}
where $\Lambda_R$ is the highest weight of the representation $R$, $\rho$ is the Weyl vector, and the 
product is over the positive roots of $G$ . We also defined the
following $q$-number:
\beq
[x]=q^{x/2} - q^{-x/2}.
\eeql{qnumber}
Another important object is 
\beq
W_{R_1 R_2} (q) = s_{R_1} ( q^{\rho}) s_{R_2} (q^{\ell(R_1) + \rho} ),
\eeql{hopfschur}
where $s_{R_2} (q^{\ell(R_2) + \rho}) = s_{R_2} (x_i = q^{l_i (R_2)-i+1/2})$. This is the 
leading part (again in the sense of \cite{Aganagic:2002qg}) of the Hopf link invariant ${\cal W}_{R_1 R_2}^{U(N)}$.

The topological vertex formula derived in \cite{Aganagic:2003db} reads
\beq
C_{R_1 R_2 R_3} = q^{{1 \over 2}(\kappa_{R_2}+\kappa_{R_3})} \sum_{Q_1,Q_2, R} N_{Q_1 R}^{R_1} N_{Q_2 R}^{R_3^t} 
{W_{R_2^t Q_1} W_{R_2 Q_2} \over W_{R_2}},
\eeql{topvertexapp}
where $\kappa_R$ is defined by $\kappa_R=\sum_i l_i(R) (l_i(R) -2i +1)$. Using \refeq{unknotschur} and \refeq{hopfschur} we can express the topological vertex in terms of Schur functions (this was first done in \cite{Okounkov:2003sp})
\beq
C_{R_1 R_2 R_3} = q^{{1 \over 2}(\kappa_{R_2}+\kappa_{R_3})} s_{R_2^t} (q^\rho) \sum_{Q} s_{R_1/Q} (q^{\ell(R_2^t) + \rho}) s_{R_3^t / Q} (q^{\ell(R_2) + \rho}),
\eeql{topvertexschur}
where we have used skew Schur functions defined as
\beq
s_{R/R_1} (x)= \sum_{Q} N_{R_1 Q}^R s_Q (x).
\eeql{skewschur}

Schur functions satisfy some useful identities. First, we have
\beq
s_{R^t} (q) = q^{-\kappa_R/2} s_{R} (q) = (-1)^{\ell (R)} s_R (q),
\eeql{identityone}
and similarly
\beq
s_{R/R_1} (q)= (-1)^{\ell(R)-\ell(R_1)} s_{R^t/R_1^t} (q).
\eeql{identitytwo}
The two following formulae are also important:
\bea
\sum_R s_{R/R_1}(x) s_{R/R_2} (y) &=& \prod_{i,j \geq 1} (1-x_i y_j)^{-1} \sum_Q s_{R_2/Q} (x) s_{R_1/Q} (y),\nn\\
\sum_R s_{R/R_1}(x) s_{R^t/R_2} (y) &=& \prod_{i,j \geq 1} (1+x_i y_j) \sum_Q s_{R_2^t/Q} (x) s_{R_1^t/Q^t} (y).
\eeal{identitythree}
The following result was proved in \cite{Eguchi:2003sj}. Let us define the ``relative" hook length
\beq
h_{R_1 R_2} (i,j) = l_i (R_1) +l_j (R_2) -i -j +1,
\eeql{hook}
and the following functions
\bea
f_R(q) &=& {q \over (q-1)} \sum_{i \geq 1} (q^{l_i(R)-i} - q^{-i}),\nn\\
{\tilde f}_{R_1 R_2} (q) &=& {(q-1)^2\over q} f_{R_1} (q) f_{R_2} (q) +f_{R_1}(q) + f_{R_2} (q).
\eeal{fctschur}
Let us denote the expansion coefficients of ${\tilde f}_{R_1 R_2} (q)$ by
\beq
{\tilde f}_{R_1 R_2} (q) = \sum_k C_k (R_1, R_2) q^k.
\eeql{expcoeff}
Alternatively, 
\beq
\sum_k C_k (R_1, R_2) q^k={ W_{R_1 \yng(1)} \over W_{R_1}}
{ W_{R_2 \yng(1)} \over W_{R_2}}- W_{\yng(1)}^2.
\eeql{coeffs}
Then it was proved that
\beq
\prod_{i,j \geq 1} (1-Q q^{h_{R_1 R_2} (i,j)} ) = \prod_{k=1}^{\infty} (1-Q q^k)^k \prod_k (1-Q q^k)^{C_k (R_1, R_2)}.
\eeql{identityfour}

Let us now present a useful result proved by Littlewood \cite{Littlewood:1940, Macdonald:1995}:
\beq
\sum_{R=R^T} s_R (x) (-1)^{{1\over2} (\ell(R)\mp r(R))} = \prod_{i=1}^{\infty} (1 \pm x_i) \prod_{1 \leq i < j < \infty} (1-x_i x_j),
\eeql{littlewood}
where $r(R)$ is the rank of $R$. The final formula that we will need reads as follows \cite{Hollowood:2003cv}
\beq
\prod_{i,j} (1-Q x_i y_j) = \exp \Big{[}- \sum_{n=1}^{\infty} {Q^n \over n} \sum_{i,j} x_i^n y_j^n\Big{]},
\eeql{exponential}
from which we can deduce the identities
\bea
\prod_i (1\mp Q^{1/2} q^{i-1/2}) &=& \exp \Big{[} \sum_{n=1}^{\infty} {(\pm 1)^n Q^{n/2} \over n (q^{n/2}-q^{-n/2})} \Big{]},\nn\\
\prod_{i,j} (1-Q q^{i+j-1}) &=& \exp \Big{[}- \sum_{n=1}^{\infty} {Q^{n} \over n (q^{n/2}-q^{-n/2})^2} \Big{]}.
\eeal{expident}

\chapter{Various Results}

\section[Full Classification of Tops]{Full Classification of Tops}\label{class}

The following tables provide the complete list of possible tops $\top$,
in terms of their duals $\top^*$.
The first column of each table identifies the polygon $F_0^*$ to
which $\top^*$ projects according to the numbering in figure~\ref{fig:16pol}.
The following columns give $z_k:=\zm(b_k)$, where $b_0$ is the
origin of $F_0^*$ and the other $b_k$ are the lattice points at the
boundary, starting at the `12 o'clock position' and proceeding clockwise.
The parameter $i$ takes values in $\{0,1,\ldots\}$.
For the elements of the $A$-series (last row for each choice of
$F_0^*$) the $z_k$ are assumed to satisfy the inequalities determined
by (\ref{latdist}).

The last column indicates the affine Kac-Moody algebra to which $\top$
corresponds, with the superscript ${}^{(1)}$ suppressed for the untwisted
cases.
The case $i=0$ can be special in the sense that the corresponding
Dynkin diagram does not belong to the general family (this occurs when
edges that are distinct for $i>0$ merge when $i=0$), or that
$\top$ is the same for some other family.
In either of these cases we also display the $i=0$ algebra, with
superscripts ${}^a$ or ${}^b$ for repeated non-trivial cases.

Any two duals of tops in our tables are different, with the
following exceptions.
The trivial top (case 0) may occur more than once as the $i=0$ case of
a $C_i$ series; $C_0$ always means the trivial case and its repetitions
are not separately indicated.
If a non-trivial top occurs more than once as an $i=0$ case, it gets a
superscript ${}^a$ or ${}^b$ which is the same for each occurrence.
For the $A$-cases, we have not eliminated the equivalences coming from
automorphisms of $F_0^*$.

\newpage
${}$
\vfill

%\bigskip\bigskip
%\bigskip\bigskip

\begin{table}[!b]
\begin{center}
\begin{footnotesize}
\begin{tabular}{|c|cccccc|c|}\hline
$F_0^*$&$z_0$&$z_1$&$z_2$&$z_3$&$z_4$&$z_5$&AKMA\\[-.5mm]
\hline\hline

1&$-1/3$&$-1$&$-1$&$1$&-&-&$E\sst {6}$\\[-.5mm]
\cline{2-8}
&$-1$&$-1$&$-1$&$-1$&-&-&$C\sst {0}$\\[-.5mm]
\cline{2-8}
&$-1$&$-1$&$-1$&$z_3$&-&-&$A\sst {3z_3+2}$\\[-.5mm]
\hline\hline

2&$-1/2$&$-1$&$-1$&$0$&$i$&-&$D\sst {2i+4}$\\[-.5mm]
\cline{2-8}
&$-1$&$-1$&$-1$&$-1$&$i-1$&-&$C\sst i$\\[-.5mm]
\cline{2-8}
&$-1$&$-1$&$-1$&$z_3$&$z_4$&-&$A\sst {2z_3+2z_4+3}$\\[-.5mm]
\hline\hline

3&$-1/3$&$-1$&$-1$&$1$&$1$&-&$E\sst 6$\\[-.5mm]
\cline{2-8}
&$-1/2$&$-1$&$-1$&$0$&$i+1$&-&$D\sst {2i+5}$\\[-.5mm]
\cline{2-8}
&$-1$&$-1$&$-1$&$-1$&$i-1$&-&$C\sst i$\\[-.5mm]
\cline{2-8}
&$-1$&$-1$&$-1$&$z_3$&$z_4$&-&$A\sst {z_3+2z_4+2}$\\[-.5mm]
\hline\hline

4&$-1/4$&$-1$&$-1$&$1/2$&$2$&-&$E\sst 7$\\[-.5mm]
\cline{2-8}
&$-1/2$&$-1$&$-1$&$0$&$i+1$&-&$D\sst {2i+4} / B\sst 3$\\[-.5mm]
\cline{2-8}
&$-1$&$-1$&$-1$&$-1$&$i-1$&-&$C\sst i$\\[-.5mm]
\cline{2-8}
&$-1$&$-1$&$-1$&$z_3$&$z_4$&-&$A\sst {2z_4+1}$\\[-.5mm]
\hline\hline

5&$-1/3$&$-1$&$-1$&$1$&$1$&$-1$&$E\sst 6$\\[-.5mm]
\cline{2-8}
&$-1/2$&$-1$&$-1$&$0$&$i$&$i$&$B\sst {2i+4}$\\[-.5mm]
\cline{2-8}
&$-1/2$&$-1$&$-1$&$0$&$i + 1$&$i$&$B\sst {2i+5}$\\[-.5mm]
\cline{2-8}
&$-1$&$-1$&$-1$&$-1$&$i-1$&$i-1$&$C\sst i$\\[-.5mm]
\cline{2-8}
&$-1$&$-1$&$-1$&$z_3$&$z_4$&$z_5$&$A\sst {2z_3+z_4+z_5+3}$\\[-.5mm]

\hline\hline

6&$-1/4$&$-1$&$-1$&$1/2$&$2$&$2$&$E\sst 7$\\[-.5mm]
\cline{2-8}
&$-1/3$&$-1$&$-1$&$1/2$&$2$&$1$&$E\sst 6$\\[-.5mm]
\cline{2-8}
&$-1/3$&$-1$&$-1$&$1$&$3$&$1$&$E\sst 6$\\[-.5mm]
\cline{2-8}
&$-1/2$&$-1$&$-1$&$0$&$i+1$&$i+1$&$D\sst {2i+4} / B\sst {3}$\\[-.5mm]
\cline{2-8}
&$-1/2$&$-1$&$-1$&$0$&$i+2$&$i+1$&$D\sst {2i+5}$\\[-.5mm]
\cline{2-8}
&$-1$&$-1$&$-1$&$-1$&$i-1$&$i-1$&$C\sst i$\\[-.5mm]
\cline{2-8}
&$-1$&$-1$&$-1$&$z_3$&$z_4$&$z_5$&$A\sst {z_4+z_5+1}$\\[-.5mm]
\hline\hline

\end{tabular}
\caption{Duals $\top^*$ of tops with $F_0^*$ one of the polygons
$1,\ldots,6$ of figure~\ref{fig:16pol}.}
\label{tab:results1to6}
\end{footnotesize}
\end{center}
\end{table}

%\vfill
%${}$

\newpage
${}$
\vfill

%\renewcommand{\baselinestretch}{1.1}

%${}$
%\vfill

\begin{table}[!b]
\begin{center}
\begin{footnotesize}
\begin{tabular}{|c|ccccccc|c|}
\hline

$F_0^*$&$z_0$&$z_1$&$z_2$&$z_3$&$z_4$&$z_5$&$z_6$&AKMA\\[-.3mm]
\hline\hline

7&$-1/3$&$-1$&$-1$&$1$&$1$&$1$&$-1$&$E\sst 6$\\[-.3mm]
\cline{2-9}
&$-1/2$&$-1$&$-1$&$0$&$0$&$i$&$i$&$D\sst {2i+4}$\\[-.3mm]
\cline{2-9}
&$-1/2$&$-1$&$-1$&$0$&$0$&$i+1$&$i$&$D\sst {2i+5}$\\[-.3mm]
\cline{2-9}
&$-1$&$-1$&$-1$&$-1$&$-1$&$i-1$&$i-1$&$C\sst i$\\[-.3mm]
\cline{2-9}

&$-1$&$-1$&$-1$&$z_3$&$z_4$&$z_5$&$z_6$&$A\sst {z_3+z_4+z_5+z_6+3}$\\[-.3mm]
\hline\hline

8&$-1/4$&$-1$&$-1$&$1/2$&$2$&$3/2$&$1$&$E\sst 7$\\[-.3mm]
\cline{2-9}
&$-1/4$&$-1$&$-1$&$1/2$&$2$&$2$&$2$&$E\sst 7$\\[-.3mm]
\cline{2-9}
&$-1/3$&$-1$&$-1$&$1/2$&$2$&$1$&$1$&$E\sst 6$\\[-.3mm]
\cline{2-9}
&$-1/3$&$-1$&$-1$&$1$&$3$&$1$&$1$&$E\sst 6$\\[-.3mm]
\cline{2-9}

&$-1/2$&$-1$&$-1$&$0$&$i+1$&$i+1/2$&$i$&$D\sst {2i+4} \yy B\sst {3}^a\xx$\\[-.3mm]
\cline{2-9}
&$-1/2$&$-1$&$-1$&$0$&$i+2$&$i+1$&$i$&$D\sst {2i+4} \yy D\sst 4^a\xx$\\[-.3mm]
\cline{2-9}
&$-1/2$&$-1$&$-1$&$0$&$i+1$&$i+1$&$i+1$&$B\sst {2i+4} \yy B\sst 3\xx$\\[-.3mm]
\cline{2-9}
&$-1/2$&$-1$&$-1$&$i$&$2i+2$&$i+1$&$0$&$B\sst {2i+4} \yy D\sst 4^a\xx$\\[-.3mm]
\cline{2-9}
&$-1/2$&$-1$&$-1$&$0$&$i+2$&$i+1$&$i+1$&$B\sst {2i+5}$\\[-.3mm]
\cline{2-9}
&$-1/2$&$-1$&$-1$&$i+1$&$2i+3$&$i+1$&$0$&$B\sst {2i+5}$\\[-.3mm]
\cline{2-9}
&$-1/2$&$-1$&$-1$&$i/2$&$i+1$&$(i+1)/2$&$0$&$B\sst {i+3} \yy B\sst {3}^a\xx$\\[-.3mm]
\cline{2-9}
&$-1$&$-1$&$-1$&$-1$&$i-1$&$i-1$&$i-1$&$C\sst i$\\[-.3mm]
\cline{2-9}
&$-1$&$-1$&$-1$&$i-1$&$2i-1$&$i-1$&$-1$&$C\sst i$\\[-.3mm]
\cline{2-9}
&$-1$&$-1$&$-1$&$z_3$&$z_4$&$z_5$&$z_6$&$A\sst {z_4+z_6+1}$\\[-.3mm]
\hline\hline

9&$-1/4$&$2$&$-1$&$-1$&$-1$&$1/2$&$2$&$E\sst 7$\\[-.3mm]
\cline{2-9}
&$-1/3$&$-1$&$-1$&$1$&$1$&$1/2$&$0$&$E\sst 6$\\[-.3mm]
\cline{2-9}
&$-1/3$&$-1$&$-1$&$1$&$1$&$1$&$1$&$E\sst 6$\\[-.3mm]
\cline{2-9}

&$-1/2$&$-1$&$-1$&$0$&$i + 1$&$i$&$i$&$D\sst {2i+4} \yy D\sst 4^a\xx$\\[-.3mm]
\cline{2-9}
&$-1/2$&$-1$&$-1$&$i$&$i + 1$&$0$&$0$&$D\sst {2i+4} \yy D\sst 4^a\xx$\\[-.3mm]
\cline{2-9}
&$-1/2$&$-1$&$-1$&$i+1$&$i+1$&$0$&$-1$&$D\sst {2i+4} \yy B\sst 3\xx$\\[-.3mm]
\cline{2-9}
&$-1/2$&$-1$&$-1$&$0$&$i+1$&$i+1/2$&$i$&$D\sst {2i+5}$\\[-.3mm]
\cline{2-9}
&$-1/2$&$-1$&$-1$&$0$&$i+1$&$i+1$&$i+1$&$D\sst {2i+5}$\\[-.3mm]
\cline{2-9}
&$-1/2$&$-1$&$-1$&$i+1$&$i+1$&$0$&$0$&$D\sst {2i+5}$\\[-.3mm]
\cline{2-9}

&$-1$&$-1$&$-1$&$-1$&$i-1$&$i-1$&$i-1$&$C\sst i$\\[-.3mm]
\cline{2-9}
&$-1$&$-1$&$-1$&$i-1$&$i-1$&$-1$&$-1$&$C\sst i$\\[-.3mm]
\cline{2-9}
&$-1$&$-1$&$-1$&$z_3$&$z_4$&$z_5$&$z_6$&$A\sst {z_3+z_4+z_6+2}$\\[-.3mm]
\hline\hline

10&$-1/6$&$-1$&$-1$&$2/3$&$7/3$&$4$&$3/2$&$E\sst 8$\\[-.3mm]
\cline{2-9}
&$-1/4$&$-1$&$-1$&$1/2$&$2$&$4$&$3/2$&$E\sst 7$\\[-.3mm]
\cline{2-9}
&$-1/4$&$-1$&$-1$&$1/2$&$2$&$5$&$2$&$E\sst 7$\\[-.3mm]
\cline{2-9}
&$-1/3$&$-1$&$-1$&$1/3$&$5/3$&$3$&$1$&$F\sst 4$\\[-.3mm]
\cline{2-9}
&$-1/3$&$-1$&$-1$&$1/2$&$2$&$4$&$1$&$E\sst 6$\\[-.3mm]
\cline{2-9}
&$-1/3$&$-1$&$-1$&$1$&$3$&$5$&$1$&$E\sst 6$\\[-.3mm]
\cline{2-9}

&$-1/2$&$-1$&$-1$&$0$&$i+1$&$2i+3$&$i+1$&$D\sst {2i+4} \yy B\sst 3\xx$\\[-.3mm]
\cline{2-9}
&$-1/2$&$-1$&$-1$&$0$&$i +2$&$2i+4$&$i+1$&$D\sst {2i+5}$\\[-.3mm]
\cline{2-9}
&$-1/2$&$-1$&$-1$&$0$&$i/2+1$&$i+2$&$(i+1)/2$&$B\sst {i+3} \yy G\sst 2\xx$\\[-.3mm]
\cline{2-9}

&$-1$&$-1$&$-1$&$-1$&$i-1$&$2i-1$&$i-1$&$C\sst i$\\[-.3mm]
\cline{2-9}

&$-1$&$-1$&$-1$&$z_3$&$z_4$&$z_5$&$z_6$&$A\sst {z_5}$\\[-.3mm]
\hline\hline

\end{tabular}
\caption{Duals $\top^*$ of tops with $F_0^*$ one of the polygons
$7,8,9,10$ of figure~\ref{fig:16pol}.}\label{tab:results7to10}
\end{footnotesize}
\end{center}
\end{table}

%\newpage
%\clearpage
%\renewcommand{\baselinestretch}{1.05}

%\renewcommand{\baselinestretch}{1.05}

\newpage
${}$
\vfill

\begin{table}[!b]
\begin{center}
\begin{footnotesize}
\begin{tabular}{|c|cccccccc|c|}
\hline

\!$F_0^*$\!\!&\!\!$z_0$\!\!&\!\!$z_1$\!\!&\!\!$z_2$\!\!&\!\!$z_3$\!\!&\!\!$z_4$\!\!&\!\!$z_5$\!\!&\!\!$z_6$\!\!&\!\!$z_7$\!&\!AKMA\!\\[-.4mm]
\hline\hline

\!11\!&\!$-1/6$\!\!&\!\!$-1$\!\!&\!\!$-1$\!\!&\!\!$-1$\!\!&\!\!$2/3$\!\!&\!\!$7/3$\!\!&\!\!$4$\!\!&\!\!$3/2$\!\!&\!\!$E\sst 8$\!\\[-.4mm]
\cline{2-10}
\!&\!$-1/4$\!\!&\!\!$-1$\!\!&\!\!$-1$\!\!&\!\!$-1$\!\!&\!\!$1/2$\!\!&\!\!$2$\!\!&\!\!$4$\!\!&\!\!$3/2$\!\!&\!\!$E\sst 7$\!\\[-.4mm]
\cline{2-10}
\!&\!$-1/4$\!\!&\!\!$-1$\!\!&\!\!$-1$\!\!&\!\!$-1$\!\!&\!\!$1/2$\!\!&\!\!$2$\!\!&\!\!$5$\!\!&\!\!$2$\!\!&\!\!$E\sst 7$\!\\[-.4mm]
\cline{2-10}
\!&\!$-1/4$\!\!&\!\!$-1$\!\!&\!\!$-1$\!\!&\!\!$0$\!\!&\!\!$2/3$\!\!&\!\!$4/3$\!\!&\!\!$2$\!\!&\!\!$1/2$\!\!&\!\!$E\sst 7$\!\\[-.4mm]
\cline{2-10}
\!&\!$-1/4$\!\!&\!\!$-1$\!\!&\!\!$-1$\!\!&\!\!$1$\!\!&\!\!$1$\!\!&\!\!$3/2$\!\!&\!\!$2$\!\!&\!\!$1/2$\!\!&\!\!$E\sst 7$\!\\[-.4mm]
\cline{2-10}
\!&\!$-1/4$\!\!&\!\!$-1$\!\!&\!\!$-1$\!\!&\!\!$2$\!\!&\!\!$2$\!\!&\!\!$2$\!\!&\!\!$2$\!\!&\!\!$1/2$\!\!&\!\!$E\sst 7$\!\\[-.4mm]
\cline{2-10}
\!&\!$-1/3$\!\!&\!\!$-1$\!\!&\!\!$-1$\!\!&\!\!$-1$\!\!&\!\!$1/3$\!\!&\!\!$5/3$\!\!&\!\!$3$\!\!&\!\!$1$\!\!&\!\!$F\sst 4$\!\\[-.4mm]
\cline{2-10}
\!&\!$-1/3$\!\!&\!\!$-1$\!\!&\!\!$1$\!\!&\!\!$1$\!\!&\!\!$1/2$\!\!&\!\!$0$\!\!&\!\!$0$\!\!&\!\!$-1$\!\!&\!\!$E\sst 6$\!\\[-.4mm]
\cline{2-10}
\!&\!$-1/3$\!\!&\!\!$-1$\!\!&\!\!$1$\!\!&\!\!$1$\!\!&\!\!$1$\!\!&\!\!$1$\!\!&\!\!$1$\!\!&\!\!$-1$\!\!&\!\!$E\sst 6$\!\\[-.4mm]
\cline{2-10}
\!&\!$-1/3$\!\!&\!\!$-1$\!\!&\!\!$-1$\!\!&\!\!$0$\!\!&\!\!$1/2$\!\!&\!\!$1$\!\!&\!\!$2$\!\!&\!\!$1/2$\!\!&\!\!$E\sst 6$\!\\[-.4mm]
\cline{2-10}
\!&\!$-1/3$\!\!&\!\!$-1$\!\!&\!\!$-1$\!\!&\!\!$1$\!\!&\!\!$1$\!\!&\!\!$1$\!\!&\!\!$2$\!\!&\!\!$1/2$\!\!&\!\!$E\sst 6$\!\\[-.4mm]
\cline{2-10}
\!&\!$-1/3$\!\!&\!\!$-1$\!\!&\!\!$-1$\!\!&\!\!$0$\!\!&\!\!$1/2$\!\!&\!\!$1$\!\!&\!\!$3$\!\!&\!\!$1$\!\!&\!\!$E\sst 6$\!\\[-.4mm]
\cline{2-10}
\!&\!$-1/3$\!\!&\!\!$-1$\!\!&\!\!$-1$\!\!&\!\!$1$\!\!&\!\!$1$\!\!&\!\!$1$\!\!&\!\!$3$\!\!&\!\!$1$\!\!&\!\!$E\sst 6$\!\\[-.4mm]
\cline{2-10}

\!&\!$-1/2$\!\!&\!\!$-1$\!\!&\!\!$-1$\!\!&\!\!$0$\!\!&\!\!$0$\!\!&\!\!$i+1$\!\!&\!\!$2i+2$\!\!&\!\!$i$\!\!&\!\!$D\sst {2i+4} \yy D\sst 4^a\xx$\!\\[-.4mm]
\cline{2-10}
\!&\!$-1/2$\!\!&\!\!$-1$\!\!&\!\!$-1$\!\!&\!\!$-1$\!\!&\!\!$0$\!\!&\!\!$i+1$\!\!&\!\!$2i+3$\!\!&\!\!$i+1$\!\!&\!\!$D\sst {2i+4} \yy B\sst 3\xx$\!\\[-.4mm]
\cline{2-10}
\!&\!$-1/2$\!\!&\!\!$-1$\!\!&\!\!$-1$\!\!&\!\!$i$\!\!&\!\!$i$\!\!&\!\!$i+1/2$\!\!&\!\!$i+1$\!\!&\!\!$0$\!\!&\!\!$D\sst {2i+4} \yy B\sst 3^a\xx$\!\\[-.4mm]
\cline{2-10}
\!&\!$-1/2$\!\!&\!\!$-1$\!\!&\!\!$-1$\!\!&\!\!$i$\!\!&\!\!$i$\!\!&\!\!$i+1$\!\!&\!\!$i+2$\!\!&\!\!$0$\!\!&\!\!$D\sst {2i+4} \yy D\sst 4^a\xx$\!\\[-.4mm]
\cline{2-10}
\!&\!$-1/2$\!\!&\!\!$-1$\!\!&\!\!$-1$\!\!&\!\!$i+1$\!\!&\!\!$i+1$\!\!&\!\!$i+1$\!\!&\!\!$i+1$\!\!&\!\!$0$\!\!&\!\!$D\sst {2i+4} \yy B\sst 3\xx$\!\\[-.4mm]
\cline{2-10}

\!&\!$-1/2$\!\!&\!\!$-1$\!\!&\!\!$-1$\!\!&\!\!$0$\!\!&\!\!$0$\!\!&\!\!$i+1$\!\!&\!\!$2i+3$\!\!&\!\!$i+1$\!\!&\!\!$D\sst {2i+5}$\!\\[-.4mm]
\cline{2-10}
\!&\!$-1/2$\!\!&\!\!$-1$\!\!&\!\!$-1$\!\!&\!\!$-1$\!\!&\!\!$0$\!\!&\!\!$i +2$\!\!&\!\!$2i+4$\!\!&\!\!$i+1$\!\!&\!\!$D\sst {2i+5}$\!\\[-.4mm]
\cline{2-10}
\!&\!$-1/2$\!\!&\!\!$-1$\!\!&\!\!$-1$\!\!&\!\!$i$\!\!&\!\!$i+1/2$\!\!&\!\!$i+1$\!\!&\!\!$i+2$\!\!&\!\!$0$\!\!&\!\!$D\sst {2i+5}$\!\\[-.4mm]
\cline{2-10}
\!&\!$-1/2$\!\!&\!\!$-1$\!\!&\!\!$-1$\!\!&\!\!$i$\!\!&\!\!$i+1$\!\!&\!\!$i+2$\!\!&\!\!$i+3$\!\!&\!\!$0$\!\!&\!\!$D\sst {2i+5}$\!\\[-.4mm]
\cline{2-10}
\!&\!$-1/2$\!\!&\!\!$-1$\!\!&\!\!$-1$\!\!&\!\!$i+1$\!\!&\!\!$i+1$\!\!&\!\!$i+1$\!\!&\!\!$i+2$\!\!&\!\!$0$\!\!&\!\!$D\sst {2i+5}$\!\\[-.4mm]
\cline{2-10}

\!&\!$-1/2$\!\!&\!\!$-1$\!\!&\!\!$-1$\!\!&\!\!$0$\!\!&\!\!$0$\!\!&\!\!$(i+1)/2$\!\!&\!\!$i+1$\!\!&\!\!$i/2$\!\!&\!\!$B\sst {i+3} \yy B\sst 3^a\xx$\!\\[-.4mm]
\cline{2-10}
\!&\!$-1/2$\!\!&\!\!$-1$\!\!&\!\!$-1$\!\!&\!\!$-1$\!\!&\!\!$0$\!\!&\!\!$i/2+1$\!\!&\!\!$i+2$\!\!&\!\!$(i+1)/2$\!\!&\!\!$B\sst {i+3} \yy G\sst 2\xx$\!\\[-.4mm]
\cline{2-10}

\!&\!$-1$\!\!&\!\!$-1$\!\!&\!\!$-1$\!\!&\!\!$i-1$\!\!&\!\!$i-1$\!\!&\!\!$i-1$\!\!&\!\!$i-1$\!\!&\!\!$-1$\!\!&\!\!$C\sst i$\!\\[-.4mm]
\cline{2-10}
\!&\!$-1$\!\!&\!\!$-1$\!\!&\!\!$-1$\!\!&\!\!$-1$\!\!&\!\!$-1$\!\!&\!\!$i-1$\!\!&\!\!$2i-1$\!\!&\!\!$i-1$\!\!&\!\!$C\sst i$\!\\[-.4mm]
\cline{2-10}

\!&\!$-1$\!\!&\!\!$-1$\!\!&\!\!$-1$\!\!&\!\!$z_3$\!\!&\!\!$z_4$\!\!&\!\!$z_5$\!\!&\!\!$z_6$\!\!&\!\!$z_7$\!\!&\!\!$A\sst {z_3+z_6+1}$\!\\[-.4mm]
\hline\hline

\!12\!\!&\!\!$-1/4$\!\!&\!\!$-3/2$\!\!&\!\!$-2$\!\!&\!\!$-1$\!\!&\!\!$2$\!\!&\!\!$2$\!\!&\!\!$1/2$\!\!&\!\!$-1$\!\!&\!\!$E\sst 7$\!\\[-.4mm]
\cline{2-10}
\!\!&\!\!$-1/4$\!\!&\!\!$-1$\!\!&\!\!$-1$\!\!&\!\!$-1$\!\!&\!\!$2$\!\!&\!\!$2$\!\!&\!\!$1/2$\!\!&\!\!$-1$\!\!&\!\!$E\sst 7$\!\\[-.4mm]
\cline{2-10}
\!\!&\!\!$-1/3$\!\!&\!\!$-3/2$\!\!&\!\!$-2$\!\!&\!\!$-1$\!\!&\!\!$1$\!\!&\!\!$2$\!\!&\!\!$1/2$\!\!&\!\!$-1$\!\!&\!\!$E\sst 6$\!\\[-.4mm]
\cline{2-10}
\!\!&\!\!$-1/3$\!\!&\!\!$-3/2$\!\!&\!\!$-2$\!\!&\!\!$-1$\!\!&\!\!$1$\!\!&\!\!$3$\!\!&\!\!$1$\!\!&\!\!$-1$\!\!&\!\!$E\sst 6$\!\\[-.4mm]
\cline{2-10}
\!\!&\!\!$-1/3$\!\!&\!\!$-1$\!\!&\!\!$-1$\!\!&\!\!$-1$\!\!&\!\!$1$\!\!&\!\!$3$\!\!&\!\!$1$\!\!&\!\!$-1$\!\!&\!\!$E\sst 6$\!\\[-.4mm]
\cline{2-10}
\!\!&\!\!$-1/3$\!\!&\!\!$-1$\!\!&\!\!$-1$\!\!&\!\!$-1$\!\!&\!\!$1$\!\!&\!\!$1$\!\!&\!\!$1/2$\!\!&\!\!$0$\!\!&\!\!$E\sst 6$\!\\[-.4mm]
\cline{2-10}
\!\!&\!\!$-1/3$\!\!&\!\!$-1$\!\!&\!\!$-1$\!\!&\!\!$-1$\!\!&\!\!$1$\!\!&\!\!$1$\!\!&\!\!$1$\!\!&\!\!$1$\!\!&\!\!$E\sst 6$\!\\[-.4mm]
\cline{2-10}

\!\!&\!\!$-1/2$\!\!&\!\!$-1$\!\!&\!\!$-1$\!\!&\!\!$0$\!\!&\!\!$0$\!\!&\!\!$i$\!\!&\!\!$i-1/2$\!\!&\!\!$i-1$\!\!&\!\!$D\sst {2i+4}\yy B\sst 3^a\xx$\!\\[-.4mm]
\cline{2-10}
\!\!&\!\!$-1/2$\!\!&\!\!$-1$\!\!&\!\!$-1$\!\!&\!\!$0$\!\!&\!\!$0$\!\!&\!\!$i$\!\!&\!\!$i$\!\!&\!\!$i$\!\!&\!\!$D\sst {2i+4}\yy D\sst 4^a\xx$\!\\[-.4mm]
\cline{2-10}
\!\!&\!\!$-1/2$\!\!&\!\!$-1$\!\!&\!\!$-1$\!\!&\!\!$0$\!\!&\!\!$0$\!\!&\!\!$i+1$\!\!&\!\!$i$\!\!&\!\!$i-1$\!\!&\!\!$D\sst {2i+4}\yy B\sst 3\xx$\!\\[-.4mm]
\cline{2-10}
\!\!&\!\!$-1/2$\!\!&\!\!$-1$\!\!&\!\!$-1$\!\!&\!\!$-1$\!\!&\!\!$0$\!\!&\!\!$i+1$\!\!&\!\!$i$\!\!&\!\!$i$\!\!&\!\!$D\sst {2i+4}\yy D\sst 4\xx$\!\\[-.4mm]
\cline{2-10}
\!\!&\!\!$-1/2$\!\!&\!\!$i$\!\!&\!\!$-1$\!\!&\!\!$-1$\!\!&\!\!$0$\!\!&\!\!$0$\!\!&\!\!$i$\!\!&\!\!$2i+1$\!\!&\!\!$D\sst {2i+4}\yy D\sst 4^a\xx$\!\\[-.4mm]
\cline{2-10}
\!\!&\!\!$-1/2$\!\!&\!\!$-1$\!\!&\!\!$-1$\!\!&\!\!$0$\!\!&\!\!$0$\!\!&\!\!$i+1$\!\!&\!\!$i$\!\!&\!\!$i$\!\!&\!\!$D\sst {2i+5}$\!\\[-.4mm]
\cline{2-10}
\!\!&\!\!$-1/2$\!\!&\!\!$-1$\!\!&\!\!$-1$\!\!&\!\!$-1$\!\!&\!\!$0$\!\!&\!\!$i+1$\!\!&\!\!$i+1/2$\!\!&\!\!$i$\!\!&\!\!$D\sst {2i+5}$\!\\[-.4mm]
\cline{2-10}
\!\!&\!\!$-1/2$\!\!&\!\!$-1$\!\!&\!\!$-1$\!\!&\!\!$-1$\!\!&\!\!$0$\!\!&\!\!$i+1$\!\!&\!\!$i+1$\!\!&\!\!$i+1$\!\!&\!\!$D\sst {2i+5}$\!\\[-.4mm]
\cline{2-10}
\!\!&\!\!$-1/2$\!\!&\!\!$-1$\!\!&\!\!$-1$\!\!&\!\!$-1$\!\!&\!\!$0$\!\!&\!\!$i+2$\!\!&\!\!$i+1$\!\!&\!\!$i$\!\!&\!\!$D\sst {2i+5}$\!\\[-.4mm]
\cline{2-10}
\!\!&\!\!$-1/2$\!\!&\!\!$i$\!\!&\!\!$-1$\!\!&\!\!$-1$\!\!&\!\!$0$\!\!&\!\!$0$\!\!&\!\!$i+1$\!\!&\!\!$2i+2$\!\!&\!\!$D\sst {2i+5}$\!\\[-.4mm]
\cline{2-10}
\!\!&\!\!$-1/2$\!\!&\!\!$(i-1)/2$\!\!&\!\!$-1$\!\!&\!\!$-1$\!\!&\!\!$0$\!\!&\!\!$0$\!\!&\!\!$i/2$\!\!&\!\!$i$\!\!&\!\!$B\sst {i+3}\yy B\sst 3^a\xx$\!\\[-.4mm]
\cline{2-10}

\!\!&\!\!$-1$\!\!&\!\!$-1$\!\!&\!\!$-1$\!\!&\!\!$-1$\!\!&\!\!$-1$\!\!&\!\!$i-1$\!\!&\!\!$i-1$\!\!&\!\!$i-1$\!\!&\!\!$C\sst i$\!\\[-.4mm]
\cline{2-10}
\!\!&\!\!$-1$\!\!&\!\!$i-1$\!\!&\!\!$-1$\!\!&\!\!$-1$\!\!&\!\!$-1$\!\!&\!\!$-1$\!\!&\!\!$i-1$\!\!&\!\!$2i-1$\!\!&\!\!$C\sst i$\!\\[-.4mm]
\cline{2-10}

\!\!&\!\!$-1$\!\!&\!\!$-1$\!\!&\!\!$-1$\!\!&\!\!$z_3$\!\!&\!\!$z_4$\!\!&\!\!$z_5$\!\!&\!\!$z_6$\!\!&\!\!$z_7$\!\!&\!\!$A\sst {z_3+z_4+z_5+z_7+3}$\!\\[-.4mm]
\hline\hline

\end{tabular}
\caption{Duals $\top^*$ of tops with $F_0^*$ one of the polygons $11,12$ of figure~\ref{fig:16pol}.}\label{tab:results1112}
\end{footnotesize}
%\end{adjustwidth}
\end{center}
\end{table}

%\newpage

%\clearpage
%\renewcommand{\baselinestretch}{1.2}

%\renewcommand{\baselinestretch}{1.2}

\newpage
${}$
\vfill

\begin{table}[!b]
\begin{center}
%\beginadjustwidth}{-0.2in}{-0.2in}
\begin{footnotesize}
\begin{tabular}{|c|ccccccccc|c|}
\hline
%$F_0^*$\!\!\!&\!\!\!\multicolumn{9}{c}{Minimal points}\vline\!\!\!&\!\!\!Lie\!\\[-.5mm]
%\cline{2-10}
$F_0^*$\!\!\!&\!\!\!$z_0$\!\!\!&\!\!\!$z_1$\!\!\!&\!\!\!$z_2$\!\!\!&\!\!\!$z_3$\!\!\!&\!\!\!$z_4$\!\!\!&\!\!\!$z_5$\!\!\!&\!\!\!$z_6$\!\!\!&\!\!\!$z_7$\!\!\!&\!\!\!$z_8$\!\!\!&\!\!\!AKMA\!\\[-.5mm]
\hline\hline

13\!\!\!&\!\!\!$-1/6$\!\!\!&\!\!\!$3/2$\!\!\!&\!\!\!$-1$\!\!\!&\!\!\!$-3/2$\!\!\!&\!\!\!$-2$\!\!\!&\!\!\!$-1$\!\!\!&\!\!\!$2/3$\!\!\!&\!\!\!$7/3$\!\!\!&\!\!\!$4$\!\!\!&\!\!\!$E\sst 8$\!\\[-.5mm]
\cline{2-11}
\!\!\!&\!\!\!$-1/6$\!\!\!&\!\!\!$3/2$\!\!\!&\!\!\!$-1$\!\!\!&\!\!\!$-1$\!\!\!&\!\!\!$-1$\!\!\!&\!\!\!$-1$\!\!\!&\!\!\!$2/3$\!\!\!&\!\!\!$7/3$\!\!\!&\!\!\!$4$\!\!\!&\!\!\!$E\sst 8$\!\\[-.5mm]
\cline{2-11}
\!\!\!&\!\!\!$-1/4$\!\!\!&\!\!\!$1/2$\!\!\!&\!\!\!$2$\!\!\!&\!\!\!$-1$\!\!\!&\!\!\!$-4$\!\!\!&\!\!\!$-13/4$\!\!\!&\!\!\!$-5/2$\!\!\!&\!\!\!$-7/4$\!\!\!&\!\!\!$-1$\!\!\!&\!\!\!$E\sst 6^{(2)}$\!\\[-.5mm]
\cline{2-11}
\!\!\!&\!\!\!$-1/4$\!\!\!&\!\!\!$3/2$\!\!\!&\!\!\!$-1$\!\!\!&\!\!\!$-3/2$\!\!\!&\!\!\!$-2$\!\!\!&\!\!\!$-1$\!\!\!&\!\!\!$1/2$\!\!\!&\!\!\!$2$\!\!\!&\!\!\!$4$\!\!\!&\!\!\!$E\sst 7$\!\\[-.5mm]
\cline{2-11}
\!\!\!&\!\!\!$-1/4$\!\!\!&\!\!\!$3/2$\!\!\!&\!\!\!$-1$\!\!\!&\!\!\!$-1$\!\!\!&\!\!\!$-1$\!\!\!&\!\!\!$-1$\!\!\!&\!\!\!$1/2$\!\!\!&\!\!\!$2$\!\!\!&\!\!\!$4$\!\!\!&\!\!\!$E\sst 7$\!\\[-.5mm]
\cline{2-11}
\!\!\!&\!\!\!$-1/4$\!\!\!&\!\!\!$2$\!\!\!&\!\!\!$-1$\!\!\!&\!\!\!$-1$\!\!\!&\!\!\!$-1$\!\!\!&\!\!\!$-1$\!\!\!&\!\!\!$1/2$\!\!\!&\!\!\!$2$\!\!\!&\!\!\!$5$\!\!\!&\!\!\!$E\sst 7$\!\\[-.5mm]
\cline{2-11}
\!\!\!&\!\!\!$-1/4$\!\!\!&\!\!\!$1/2$\!\!\!&\!\!\!$2$\!\!\!&\!\!\!$-1$\!\!\!&\!\!\!$-3$\!\!\!&\!\!\!$-3$\!\!\!&\!\!\!$-7/3$\!\!\!&\!\!\!$-5/3$\!\!\!&\!\!\!$-1$\!\!\!&\!\!\!$E\sst 7$\!\\[-.5mm]
\cline{2-11}
\!\!\!&\!\!\!$-1/4$\!\!\!&\!\!\!$1/2$\!\!\!&\!\!\!$2$\!\!\!&\!\!\!$-1$\!\!\!&\!\!\!$-3$\!\!\!&\!\!\!$-5/2$\!\!\!&\!\!\!$-2$\!\!\!&\!\!\!$-3/2$\!\!\!&\!\!\!$-1$\!\!\!&\!\!\!$E\sst 7$\!\\[-.5mm]
\cline{2-11}
\!\!\!&\!\!\!$-1/4$\!\!\!&\!\!\!$1/2$\!\!\!&\!\!\!$2$\!\!\!&\!\!\!$-1$\!\!\!&\!\!\!$-2$\!\!\!&\!\!\!$-2$\!\!\!&\!\!\!$-2$\!\!\!&\!\!\!$-3/2$\!\!\!&\!\!\!$-1$\!\!\!&\!\!\!$E\sst 7$\!\\[-.5mm]
\cline{2-11}
\!\!\!&\!\!\!$-1/4$\!\!\!&\!\!\!$1/2$\!\!\!&\!\!\!$2$\!\!\!&\!\!\!$-1$\!\!\!&\!\!\!$-1$\!\!\!&\!\!\!$-1$\!\!\!&\!\!\!$-1$\!\!\!&\!\!\!$-1$\!\!\!&\!\!\!$-1$\!\!\!&\!\!\!$E\sst 7$\!\\[-.5mm]
\cline{2-11}
\!\!\!&\!\!\!$-1/3$\!\!\!&\!\!\!$1/2$\!\!\!&\!\!\!$-1$\!\!\!&\!\!\!$-1$\!\!\!&\!\!\!$-1$\!\!\!&\!\!\!$-1/3$\!\!\!&\!\!\!$1/3$\!\!\!&\!\!\!$1$\!\!\!&\!\!\!$2$\!\!\!&\!\!\!$F\sst 4$\!\\[-.5mm]
\cline{2-11}
\!\!\!&\!\!\!$-1/3$\!\!\!&\!\!\!$1$\!\!\!&\!\!\!$-1$\!\!\!&\!\!\!$-1$\!\!\!&\!\!\!$-1$\!\!\!&\!\!\!$-1/3$\!\!\!&\!\!\!$1/3$\!\!\!&\!\!\!$1$\!\!\!&\!\!\!$3$\!\!\!&\!\!\!$F\sst 4$\!\\[-.5mm]
\cline{2-11}
\!\!\!&\!\!\!$-1/3$\!\!\!&\!\!\!$1/2$\!\!\!&\!\!\!$-1$\!\!\!&\!\!\!$-1$\!\!\!&\!\!\!$0$\!\!\!&\!\!\!$0$\!\!\!&\!\!\!$1/2$\!\!\!&\!\!\!$1$\!\!\!&\!\!\!$2$\!\!\!&\!\!\!$E\sst 6$\!\\[-.5mm]
\cline{2-11}
\!\!\!&\!\!\!$-1/3$\!\!\!&\!\!\!$1/2$\!\!\!&\!\!\!$-1$\!\!\!&\!\!\!$-1$\!\!\!&\!\!\!$1$\!\!\!&\!\!\!$1$\!\!\!&\!\!\!$1$\!\!\!&\!\!\!$1$\!\!\!&\!\!\!$2$\!\!\!&\!\!\!$E\sst 6$\!\\[-.5mm]
\cline{2-11}
\!\!\!&\!\!\!$-1/3$\!\!\!&\!\!\!$1$\!\!\!&\!\!\!$-1$\!\!\!&\!\!\!$-1$\!\!\!&\!\!\!$0$\!\!\!&\!\!\!$0$\!\!\!&\!\!\!$1/2$\!\!\!&\!\!\!$1$\!\!\!&\!\!\!$3$\!\!\!&\!\!\!$E\sst 6$\!\\[-.5mm]
\cline{2-11}
\!\!\!&\!\!\!$-1/3$\!\!\!&\!\!\!$1$\!\!\!&\!\!\!$-1$\!\!\!&\!\!\!$-1$\!\!\!&\!\!\!$1$\!\!\!&\!\!\!$1$\!\!\!&\!\!\!$1$\!\!\!&\!\!\!$1$\!\!\!&\!\!\!$3$\!\!\!&\!\!\!$E\sst 6$\!\\[-.5mm]

\cline{2-11}
\!\!\!&\!\!\!$-1/2$\!\!\!&\!\!\!$i$\!\!\!&\!\!\!$-1$\!\!\!&\!\!\!$-1$\!\!\!&\!\!\!$0$\!\!\!&\!\!\!$0$\!\!\!&\!\!\!$0$\!\!\!&\!\!\!$i+1$\!\!\!&\!\!\!$2i+2$\!\!\!&\!\!\!$D\sst {2i+4}\yy D\sst 4^a\xx$\!\\[-.5mm]
\cline{2-11}
\!\!\!&\!\!\!$-1/2$\!\!\!&\!\!\!$i+1$\!\!\!&\!\!\!$-1$\!\!\!&\!\!\!$-1$\!\!\!&\!\!\!$-1$\!\!\!&\!\!\!$-1$\!\!\!&\!\!\!$0$\!\!\!&\!\!\!$i+1$\!\!\!&\!\!\!$2i+3$\!\!\!&\!\!\!$D\sst {2i+4}\yy B\sst 3\xx$\!\\[-.5mm]
\cline{2-11}
\!\!\!&\!\!\!$-1/2$\!\!\!&\!\!\!$-1$\!\!\!&\!\!\!$-1$\!\!\!&\!\!\!$0$\!\!\!&\!\!\!$i+1$\!\!\!&\!\!\!$i+1/2$\!\!\!&\!\!\!$i$\!\!\!&\!\!\!$i-1/2$\!\!\!&\!\!\!$i-1$\!\!\!&\!\!\!$D\sst {2i+4}\yy D\sst 3^{(2)a}\xx$\!\\[-.5mm]
\cline{2-11}
\!\!\!&\!\!\!$-1/2$\!\!\!&\!\!\!$-1$\!\!\!&\!\!\!$-1$\!\!\!&\!\!\!$0$\!\!\!&\!\!\!$i+1$\!\!\!&\!\!\!$i+1$\!\!\!&\!\!\!$i+1$\!\!\!&\!\!\!$i+1$\!\!\!&\!\!\!$i+1$\!\!\!&\!\!\!$D\sst {2i+4}\yy B\sst 3\xx$\!\\[-.5mm]
\cline{2-11}
\!\!\!&\!\!\!$-1/2$\!\!\!&\!\!\!$-1$\!\!\!&\!\!\!$-1$\!\!\!&\!\!\!$0$\!\!\!&\!\!\!$i+2$\!\!\!&\!\!\!$i+1$\!\!\!&\!\!\!$i$\!\!\!&\!\!\!$i$\!\!\!&\!\!\!$i$\!\!\!&\!\!\!$D\sst {2i+4}\yy D\sst 4^a\xx$\!\\[-.5mm]
\cline{2-11}
\!\!\!&\!\!\!$-1/2$\!\!\!&\!\!\!$-1$\!\!\!&\!\!\!$-1$\!\!\!&\!\!\!$0$\!\!\!&\!\!\!$i+1$\!\!\!&\!\!\!$i+1/2$\!\!\!&\!\!\!$i$\!\!\!&\!\!\!$i$\!\!\!&\!\!\!$i$\!\!\!&\!\!\!$D\sst {2i+4}\yy B\sst 3^a\xx$\!\\[-.5mm]
\cline{2-11}
\!\!\!&\!\!\!$-1/2$\!\!\!&\!\!\!$i+1$\!\!\!&\!\!\!$-1$\!\!\!&\!\!\!$-1$\!\!\!&\!\!\!$0$\!\!\!&\!\!\!$0$\!\!\!&\!\!\!$0$\!\!\!&\!\!\!$i+1$\!\!\!&\!\!\!$2i+3$\!\!\!&\!\!\!$D\sst {2i+5}$\!\\[-.5mm]
\cline{2-11}
\!\!\!&\!\!\!$-1/2$\!\!\!&\!\!\!$-1$\!\!\!&\!\!\!$-1$\!\!\!&\!\!\!$0$\!\!\!&\!\!\!$i+2$\!\!\!&\!\!\!$i+1$\!\!\!&\!\!\!$i+1/2$\!\!\!&\!\!\!$i$\!\!\!&\!\!\!$i$\!\!\!&\!\!\!$D\sst {2i+5}$\!\\[-.5mm]
\cline{2-11}
\!\!\!&\!\!\!$-1/2$\!\!\!&\!\!\!$-1$\!\!\!&\!\!\!$-1$\!\!\!&\!\!\!$0$\!\!\!&\!\!\!$i+2$\!\!\!&\!\!\!$i+1$\!\!\!&\!\!\!$i+1$\!\!\!&\!\!\!$i+1$\!\!\!&\!\!\!$i+1$\!\!\!&\!\!\!$D\sst {2i+5}$\!\\[-.5mm]
\cline{2-11}

\!\!\!&\!\!\!$-1/2$\!\!\!&\!\!\!$i/2$\!\!\!&\!\!\!$-1$\!\!\!&\!\!\!$-1$\!\!\!&\!\!\!$0$\!\!\!&\!\!\!$0$\!\!\!&\!\!\!$0$\!\!\!&\!\!\!$(i+1)/2$\!\!\!&\!\!\!$i+1$\!\!\!&\!\!\!$B\sst {i+3}\yy B\sst 3^a\xx$\!\\[-.5mm]
\cline{2-11}
\!\!\!&\!\!\!$-1/2$\!\!\!&\!\!\!$(i+1)/2$\!\!\!&\!\!\!$-1$\!\!\!&\!\!\!$-1$\!\!\!&\!\!\!$-1$\!\!\!&\!\!\!$-1$\!\!\!&\!\!\!$0$\!\!\!&\!\!\!$i/2+1$\!\!\!&\!\!\!$i+2$\!\!\!&\!\!\!$B\sst {i+3}\yy G\sst 2\xx$\!\\[-.5mm]
\cline{2-11}
\!\!\!&\!\!\!$-1/2$\!\!\!&\!\!\!$i+1/2$\!\!\!&\!\!\!$-1$\!\!\!&\!\!\!$-1$\!\!\!&\!\!\!$-1$\!\!\!&\!\!\!$-1/2$\!\!\!&\!\!\!$0$\!\!\!&\!\!\!$i+1$\!\!\!&\!\!\!$2i+2$\!\!\!&\!\!\!$D\sst {2i+4}^{(2)}$\!\\[-.5mm]
\cline{2-11}
\!\!\!&\!\!\!$-1/2$\!\!\!&\!\!\!$i$\!\!\!&\!\!\!$-1$\!\!\!&\!\!\!$-1$\!\!\!&\!\!\!$-1$\!\!\!&\!\!\!$-1/2$\!\!\!&\!\!\!$0$\!\!\!&\!\!\!$i+1/2$\!\!\!&\!\!\!$2i+1$\!\!\!&\!\!\!$D\sst {2i+3}^{(2)}\yy D\sst 3^{(2)a}\xx$\!\\[-.5mm]
\cline{2-11}
\!\!\!&\!\!\!$-1/2$\!\!\!&\!\!\!$i-1/2$\!\!\!&\!\!\!$-1$\!\!\!&\!\!\!$-1/2$\!\!\!&\!\!\!$0$\!\!\!&\!\!\!$0$\!\!\!&\!\!\!$0$\!\!\!&\!\!\!$i$\!\!\!&\!\!\!$2i$\!\!\!&\!\!\!$D\sst
{2i+3}^{(2)}\yy A\sst 2^{(2)}\xx$\!\\[-.5mm]
\cline{2-11}

\!\!\!&\!\!\!$-1$\!\!\!&\!\!\!$i-1$\!\!\!&\!\!\!$-1$\!\!\!&\!\!\!$-1$\!\!\!&\!\!\!$-1$\!\!\!&\!\!\!$-1$\!\!\!&\!\!\!$-1$\!\!\!&\!\!\!$i-1$\!\!\!&\!\!\!$2i-1$\!\!\!&\!\!\!$C\sst i$\!\\[-.5mm]
\cline{2-11}
\!\!\!&\!\!\!$-1$\!\!\!&\!\!\!$-1$\!\!\!&\!\!\!$-1$\!\!\!&\!\!\!$-1$\!\!\!&\!\!\!$i-1$\!\!\!&\!\!\!$i-1$\!\!\!&\!\!\!$i-1$\!\!\!&\!\!\!$i-1$\!\!\!&\!\!\!$i-1$\!\!\!&\!\!\!$C\sst i$\!\\[-.5mm]
\cline{2-11}

\!\!\!&\!\!\!$-1$\!\!\!&\!\!\!$-1$\!\!\!&\!\!\!$-1$\!\!\!&\!\!\!$z_3$\!\!\!&\!\!\!$z_4$\!\!\!&\!\!\!$z_5$\!\!\!&\!\!\!$z_6$\!\!\!&\!\!\!$z_7$\!\!\!&\!\!\!$z_8$\!\!\!&\!\!\!$A\sst {z_4+z_8+1}$\!\\[-.5mm]

\hline\hline

\end{tabular}
\caption{Duals $\top^*$ of tops with $F_0^*$ being polygon $13$ of figure~\ref{fig:16pol}.}\label{tab:results13}
\end{footnotesize}
%\end{adjustwidth}
\end{center}
\end{table}

%\newpage

\newpage
%\clearpage

${}$
\vfill

\begin{table}[!b]
\begin{center}
%\beginadjustwidth}{-0.2in}{-0.2in}
\begin{footnotesize}
\begin{tabular}{|c|ccccccccc|c|}
\hline
%$F_0^*$\!&\!\multicolumn{9}{c}{Minimal points}\vline\!&\!Lie\!\\[-.5mm]
%\cline{2-10}\!\!
$\!F_0^*$\!&\!$z_0$\!&\!$z_1$\!&\!$z_2$\!&\!$z_3$\!&\!$z_4$\!&\!$z_5$\!&\!$z_6$\!&\!$z_7$\!&\!$z_8$\!&\!AKMA\!\\[-.5mm]
\hline\hline

\!14\!&\!$-1/6$\!&\!$-1$\!&\!$-2$\!&\!$-3/2$\!&\!$-1$\!&\!$2/3$\!&\!$7/3$\!&\!$4$\!&\!$3/2$\!&\!$E\sst 8$\!\\[-.5mm]
\cline{2-11}
\!&\!$-1/6$\!&\!$-1$\!&\!$-1$\!&\!$-1$\!&\!$-1$\!&\!$2/3$\!&\!$7/3$\!&\!$4$\!&\!$3/2$\!&\!$E\sst 8$\!\\[-.5mm]
\cline{2-11}
\!&\!$-1/4$\!&\!$2$\!&\!$2$\!&\!$1/2$\!&\!$-1$\!&\!$-5/3$\!&\!$-7/3$\!&\!$-3$\!&\!$-1$\!&\!$E\sst 7$\!\\[-.5mm]
\cline{2-11}
\!&\!$-1/4$\!&\!$2$\!&\!$2$\!&\!$1/2$\!&\!$-1$\!&\!$-3/2$\!&\!$-2$\!&\!$-2$\!&\!$-1$\!&\!$E\sst 7$\!\\[-.5mm]
\cline{2-11}
\!&\!$-1/4$\!&\!$2$\!&\!$2$\!&\!$1/2$\!&\!$-1$\!&\!$-1$\!&\!$-1$\!&\!$-1$\!&\!$-1$\!&\!$E\sst 7$\!\\[-.5mm]
\cline{2-11}
\!&\!$-1/4$\!&\!$-1$\!&\!$-2$\!&\!$-3/2$\!&\!$-1$\!&\!$1/2$\!&\!$2$\!&\!$4$\!&\!$3/2$\!&\!$E\sst 7$\!\\[-.5mm]
\cline{2-11}
\!&\!$-1/4$\!&\!$-1$\!&\!$-2$\!&\!$-3/2$\!&\!$-1$\!&\!$1/2$\!&\!$2$\!&\!$5$\!&\!$2$\!&\!$E\sst 7$\!\\[-.5mm]
\cline{2-11}
\!&\!$-1/4$\!&\!$-1$\!&\!$-1$\!&\!$-1$\!&\!$-1$\!&\!$1/2$\!&\!$2$\!&\!$4$\!&\!$3/2$\!&\!$E\sst 7$\!\\[-.5mm]
\cline{2-11}
\!&\!$-1/4$\!&\!$-1$\!&\!$-1$\!&\!$-1$\!&\!$-1$\!&\!$1/2$\!&\!$2$\!&\!$5$\!&\!$2$\!&\!$E\sst 7$\!\\[-.5mm]
\cline{2-11}
\!&\!$-1/3$\!&\!$-1$\!&\!$0$\!&\!$1/2$\!&\!$1$\!&\!$1/3$\!&\!$-1/3$\!&\!$-1$\!&\!$-1$\!&\!$F\sst 4$\!\\[-.5mm]
\cline{2-11}
\!&\!$-1/3$\!&\!$-1$\!&\!$1$\!&\!$1$\!&\!$1$\!&\!$1/3$\!&\!$-1/3$\!&\!$-1$\!&\!$-1$\!&\!$F\sst 4$\!\\[-.5mm]
\cline{2-11}
\!&\!$-1/3$\!&\!$-1$\!&\!$0$\!&\!$1/2$\!&\!$1$\!&\!$1/2$\!&\!$0$\!&\!$0$\!&\!$-1$\!&\!$E\sst 6$\!\\[-.5mm]
\cline{2-11}
\!&\!$-1/3$\!&\!$-1$\!&\!$0$\!&\!$1/2$\!&\!$1$\!&\!$1$\!&\!$1$\!&\!$1$\!&\!$-1$\!&\!$E\sst 6$\!\\[-.5mm]
\cline{2-11}
\!&\!$-1/3$\!&\!$-1$\!&\!$1$\!&\!$1$\!&\!$1$\!&\!$1/2$\!&\!$0$\!&\!$0$\!&\!$-1$\!&\!$E\sst 6$\!\\[-.5mm]
\cline{2-11}
\!&\!$-1/3$\!&\!$-1$\!&\!$1$\!&\!$1$\!&\!$1$\!&\!$1$\!&\!$1$\!&\!$1$\!&\!$-1$\!&\!$E\sst 6$\!\\[-.5mm]
\cline{2-11}
\!&\!$-1/3$\!&\!$-1$\!&\!$-1$\!&\!$-1$\!&\!$0$\!&\!$1/2$\!&\!$1$\!&\!$2$\!&\!$1/2$\!&\!$E\sst 6$\!\\[-.5mm]
\cline{2-11}
\!&\!$-1/3$\!&\!$-1$\!&\!$-1$\!&\!$-1$\!&\!$1$\!&\!$1$\!&\!$1$\!&\!$2$\!&\!$1/2$\!&\!$E\sst 6$\!\\[-.5mm]
\cline{2-11}
\!&\!$-1/3$\!&\!$-1$\!&\!$-1$\!&\!$-1$\!&\!$0$\!&\!$1/2$\!&\!$1$\!&\!$3$\!&\!$1$\!&\!$E\sst 6$\!\\[-.5mm]
\cline{2-11}
\!&\!$-1/3$\!&\!$-1$\!&\!$-1$\!&\!$-1$\!&\!$1$\!&\!$1$\!&\!$1$\!&\!$3$\!&\!$1$\!&\!$E\sst 6$\!\\[-.5mm]
\cline{2-11}

\!&\!$-1/2$\!&\!$-1$\!&\!$-1$\!&\!$-1$\!&\!$i$\!&\!$i$\!&\!$i+1/2$\!&\!$i+1$\!&\!$0$\!&\!$D\sst {2i+4}\yy B\sst 3^a\xx$\!\\[-.5mm]
\cline{2-11}
\!&\!$-1/2$\!&\!$-1$\!&\!$-1$\!&\!$-1$\!&\!$i$\!&\!$i$\!&\!$i+1$\!&\!$i+2$\!&\!$0$\!&\!$D\sst {2i+4}\yy D\sst 4^a\xx$\!\\[-.5mm]
\cline{2-11}
\!&\!$-1/2$\!&\!$-1$\!&\!$-1$\!&\!$-1$\!&\!$i+1$\!&\!$i+1$\!&\!$i+1$\!&\!$i+1$\!&\!$0$\!&\!$D\sst {2i+4}\yy B\sst 3\xx$\!\\[-.5mm]
\cline{2-11}
\!&\!$-1/2$\!&\!$-1$\!&\!$-1$\!&\!$-1/2$\!&\!$0$\!&\!$0$\!&\!$i$\!&\!$2i+1$\!&\!$i$\!&\!$D\sst {2i+4}\yy B\sst 3^b\xx$\!\\[-.5mm]
\cline{2-11}
\!&\!$-1/2$\!&\!$-1$\!&\!$-1$\!&\!$0$\!&\!$1$\!&\!$0$\!&\!$i$\!&\!$2i+1$\!&\!$i$\!&\!$D\sst {2i+4}$\!\\[-.5mm]
\cline{2-11}
\!&\!$-1/2$\!&\!$-1$\!&\!$-1$\!&\!$-1$\!&\!$0$\!&\!$0$\!&\!$i+1$\!&\!$2i+2$\!&\!$i$\!&\!$D\sst {2i+4}\yy D\sst 4^a\xx$\!\\[-.5mm]
\cline{2-11}
\!&\!$-1/2$\!&\!$-1$\!&\!$-1$\!&\!$-1$\!&\!$-1$\!&\!$0$\!&\!$i+1$\!&\!$2i+3$\!&\!$i+1$\!&\!$D\sst {2i+4}\yy B\sst 3\xx$\!\\[-.5mm]
\cline{2-11}
\!&\!$-1/2$\!&\!$-1$\!&\!$-1$\!&\!$-1$\!&\!$i$\!&\!$i+1/2$\!&\!$i+1$\!&\!$i+2$\!&\!$0$\!&\!$D\sst {2i+5}$\!\\[-.5mm]
\cline{2-11}
\!&\!$-1/2$\!&\!$-1$\!&\!$-1$\!&\!$-1$\!&\!$i$\!&\!$i+1$\!&\!$i+2$\!&\!$i+3$\!&\!$0$\!&\!$D\sst {2i+5}$\!\\[-.5mm]
\cline{2-11}
\!&\!$-1/2$\!&\!$-1$\!&\!$-1$\!&\!$-1$\!&\!$i+1$\!&\!$i+1$\!&\!$i+1$\!&\!$i+2$\!&\!$0$\!&\!$D\sst {2i+5}$\!\\[-.5mm]
\cline{2-11}
\!&\!$-1/2$\!&\!$-1$\!&\!$-1$\!&\!$-1/2$\!&\!$0$\!&\!$0$\!&\!$i+1$\!&\!$2i+2$\!&\!$i$\!&\!$D\sst {2i+5}$\!\\[-.5mm]
\cline{2-11}
\!&\!$-1/2$\!&\!$-1$\!&\!$-1$\!&\!$0$\!&\!$1$\!&\!$0$\!&\!$i+1$\!&\!$2i+2$\!&\!$i$\!&\!$D\sst {2i+5}$\!\\[-.5mm]
\cline{2-11}
\!&\!$-1/2$\!&\!$-1$\!&\!$-1$\!&\!$-1$\!&\!$0$\!&\!$0$\!&\!$i+1$\!&\!$2i+3$\!&\!$i+1$\!&\!$D\sst {2i+5}$\!\\[-.5mm]
\cline{2-11}
\!&\!$-1/2$\!&\!$-1$\!&\!$-1$\!&\!$-1$\!&\!$-1$\!&\!$0$\!&\!$i+2$\!&\!$2i+4$\!&\!$i+1$\!&\!$D\sst {2i+5}$\!\\[-.5mm]
\cline{2-11}
\!&\!$-1/2$\!&\!$-1$\!&\!$-1$\!&\!$-1/2$\!&\!$0$\!&\!$0$\!&\!$i/2$\!&\!$i$\!&\!$(i-1)/2$\!&\!$B\sst {i+3}\yy G\sst 2\xx$\!\\[-.5mm]
\cline{2-11}
\!&\!$-1/2$\!&\!$-1$\!&\!$-1$\!&\!$0$\!&\!$1$\!&\!$0$\!&\!$i/2$\!&\!$i$\!&\!$(i-1)/2$\!&\!$B\sst {i+3}\yy B\sst 3^b\xx$\!\\[-.5mm]
\cline{2-11}
\!&\!$-1/2$\!&\!$-1$\!&\!$-1$\!&\!$-1$\!&\!$0$\!&\!$0$\!&\!$(i+1)/2$\!&\!$i+1$\!&\!$i/2$\!&\!$B\sst {i+3}\yy B\sst 3^a\xx$\!\\[-.5mm]
\cline{2-11}
\!&\!$-1/2$\!&\!$-1$\!&\!$-1$\!&\!$-1$\!&\!$-1$\!&\!$0$\!&\!$i/2+1$\!&\!$i+2$\!&\!$(i+1)/2$\!&\!$B\sst {i+3}\yy G\sst 2\xx$\!\\[-.5mm]
\cline{2-11}

\!&\!$-1$\!&\!$-1$\!&\!$-1$\!&\!$-1$\!&\!$-1$\!&\!$-1$\!&\!$i-1$\!&\!$2i-1$\!&\!$i-1$\!&\!$C\sst i$\!\\[-.5mm]
\cline{2-11}
\!&\!$-1$\!&\!$-1$\!&\!$-1$\!&\!$-1$\!&\!$i-1$\!&\!$i-1$\!&\!$i-1$\!&\!$i-1$\!&\!$-1$\!&\!$C\sst i$\!\\[-.5mm]
\cline{2-11}

\!&\!$-1$\!&\!$-1$\!&\!$-1$\!&\!$z_3$\!&\!$z_4$\!&\!$z_5$\!&\!$z_6$\!&\!$z_7$\!&\!$z_8$\!&\!$A\sst {z_4+z_7+1}$\!\\[-.5mm]

\hline\hline

\end{tabular}
\caption{Duals $\top^*$ of tops with $F_0^*$ being polygon $14$ of figure~\ref{fig:16pol}.}\label{tab:results14}
\end{footnotesize}
%\end{adjustwidth}
\end{center}
\end{table}

%\newpage
%\renewcommand{\baselinestretch}{1.05}

\newpage
%\clearpage

%\renewcommand{\baselinestretch}{1.05}

${}$
\vfill

\begin{table}[!b]
\begin{center}
%\beginadjustwidth}{-0.7in}{-0.7in}
\begin{footnotesize}
\begin{tabular}{|c|cccccccccc|c|}
\hline
%$F_0^*$\!\!\!&\!\!\!\multicolumn{10}{c}{Minimal points}\vline\!\!\!&\!\!\!Lie\!\\[-.5mm]
%\cline{2-11}
$F_0^*$\!\!\!&\!\!\!$z_0$\!\!\!&\!\!\!$z_1$\!\!\!&\!\!\!$z_2$\!\!\!&\!\!\!$z_3$\!\!\!&\!\!\!$z_4$\!\!\!&\!\!\!$z_5$\!\!\!&\!\!\!$z_6$\!\!\!&\!\!\!$z_7$\!\!\!&\!\!\!$z_8$\!\!\!&\!\!\!$z_9$\!\!\!&\!\!\!AKMA\!\\[-.5mm]
\hline\hline

15\!\!\!&\!\!\!$-1/4$\!\!\!&\!\!\!$-1$\!\!\!&\!\!\!$1$\!\!\!&\!\!\!$3/2$\!\!\!&\!\!\!$2$\!\!\!&\!\!\!$1/2$\!\!\!&\!\!\!$-1$\!\!\!&\!\!\!$-3/2$\!\!\!&\!\!\!$-2$\!\!\!&\!\!\!-\!\!\!&\!\!\!$E\sst 7$\!\\[-.5mm]
\cline{2-12}
\!\!\!&\!\!\!$-1/4$\!\!\!&\!\!\!$-1$\!\!\!&\!\!\!$1$\!\!\!&\!\!\!$3/2$\!\!\!&\!\!\!$2$\!\!\!&\!\!\!$1/2$\!\!\!&\!\!\!$-1$\!\!\!&\!\!\!$-1$\!\!\!&\!\!\!$-1$\!\!\!&\!\!\!-\!\!\!&\!\!\!$E\sst 7$\!\\[-.5mm]
\cline{2-12}
\!\!\!&\!\!\!$-1/4$\!\!\!&\!\!\!$-1$\!\!\!&\!\!\!$2$\!\!\!&\!\!\!$2$\!\!\!&\!\!\!$2$\!\!\!&\!\!\!$1/2$\!\!\!&\!\!\!$-1$\!\!\!&\!\!\!$-1$\!\!\!&\!\!\!$-1$\!\!\!&\!\!\!-\!\!\!&\!\!\!$E\sst 7$\!\\[-.5mm]
\cline{2-12}
\!\!\!&\!\!\!$-1/3$\!\!\!&\!\!\!$-1$\!\!\!&\!\!\!$-1$\!\!\!&\!\!\!$-1$\!\!\!&\!\!\!$0$\!\!\!&\!\!\!$1/2$\!\!\!&\!\!\!$1$\!\!\!&\!\!\!$1/2$\!\!\!&\!\!\!$0$\!\!\!&\!\!\!-\!\!\!&\!\!\!$E\sst 6$\!\\[-.5mm]
\cline{2-12}
\!\!\!&\!\!\!$-1/3$\!\!\!&\!\!\!$-1$\!\!\!&\!\!\!$-1$\!\!\!&\!\!\!$-1$\!\!\!&\!\!\!$0$\!\!\!&\!\!\!$1/2$\!\!\!&\!\!\!$1$\!\!\!&\!\!\!$1$\!\!\!&\!\!\!$1$\!\!\!&\!\!\!-\!\!\!&\!\!\!$E\sst 6$\!\\[-.5mm]
\cline{2-12}
\!\!\!&\!\!\!$-1/3$\!\!\!&\!\!\!$-1$\!\!\!&\!\!\!$-1$\!\!\!&\!\!\!$-1$\!\!\!&\!\!\!$1$\!\!\!&\!\!\!$1$\!\!\!&\!\!\!$1$\!\!\!&\!\!\!$1$\!\!\!&\!\!\!$1$\!\!\!&\!\!\!-\!\!\!&\!\!\!$E\sst 6$\!\\[-.5mm]
\cline{2-12}

\!\!\!&\!\!\!$-1/2$\!\!\!&\!\!\!$-1$\!\!\!&\!\!\!$-1$\!\!\!&\!\!\!$-1/2$\!\!\!&\!\!\!$0$\!\!\!&\!\!\!$0$\!\!\!&\!\!\!$i$\!\!\!&\!\!\!$i-1/2$\!\!\!&\!\!\!$i-1$\!\!\!&\!\!\!-\!\!\!&\!\!\!$D\sst {2i+4}\yy D\sst 3^{(2)a}\xx$\!\\[-.5mm]
\cline{2-12}
\!\!\!&\!\!\!$-1/2$\!\!\!&\!\!\!$-1$\!\!\!&\!\!\!$-1$\!\!\!&\!\!\!$-1/2$\!\!\!&\!\!\!$0$\!\!\!&\!\!\!$0$\!\!\!&\!\!\!$i$\!\!\!&\!\!\!$i$\!\!\!&\!\!\!$i$\!\!\!&\!\!\!-\!\!\!&\!\!\!$D\sst {2i+4}\yy B\sst 3^a\xx$\!\\[-.5mm]
\cline{2-12}
\!\!\!&\!\!\!$-1/2$\!\!\!&\!\!\!$-1$\!\!\!&\!\!\!$-1$\!\!\!&\!\!\!$0$\!\!\!&\!\!\!$1$\!\!\!&\!\!\!$0$\!\!\!&\!\!\!$i$\!\!\!&\!\!\!$i$\!\!\!&\!\!\!$i$\!\!\!&\!\!\!-\!\!\!&\!\!\!$D\sst {2i+4}\yy D\sst 4^a\xx$\!\\[-.5mm]
\cline{2-12}
\!\!\!&\!\!\!$-1/2$\!\!\!&\!\!\!$-1$\!\!\!&\!\!\!$-1$\!\!\!&\!\!\!$0$\!\!\!&\!\!\!$1$\!\!\!&\!\!\!$0$\!\!\!&\!\!\!$i+1$\!\!\!&\!\!\!$i$\!\!\!&\!\!\!$i-1$\!\!\!&\!\!\!-\!\!\!&\!\!\!$D\sst {2i+4}\yy B\sst 3\xx$\!\\[-.5mm]
\cline{2-12}
\!\!\!&\!\!\!$-1/2$\!\!\!&\!\!\!$-1$\!\!\!&\!\!\!$-1$\!\!\!&\!\!\!$-1$\!\!\!&\!\!\!$0$\!\!\!&\!\!\!$0$\!\!\!&\!\!\!$i+1$\!\!\!&\!\!\!$i$\!\!\!&\!\!\!$i$\!\!\!&\!\!\!-\!\!\!&\!\!\!$D\sst {2i+4}$\!\\[-.5mm]
\cline{2-12}
\!\!\!&\!\!\!$-1/2$\!\!\!&\!\!\!$i$\!\!\!&\!\!\!$-1$\!\!\!&\!\!\!$-1$\!\!\!&\!\!\!$0$\!\!\!&\!\!\!$0$\!\!\!&\!\!\!$0$\!\!\!&\!\!\!$i$\!\!\!&\!\!\!$2i+1$\!\!\!&\!\!\!-\!\!\!&\!\!\!$D\sst {2i+4}\yy D\sst 4^a\xx$\!\\[-.5mm]
\cline{2-12}
\!\!\!&\!\!\!$-1/2$\!\!\!&\!\!\!$-1$\!\!\!&\!\!\!$-1$\!\!\!&\!\!\!$-1/2$\!\!\!&\!\!\!$0$\!\!\!&\!\!\!$0$\!\!\!&\!\!\!$i+1$\!\!\!&\!\!\!$i$\!\!\!&\!\!\!$i$\!\!\!&\!\!\!-\!\!\!&\!\!\!$D\sst {2i+5}$\!\\[-.5mm]
\cline{2-12}
\!\!\!&\!\!\!$-1/2$\!\!\!&\!\!\!$-1$\!\!\!&\!\!\!$-1$\!\!\!&\!\!\!$0$\!\!\!&\!\!\!$1$\!\!\!&\!\!\!$0$\!\!\!&\!\!\!$i+1$\!\!\!&\!\!\!$i$\!\!\!&\!\!\!$i$\!\!\!&\!\!\!-\!\!\!&\!\!\!$D\sst {2i+5}$\!\\[-.5mm]
\cline{2-12}
\!\!\!&\!\!\!$-1/2$\!\!\!&\!\!\!$i$\!\!\!&\!\!\!$-1$\!\!\!&\!\!\!$-1$\!\!\!&\!\!\!$0$\!\!\!&\!\!\!$0$\!\!\!&\!\!\!$0$\!\!\!&\!\!\!$i+1$\!\!\!&\!\!\!$2i+2$\!\!\!&\!\!\!-\!\!\!&\!\!\!$D\sst {2i+5}$\!\\[-.5mm]
\cline{2-12}
\!\!\!&\!\!\!$-1/2$\!\!\!&\!\!\!$(i-1)/2$\!\!\!&\!\!\!$-1$\!\!\!&\!\!\!$-1$\!\!\!&\!\!\!$0$\!\!\!&\!\!\!$0$\!\!\!&\!\!\!$0$\!\!\!&\!\!\!$i/2$\!\!\!&\!\!\!$i$\!\!\!&\!\!\!-\!\!\!&\!\!\!$B\sst {i+3}\yy B\sst 3^a\xx$\!\\[-.5mm]
\cline{2-12}
\!\!\!&\!\!\!$-1/2$\!\!\!&\!\!\!$(i-1)/2$\!\!\!&\!\!\!$-1$\!\!\!&\!\!\!$-1$\!\!\!&\!\!\!$-1$\!\!\!&\!\!\!$-1/2$\!\!\!&\!\!\!$0$\!\!\!&\!\!\!$i/2$\!\!\!&\!\!\!$i$\!\!\!&\!\!\!-\!\!\!&\!\!\!$D\sst {i+3}^{(2)}\yy D\sst 3^{(2)a}\xx$\!\\[-.5mm]
\cline{2-12}

\!\!\!&\!\!\!$-1$\!\!\!&\!\!\!$-1$\!\!\!&\!\!\!$-1$\!\!\!&\!\!\!$-1$\!\!\!&\!\!\!$-1$\!\!\!&\!\!\!$-1$\!\!\!&\!\!\!$i-1$\!\!\!&\!\!\!$i-1$\!\!\!&\!\!\!$i-1$\!\!\!&\!\!\!-\!\!\!&\!\!\!$C\sst i$\!\\[-.5mm]
\cline{2-12}
\!\!\!&\!\!\!$-1$\!\!\!&\!\!\!$i-1$\!\!\!&\!\!\!$-1$\!\!\!&\!\!\!$-1$\!\!\!&\!\!\!$-1$\!\!\!&\!\!\!$-1$\!\!\!&\!\!\!$-1$\!\!\!&\!\!\!$i-1$\!\!\!&\!\!\!$2i-1$\!\!\!&\!\!\!-\!\!\!&\!\!\!$C\sst i$\!\\[-.5mm]
\cline{2-12}

\!\!\!&\!\!\!$-1$\!\!\!&\!\!\!$-1$\!\!\!&\!\!\!$-1$\!\!\!&\!\!\!$z_3$\!\!\!&\!\!\!$z_4$\!\!\!&\!\!\!$z_5$\!\!\!&\!\!\!$z_6$\!\!\!&\!\!\!$z_7$\!\!\!&\!\!\!$z_8$\!\!\!&\!\!\!-\!\!\!&\!\!\!$A\sst {z_4+z_6+z_8+2}$\!\\[-.5mm]
\hline\hline
16\!\!\!&\!\!\!$-1/6$\!\!\!&\!\!\!$-1$\!\!\!&\!\!\!$-3$\!\!\!&\!\!\!$-2$\!\!\!&\!\!\!$-3/2$\!\!\!&\!\!\!$-1$\!\!\!&\!\!\!$2/3$\!\!\!&\!\!\!$7/3$\!\!\!&\!\!\!$4$\!\!\!&\!\!\!$3/2$\!\!\!&\!\!\!$E\sst 8$\!\\[-.5mm]
\cline{2-12}
\!\!\!&\!\!\!$-1/6$\!\!\!&\!\!\!$-1$\!\!\!&\!\!\!$-3$\!\!\!&\!\!\!$-7/3$\!\!\!&\!\!\!$-5/3$\!\!\!&\!\!\!$-1$\!\!\!&\!\!\!$2/3$\!\!\!&\!\!\!$7/3$\!\!\!&\!\!\!$4$\!\!\!&\!\!\!$3/2$\!\!\!&\!\!\!$E\sst 8$\!\\[-.5mm]
\cline{2-12}
\!\!\!&\!\!\!$-1/6$\!\!\!&\!\!\!$-1$\!\!\!&\!\!\!$-1$\!\!\!&\!\!\!$-1$\!\!\!&\!\!\!$-1$\!\!\!&\!\!\!$-1$\!\!\!&\!\!\!$2/3$\!\!\!&\!\!\!$7/3$\!\!\!&\!\!\!$4$\!\!\!&\!\!\!$3/2$\!\!\!&\!\!\!$E\sst 8$\!\\[-.5mm]
\cline{2-12}

\!\!\!&\!\!\!$-1/4$\!\!\!&\!\!\!$-1$\!\!\!&\!\!\!$-3$\!\!\!&\!\!\!$-7/3$\!\!\!&\!\!\!$-5/3$\!\!\!&\!\!\!$-1$\!\!\!&\!\!\!$1/2$\!\!\!&\!\!\!$2$\!\!\!&\!\!\!$4$\!\!\!&\!\!\!$3/2$\!\!\!&\!\!\!$E\sst 7$\!\\[-.5mm]
\cline{2-12}
\!\!\!&\!\!\!$-1/4$\!\!\!&\!\!\!$-1$\!\!\!&\!\!\!$-3$\!\!\!&\!\!\!$-7/3$\!\!\!&\!\!\!$-5/3$\!\!\!&\!\!\!$-1$\!\!\!&\!\!\!$1/2$\!\!\!&\!\!\!$2$\!\!\!&\!\!\!$5$\!\!\!&\!\!\!$2$\!\!\!&\!\!\!$E\sst 7$\!\\[-.5mm]
\cline{2-12}
\!\!\!&\!\!\!$-1/4$\!\!\!&\!\!\!$-1$\!\!\!&\!\!\!$-2$\!\!\!&\!\!\!$-2$\!\!\!&\!\!\!$-3/2$\!\!\!&\!\!\!$-1$\!\!\!&\!\!\!$1/2$\!\!\!&\!\!\!$2$\!\!\!&\!\!\!$4$\!\!\!&\!\!\!$3/2$\!\!\!&\!\!\!$E\sst 7$\!\\[-.5mm]
\cline{2-12}
\!\!\!&\!\!\!$-1/4$\!\!\!&\!\!\!$-1$\!\!\!&\!\!\!$-2$\!\!\!&\!\!\!$-2$\!\!\!&\!\!\!$-3/2$\!\!\!&\!\!\!$-1$\!\!\!&\!\!\!$1/2$\!\!\!&\!\!\!$2$\!\!\!&\!\!\!$5$\!\!\!&\!\!\!$2$\!\!\!&\!\!\!$E\sst 7$\!\\[-.5mm]
\cline{2-12}
\!\!\!&\!\!\!$-1/4$\!\!\!&\!\!\!$-1$\!\!\!&\!\!\!$-1$\!\!\!&\!\!\!$-1$\!\!\!&\!\!\!$-1$\!\!\!&\!\!\!$-1$\!\!\!&\!\!\!$1/2$\!\!\!&\!\!\!$2$\!\!\!&\!\!\!$4$\!\!\!&\!\!\!$3/2$\!\!\!&\!\!\!$E\sst 7$\!\\[-.5mm]
\cline{2-12}
\!\!\!&\!\!\!$-1/4$\!\!\!&\!\!\!$-1$\!\!\!&\!\!\!$-1$\!\!\!&\!\!\!$-1$\!\!\!&\!\!\!$-1$\!\!\!&\!\!\!$-1$\!\!\!&\!\!\!$1/2$\!\!\!&\!\!\!$2$\!\!\!&\!\!\!$5$\!\!\!&\!\!\!$2$\!\!\!&\!\!\!$E\sst 7$\!\\[-.5mm]
\cline{2-12}
\!\!\!&\!\!\!$-1/3$\!\!\!&\!\!\!$-1$\!\!\!&\!\!\!$-1$\!\!\!&\!\!\!$-1/3$\!\!\!&\!\!\!$1/3$\!\!\!&\!\!\!$1$\!\!\!&\!\!\!$1/3$\!\!\!&\!\!\!$-1/3$\!\!\!&\!\!\!$-1$\!\!\!&\!\!\!$-1$\!\!\!&\!\!\!$D\sst 4^{(3)}$\!\\[-.5mm]
\cline{2-12}
\!\!\!&\!\!\!$-1/3$\!\!\!&\!\!\!$-1$\!\!\!&\!\!\!$-1$\!\!\!&\!\!\!$-1/3$\!\!\!&\!\!\!$1/3$\!\!\!&\!\!\!$1$\!\!\!&\!\!\!$1/2$\!\!\!&\!\!\!$0$\!\!\!&\!\!\!$0$\!\!\!&\!\!\!$-1$\!\!\!&\!\!\!$F\sst 4$\!\\[-.5mm]
\cline{2-12}
\!\!\!&\!\!\!$-1/3$\!\!\!&\!\!\!$-1$\!\!\!&\!\!\!$-1$\!\!\!&\!\!\!$-1/3$\!\!\!&\!\!\!$1/3$\!\!\!&\!\!\!$1$\!\!\!&\!\!\!$1$\!\!\!&\!\!\!$1$\!\!\!&\!\!\!$1$\!\!\!&\!\!\!$-1$\!\!\!&\!\!\!$F\sst 4$\!\\[-.5mm]
\cline{2-12}
\!\!\!&\!\!\!$-1/3$\!\!\!&\!\!\!$-1$\!\!\!&\!\!\!$0$\!\!\!&\!\!\!$0$\!\!\!&\!\!\!$1/2$\!\!\!&\!\!\!$1$\!\!\!&\!\!\!$1/2$\!\!\!&\!\!\!$0$\!\!\!&\!\!\!$0$\!\!\!&\!\!\!$-1$\!\!\!&\!\!\!$E\sst 6$\!\\[-.5mm]
\cline{2-12}
\!\!\!&\!\!\!$-1/3$\!\!\!&\!\!\!$-1$\!\!\!&\!\!\!$0$\!\!\!&\!\!\!$0$\!\!\!&\!\!\!$1/2$\!\!\!&\!\!\!$1$\!\!\!&\!\!\!$1$\!\!\!&\!\!\!$1$\!\!\!&\!\!\!$1$\!\!\!&\!\!\!$-1$\!\!\!&\!\!\!$E\sst 6$\!\\[-.5mm]
\cline{2-12}
\!\!\!&\!\!\!$-1/3$\!\!\!&\!\!\!$-1$\!\!\!&\!\!\!$1$\!\!\!&\!\!\!$1$\!\!\!&\!\!\!$1$\!\!\!&\!\!\!$1$\!\!\!&\!\!\!$1$\!\!\!&\!\!\!$1$\!\!\!&\!\!\!$1$\!\!\!&\!\!\!$-1$\!\!\!&\!\!\!$E\sst 6$\!\\[-.5mm]
\cline{2-12}
\!\!\!&\!\!\!$-1/3$\!\!\!&\!\!\!$-1$\!\!\!&\!\!\!$-2$\!\!\!&\!\!\!$-3/2$\!\!\!&\!\!\!$-1$\!\!\!&\!\!\!$0$\!\!\!&\!\!\!$1/2$\!\!\!&\!\!\!$1$\!\!\!&\!\!\!$2$\!\!\!&\!\!\!$1/2$\!\!\!&\!\!\!$E\sst 6$\!\\[-.5mm]
\cline{2-12}
\!\!\!&\!\!\!$-1/3$\!\!\!&\!\!\!$-1$\!\!\!&\!\!\!$-2$\!\!\!&\!\!\!$-3/2$\!\!\!&\!\!\!$-1$\!\!\!&\!\!\!$1$\!\!\!&\!\!\!$1$\!\!\!&\!\!\!$1$\!\!\!&\!\!\!$2$\!\!\!&\!\!\!$1/2$\!\!\!&\!\!\!$E\sst 6$\!\\[-.5mm]
\cline{2-12}
\!\!\!&\!\!\!$-1/3$\!\!\!&\!\!\!$-1$\!\!\!&\!\!\!$-2$\!\!\!&\!\!\!$-3/2$\!\!\!&\!\!\!$-1$\!\!\!&\!\!\!$1$\!\!\!&\!\!\!$1$\!\!\!&\!\!\!$1$\!\!\!&\!\!\!$3$\!\!\!&\!\!\!$1$\!\!\!&\!\!\!$E\sst 6$\!\\[-.5mm]
\cline{2-12}
\!\!\!&\!\!\!$-1/3$\!\!\!&\!\!\!$-1$\!\!\!&\!\!\!$-1$\!\!\!&\!\!\!$-1$\!\!\!&\!\!\!$-1$\!\!\!&\!\!\!$1$\!\!\!&\!\!\!$1$\!\!\!&\!\!\!$1$\!\!\!&\!\!\!$3$\!\!\!&\!\!\!$1$\!\!\!&\!\!\!$E\sst 6$\!\\[-.5mm]
\cline{2-12}

\!\!\!&\!\!\!$-1/2$\!\!\!&\!\!\!$-1$\!\!\!&\!\!\!$-1$\!\!\!&\!\!\!$-1/2$\!\!\!&\!\!\!$0$\!\!\!&\!\!\!$1$\!\!\!&\!\!\!$0$\!\!\!&\!\!\!$i$\!\!\!&\!\!\!$2i$\!\!\!&\!\!\!$i-1$\!\!\!&\!\!\!$D\sst {2i+4}\yy B\sst 3^a\xx$\!\\[-.5mm]
\cline{2-12}
\!\!\!&\!\!\!$-1/2$\!\!\!&\!\!\!$-1$\!\!\!&\!\!\!$-1$\!\!\!&\!\!\!$0$\!\!\!&\!\!\!$1$\!\!\!&\!\!\!$2$\!\!\!&\!\!\!$0$\!\!\!&\!\!\!$i$\!\!\!&\!\!\!$2i$\!\!\!&\!\!\!$i-1$\!\!\!&\!\!\!$D\sst {2i+4}\yy B\sst 3\xx$\!\\[-.5mm]
\cline{2-12}
\!\!\!&\!\!\!$-1/2$\!\!\!&\!\!\!$-1$\!\!\!&\!\!\!$-1$\!\!\!&\!\!\!$-1$\!\!\!&\!\!\!$0$\!\!\!&\!\!\!$1$\!\!\!&\!\!\!$0$\!\!\!&\!\!\!$i$\!\!\!&\!\!\!$2i+1$\!\!\!&\!\!\!$i$\!\!\!&\!\!\!$D\sst {2i+4}$\!\\[-.5mm]
\cline{2-12}
\!\!\!&\!\!\!$-1/2$\!\!\!&\!\!\!$-1$\!\!\!&\!\!\!$-1$\!\!\!&\!\!\!$-1/2$\!\!\!&\!\!\!$0$\!\!\!&\!\!\!$1$\!\!\!&\!\!\!$0$\!\!\!&\!\!\!$i$\!\!\!&\!\!\!$2i+1$\!\!\!&\!\!\!$i$\!\!\!&\!\!\!$D\sst {2i+5}$\!\\[-.5mm]
\cline{2-12}
\!\!\!&\!\!\!$-1/2$\!\!\!&\!\!\!$-1$\!\!\!&\!\!\!$-1$\!\!\!&\!\!\!$0$\!\!\!&\!\!\!$1$\!\!\!&\!\!\!$2$\!\!\!&\!\!\!$0$\!\!\!&\!\!\!$i$\!\!\!&\!\!\!$2i+1$\!\!\!&\!\!\!$i$\!\!\!&\!\!\!$D\sst {2i+5}$\!\\[-.5mm]
\cline{2-12}
\!\!\!&\!\!\!$-1/2$\!\!\!&\!\!\!$-1$\!\!\!&\!\!\!$-1$\!\!\!&\!\!\!$-1$\!\!\!&\!\!\!$0$\!\!\!&\!\!\!$1$\!\!\!&\!\!\!$0$\!\!\!&\!\!\!$i+1$\!\!\!&\!\!\!$2i+2$\!\!\!&\!\!\!$i$\!\!\!&\!\!\!$D\sst {2i+5}$\!\\[-.5mm]
\cline{2-12}
\!\!\!&\!\!\!$-1/2$\!\!\!&\!\!\!$-1$\!\!\!&\!\!\!$-1$\!\!\!&\!\!\!$-1/2$\!\!\!&\!\!\!$0$\!\!\!&\!\!\!$1$\!\!\!&\!\!\!$0$\!\!\!&\!\!\!$(i-1)/2$\!\!\!&\!\!\!$i-1$\!\!\!&\!\!\!$i/2-1$\!\!\!&\!\!\!$B\sst {i+3}\yy G\sst 2\xx$\!\\[-.5mm]
\cline{2-12}
\!\!\!&\!\!\!$-1/2$\!\!\!&\!\!\!$-1$\!\!\!&\!\!\!$-1$\!\!\!&\!\!\!$0$\!\!\!&\!\!\!$1$\!\!\!&\!\!\!$2$\!\!\!&\!\!\!$0$\!\!\!&\!\!\!$(i-1)/2$\!\!\!&\!\!\!$i-1$\!\!\!&\!\!\!$i/2-1$\!\!\!&\!\!\!$B\sst {i+3}\yy G\sst 2\xx$\!\\[-.5mm]
\cline{2-12}
\!\!\!&\!\!\!$-1/2$\!\!\!&\!\!\!$-1$\!\!\!&\!\!\!$-1$\!\!\!&\!\!\!$-1$\!\!\!&\!\!\!$0$\!\!\!&\!\!\!$1$\!\!\!&\!\!\!$0$\!\!\!&\!\!\!$i/2$\!\!\!&\!\!\!$i$\!\!\!&\!\!\!$(i-1)/2$\!\!\!&\!\!\!$B\sst {i+3}\yy B\sst 3^a\xx$\!\\[-.5mm]
\cline{2-12}

\!\!\!&\!\!\!$-1$\!\!\!&\!\!\!$-1$\!\!\!&\!\!\!$-1$\!\!\!&\!\!\!$-1$\!\!\!&\!\!\!$-1$\!\!\!&\!\!\!$-1$\!\!\!&\!\!\!$-1$\!\!\!&\!\!\!$i-1$\!\!\!&\!\!\!$2i-1$\!\!\!&\!\!\!$i-1$\!\!\!&\!\!\!$C\sst i$\!\\[-.5mm]
\cline{2-12}

\!\!\!&\!\!\!$-1$\!\!\!&\!\!\!$-1$\!\!\!&\!\!\!$-1$\!\!\!&\!\!\!$z_3$\!\!\!&\!\!\!$z_4$\!\!\!&\!\!\!$z_5$\!\!\!&\!\!\!$z_6$\!\!\!&\!\!\!$z_7$\!\!\!&\!\!\!$z_8$\!\!\!&\!\!\!$z_9$\!\!\!&\!\!\!$A\sst {z_5+z_8+1}$
\!\\[-.5mm]

\hline\hline
\end{tabular}
\caption{Duals $\top^*$ of tops with $F_0^*$ one of the polygons $15,16$
of figure~\ref{fig:16pol}.}
\label{tab:results16}
\end{footnotesize}
%\end{adjustwidth}
\end{center}
\end{table}

\section{Full Results for the Closed Topological String Amplitudes}\label{resultfunct}

Here we present the results for the full generating functionals given by \refeq{functionalsi}. The $\pm$ sign corresponds to $Sp$ and $SO$, 
respectively. Of course, the oriented contribution for $Q=0$ agrees with previous results for the local del Pezzo $dP_3$ with one 
K\"ahler parameter sent to infinity \cite{Aganagic:2002qg,Chiang:1999tz}, and if we set $q_{1,2}=0$ we recover the results presented in tables \ref{tabfunct1}--\ref{tabfunct16} (taking 
into account the $1/2$ factor in the definition of the $c=0$ generating functional).

We computed the results up to degree 5 in $e^{-t}$, but we will present only the results up to degree 3 as the higher degree results are rather cumbersome. Note that $\CF_0^{0,2}$, $\CF_1^{0,2}$, $\CF_2^{0,2}$ and $\CF_3^{1,2}$ are $0$ (but not $\CF_3^{0,2}$), and therefore we omit them.

\bea
\CF_0^{0,0}&=&~q_1+q_2+{1 \over 2}Q,\nn\\
\CF_0^{0,1}&=& \pm [Q^{1/2}],\nn\\
\CF_1^{0,0}&=&~3-2(q_1+q_2+Q)+(q_1q_2+q_2Q+q_1Q),\nn\\
\CF_1^{0,1}&=&\pm [-2Q^{1/2}+( q_1 Q^{1/2}+q_2Q^{1/2})],\nn\\
\CF_2^{0,0}&=&-6+5(q_1+q_2)+7Q-4q_1q_2-6(q_1Q+q_2Q)+4q_1q_2Q+{1 \over 2} (q_1^2 Q + q_2^2 Q)\nn\\
&&-Q^2+(q_1 Q^2+q_2 Q^2) - q_1 q_2 Q^2,\nn\\
\CF_2^{0,1}&=&\pm[5 Q^{1/2} -4(q_1Q^{1/2}+q_2Q^{1/2}) +3 q_1 q_2 Q^{1/2} -3Q^{3/2}\nn\\
&&+2(q_1 Q^{3/2} + q_2 Q^{3/2}) - q_1 q_2 Q^{3/2}],
\nn\\
\CF_3^{0,0}&=&~27 - 32(q_1 +q_2) - 42Q + 35 q_1 q_2 + 48(q_1 Q + q_2 Q)-50 q_1 q_2 Q + 7(q_1^2 + q_2^2)\nn\\
&&+ 15 Q^2 - 6(q_1^2 q_2 + q_1 q_2^2)- 10(q_1^2 Q + q_2^2 Q)-16(q_1 Q^2 + q_2 Q^2) \nn\\
&&+8(q_1^2 q_2 Q + q_1 q_2^2 Q)+ 3 (q_1^2 Q^2+q_2^2 Q^2) - 2(q_1^2 q_2 Q^2 + q_1 q_2^2 Q^2) + 15 q_1 q_2 Q^2,\nn\\
\CF_3^{0,1}&=&\pm[-32 Q^{1/2} +35(q_1 Q^{1/2} + q_2 Q^{1/2})-36q_1q_2Q^{1/2}-6(q_1^2 Q^{1/2} + q_2^2 Q^{1/2})\nn\\
&&+(q_1^2 q_2 Q^{1/2}+q_1 q_2^2 Q^{1/2}) +30 Q^{3/2}-30(q_1 Q^{3/2} + q_2 Q^{3/2})\nn\\
&& +4(q_1^2 Q^{3/2} + q_2^2 Q^{3/2})+28 q_1 q_2 Q^{3/2}-3(q_1^2 q_2 Q^{3/2} + q_1 q_2^2 Q^{3/2}) -4 Q^{5/2} \nn\\
&&+3(q_1 Q^{5/2}+ q_2 Q^{5/2})-2 q_1 q_2 Q^{5/2}],\nn\\
\CF_3^{0,2}&=&~Q^2 -(q_1 Q^2 +q_2 Q^2) + q_1 q_2 Q^2,\nn\\
\CF_3^{1,0}&=&~10-9 (q_1+q_2+Q)+8(q_1q_2 +q_1Q +q_2Q)-7q_1q_2Q,\nn\\
\CF_3^{1,1}&=&\pm[-9 Q^{1/2} +8 (q_1 Q^{1/2} + q_2 Q^{1/2}) - 7 q_1 q_2 Q^{1/2}+7 Q^{3/2}\nn\\
&& - 6(q_1 Q^{3/2} + q_2 Q^{3/2}) +5 q_1 q_2 Q^{3/2}].\nn
\eeal{fullresultii}

\section{BPS Invariants for the Trefoil Knot}\label{trefoilapp}

In this appendix, we list the BPS invariants $N^{c=1}_{R,g,\beta}$ for the trefoil knot, for representations $R$ with three boxes.

\begin{table}[!b]
\begin{center}
\begin{footnotesize}
\begin{tabular}{|c|rrrrrrrr|}\hline
&$\beta=2$&$3$&$4$&$5$&$6$&$7$&$8$&$9$
\\
\hline
$g=0$&$18$&$-270$&$1185$&$-2380$&$2430$&$-1188$&$175$&$30$
\\
$1$&$21$&$-753$&$4924$&$-12209$&$13203$&$-4856$&$-1300$&$970$
\\
$2$&$8$&$-1007$&$10374$&$-31348$&$31419$&$4028$&$-22155$&$8681$
\\
$3$&$1$&$-793$&$13920$&$-50383$&$30636$&$84956$&$-117415$&$39078$
\\
$4$&$0$&$-378$&$12688$&$-54222$&$-24584$&$305272$&$-343318$&$104542$
\\
$5$&$0$&$-106$&$8006$&$-40151$&$-118255$&$609701$&$-639896$&$180701$
\\
$6$&$0$&$-16$&$3486$&$-20657$&$-178503$&$797521$&$-813994$&$212163$
\\
$7$&$0$&$-1$&$1024$&$-7353$&$-161931$&$728309$&$-734484$&$174436$
\\
$8$&$0$&$0$&$193$&$-1773$&$-98947$&$478948$&$-480509$&$102088$
\\
$9$&$0$&$0$&$21$&$-276$&$-42205$&$229955$&$-230209$&$42714$
\\
$10$&$0$&$0$&$1$&$-25$&$-12624$&$80705$&$-80729$&$12672$
\\
$11$&$0$&$0$&$0$&$-1$&$-2599$&$20474$&$-20475$&$2601$
\\
$12$&$0$&$0$&$0$&$0$&$-351$&$3654$&$-3654$&$351$
\\
$13$&$0$&$0$&$0$&$0$&$-28$&$435$&$-435$&$28$
\\
$14$&$0$&$0$&$0$&$0$&$-1$&$31$&$-31$&$1$
\\
$15$&$0$&$0$&$0$&$0$&$0$&$1$&$-1$&$0$
\\
\hline
\end{tabular}
\caption{BPS invariants $N_{(3),g,\beta}^{c=1}$ for the trefoil knot.}
\label{trefoil4}
\end{footnotesize}
\end{center}
\end{table}

\begin{table}[!b]
\begin{center}
\begin{footnotesize}
\begin{tabular}{|c|rrrrrrrr|}\hline
&$\beta=2$&$3$&$4$&$5$&$6$&$7$&$8$&$9$
\\
\hline
$g=0$&$99$&$-1125$&$4359$&$-8096$&$7828$&$-3699$&$563$&$72$
\\
$1$&$201$&$-4194$&$22748$&$-51475$&$53807$&$-21649$&$-2204$&$2766$
\\
$2$&$164$&$-7702$&$60811$&$-165827$&$171590$&$-19997$&$-68978$&$29939$
\\
$3$&$66$&$-8701$&$104757$&$-338906$&$282625$&$264688$&$-468878$&$164349$
\\
$4$&$13$&$-6395$&$125047$&$-472907$&$124226$&$1398430$&$-1710505$&$542091$
\\
$5$&$1$&$-3092$&$106648$&$-466523$&$-477321$&$3645201$&$-3976290$&$1171376$
\\
$6$&$0$&$-971$&$65795$&$-331606$&$-1232410$&$6113672$&$-6363573$&$1749093$
\\
$7$&$0$&$-190$&$29358$&$-171307$&$-1590490$&$7192295$&$-7328205$&$1868539$
\\
$8$&$0$&$-21$&$9358$&$-64261$&$-1351903$&$6186865$&$-6240225$&$1460187$
\\
$9$&$0$&$-1$&$2072$&$-17298$&$-815116$&$3979137$&$-3994110$&$845316$
\\
$10$&$0$&$0$&$302$&$-3252$&$-358192$&$1934294$&$-1937220$&$364068$
\\
$11$&$0$&$0$&$26$&$-405$&$-115397$&$712126$&$-712504$&$116154$
\\
$12$&$0$&$0$&$1$&$-30$&$-26996$&$197286$&$-197315$&$27054$
\\
$13$&$0$&$0$&$0$&$-1$&$-4465$&$40454$&$-40455$&$4467$
\\
$14$&$0$&$0$&$0$&$0$&$-495$&$5952$&$-5952$&$495$
\\
$15$&$0$&$0$&$0$&$0$&$-33$&$594$&$-594$&$33$
\\
$16$&$0$&$0$&$0$&$0$&$-1$&$36$&$-36$&$1$
\\
$17$&$0$&$0$&$0$&$0$&$0$&$1$&$-1$&$0$
\\
\hline
\end{tabular}
\caption{BPS invariants $N_{(2,1),g,\beta}^{c=1}$ for the trefoil knot.}
\label{trefoil5}
\end{footnotesize}
\end{center}
\end{table}

\begin{table}[!b]
\begin{center}
\begin{footnotesize}
\begin{tabular}{|c|rrrrrrrr|}\hline
&$\beta=2$&$3$&$4$&$5$&$6$&$7$&$8$&$9$
\\
\hline
$g=0$&$108$&$-1044$&$3705$&$-6484$&$6000$&$-2754$&$427$&$42$
\\
$1$&$306$&$-4818$&$23074$&$-48785$&$49436$&$-20669$&$-448$&$1904$
\\
$2$&$366$&$-11012$&$73663$&$-186538$&$193691$&$-44683$&$-49616$&$24129$
\\
$3$&$230$&$-15636$&$151596$&$-453623$&$421630$&$161750$&$-421269$&$155322$
\\
$4$&$79$&$-14720$&$216949$&$-756616$&$429479$&$1359478$&$-1836601$&$601952$
\\
$5$&$14$&$-9381$&$223615$&$-898781$&$-235791$&$4434624$&$-5047078$&$1532778$
\\
$6$&$1$&$-4047$&$168943$&$-777340$&$-1531480$&$8961515$&$-9525899$&$2708307$
\\
$7$&$0$&$-1160$&$94128$&$-495542$&$-2661004$&$12577678$&$-12957296$&$3443196$
\\
$8$&$0$&$-211$&$38523$&$-233794$&$-2843448$&$12900213$&$-13087921$&$3226638$
\\
$9$&$0$&$-22$&$11409$&$-81283$&$-2124814$&$9936047$&$-10004126$&$2262789$
\\
$10$&$0$&$-1$&$2373$&$-20525$&$-1160684$&$5832726$&$-5850601$&$1196712$
\\
$11$&$0$&$0$&$328$&$-3656$&$-470990$&$2625946$&$-2629249$&$477621$
\\
$12$&$0$&$0$&$27$&$-435$&$-142042$&$905758$&$-906165$&$142857$
\\
$13$&$0$&$0$&$1$&$-31$&$-31433$&$237305$&$-237335$&$31493$
\\
$14$&$0$&$0$&$0$&$-1$&$-4959$&$46375$&$-46376$&$4961$
\\
$15$&$0$&$0$&$0$&$0$&$-528$&$6545$&$-6545$&$528$
\\
$16$&$0$&$0$&$0$&$0$&$-34$&$630$&$-630$&$34$
\\
$17$&$0$&$0$&$0$&$0$&$-1$&$37$&$-37$&$1$
\\
$18$&$0$&$0$&$0$&$0$&$0$&$1$&$-1$&$0$
\\
\hline
\end{tabular}
\caption{BPS invariants $N_{(1,1,1),g,\beta}^{c=1}$ for the trefoil knot.}
\label{trefoil6}
\end{footnotesize}
\end{center}
\end{table}

%Bibliography
\addcontentsline{toc}{chapter}{Bibliography}
\bibliography{../BibTex/refs}

\end{document}%